\pdfoutput=1
\documentclass[a4paper,12pt,twoside,openright]{report}

\usepackage[a4paper,left=27mm,right=27mm,top=28mm,bottom=28mm]{geometry}

\usepackage[utf8]{inputenc}
\usepackage[english]{babel}
\usepackage[T1]{fontenc}
\usepackage{polski}
\usepackage{csquotes}
\usepackage{xpatch}
\usepackage{amsthm}
\usepackage{amssymb}
\usepackage{latexsym}
\usepackage{amsmath}
\usepackage{amssymb}
\usepackage{amsfonts}
\usepackage{mathrsfs}
\usepackage{bbm}
\usepackage{slashed}
\usepackage{amstext}
\usepackage{enumerate}
\usepackage{environ}
\usepackage{mathtools}
\mathtoolsset{showonlyrefs}
\usepackage{enumitem}
\usepackage{stmaryrd}
\usepackage{hyphenat}
\usepackage[mathscr]{eucal}
\usepackage{afterpage}

\selectfont

\DeclareFontFamily{OT1}{pzc}{}
\DeclareFontShape{OT1}{pzc}{m}{bf}{<-> s * [1.1] pzcmi7t}{}
\DeclareMathAlphabet{\mathpzc}{OT1}{pzc}{m}{bf}

\usepackage{marginnote}

\usepackage[maxbibnames=99,maxcitenames=5,maxalphanames=5,given
inits=true,backend=biber,style=alphabetic,citestyle=alphabetic,backref=true]{biblatex}
\renewbibmacro{in:}{%
  \ifentrytype{article}{}{\printtext{\bibstring{in}\intitlepunct}}}

\renewbibmacro*{volume+number+eid}{%
  {\bf\printfield{volume}}%
  \printfield{number}%
  \setunit{\addcomma\space}%
  \printfield{eid}}
\DeclareFieldFormat[article]{number}{\mkbibparens{#1}}
\DeclareFieldFormat{pages}{#1}

\xpatchbibdriver{article}
  {\newunit
   \usebibmacro{note+pages}}
  {}{}{}
  
\renewbibmacro*{journal+issuetitle}{%
  \usebibmacro{journal}%
  \setunit*{\addspace}%
  \iffieldundef{series}
    {}
    {\newunit
     \printfield{series}%
     \setunit{\addspace}}%
  \usebibmacro{volume+number+eid}%
  \newunit
  \usebibmacro{note+pages}%
  \setunit{\addspace}%
  \usebibmacro{issue+date}%
  \setunit{\addcolon\space}%
  \usebibmacro{issue}%
  \newunit}  

\xpatchbibdriver{incollection}
  {\usebibmacro{byeditor+others}}
  {\nopunct\usebibmacro{byeditor+others}}
  {}{\typeout{failed to patch bibmacro incollection to remove punctuation before byeditors}}

\xpatchbibdriver{incollection}
  {\newunit\newblock
   \usebibmacro{chapter+pages}}
  {}
  {}{\typeout{failed to patch bibmacro incollection to remove page macro}}

\xpatchbibdriver{incollection}
  {\usebibmacro{series+number}}
  {\usebibmacro{series+number}
   \newunit\newblock
   \usebibmacro{chapter+pages}}
  {}{\typeout{failed to patch bibmacro incollection to re-add page macro}}  

\DeclareFieldFormat[incollection]{year}{\mkbibparens{#1}}

\usepackage{color,hyperref}
\definecolor{darkblue}{rgb}{0.0,0.0,0.5}
\hypersetup{
  linkcolor  = darkblue,
  citecolor  = darkblue,
  urlcolor   = darkblue,
  colorlinks = true,
}

\makeatletter
\AtBeginDocument{
  \hypersetup{
    pdftitle = {\@title},
    pdfauthor = {\@author}
  }
}
\makeatother

\usepackage{smartref}
\addtoreflist{section}

\newcommand{\N}{\mathbb{N}}

\newcommand{\C}{\mathbb{C}}
\newcommand{\Z}{\mathbb{Z}}
\newcommand{\R}{\mathbb{R}}

\newcommand{\ri}{\mathrm{i}}
\newcommand{\rd}{\mathrm{d}}

\newcommand{\Span}{\mathrm{span}}

\newcommand{\ret}{\mathrm{ret}}
\newcommand{\adv}{\mathrm{adv}}

\newcommand{\rEG}{\mathrm{EG}}
\newcommand{\rint}{\mathrm{int}}
\newcommand{\const}{\mathrm{const}}

\DeclareMathOperator{\sgn}{sgn}
\DeclareMathOperator{\Texp}{Texp}

\DeclareMathOperator{\T}{T}

\DeclareMathOperator{\aT}{\overline{T}}
\DeclareMathOperator{\Ret}{R}

\DeclareMathOperator{\Adv}{A}

\DeclareMathOperator{\Dif}{D}
\DeclareMathOperator{\Gre}{G}
\DeclareMathOperator{\Wig}{W}

\DeclareMathOperator{\sd}{sd}

\DeclareMathOperator{\ext}{e}
\DeclareMathOperator{\der}{d}
\DeclareMathOperator{\CC}{\mathbf{c}}
\newcommand{\Fa}{\mathcal{F}}
\newcommand{\Fh}{\mathcal{F}^\mathrm{hom}}
\newcommand{\id}{\mathbbm{1}}

\newcommand{\supp}{\mathrm{supp}}
\newcommand{\zerop}{0}
\newcommand{\normord}[1]{:\mathrel{#1}:}
\newcommand{\mP}[1]{\frac{\rd^4 #1}{(2\pi)^4}}
\newcommand{\mH}[2]{\rd\mu_{#1}(#2)}
\newcommand{\F}[1]{\tilde{#1}}
\newcommand{\FF}[1]{\widetilde{#1}}

\makeatletter
\newcommand{\leqnomode}{\tagsleft@true}
\newcommand{\reqnomode}{\tagsleft@false}
\makeatother

\newtheorem{thm}{Theorem}
\newtheorem{dfn}{Definition}
\newtheorem{asm}{Assumption}
\newtheorem{lem}[thm]{Lemma}

\newtheorem{rem}{Remark}

\numberwithin{equation}{section}
\numberwithin{thm}{section}
\numberwithin{dfn}{section}
\numberwithin{rem}{section}
\numberwithin{asm}{section}

\newenvironment{changemargin}[2]{%
\begin{list}{}{%
\setlength{\topsep}{0pt}%
\setlength{\leftmargin}{#1}%
\setlength{\rightmargin}{#2}%
\setlength{\listparindent}{\parindent}%
\setlength{\itemindent}{\parindent}%
\setlength{\parsep}{\parskip}%
}%
\item[]}{\end{list}}

\begin{document}
\pagenumbering{gobble}
\begin{titlepage}
\begin{center}

\includegraphics[width=0.3\textwidth]{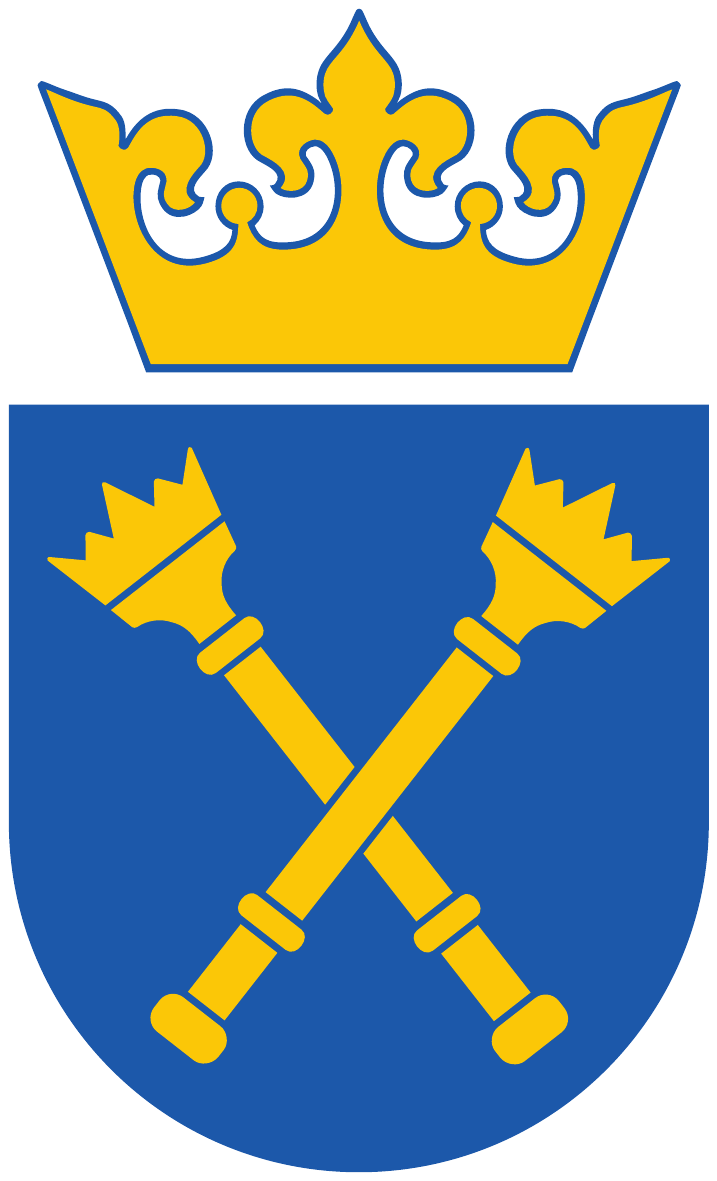} 

\vspace{-1cm}

{\scshape\LARGE Jagiellonian University \par}

\vspace{0.7cm}

\textsc{\Large Doctoral Thesis}

\vspace{0.7cm}

\rule{\textwidth}{0.4mm}

\vspace{0.4cm}

{\huge \bfseries Massless fields and adiabatic limit in quantum field theory \par}

\vspace{0.4cm} 

\rule{\textwidth}{0.4mm}

\vspace{2cm}

{\LARGE Paweł Duch}

\vspace{2cm}

{\Large Supervisor: prof. Andrzej Herdegen}

\vspace{2cm}

\vfill

{\large Krak\'ow, June 2017}\\

\end{center}
\end{titlepage}

\vspace{1mm}

\begin{changemargin}{0.1cm}{0.1cm}

\vspace{1cm}
\noindent
\begin{flushleft}
Wydział Fizyki, Astronomii i Informatyki Stosowanej
\\
Uniwersytet Jagielloński
\end{flushleft}

\vspace{1cm}

\begin{Large}
\begin{center}
 Oświadczenie
\end{center}
\end{Large}

\vspace{1cm}

Ja niżej podpisany Paweł Duch (nr indeksu: 1098445) doktorant Wydziału Fizyki, Astronomii i~Informatyki Stosowanej Uniwersytetu Jagiellońskiego oświadczam, że przed\-łożona przeze mnie rozprawa doktorska pt. "Massless fields and adiabatic limit in quantum field theory" jest oryginalna i~przedstawia wyniki badań wykonanych przeze mnie osobiście, pod kierunkiem prof. dr hab. Andrzeja Herdegena. Pracę napisałem samodzielnie.

Oświadczam, że moja rozprawa doktorska została opracowana zgodnie z~Ustawą o~prawie autorskim i prawach pokrewnych z dnia 4 lutego 1994 r. (Dziennik Ustaw 1994 nr 24 poz. 83 wraz z późniejszymi zmianami). Jestem świadom, że niezgodność niniejszego oświadczenia z prawdą ujawniona w~dowolnym czasie, niezależnie od skutków prawnych wynikających z~ww.~ustawy, może spowodować unieważnienie stopnia nabytego na podstawie tej rozprawy.

\vspace{2cm}

\noindent
\begin{tabular}[t]{@{}l} 
  Kraków, dnia 30.06.2017
\end{tabular}
\hfill% move it to the right
\begin{tabular}[t]{l@{}}
 ...................................................
\end{tabular}

\end{changemargin}

\begin{abstract}
The thesis is devoted to a rigorous construction of the Wightman and Green functions in models of the perturbative quantum field theory in the four-dimensional Minkowski spacetime in the framework of the causal perturbation theory developed by Epstein and Glaser. In this approach each interaction term is multiplied by a switching function which vanishes rapidly at infinity, and thus plays the role of an infrared regularization. In order to obtain the physical Wightman and Green functions, the scattering matrix or the interacting fields one has to remove this regularization by taking the adiabatic limit.

In the first part of the thesis we give an overview of the Epstein-Glaser approach to the perturbative quantum field theory. We outline the construction of the time-ordered products which is applicable also when fermionic fields are present. Next, we recall the definition of the Wightman and Green functions, the scattering matrix and the interacting fields with the infrared regularization. Subsequently, we discuss the method of removing this regularization by means of the adiabatic limit.

In the second part we construct the Wightman and Green functions in a large class of models, generalizing the result due to Blanchard and Seneor. To this end, we show the existence of the so-called weak adiabatic limit. The proof of the existence of this limit is valid under the assumption that the time-ordered products satisfy certain normalization condition. We show that this normalization condition may be imposed in all models with interaction vertices of the canonical dimension equal to four as well as in all models with interaction vertices of the canonical dimension equal to three provided each of them contains at least one massive field. Moreover, we prove that the above-mentioned normalization condition is compatible with all the standard normalization conditions which are usually imposed on the time-ordered products. We consider in detail the case of the quantum electrodynamics with a massive or massless, spinor or scalar charged field and certain model of interacting scalar fields with the interaction vertex of dimension three which we call the scalar model. Our result is also applicable to non-abelian Yang-Mills theories. Using the method developed in the proof of the existence of the weak adiabatic limit, we also show the existence of the central splitting solution in the quantum electrodynamics with a massive spinor field.

\end{abstract}

\renewcommand{\abstractname}{Streszczenie}
\begin{abstract}
Rozprawa jest poświęcona ścisłej konstrukcji funkcji Wightmana oraz Greena w modelach perturbacyjnej kwantowej teorii pola w czterowymiarowej czasoprzestrzeni Min\-kow\-skie\-go w ramach przyczynowej teorii zaburzeń rozwiniętej przez Epsteina i Glasera. W tym podejściu każdy z wierzchołków oddziaływania jest przemnożony przez funkcję włączającą, która znika szybko w nieskończoności i w związku z tym odgrywa rolę regularyzacji w podczerwieni.  W celu uzyskania fizycznych funkcji Wightmana i Greena, macierzy rozpraszania lub pól oddziałujących konieczne jest usunięcie tej regularyzacji poprzez wzięcie granicy adiabatycznej.

W pierwszej części pracy przedstawiamy zarys podejścia Epsteina i Glasera do perturbacyjnej kwantowej teorii pola. Podajemy konstrukcję iloczynów uporządkowanych chronologicznie, która może być zastosowana również w przypadku obecności pól fer\-mio\-no\-wych. W dalszej części przypominamy definicje funkcji Wightmana oraz Greena, macierzy rozpraszania oraz pól oddziałujących z regularyzacją w podczerwieni. Następnie omawiamy metodę usunięcia tej regularyzacji przy wykorzystaniu granicy adiabatycznej. 

W drugiej części rozprawy konstruujemy funkcje Wightmana oraz Greena w szerokiej klasie modeli, uogólniając wcześniejsze wyniki Blancharda i Seneora. W tym celu pokazujemy istnienie tzw. słabej granicy adiabatycznej. Przeprowadzenie dowodu istnienia tej granicy wymaga założenia, że iloczyny uporządkowane chronologicznie spełniają pewien warunek normalizacji. Pokazujemy, że wspomniany warunek normalizacji może być narzucony w dowolnym modelu o wierzchołkach oddziaływania o wymiarze kanonicznym równym cztery oraz w dowolnym modelu o wierzchołkach oddziaływania o wymiarze kanonicznym równym trzy pod warunkiem, że każdy z nich zawiera przynajmniej jedno pole masywne. Ponadto dowodzimy, że powyższy warunek normalizacji jest kompatybilny ze wszystkimi standardowymi warunkami normalizacji, które są zazwyczaj narzucane na iloczyny uporządkowane chronologicznie. Rozważamy szczegółowo przypadek elektrodynamiki kwantowej z masywnym lub bezmasowym, spinorowym lub skalarnym polem naładowanym oraz pewien model oddziałujących pól skalarnych z wierzchołkiem oddziaływania o wymiarze trzy, który nazywamy modelem skalarnym. Nasz wynik stosuje się również do nieabelowych teorii Yanga-Millsa. Wykorzystując metodę opracowaną przy dowodzeniu istnienia słabej granicy adiabatycznej pokazujemy ponadto istnienie centralnego rozwiązania problemu podziału w elektrodynamice kwantowej z masywnym polem spino\-rowym.
\end{abstract}

{\Large \bf Acknowledgement}

\vspace{4mm}

I express my deep gratitude to my supervisor prof. Andrzej Herdegen for numerous discussions, his constant interest in this work and many helpful suggestions.

W sposób szczególny dziękuję moim rodzicom oraz bratu za ich wsparcie w każdym momencie mojego życia. 

\newpage

\pagenumbering{arabic}
\setcounter{tocdepth}{1}

\tableofcontents

%===================================================================================================
%++++++++++++++++++++++++++++++++++++++INTRODUCTION+++++++++++++++++++++++++++++++++++++++++++++++++
%===================================================================================================

\chapter{Introduction}

The relativistically covariant perturbative quantum field theory (QFT) in the Minkowski spacetime provides a general framework which allows to describe three of the four fundamental forces of nature -- the electromagnetic, the strong, and the weak interactions. Its applicability is restricted by its perturbative character and incompatibility with the general theory of relativity. Nevertheless, because of the accuracy of the predictions of its models, especially the Standard Model, which have been repeatedly confirmed, it is one of the most successful modern physical theories. 

The relativistic QFT was born in 1920s as an attempt to reconcile the quantum mechanics and the special theory of relativity which was essential for the quantum mechanical description of the electromagnetic field. The main obstacle in its initial development were the so-called ultraviolet (UV) divergences which appeared in calculations of the scattering amplitudes starting from the second-order corrections. In the middle of the 20th century the UV problem was understood with the invention of the renormalization by the pioneering work of Feynman, Schwinger and Tomonaga. The process of renormalization was put on mathematically sound footing about 50 years ago by Bogoliubov, Parasiuk, Hepp and Zimmermann (BPHZ) \cite{bogoliubov1957multiplication,hepp1966proof,zimmermann1969convergence} and Epstein and Glaser  (EG) \cite{epstein1973role}.  However, the mathematical foundations of the perturbative QFT are still a subject of an ongoing research. Let us mention only two recent developments: the discovery \cite{kreimer1998hopf} that the BPHZ renormalization procedure can be understood in terms of Hopf algebras, which stimulated the investigation of the perturbative QFT in the mathematical community, and the successful development of the QFT on curved spacetime \cite{brunetti2000microlocal,hollands2001local,hollands2002existence}, which revived the interest in the EG approach and led to many new structural results of importance in the case of the Minkowski spacetime \cite{dutsch1999local,dutsch2001algebraic,dutsch2001perturbative,dutsch2004causal,hollands2004algebraic, brunetti2009perturbative}. The investigations of the mathematical foundations of the perturbative QFT have also contributed recently to the development of new practical methods for calculations \cite{gracia2003improved,gracia2014improved,dutsch2014dimensional,nikolov2014renormalization,gracia2016systematic,todorov2017renormalization}.

The fundamental objects of interest in the QFT in the flat spacetime are the vacuum expectation values (VEVs) of the products, and the time-ordered products, of the interacting fields called the Wightman and Green functions, respectively. Conceptually the Wightman functions are of more fundamental character. They are used for examining the basic physical structure of a theory. Their properties: the Poincar{\'e} covariance, the spectrum condition, the Hermiticity, the locality, the positivity  and the clustering reflect the basic assumptions of the relativistic QFT. Moreover, by the Wightman reconstruction theorem \cite{streater2000pct} their knowledge allows to obtain the operator formulation of the theory. On the other hand, the Green functions are of more practical importance and it is usually easier to obtain them in the perturbation theory. Furthermore, by the LSZ reduction formula \cite{lehmann1955formulierung} they are directly related to the S-matrix elements. Because of this they have been studied in the physical literature more extensively than the Wightman functions.

The main result of the thesis is the \emph{perturbative} construction of the Wightman and Green functions in models of the quantum field theory with both massive and massless particles in the framework of causal perturbation theory developed by Epstein and Glaser \cite{epstein1973role}. Before we describe our results in detail let us give a brief overview of the perturbative definitions of the Wightman and Green functions existing in the literature. 

\section{Definitions of Wightman and Green functions}

\subsubsection{Axiomatic approach}\label{sec:axiomatic_approach}

Assume that a model of the quantum field theory under consideration satisfies the Wightman axioms \cite{streater2000pct} generalized appropriately to the perturbative context.
\begin{enumerate}[leftmargin=*,label={(\arabic*)}]
 \item  There is a linear space $\mathcal{D}$ with a positive-definite inner product denoted by $(\cdot|\cdot)$. The continuous unitary representation of the inhomogeneous $SL(2,\C)$ group which is the universal cover of the Poincar{\'e} group is defined on $\mathcal{D}$
 \begin{equation}
  \R^4\rtimes SL(2,\C) \ni (a,\Lambda) \mapsto U(a,\Lambda) \in L(\mathcal{D}),
 \end{equation}
 where $\R^4\rtimes SL(2,\C)$ is the inhomogeneous $SL(2,\C)$ group and $L(\mathcal{D})$ is the space of operators mapping $\mathcal{D}$ into $\mathcal{D}$. 
 
 There exists a state $\Omega\in\mathcal{D}$ called the vacuum which is Poincar{\'e} invariant, i.e.
 \begin{equation}
 \forall_{(a,\Lambda)\in \R^4\rtimes SL(2,\C)}~~ U(a,\Lambda)\Omega = \Omega.
 \end{equation}

 Moreover, for any $\Psi_1,\Psi_2\in\mathcal{D}$ the support of the Fourier transform of 
 \begin{equation}
  \R^4\ni a\mapsto (\Psi_1|U(a,\id)\Psi_2)\in\C
 \end{equation}
 is contained in the closed future light cone $\overline{V}^+:=\{p\in\R^4\,:\, p^2 \geq 0,~p^0\geq 0\}$ (the relativistic spectral condition).

 \item For any Schwartz function $f$ the interacting fields $B_{1,\rint}(f),\ldots,B_{n,\rint}(f)$ are formal power series in the coupling constant with coefficients in $L(\mathcal{D})$. For each interacting field $B_{i,\rint}(f)$ its Hermitian adjoint  $B_{i,\rint}(f)^*$ is also included in the list $B_{1,\rint}(f),\ldots,B_{n,\rint}(f)$. For any $i$ and $\Psi_1,\Psi_2\in\mathcal{D}$ the distribution $\mathcal{S}(\R^4)\ni f \mapsto (\Psi_1|B_{i,\rint}(f)\Psi_2)$ is a formal power series in the coupling constant with coefficients belonging to the space of Schwartz distributions. 
 
 \item The fields $B_{1,\rint}(x),\ldots,B_{n,\rint}(x)$ are Poincar{\'e} covariant.
 
 \item The axiom of local (anti)commutativity is satisfied, i.e. if the supports of $f$ and $g$ are spacelike separated then $B_{i,\rint}(f)$ commutes or anticommutes with $B_{j,\rint}(g)$ for any $i$ and $j$.  
\end{enumerate}
The Wightman functions are the vacuum expectation values of the interacting fields $B_{1,\rint}(x),\ldots,B_{n,\rint}(x)$ introduced above, i.e.
\begin{equation}\label{eq:wig}
  \Wig(C_1(x_1),\ldots,C_m(x_m)) := (\Omega|C_{1,\rint}(x_1),\ldots,C_{m,\rint}(x_m)\Omega),
\end{equation}
where $C_{1,\rint}(x_1),\ldots,C_{m,\rint}(x_m)$ are any of the fields $B_{1,\rint}(x),\ldots,B_{n,\rint}(x)$. By the nuclear theorem the Wightman functions are formal power series in the coupling constant with coefficients in $S'(\R^{4m})$. They satisfy a number of conditions: 
\begin{enumerate}[leftmargin=*,label={(\arabic*)}]
 \item Poincar{\'e} covariance, 
 \item relativistic spectral condition,
 \item Hermiticity,
 \item local (anti)commutativity and
 \item positive definiteness condition.
\end{enumerate}
The Green functions are the vacuum expectation values of the time-ordered products of the interacting fields $B_{1,\rint}(x),\ldots,B_{n,\rint}(x)$, i.e.
\begin{equation}\label{eq:gre}
  \Gre(C_1(x_1),\ldots,C_m(x_m)) := (\Omega|\T(C_{1,\rint}(x_1),\ldots,C_{m,\rint}(x_m))\Omega),
\end{equation}
where $C_{1,\rint}(x),\ldots,C_{m,\rint}(x)$ are any of the fields $B_{1,\rint}(x),\ldots,B_{n,\rint}(x)$. The time-ordering operator $\T$ is not determined uniquely in the axiomatic setting. We assume that it is defined in such a way that the resulting Green functions are formal power series in the coupling constant with coefficients in $S'(\R^{4m})$ which have the following properties:
\begin{enumerate}[leftmargin=*,label={(\arabic*)}]
 \item Poincar{\'e} covariance,
 \item symmetry (or graded-symmetry in the presence of fermionic fields) under the permutation of its arguments and
 \item causality (if for all $j\in\{1,\ldots,m-1\}$ the point $x_{j}$ is not in the causal past of any of the points $x_{j+1},\ldots,x_m$ then the Green function $ \Gre(C_1(x_1),\ldots,C_m(x_m))$ coincides with the Wightman function $ \Wig(C_1(x_1),\ldots,C_m(x_m))$).
\end{enumerate}
By the last two conditions the Green functions are determined uniquely for non-coinciding points in terms of the Wightman functions. For the precise form of all of the above properties of the Wightman and Green functions we refer the reader to Section~\ref{sec:properties}.

The Wightman and Green functions are always defined in the literature with the use of indirect methods. The starting point are formulas which have been regularized in the infrared regime and only formally correspond to the expressions \eqref{eq:wig} and \eqref{eq:gre}. The Wightman or Green functions are obtained by taking an appropriate limit in which this regularization is removed. The method of defining the Wightman and Green functions due to Epstein and Glaser which is used in the thesis is also of this type. Because of the use of an infrared regularization in the intermediate steps of the construction of the Wightman and Green functions it is not clear whether the resulting functions have anything in common with the vacuum expectation values of the products or the time-order products of the interacting fields. In order to address this objection and justify the correctness of our definition of the Wightman and Green functions we prove that these functions have all the properties listed above. Unfortunately, this still does not guarantee that the obtained Wightman and Green functions are given by formulas \eqref{eq:wig} and \eqref{eq:gre} for interacting fields of some perturbative model satisfying the generalized Wightman axioms. Note that the Wightman reconstruction theorem \cite{streater2000pct} is not applicable in the perturbative setting. 

The existence of the perturbative models (in an operator formulation) satisfying the generalized Wightman axioms has not been established in the literature. We expect that it is possible to generalize the results of \cite{epstein1976adiabatic} to show that any purely massive model fulfills these axioms. Because of the infrared problem the proof of the analogous statement in the case of theories with massless particles is much more difficult task since there are no natural candidates for the linear space $\mathcal{D}$ and the representation of the Poincar{\'e} group $U(a,\Lambda)$. These problems will, however, not be investigated in the thesis. We stated the generalized Wightman axioms only in order to provide a motivation for the properties of the Wightman and Green functions listed above.

\subsubsection{BPHZ approach}

The first complete proof of the existence of the Green functions was given by Hepp \cite{hepp1966proof} based on the earlier work of Bogoliubov and Parasiuk \cite{bogoliubov1957multiplication}. The proof was subsequently simplified and generalized by Zimmermann \cite{zimmermann1969convergence} with the use of the Forest Formula which explicitly solves the Bogoliubov's recurrence and allows to define the time-ordered products without introducing any ultraviolet (UV) regularization. The method of renormalization developed in the above papers is known as the BPHZ approach. Let us describe its main steps.

The starting point is the formal formula due to Gell-Mann and Low \cite{gell1951bound}
\begin{multline}\label{eq:gellmann_low}
 \Gre(C_1(x_1),\ldots,C_m(x_m)) =
 \\
 \frac{\sum_{n=0}^\infty \frac{\ri^ne^n}{n!} \int\rd^4 y_1\ldots\rd^4 y_n\,(\Omega|\T(\mathcal{L}(y_1),\ldots,\mathcal{L}(y_n),C_1(x_1),\ldots,C_m(x_m))\Omega)}{\sum_{n=0}^\infty \frac{\ri^ne^n}{n!} \int\rd^4 y_1\ldots\rd^4 y_n\,(\Omega|\T(\mathcal{L}(y_1),\ldots,\mathcal{L}(y_n))\Omega)},
\end{multline}
which allows to express the Green function $\Gre(C_1(x_1),\ldots,C_m(x_m))$ of the interacting fields $C_{1,\rint},\ldots,C_{m,\rint}$ as a formal power series in the coupling constant $e$ with coefficients given in terms of the Green functions of the free fields $C_1,\ldots,C_m$ and the interaction vertex $\mathcal{L}$ (for simplicity we assume that there is only one interaction vertex; the method is also applicable when there are more interaction vertices). Note that, even if the vacuum expectation values of the time-ordered products are correctly defined as Schwartz distributions, i.e. the ultraviolet (UV) problem is solved, one still has to deal with the infrared (IR) problem as the integrals on the RHS of the Gell-Mann and Low formula are usually meaningless. 

Using Wick's theorem the RHS of Equation \eqref{eq:gellmann_low} is formally expanded into a sum over Feynman diagrams. To handle the IR problem one replaces the standard general expression for the propagator in the momentum representation 
\begin{equation}
 \F{D}_{ij}(l)=\frac{\ri P_{ij}(l)}{l^2-m^2+\ri 0},
\end{equation}
where $P_{ij}(l)$ are polynomials in the four-momentum $l$, by the covariant regularized propagator
\begin{equation}\label{eq:propagator_reg_cov}
 \F{D}^\textrm{cov}_{ij}(\epsilon;l)=\frac{\ri P_{ij}(l)}{l^2-m^2+\ri \epsilon},~~~~~\epsilon>0
\end{equation}
or the noncovaraiant regularized propagator
\begin{equation}\label{eq:propagator_reg}
 \F{D}_{ij}(\epsilon;l)=\frac{\ri P_{ij}(l)}{l^2-m^2+\ri \epsilon (\vec{l}^2+m^2)},~~~~~\epsilon>0,
\end{equation}
where $\vec{l}^2 = l_1^2+l_2^2+l_3^2$. The covariant regularized propagator is used in the method due to Bogoliubov, Parasiuk and Hepp \cite{hepp1966proof}. This method requires the introduction of some UV regularization, e.g. the Pauli-Villars regularization, in the intermediate steps. The advantage of the noncovaraiant regularized propagator is its better UV behavior. It turns out that for any $\epsilon>0$ it satisfies the following bound 
\begin{equation}
 |\F{D}_{ij}(\epsilon;l)| \leq \const~ \frac{|P_{ij}(l)|}{|l|^2+m^2},
\end{equation}
where $|l|^2=l_0^2+\vec{l}^2$ and the constant above depends on $\epsilon$. The validity of the above estimate, which is violated by the covariant regularized propagator $\F{D}^\textrm{cov}_{ij}(\epsilon;l)$, is indispensable in the Zimmermann's method. In this method the UV regularization is not needed. Moreover, as shown by Lowenstein it can be generalized to the case of models with massless particles. In what follows we give an overview of the Zimmermann's construction of the Green functions in purely massive models and subsequently outline its generalization due to Lowenstein.

In order to solve the UV problem the integrand obtained by the application of the Feynman rules is modified by subtracting Taylor polynomial according to the Zimmermann's Forest Formula. The expression for the connected diagram $\Gamma$ obtained in this way has the following form
\begin{equation}\label{eq:BPHZ_R}
 R_\Gamma(\epsilon;k_1,\ldots,k_L;p_1,\ldots,p_M)=
 \frac{P(k,p,\epsilon)}{\prod_{i=1}^K [l_i(k,p)^2-m_i^2-\ri \epsilon (\vec{l}_i(k,p)^2+m_i^2)]},
\end{equation}
where $m_i> 0$ are mass parameters, $P$ is a polynomial in $k$, $p$, $\epsilon$ and the  four-vectors $l_j(k,p)$ are linear functions of the integration variables $k_1,\ldots,k_L$ and the external momenta $p_1,\ldots,p_M$. For any $\epsilon>0$ the integral of the RHS of \eqref{eq:BPHZ_R} is bounded from above by the corresponding Euclidean integral -- this is the reason for the non-covariant regularization \eqref{eq:propagator_reg} of propagators. By the above property the absolute convergence of
\begin{equation}\label{eq:bphz_integrated}
 \int\frac{\rd^4k_1}{(2\pi)^4}\ldots \frac{\rd^4k_L}{(2\pi)^4}\, R_\Gamma(\epsilon;k_1,\ldots,k_L;p_1,\ldots,p_M)
\end{equation}
for any $\epsilon>0$ and $p_1,\ldots,p_M\in\R^4$ follows from the Weinberg's power counting theorem \cite{weinberg1960high}. It was shown in \cite{zimmermann1968power} (based on the results of \cite{hepp1966proof}) that the expression \eqref{eq:bphz_integrated} converges in $\mathcal{S}'(\R^{4M})$ for $\epsilon\searrow 0$ to a Lorentz covariant Schwartz distribution. The connected Green function $ \Gre_{\textrm{c}}(C_1(x_1),\ldots,C_m(x_m))$ is given by
\begin{multline}\label{eq:bphz_epsilon_lim}
 \int\rd^4 x_1\ldots\rd^4 x_m\, \Gre_{\textrm{c}}(C_1(x_1),\ldots,C_m(x_m)) f(x_1,\ldots,x_m):=
 \\
 \sum_{\Gamma} \lim_{\epsilon\searrow0}\int\frac{\rd^4k_1}{(2\pi)^4}\ldots \frac{\rd^4k_L}{(2\pi)^4}\frac{\rd^4p_1}{(2\pi)^4}\ldots \frac{\rd^4p_m}{(2\pi)^4}\, \F{f}(-p_1,\ldots,-p_m)
 \\
 \times (2\pi)^4\delta(p_1+\ldots+p_m)\,R_\Gamma(\epsilon;k_1,\ldots,k_L;p_1,\ldots,p_{m-1}) 
\end{multline}
where the sum is over Feynman diagrams contributing to the connected Green function under consideration (it is a formal power series in the coupling constant $e$).

The BPHZ method described above was subsequently generalized to the case when some fields are massless by Lowenstein \cite{lowenstein1976convergence} (based on his earlier results obtained in collaboration with Zimmermann \cite{lowenstein1975formulation}). The key ingredient is the modified Forest Formula. The Taylor subtractions are made  again at the vanishing external momenta of the (sub)diagram under consideration. However, to avoid the IR problem, in some of the subtracted terms the massless propagators are replaced by the massive ones. In this way the Lorentz covariance is preserved (after taking the limit $\epsilon\searrow0$) at the cost of introducing additional energy scale in the theory. By the power-counting theorem \cite{lowenstein1975power} generalized to the case when some propagators are massless the following integral
\begin{equation}
\int\frac{\rd^4k_1}{(2\pi)^4}\ldots \frac{\rd^4k_L}{(2\pi)^4}\frac{\rd^4p_1}{(2\pi)^4}\ldots \frac{\rd^4p_M}{(2\pi)^4}\, f(p_1,\ldots,p_M) R_\Gamma(\epsilon;k_1,\ldots,k_L;p_1,\ldots,p_M)
\end{equation}
is absolutely convergent for any $\epsilon>0$ and $g\in\mathcal{S}(\R^{4M})$. The expression \eqref{eq:bphz_epsilon_lim} converges in the limit $\epsilon\searrow0$ to a covariant Schwartz distribution as shown in \cite{lowenstein1976distributional}. This proves the existence of the Green functions for theories with massless particles. 

More precisely, the method of Lowenstein applies to all models with interaction vertices of the infrared dimension equal at least $4$. Each interaction vertex is a polynomial in the free fields and their derivatives. The infrared dimension is the additive number calculated by assigning dimension $2$ to massive fields, dimension $\frac{3}{2}$ to massless Dirac fields and dimension $1$ to massless boson fields and each derivative. In contrast, the canonical dimension is the additive number calculated by assigning dimension $1$ to each derivative as well as scalar and vector fields\footnote{In the thesis we only consider vector fields defined with the use of the Gupta-Bleuler approach.} and dimension $\frac{3}{2}$ to Dirac fields. The canonical dimension is used in the classification of models of the quantum field theory into super-renormalizable, renormalizable and non-renormalizable models. Since the infrared dimension is not lower then the canonical dimension the Green functions exist in particular in all models with interaction vertices of the canonical dimension equal to~$4$ (all important physical models satisfy this property).

The BPHZ method of renormalization (or rather, its variants) are described in detail in the monographs \cite{manoukian1983renormalization} and \cite{zavialov2012renormalized} (see also the lecture notes by Zimmermann and Lowenstein \cite{zimmermann1970local,lowenstein1976bphz}). Note that the Wightman functions have not been defined in the BPHZ approach.

\subsubsection{Dimensional regularization}

Other definition of the Green functions was given by Breitenlohner and Maison in a series of papers \cite{breitenlohner1977dimensional,breitenlohner1977dimensionally1,breitenlohner1977dimensionally2} with the use of the dimensional regularization \cite{bollini1972dimensional,veltman1972regularization}. Like the BPHZ method, this method is also based on the formal representation of the Green functions in terms of the Feynman graphs with covariant regularized propagators
\begin{equation}
 \F{D}^\textrm{cov}_{ij}(\epsilon;l)=\frac{\ri P_{ij}(l)}{l^2-m^2+\ri \epsilon},~~~~~\epsilon>0.
\end{equation}
After introducing the Schwinger parameters the momentum space integrals over internal momenta of a given (sub)graph are performed. The resulting expressions $$I_\gamma(\epsilon;d;\alpha_1,\ldots,\alpha_n;p_1,\ldots,p_m)$$  depend on the Schwinger parameters $\alpha_1,\ldots,\alpha_n$ and the external momenta $p_1,\ldots,p_m$ of the considered (sub)graph $\gamma$ and are meromorphic functions of the dimension $d$ of the spacetime. The UV problem is related to the divergence of the integral of $I_\gamma$ over $\alpha_1,\ldots,\alpha_n$ and is solved by the method analogous to the one used in the BPHZ approach. The difference is that instead of considering the Taylor expansion of $I_\gamma$ in the external momenta at the origin one subtracts the principal part of its Laurent series in $d$ at $d=4$ (it is a polynomial in the external momenta of the (sub)graph $\gamma$). After that one shows the existence of the $\epsilon\searrow0$ limit in the sense of Schwartz distributions. This limit always exists in purely massive theories and purely massless theories, in the later case provided that all interaction vertices have the canonical dimension $4$. If both massive and massless fields are present then the limit $\epsilon\searrow0$ exists provided the vertex functions are renormalized appropriately. Since dimensional regularization preserves most of the symmetries, in particular the local gauge symmetry, this method is well-suited for practical calculations.

\subsubsection{EG approach}

A completely different method for the construction of the Green and Wightman functions was proposed by Epstein and Glaser in their seminal paper \cite{epstein1973role} in which they formulated and developed the causal perturbation theory. Their approach is based on the ideas due to St{\"u}ckelberg \cite{stuckelberg1950propos} and Bogoliubov \cite{bogoliubov1959introduction}. The time-ordered products are defined as (unbounded) operator-valued Schwartz distributions which satisfy a set of axioms formulated in the position representation. The key observation which allows their inductive construction is that the time-ordered product of $n+1$ local fields at points $x_1,\ldots,x_{n+1}$ is determined uniquely by the time-ordered products of at most $n$ fields for all test functions with support away from the main diagonal $x_1=\ldots=x_{n+1}$. The solution of the UV problem in the EG approach amounts to finding extensions to the space of all test functions which are compatible with the axioms. Note that one neither introduces any regularization nor performs any subtractions. The existing freedom in the definition of the time-ordered products is described by the St{\"u}ckelberg-Petermann renormalization group \cite{dutsch2004causal,brunetti2009perturbative}. The EG method is particularly well-suited for the investigation of the structural properties of the theory. In contrast to the BPHZ approach, it enables very transparent  formulations of the causality and unitarity conditions.

A particular model of the causal perturbation theory is specified by the list of its interaction vertices
\begin{equation}
 \mathcal{L}_1,\ldots,\mathcal{L}_{\mathrm{q}},
\end{equation}
which are polynomials in free fields and their derivatives. To each interaction vertex $\mathcal{L}_l$ we associate a coupling constant $e_l$ and a switching function $g_l\in\mathcal{S}(\R^4)$. For example, the scattering matrix of the model is by definition
\begin{equation}
 S(\mathpzc{g}) = \Texp\left(  \ri \int \rd^4 x \, \sum_{l=1}^\mathrm{q} e_l g_l(x)\mathcal{L}_l(x) \right),
\end{equation}
where $\mathpzc{g}$ is the list of the switching functions $(g_1,\ldots,g_{\mathrm{q}})$. The time-ordered exponential is given in terms of the the time-ordered products of free fields which are operator-valued Schwartz distribution. Consequently, the scattering matrix $S(\mathpzc{g})$ is a well-defined operator-valued formal power series in the coupling constants as long as all the switching functions belong to the Schwartz class. The functions $g_1,\ldots,g_\mathrm{q}$ switch off the interaction as $|x|\to\infty$ and are introduced in order to avoid the IR problem in the definition of the scattering matrix and the interacting fields. In order to remove the above IR regularization and obtain the physical scattering matrix or the physical interacting fields one performs the adiabatic limit.

The interacting advanced field with the IR regularization is given by the Bogoliubov formula
\begin{align}
 C_\adv(\mathpzc{g};x) &:= (-\ri)\frac{\delta}{\delta h(x)} S(\mathpzc{g};h)S(\mathpzc{g})^{-1}\bigg|_{h=0},
%  \\
%  C_\ret(g;x) &:= (-\ri)\frac{\delta}{\delta h(x)}S(g)^{-1}S(g;h)\bigg|_{h=0},
\end{align}
where $C$ is a polynomial in free fields and their derivatives and
\begin{equation}
 S(\mathpzc{g};h):=\Texp\left( \ri \int \rd^4 x \, \sum_{l=1}^\mathrm{q} e_l g_l(x)\mathcal{L}_l(x) +  \ri \int \rd^4 x \, h(x)C(x)\right).
\end{equation}
Similarly, the time-ordered products of the interacting advanced fields with the IR regularization are given by
\begin{align}
 \T(C_{1,\adv}(\mathpzc{g};x_1)\ldots C_{m,\adv}(\mathpzc{g};x_m))  
 := (-\ri)^{m} \frac{\delta}{\delta h_m(x_m)}\ldots\frac{\delta}{\delta h_1(x_1)} 
 S(\mathpzc{g};\mathpzc{h})S(\mathpzc{g})^{-1}\bigg|_{\mathpzc{h}=0},
%  \\
%  \T(C_1(x_1),\ldots,C_{m}(x_m))   
%  := (-\ri)^{m} \frac{\delta}{\delta h_m(x_m)}\ldots\frac{\delta}{\delta h_1(x_1)} 
%  S(g)^{-1}S(g;h_1,\ldots,h_m)\bigg|_{h=0},
\end{align}
where 
\begin{equation}
 S(\mathpzc{g};\mathpzc{h}):=\Texp\left( \ri \int \rd^4 x \, \sum_{l=1}^\mathrm{q} e_l g_l(x)\mathcal{L}_l(x) +  \ri \int \rd^4 x \, \sum_{j=1}^m h_j(x)C_j(x)\right),
\end{equation}
$\mathpzc{h}:=(h_1,\ldots,h_m)$, $h_j\in\mathcal{S}(\R^4)$ and $C_1,\ldots,C_m$ are polynomials in free fields and their derivatives. Note that the interacting fields and their time-ordered products are formal power series in the coupling constants $e_1,\ldots,e_\mathrm{q}$.

The physical Wightman and Green functions are obtained as the following limits of the vacuum expectation values of the products or the time-ordered products of the interacting fields
\begin{equation}\label{eq:intro_W_G}
\begin{aligned}
 \Wig(C_1(x_1),\ldots,C_m(x_m)) &:=
 \lim_{\epsilon\searrow 0}~ (\Omega|C_{1,\adv}(\mathpzc{g}_\epsilon;x_1)\ldots C_{m,\adv}(\mathpzc{g}_\epsilon;x_m)\Omega),
 \\
 \Gre(C_1(x_1),\ldots,C_m(x_m)) &:=
 \lim_{\epsilon\searrow 0}~ (\Omega|\T(C_{1,\adv}(\mathpzc{g}_\epsilon;x_1)\ldots C_{m,\adv}(\mathpzc{g}_\epsilon;x_m))\Omega).
\end{aligned}
\end{equation}
By definition $\mathpzc{g}_\epsilon=(g_{1,\epsilon},\ldots,g_{\mathrm{q},\epsilon})$ and $g_{l,\epsilon}(x):=g_l(\epsilon x)$, where $g_l$ is a Schwartz function such that $g_l(0)=1$ for all $l\in\{1,\ldots,\mathrm{q}\}$. Note that in the limit $\epsilon\searrow0$ the interaction is adiabatically turned on and off. The above limit is called the weak adiabatic limit. Note that its existence does not imply that the Wightman and Green functions are the vacuum expectation values of some operator-valued Schwartz distribution. It would be indeed the case if e.g. the limit
\begin{equation}
 C_{\adv}(f)\Psi := \lim_{\epsilon\searrow 0} C_{\adv}(\mathpzc{g}_\epsilon;f)\Psi \in \mathcal{D}_1,
\end{equation}
called the strong adiabatic limit, existed for all polynomials $C$, $f\in\mathcal{S}(\R^4)$ and $\Psi\in\mathcal{D}_1$, where $\mathcal{D}_1$ is certain domain in the Fock space containing the vacuum vector. However, it is known that the strong adiabatic limit does not exist in many theories with massless particles such as e.g. QED.

The existence of the weak adiabatic limit for purely massive theories was shown by Epstein and Glaser in \cite{epstein1973role}. This result was subsequently extended to the case of the quantum electrodynamics and the massless $\varphi^4$ theory by Blanchard and Seneor \cite{blanchard1975green}. The proof of the existence of the weak adiabatic limit in a more general class of models has not been given in the literature.

Let us mention other related results. First, the existence of the strong adiabatic limit  $\lim_{\epsilon\searrow 0} S(\mathpzc{g}_\epsilon)$, which provides the definition of the physical S-matrix, was shown in \cite{epstein1976adiabatic} for purely massive theories. Adiabatic limit of the inclusive cross-sections in the low orders of the perturbation theory of QED was considered in \cite{scharf2014} (see also \cite{dutsch1993infrared,dutsch1993vertex}). Finally, the existence of expectation values of the products of the interacting fields in thermal states has been recently proved by Fredenhagen and Lindner in the EG framework with the use of the time-slice axiom and the time-averaged Hamiltonian \cite{fredenhagen2014constructionerratum} (see also \cite{lindner2013perturbative,drago2015generalised}).

\subsubsection{Steinmann's approach}

Let us also mention Steinmann's definition of the Wightman and Green functions of the basic fields (products of the basic fields at a point are excluded) in the case of QED \cite{steinmann2013} and the massive or massless $\varphi^4(x)$ theory \cite{steinmann1993perturbation}. The author's construction is based on the set of axioms imposed on the Wightman functions and is performed mostly in the position representation. The UV problem is, however, solved in the momentum space with the use of BPHZ approach. The Steinmann's method is quite complicated and not very transparent because it requires manipulation of ill-defined distributions in the intermediate steps of the construction. Only the final expression for the Wightman and Green functions is free from the infrared problem.

\section{Summary of results}

\vspace{-2mm}
\subsubsection{\hyperref[Sec:existence_wAL]{Wightman and Green functions}}
\vspace{-2mm}

Our main result is the proof of the existence of the weak adiabatic limit in a large class of models. The result provides a method for the construction of the Wightman and Green functions in the Epstein and Glaser approach. It is a generalization of the earlier results due to Epstein and Glaser \cite{epstein1973role} and Blanchard and Seneor \cite{blanchard1975green} where the existence of the adiabatic limit in the massive models, the quantum electrodynamics and the massless $\varphi^4$ theory was shown. Let $\mathcal{L}_1,\ldots,\mathcal{L}_\mathrm{q}$ be the interaction vertices. Our proof of the existence of the weak adiabatic limit applies to all models which satisfy one of the following conditions:
\begin{enumerate}[leftmargin=*,label={(\arabic*)}]
\item $\dim(\mathcal{L}_l)=3$ and $\mathcal{L}_l$ contains at least one massive field for all $l\in\{1,\ldots,\mathrm{q}\}$,
\item $\dim(\mathcal{L}_l)=4$ for all $l\in\{1,\ldots,\mathrm{q}\}$,
\end{enumerate}
where $\dim(B)$ is the canonical dimension of the polynomial $B$. In the case (1) we impose a non-standard bound on the Steinmann scaling degree of the vacuum expectation values of the time-ordered products which allows more freedom in their definition. The resulting Wightman and Green functions fulfill all the conditions following from the modified Wightman axioms which were mentioned in Section~\ref{sec:axiomatic_approach}. Note that the Green functions have already been constructed in the above-mentioned models with the use of momentum-space methods as descried in the previous section. However, the general construction of the Wightman functions has not been given in the literature.

The existence of the Wightman functions can be used to define a real and Poincar\'e invariant functional on the algebra of interacting fields obtained by means of the algebraic adiabatic limit. In the case of models without vector fields this functional is positive, and thus, may be interpreted as the vacuum state. In QED and other theories which contain vector fields it is expected that the above-mentioned functional can be used to define a vacuum state on the algebra of interacting observables. 

\vspace{-4mm}
\subsubsection{\hyperref[sec:generalized_central]{Central normalization condition}}
\vspace{-2mm}

We also formulate a normalization condition of the time-ordered products which we call the central normalization condition. Using the technique of the proof of the existence of the weak adiabatic limit, we show that it is possible to define the time-ordered products which fulfill this condition. The central normalization condition significantly restricts the ambiguity in the definition of the time-ordered products and is compatible with the standard normalization conditions. In the case of the quantum electrodynamics (with a massive spinor field) it fixes uniquely all the time-ordered products of sub-polynomials of the interaction vertex and in particular implies the Ward identities.

\section{Structure of the thesis}

\vspace{-2mm}
\subsubsection{\hyperref[Sec:causal]{Outline of causal perturbation theory}}
\vspace{-2mm}

In Chapter~\ref{Sec:causal} we give an overview of the approach to the perturbative quantum field theory developed by Epstein and Glaser. We recall main steps of the construction of the time-ordered products of the free fields, which are the basic objects of this approach, introduce the notation which is used throughout the thesis and present some auxiliary results. 

\vspace{-4mm}
\subsubsection{\hyperref[sec:models]{Interacting models}}
\vspace{-2mm}

In Chapter~\ref{sec:models} we recall the definition of the interacting models in the Epstein-Glaser approach. We give the formulas for the scattering matrix, the advanced and retarded fields and the Wightman and Green functions with the infrared regularization. In Section~\ref{sec:W_G_IR} we collect identities which play an important role in the proof of the existence of the Wightman and Green functions.

\vspace{-4mm}
\subsubsection{\hyperref[sec:adiabatic_limits]{Adiabatic limits}}
\vspace{-2mm}

Chapter~\ref{sec:adiabatic_limits} contains a survey of different types of the adiabatic limits which have been introduced in the literature. In particular we give the precise definition of the weak adiabatic limit which is used to determine the Wightman and Green functions. We define the algebra of interacting fields and introduce the notion of a state on this algebra.

\vspace{-4mm}
\subsubsection{\hyperref[Sec:existence_wAL]{Wightman and Green functions}}
\vspace{-2mm}

In Chapter~\ref{Sec:existence_wAL} we present the main result  of the thesis i.e. the proof of the existence of the week adiabatic limit which allows to define the Wightman and Green functions. We first recall the proof in the case of the purely massive models which was given by Epstein and Glaser. Next, in Section~\ref{sec:idea} we describe the idea of the proof in the general case. In Sections \ref{sec:math}, \ref{sec:prod} and \ref{sec:split} we collect intermediate results. The proof of the existence of the week adiabatic limit in the general case is contained in Section~\ref{sec:proof_scalar}. At the end of this section the comparison of our method with the one due to Blanchard and Seneor may be found.

\vspace{-4mm}
\subsubsection{\hyperref[ch:comp]{Compatibility of normalization conditions}}
\vspace{-2mm}

In Chapter~\ref{ch:comp} we prove the compatibility of the normalization condition needed for the existence of the week adiabatic limit with the standard normalization conditions usually imposed on the time-ordered products which guarantee e.g. the unitarity and Poincar{\'e} covariance of the scattering matrix.

\vspace{-4mm}
\subsubsection{\hyperref[sec:generalized_central]{Central normalization condition}}
\vspace{-2mm}

In Chapter~\ref{sec:generalized_central} we formulate the normalization condition of the time-ordered products which in the case of the quantum electrodynamics fixes uniquely all the time-ordered products of the sub-polynomials of the interaction vertex.

\vspace{-4mm}
\subsubsection{\hyperref[app:grassmann]{Appendices}}
\vspace{-2mm}

In Appendix \ref{app:grassmann} we define the notion of the Grassmann-valued Schwartz function which is used in the Chapter~\ref{Sec:causal} to define the generating functional of the time-ordered products of fields with odd fermion number. Appendix \ref{sec:magic_formula} contains the comparison of the Green functions defined in the Epstein-Glaser framework and the Green functions obtained with the use of the Gell-Mann and Low formula. In Appendix \ref{sec:mass} we show that the correct mass normalization of all massless fields is necessary for the existence of the weak adiabatic limit.

\section{Notation}

\begin{itemize}[leftmargin=*]
 \item The Minkowski spacetime is identified with $\R^4$. It is equipped with the inner product given by $x\cdot y = x^0 y^0 - x^1 y^1-x^2 y^2-x^3 y^3$.
 
 \item The closed future and past light cones in the Minkowski spacetime are denoted by $\overline{V}^\pm=\{p\in\R^4\,:\, p^2 \geq 0,~\pm p^0\geq 0\}$, respectively.
 
 \item The invariant measure on the mass hyperboloid $H_m:=\{p\in\R^4\,:\,p^2=m^2,\,p^0\geq 0\}$ is given by $\rd\mu_m(p) := \frac{1}{(2\pi)^3} \rd^4 p\, \theta(p^0) \delta(p^2-m^2) $ for $m\geq 0$.

 \item The space of test functions with compact support and Schwartz functions on $\R^N$ are denoted by $\mathcal{D}(\R^N)$ and $\mathcal{S}(\R^N)$, respectively.
 
 \item  If $t\in\mathcal{S}'(\R^N)$ is a Schwartz distribution and $g\in\mathcal{S}(\R^N)$ then
\begin{equation}
 \int \rd^N x \, t(x) g(x).
\end{equation}
stands for the value of $t$ acting on a Schwartz function $g$.

\item The Fourier transform of the Schwartz distribution $t\in\mathcal{S}'(\R^N)$ is denoted by $\tilde{t}$. For any $g\in\mathcal{S}(\R^N)$ it holds
\begin{equation}
 \tilde{g}(q):=\int\rd^N x\, \exp(\ri q \cdot x) g(x),~~~~~~g(x)=\int\frac{\rd^N q}{(2\pi)^N}\, \exp(-\ri q \cdot x) \tilde{g}(q).
\end{equation}

\item The set of positive natural numbers is denoted by $\N_+$. By definition $\N_0:=\{0\}\cup\N_+$.

\item  The four-dimensional multi-indices which are sometimes called the quadri-indices are denoted by $\alpha,\beta,\gamma,\ldots \in \N_0^4$.

\item  We use the Einstein summation convention for Lorentz and spinor indices.

\item For $x\in\R^N$ we set $|x|:=(x_1^2+\ldots+x_N^2)^{1/2}$.

\item  The acronym \emph{VEV} stands for \emph{vacuum expectation value}.
\end{itemize}

%===================================================================================================
%+++++++++++++++++++++++++++Causal perturbation theory++++++++++++++++++++++++++++++++++++++++++++++
%===================================================================================================

\part{Epstein-Glaser approach to quantum field theory}

\chapter{Outline of causal perturbation theory}\label{Sec:causal}

The main objects of the Epstein-Glaser approach to the perturbative quantum field theory, which is also known as the causal perturbation theory, are the time-ordered products of the polynomials of free fields. In this chapter we outline the main points of their construction in the four-dimensional Minkowski spacetime. It is performed entirely in the position space and is based on causality and translation invariance. One first specifies the set of axioms which are satisfied by the time-ordered products and subsequently constructs them inductively. The solution of the UV problem consists of finding an extension of distributions which are initially defined only on a suitable subspace of the space of all test functions. The renormalization freedom is the consequence of non-uniqueness of the extension. The scattering matrix, the interacting fields and the Wightman and Green functions (with the IR regularization) are expressed in terms of the time-ordered products of the free fields. The definition of these objects is, however, postponed to the next chapter where we study the interacting models in the framework of the causal perturbation theory.

The method of Epstein and Glaser has been reformulated and generalized in a number of ways. It is the basis of QFT in curved spacetime which is used to construct interacting quantum fields propagating in an arbitrary globally hyperbolic spacetimes \cite{hollands2015quantum,benini2013quantum,brunetti2009quantum}. The ideas of Epstein and Glaser combined with the formalism of deformation quantization and functional approach led also to the foundation of the perturbative algebraic quantum filed theory (pAQFT) -- the formulation of the perturbative QFT in terms of abstract local algebras \cite{fredenhagen2015perturbative,rejzner2016perturbative}.

In the thesis we follow the traditional approach in which fields are operator-valued distributions defined on a suitable domain of the Fock space. The formalism of the pAQFT framework is used only in Chapter~\ref{sec:adiabatic_limits}. In the outline of the Epstein-Glaser method presented below we follow closely the paper \cite{epstein1973role}, generalizing its results to the case of models including fermionic fields. For details we refer the reader to \cite{epstein1973role,scharf2014,scharf2016gauge}. In Section~\ref{sec:aux} we recall some important formulas and present auxiliary results which will be needed in the proof of the existence of the weak adiabatic limit. The bound on the Steinmann scaling degree formulated in Section~\ref{sec:sd} is the generalization of the standard bound which is known in the literature. The generalization is necessary for the existence of the weak adiabatic limit in models with interaction vertices of the canonical dimension equal to $3$.

\section{Algebra \texorpdfstring{$\Fa$}{F} of symbolic fields}\label{sec:ff}

In perturbative QFT the interacting models are built with the use of the free fields. Let
\begin{equation}\label{eq:basic_gen}
 \mathcal{G}_0:=\{A_1,\ldots,A_\mathrm{p}\}\simeq\{1,\ldots,\mathrm{p}\}
\end{equation}
be the set of symbols denoting types of free fields needed for the definition of a 
model under consideration. The elements of this set are called the {\it basic 
generators}. All components of vector or spinor fields are included in this set 
as separate symbols (if fields of these types are present in the model). We assume that an 
involution denoted by ${}^*$ is defined in~$\mathcal{G}_0$. It means that if a 
charged field $A_i$ belongs to $\mathcal{G}_0$, then also its Hermitian conjugation 
denoted by $A^{*}_i$ belongs to it, i.e. $A_i^{*}=A_{i'}$ for some 
$i'\in\{1,\ldots,\mathrm{p}\}$. We will consider the models containing the 
following types of fields (massive or massless)
\begin{enumerate}[leftmargin=*,label={(\arabic*)}]
 \item real scalar fields,
 \item charged scalar fields,
 \item Dirac spinor fields,
 \item real vector fields.
\end{enumerate} 
For example, in the case of QED there are 12 basic generators: the four components of the vector potential $A_\mu=A^{*}_\mu$, and the four components of the spinor field $\psi_a$ and its Hermitian conjugate $\psi_a^*=(\overline{\psi}\gamma^0)_a$. 

The basic generators \eqref{eq:basic_gen} supplemented with the symbols corresponding to their derivatives form the set of the {\it generators}
\begin{equation}\label{eq:gen}
 \mathcal{G}:=\{\partial^\alpha\! A_i:~i\in\{1,\ldots,\mathrm{p}\}, \alpha\in\N_0^4 \}\simeq\{1,\ldots,\mathrm{p}\}\times \N_0^4,
\end{equation}
where $\alpha$ is a multi-index. Note that $\mathcal{G}$ may be identified with the set of pairs $(i,\alpha)$ where $i\in\{1,\ldots,\mathrm{p}\}$ and $\alpha\in\N_0^4$. We set $(\partial^\alpha\! A_i)^*:=\partial^\alpha\! A^*_i$. To every generator we associate the following quantum numbers
\begin{enumerate}[leftmargin=*,label={(\arabic*)}]
 \item fermion number $\mathbf{f}(\partial^\alpha\! A_i)\in\Z$, 
 \item charge number $\mathbf{q}(\partial^\alpha\! A_i)\in\Z$,
 \item canonical dimension $\dim(\partial^\alpha\! A_i)\in\frac{1}{2}\Z$.
\end{enumerate}
For example in the case of QED we have 
\begin{enumerate}[leftmargin=*,label={(\arabic*)}]
 \item $\mathbf{f}(\partial^\alpha\! A_\mu)=0$, $\mathbf{f}(\partial^\alpha\! \psi_a)=-\mathbf{f}(\partial^\alpha\! \psi^*_a)=1$,
 \item $\mathbf{q}(\partial^\alpha\! A_\mu)=0$, $\mathbf{q}(\partial^\alpha\! \psi_a)=-1$, $\mathbf{q}(\partial^\alpha\! \psi^*_a)=1$,
 \item $\dim(\partial^\alpha\! A_\mu)=1+|\alpha|$, $\dim(\partial^\alpha\! \psi_a)=\dim(\partial^\alpha\! \psi^*_a)=\frac{3}{2}+|\alpha|$.
\end{enumerate}
The canonical dimension of the scalar and vector field is equal $1$ and the canonical dimension of the Dirac spinor field is equal $3/2$. Each derivative increases the canonical dimension by $1$. The canonical dimension of fields plays a crucial role in determining the renormalizability of a given model.

Following \cite{boas2000gauge} we define the {\it algebra of symbolic fields} denoted by $\Fa$. It is a free unital graded-commutative ${}^*$-algebra over $\C$ generated by the elements of the set $\mathcal{G}$. The adjoint is defined uniquely by the following conditions: the anti-linearity and the identity $(B_1 B_2)^*=B_2^* B_1^*$ which holds for all $B_1,B_2\in\Fa$. The graded commutativity means that for any $B_1,B_2\in\Fa$, which are monomials in the generators it holds
\begin{equation}
 B_1 B_2 = (-1)^{\mathbf{f}(B_1)\mathbf{f}(B_2)} B_2 B_1. 
\end{equation}
The definition of $\mathbf{f}(B)$ is extended to all monomials $B\in\Fa$ by $\mathbf{f}(cB)=\mathbf{f}(B)$ for $c\in\C$ and the additivity $\mathbf{f}(B_1B_2)=\mathbf{f}(B_1)+\mathbf{f}(B_2)$ for any monomials $B_1,B_2\in\Fa$. The same holds for the other quantum numbers. We say that $B\in\Fa$ is a homogenous polynomial if it is a linear combination of monomials with the same quantum numbers. The set of homogenous polynomials is denoted by $\Fh$. The definition of quantum numbers is naturally extended to $\Fh$.

The {\it super-quadri-index} is by definition a map
\begin{equation}
 r:~\mathcal{G} \ni (i,\alpha) \mapsto r(i,\alpha)\equiv r(\partial^\alpha\!A_i) \in \N
\end{equation}
supported on a finite subset of $\mathcal{G}$. We say that the super-quadri-index $r$ involves only the field $A_i$ (involves only the massless fields) if $r(i',\alpha)=0$ for $i'\neq i$ (for all $i'$ such that $A_{i'}$ is a massive field). We write $r\geq s$ iff $r(i,\alpha)\geq s(i,\alpha)$ for all $(i,\alpha)\in\mathcal{G}$. 
%If $r\geq s$ then the super-quadri-index $u=r-s$ is given by $u(i,\alpha)=r(i,\alpha)-s(i,\alpha)$ for all $(i,\alpha)\in\mathcal{G}$. 
Moreover, we set
\begin{equation}
 |r| := \sum_{i=1}^{\mathrm{p}} \sum_{\alpha\in\N_0^4}r(i,\alpha),
 ~~~~~
 r!:=\prod_{i=1}^{\mathrm{p}} \prod_{\alpha\in\N_0^4}r(i,\alpha).
\end{equation}

The monomial $A^r\in\Fa$ labeled by the super-quadri-index $r$ is by definition
\begin{equation}\label{eq:wick_def}
 A^r := \prod_{i=1}^{\mathrm{p}} \prod_{\alpha\in\N_0^4}\,\,(\partial^\alpha\!A_{i})^{r(i,\alpha)}.
\end{equation}
Note that if fermionic fields are present then the order of factors in the above product matters.  To remove the ambiguity in the definition of $A^r$, we fix some linear ordering in the set $\mathcal{G}\simeq\{1,\ldots,\mathrm{p}\}\times \N_0^4$ (its precise form is irrelevant for our purposes) and always multiply the generators in this order. All monomials in $\Fa$ are proportional to $A^r$ for some super-quadri-index $r$. The correspondence between super-quadri-indices and monomials $A^r$ is one-to-one. The set 
\begin{equation}
 \{A^r\,:\,r ~\textrm{ is a super-quadri-index}\}
\end{equation}
is a linear basis of the algebra $\Fa$. For example, in the case of QED the electric current $j^\mu=\overline{\psi}\gamma^\mu\psi = \psi^*_a \gamma_{ab}^0 \gamma_{bc}^\mu \psi_c\in\Fa$ (we use the Einstein summation convention) is a combination of the monomials $\psi^*_a \psi_c$.

The derivative of $B\in\Fa$ with respect to the generator $\partial^\alpha\!A_i$ is defined as a graded derivation 
\begin{equation}
 \frac{\partial}{\partial(\partial^\alpha\!A_i)} BC = \frac{\partial B}{\partial(\partial^\alpha\!A_i)} \, C + 
 (-1)^{\mathbf{f}(B)\mathbf{f}(A_i)} \, B\, \frac{\partial C}{\partial(\partial^\alpha\!A_i)}
 ~~~~\forall_{B,C\in\Fh}
\end{equation}
such that
\begin{equation}
 \frac{\partial(\partial^{\alpha'}\!\!A_{i'})}{\partial(\partial^\alpha\!A_i)} = \delta_{ii'} \delta_{\alpha\alpha'}.
\end{equation}
Now let $s$ be a super-quadri-index. We define the linear map $\Fa\ni B\mapsto B^{(s)}\in\Fa$ by
\begin{equation}
 B^{(s)}:= \prod_{i=1}^{\mathrm{p}} \prod_{\alpha\in\N_0^4}\,\,\left(\frac{\partial}{\partial(\partial^\alpha\!A_{i})}\right)^{s(i,\alpha)} B.
\end{equation}
For example if $B=A^r$ then $B^{(s)}$ is proportional to $A^{r-s}$ if $r\geq s$ and vanishes otherwise. The polynomial $C\in\Fa$ is a {\it sub-polynomial} of the polynomial $B\in\Fa$ iff $C$ equals $B^{(s)}$ up to a multiplicative constant for some super-quadri-index $s$. A sub-polynomial $C$ of a monomial $A^r$ is a monomial of the form $A^s$ for some super-quadri-index $s$ such that $r\geq s$. If a given sub-polynomial of $B$ is a monomial then we call it a \emph{sub-monomial} of~$B$.

As an example let us consider the interaction vertex of QED
\begin{equation}
 \mathcal{L} = \overline{\psi}\slashed{A}\psi =  \psi^*_a (\gamma^0 \gamma^\mu)_{ab} \psi_b A_\mu\in\Fa.
\end{equation}
There are $8$ different sub-polynomials of $\mathcal{L}$ which are all listed below 
\begin{equation}\label{eq:qed_sub_pol}
\begin{aligned}
 &\mathcal{L},~~~~
 j^\mu=\frac{\partial \mathcal{L}}{\partial A_\mu},~~~~
 (\slashed{A}\psi)_a=\frac{\partial\mathcal{L}}{\partial\overline{\psi}_a},~~~~
 (\overline{\psi}\slashed{A})_a=\frac{\partial\mathcal{L}}{\partial\psi_a},~~~~
 (\gamma^0 \gamma^\mu\psi)_a=\frac{\partial^2\mathcal{L}}{\partial A_\mu\partial\overline{\psi}_a},
 \\
 &(\overline{\psi}\gamma^0 \gamma^\mu)_a=\frac{\partial^2\mathcal{L}}{\partial A_\mu\partial\psi_a},~~~~
 (\gamma^0 \slashed{A})_{ab} = \frac{\partial^2\mathcal{L}}{\partial\psi_b \partial\overline{\psi}_a},~~~~
 (\gamma^0 \gamma^\mu)_{ab} = \frac{\partial^3\mathcal{L}}{\partial\psi_a\partial A_\mu \partial\overline{\psi}_b}.
\end{aligned} 
\end{equation}
For example we have $\mathcal{L}=\mathcal{L}^{(s)}$, where $s=0$ and $(\gamma^0 \gamma^\mu\psi)_a = \mathcal{L}^{(s)}$, where $s(A_\mu)=1$, $s(\overline{\psi}_a)=1$ and $s(\partial^\alpha\!A_i)=0$ otherwise. The latter super-quadri-index $s$ involves only the fields $A_\mu$ and $\overline{\psi}_a$. Note that the linear span of the sub-polynomials of the interaction vertex in QED is the same as the linear span of
\begin{equation}
 \mathcal{L},~j^\mu,~ (\slashed{A}\psi)_a,~(\overline{\psi}\slashed{A})_a,~\psi_a,~
 \overline{\psi}_a,~A_\mu,~1.
\end{equation}

We introduce the representation of the $SL(2,\C)$ group, which is the covering group of the Lorentz group, acting on the algebra of symbolic fields $\Fa$ and denote it by $\rho$. First, the representation of the $SL(2,\C)$ group on the vector space whose basis is the set of generators $\mathcal{G}$ is defined in the standard way using the transformation laws of the scalar, spinor and vector fields and their derivatives. In order to define the representation of the $SL(2,\C)$ group on $\Fa$ we interpret its elements as complex-valued functions on the above vector space.

\section{Wick polynomials}\label{sec:Wick}

Let $\mathcal{D}$ be a linear space over $\C$ equipped with a sesquilinear inner product $(\cdot|\cdot)$ and let $L(\mathcal{D})$ be the space of linear maps $\mathcal{D}\to\mathcal{D}$. By an {\it operator-valued Schwartz distribution} on $\mathcal{D}$ we mean a map $T:\,\mathcal{S}(\R^N)\to L(\mathcal{D})$ such that for any $\Psi,\Psi'\in\mathcal{D}$
\begin{equation}
 \mathcal{S}(\R^N)\ni g \mapsto (\Psi|\int \rd^N\!x\, g(x)\, T(x)\Psi') \in\C
\end{equation}
is a Schwartz distribution. The space of operator-valued Schwartz distributions is denoted by $\mathcal{S}'(\R^N,L(\mathcal{D}))$. By the nuclear theorem $T(x)T'(x')\in\mathcal{S}'(\R^{N+N'},L(\mathcal{D}))$ if $T(x)\in\mathcal{S}'(\R^{N},L(\mathcal{D}))$ and $T'(x')\in\mathcal{S}'(\R^{N'},L(\mathcal{D}))$.

To every symbolic field $B\in\Fa$ we associate the {\it Wick polynomial} $\normord{B(x)}$ which is an operator-valued Schwartz distribution \cite{wightman1965fields,streater2000pct} on a suitable domain $\mathcal{D}_0$ in the Fock Hilbert space to be specified below. The map
\begin{equation}
 \Fa \ni B \mapsto \,\normord{B(x)}\, \in \mathcal{S}'(\R^4,L(\mathcal{D}_0))
\end{equation}
is linear but it does not preserve the structure of $\Fa$ as an algebra. 
For example, $\normord{A_i(x)}$ denotes one of the basic free fields and 
$\normord{\partial^\alpha\! A_i(x)}\,=\partial^\alpha\!\!\normord{A_i(x)}$ -- its 
derivative. In order to distinguish free field operators from corresponding 
symbols we always use the colons to denote the former. All Wick monomials at 
point $x\in\R^4$ are up to a multiplicative constant of the form 
\begin{equation}\label{eq:wick_monomial}
 \normord{A^{r}(x)} ~=~ \normord{ \prod_{i=1}^{\mathrm{p}} \prod_{\alpha\in\N_0^4} \,(\partial^\alpha\! A_{i}(x))^{r(i,\alpha)} }
\end{equation} 
for some super-quadri-index $r$. Observe that $\normord{B(x)}\,=0$ does not imply $B=0$. For example, we have $\normord{\square \varphi(x)}\,=\!\square\!\! \normord{\varphi(x)}\,=0$ if $\varphi$ is free field fulfilling the wave equation whereas the symbol $\square \varphi\in\Fa$ is by definition a non-zero generator.

The definition of all basic free fields $\normord{A_i(x)}$ and the Fock spaces on which they act is standard \cite{scharf2016gauge,weinberg1995quantum,derezinski2014quantum}. In the case of the vector field we use the Gupta-Bleuler approach \cite{gupta1950theory,bleuler1950neue} which is nicely summarized in the monographs \cite{scharf2016gauge,strocchi2013introduction}. Let us only mention that in this approach the vector field is defined on the Krein space with two inner products. The field is Hermitian only with respect to the covariant inner product which is not positive-definite. The positive-definite inner product is only used to define topology in the Fock space and as will become clear soon plays no role in the construction of the time-ordered products and the definition of the interacting models.

The Hilbert space $\mathcal{H}$ on which the model is defined is the tensor product of the Fock spaces on which the fields $\normord{A_1(x)},\ldots,\normord{A_\mathrm{p}(x)}$ corresponding to basic generators act. The vacuum state in the Fock space $\mathcal{H}$ is denoted by $\Omega$. We introduce the following dense subspace in $\mathcal{H}$
\begin{multline}\label{eq:dom}
 \mathcal{D}_0 :=\Span_\C\bigg\{ \int\rd^4 x_1\ldots\rd^4 x_n\,f(x_1,\ldots,x_n)\,\normord{A_{i_1}(x_1)\ldots A_{i_n}(x_n)}\Omega ~:
 \\
 ~n\in\N_0, ~i_1,\ldots,i_n\in\{1,\ldots,\mathrm{p}\},~f\in\mathcal{S}(\R^{4n}) \bigg\}.
\end{multline}
From now on, all the operators we consider are elements of $L(\mathcal{D}_0)$. As shown in \cite{wightman1965fields} Wick polynomials are are well-defined as operator-valued Schwartz distributions on~$\mathcal{D}_0$. The fact that $\mathcal{D}_0$ is embedded in the Hilbert space $\mathcal{H}$ is of no importance. The non-covariant positive-definite inner product was introduced only to make the definition of the space $\mathcal{D}_0$ easier (in fact, it is possible to define $\mathcal{D}_0$ without introducing the Hilbert space $\mathcal{H}$ at all). The space $\mathcal{D}_0$ is equipped with a Poincar{\'e} covariant and non-degenerate inner product, which is denoted by $(\cdot|\cdot)$. The product  is positive-definite unless the vector fields are present in the model. The Hermitian conjugation and the notion of unitarity are defined with respect to this product. To be more precise, for any operator $B\in L(\mathcal{D}_0)$, its Hermitian conjugation $B^* \in L(\mathcal{D}_0)$ is the unique operator fulfilling the identity $(\Psi|B\Psi')=(B^* \Psi|\Psi')$ for all $\Psi,\Psi'\in\mathcal{D}_0$. The unitary representation of the Poincar{\'e} group (or more correctly its universal covering group which is the inhomogeneous $SL(2,\C)$ group), denoted by $U(a,\Lambda)$, where $a\in\R^4$ and $\Lambda$ is a Lorentz transformation (element of the group $SL(2,\C)$), is defined on $\mathcal{D}_0$ in the standard way. In the case of pure translations we write $U(a)\equiv U(a,\id)$. We also consider the discrete symmetries: the charge conjugation, the spatial-inversion and the time-reversal which act on $\mathcal{D}_0$ as unitary or anti-unitary transformations (the details may be found e.g. in  \cite{weinberg1995quantum}).

\section{\texorpdfstring{$\Fa$}{F} products}\label{sec:F_prod}

In this section we introduce the notion of the $\Fa$ product. The time-ordered products and all other products which are used in the EG approach are examples of the $\Fa$ product. By definition the $\Fa$ product is any multi-linear map
\begin{equation}
 F:~\Fa^n \ni (B_1,\ldots,B_n) \mapsto \,F(B_1(x_1),\ldots ,B_n(x_n)) \,\in \mathcal{S}'(\R^{4n},L(\mathcal{D}_0))
\end{equation}
such that the following conditions hold for any $B_1,\ldots,B_n\in\Fh$:
\begin{enumerate}[label=\bf{A.\arabic*},leftmargin=*]
\item\label{axiom1} Translational covariance: 
\begin{equation}
 U(a)F(B_1(x_1),\ldots,B_n(x_n))U(a)^{-1} 
 \\
 = F(B_1(x_1+a),\ldots,B_n(x_n+a)).
\end{equation}
\item\label{axiom2} If $\mathbf{f}(B_1)+\ldots+\mathbf{f}(B_n)\neq0$ then
\begin{equation}
 (\Omega|F(B_1(x_1),\ldots,B_n(x_n)) \Omega) = 0.
\end{equation}
\item\label{axiom3} Wick expansion: The product $F(B_1(x_1),\ldots,B_n(x_n))$ is uniquely determined by the VEVs of the product $F$ of the sub-polynomials of $B_1,\ldots,B_n$:
\begin{multline}\label{eq:T_expansion}
 F(B_1(x_1),\ldots,B_n(x_n)) 
 =
 \\
 \sum_{s_1,\ldots,s_n} (-1)^{\mathbf{f}(s_1,\ldots,s_n)}\, 
 ~(\Omega|F(B_1^{(s_1)}(x_1),\ldots,B_n^{(s_n)}(x_n))\Omega)
 ~\frac{\normord{A^{s_1}(x_1)\ldots A^{s_n}(x_n)}}{s_1!\ldots s_n!}.
\end{multline}
\end{enumerate}

The definition of $\Fa$ product is modeled on the ordinary product of normally ordered operators
\begin{equation}
 \Fa^n \ni (B_1,\ldots,B_n) \mapsto \,\normord{B_1(x_1)}\ldots \normord{B_n(x_n)} \,\in \mathcal{S}'(\R^{4n},L(\mathcal{D}_0)),
\end{equation}
in which case the properties \ref{axiom1} and \ref{axiom2} are trivially satisfied and property \ref{axiom3} is the usual Wick expansion:
\begin{multline}
 \normord{B_1(x_1)}\ldots \normord{B_n(x_n)} 
 \,=
 \\
 \sum_{s_1,\ldots,s_n} (-1)^{\mathbf{f}(s_1,\ldots,s_n)}\, 
 ~(\Omega|\normord{B_1^{(s_1)}(x_1)}\ldots\normord{B_n^{(s_n)}(x_n)}\Omega)
 ~\frac{\normord{A^{s_1}(x_1)\ldots A^{s_n}(x_n)}}{s_1!\ldots s_n!}.
\end{multline}
The factor $(-1)^{\mathbf{f}(s_1,\ldots,s_n)}$ in \eqref{eq:T_expansion} may be read off from the above equation. In particular, if $B_1,\ldots,B_n$ have even fermion number then $(-1)^{\mathbf{f}(s_1,\ldots,s_n)}=1$. The condition \ref{axiom3} allows to express $F(B_1(x_1),\ldots,B_n(x_n)) $, which is a operator-valued distribution, in terms of numerical distributions. It also provides a relation between $F(B_1(x_1),\ldots,B_n(x_n))$ and the product $F$ of the sub-polynomials of $B_1,\ldots,B_n$. For example in the case of QED we have 
\begin{equation}
 F(j^\mu(x_1),\psi_a(x_2)) =~ \normord{j^\mu(x_1) \psi_a(x_2)} ~-~
  (\Omega|F(\,(\overline{\psi}(x_1)\gamma^\mu)_b\,,\,\psi_a(x_2)\,)\Omega) ~\normord{\psi_b(x_1)}
\end{equation}
and
\begin{multline}
 F(j^\mu(x_1),j^\nu(x_2)) = \normord{j^\mu(x_1)j^\nu(x_2)} ~+ ~
 \\
 (\Omega|F(\,(\overline{\psi}(x_1)\gamma^\mu)_a\,,\,(\gamma^\nu \psi(x_2))_b\,)\Omega) ~ \normord{\psi_a(x_1)\overline{\psi}_b(x_2)}~+~
 \\
 (\Omega|F(\,(\gamma^\nu \psi(x_2))_b\,,\,(\overline{\psi}(x_1)\gamma^\mu)_a\,)\Omega) ~ \normord{\overline{\psi}_a(x_1)\psi_b(x_2)} ~+~
  (\Omega|F(j^\mu(x_1),j^\nu(x_2))\Omega),
\end{multline}
where we also used the condition \ref{axiom2}.

Examples of the $\Fa$ products are: the above-mentioned product of normally ordered operators, the Wick product
\begin{equation}\label{eq:wick_product}
 \Fa^n \ni (B_1,\ldots,B_n) \mapsto \,\normord{B_1(x_1)\ldots B_n(x_n)} \,\in \mathcal{S}'(\R^{4n},L(\mathcal{D}_0))
\end{equation}
and the time-ordered product introduced in the next section (we will encounter many other examples of the $\Fa$ products in this thesis). Our notation
\begin{equation}\label{eq:notation_F}
 F(B_1(x_1),\ldots,B_n(x_n)),
\end{equation}
which is commonly used in physics in the case of the time-ordered product (often the commas are omitted), should be treated symbolically. From the mathematical point of view it would be more natural to write $F(B_1,\ldots,B_n)(x_1,\ldots,x_n)$ instead of \eqref{eq:notation_F}. The RHS of Equation \eqref{eq:T_expansion} is a well-defined operator-valued Schwartz distribution as a result of the following theorem.
\begin{thm}\label{thm:eg0}\emph{\cite{epstein1973role}} 
Let $t\in\mathcal{S}'(\R^{4n})$ be translationally invariant, i.e.
\begin{equation}
 t(x_1,\ldots,x_n) = t(x_1+a,\ldots,x_n+a)
\end{equation}
for all $a\in\R^4$. Then for any $B_1,\ldots,B_n\in\Fa$
\begin{equation}
 t(x_1,\ldots,x_n) \normord{B_1(x_1)\ldots B_n(x_n)}~\in \mathcal{S}'(\R^{4n},L(\mathcal{D}_0)).
\end{equation}
\end{thm}
We shall frequently use the following properties of the translationally-invariant distributions.
\begin{rem}\label{rem:transl}
For any translationally-invariant distribution $t\in\mathcal{S}'(\R^{4(n+1)})$, $n\in\N_0$ we can define the associated distribution $\underline{t}\in\mathcal{S}'(\R^{4n})$ by 
\begin{equation}\label{eq:dist_trans_inv}
 \underline{t}(x_1,\ldots,x_n):=t(x_1,\ldots,x_n,0):=\int \rd^4 y \, t(x_1+y,\ldots,x_n+y,y)\, h(y),
\end{equation}
where $h\in\mathcal{S}(\R^4)$, $\int\rd^4 y \, h(y)=1$. The RHS of the above equation is independent of $h$. The Fourier transforms of the distributions $t$ and $\underline{t}$ are related by
\begin{equation}\label{eq:dist_trans_inv_F}
 \F{t}(q_1,\ldots,q_{n+1}) = (2\pi)^4 \delta(q_1+\ldots+q_{n+1})\,\F{\underline{t}}(q_1,\ldots,q_n).
\end{equation}
%Note also that $\F{\underline{t}}(q_1,\ldots,q_n)$ is the Fourier transform of the distribution $t(x_1,\ldots,x_n,0)$.
\end{rem}

Finally, we define the product and the graded commutator of two $\Fa$ products:
\begin{equation}
\begin{aligned}\label{eq:F_Fprime}
 F:~\Fa^n \ni (B_1,\ldots,B_n) &\mapsto \,F(B_1(x_1),\ldots ,B_n(x_n)) \,\in \mathcal{S}'(\R^{4n},L(\mathcal{D}_0)),
  \\
 F':~\Fa^{n'} \ni (B'_1,\ldots,B'_{n'}) &\mapsto \,F(B'_1(x'_1),\ldots ,B'_{n'}(x'_{n'})) \,\in \mathcal{S}'(\R^{4n'},L(\mathcal{D}_0)).
\end{aligned}
\end{equation}
The product of $\Fa$ products $F$ and $F'$ is by definition the following multi-linear map
\begin{multline}\label{eq:F_product}
 \Fa^{n+n'}\ni (B_1,\ldots,B_n;B'_1,\ldots,B'_{n'})\mapsto
 \\
 F(B_1(x_1),\ldots,B_n(x_n)) F'(B'_1(x'_1),\ldots,B'_{n'}(x'_{n'})) \in \mathcal{S}'(\R^{4(n+n')},L(\mathcal{D}_0)),
\end{multline}
which, as it turns out, is again an $\Fa$ product. The only non-trivial part in the proof of this fact is the verification that the product of $F$ and $F'$ fulfills the condition \eqref{eq:T_expansion}, which follows from the properties of the Wick products (cf. Section 4 in \cite{epstein1973role}). The graded commutator of $\Fa$ products $F$ and $F'$ is the $\Fa$ product defined by
\begin{multline}\label{eq:com}
 \left[ F(B_1(x_1),\ldots,B_n(x_n)), F'(B'_1(x'_1),\ldots,B'_n(x'_{n'})) \right] :=
 \\[6pt]
 F(B_1(x_1),\ldots,B_n(x_n))~F'(B'_1(x'_1),\ldots,B'_n(x'_{n'}))
 \\[6pt]
 - (-1)^{\mathbf{f}(B_1\ldots B_n)\mathbf{f}(B'_1\ldots B'_n)}~\,
 F'(B'_1(x'_1),\ldots,B'_n(x'_{n'}))~ F(B_1(x_1),\ldots,B_n(x_n)),
\end{multline}
where $B_1\ldots,B_n,B'_1,\ldots,B'_{n'}\in\Fh$. For example in the case of QED we have
\begin{equation}
 [\,\normord{\overline{\psi}_a(x_1)}\,,\,\normord{\psi_b(x_2)}\,] ~=~ \normord{\overline{\psi}_a(x_1)}\,\normord{\psi_b(x_2)} ~+~
 \normord{\psi_b(x_2)}\,\normord{\overline{\psi}_a(x_1)}.
\end{equation}

\section{Time-ordered product}\label{sec:axioms}

The scattering matrix and the interacting fields in the perturbative QFT are given in terms of the time-ordered products. They form a~family of $\Fa$ products
\begin{equation}\label{eq:time_ordered_product}
 \Fa^n \ni (B_1,\ldots,B_n) \mapsto \,\T(B_1(x_1),\ldots,B_n(x_n)) \,\in \mathcal{S}'(\R^{4n},L(\mathcal{D}_0))
\end{equation}
indexed by $n\in\N_0$ which satisfies the following axioms (besides the conditions \ref{axiom1}-\ref{axiom3} stated in the previous section which we also call the axioms):
\begin{enumerate}[label=\bf{A.\arabic*},leftmargin=*]
\setcounter{enumi}{3}
\item\label{axiom4} $\T(\emptyset)=\id$, $\T(B(x))=\,\,\normord{B(x)}$,
\begin{equation}
\T(B_1(x_1),\ldots,B_n(x_n),1(x_{n+1})) = \T(B_1(x_1),\ldots,B_n(x_n)),
\end{equation}
where $1$ on the LHS of the above equality is the unity in $\Fa$.
\item\label{axiom5} Graded symmetry: For any $B_1,\ldots,B_n\in\Fh$ it holds
\begin{equation}\label{eq:T_graded}
 \T(B_1(x_1),\ldots,B_n(x_n)) = (-1)^{\mathbf{f}(\pi)}\T(B_{\pi(1)}(x_{\pi(1)}),\ldots,B_{\pi(n)}(x_{\pi(n)})),
\end{equation}
where $\pi$ is any permutation of the set $\{1,\ldots,n\}$ and $\mathbf{f}(\pi)\in\Z/2\Z$ is the number of transpositions in $\pi$ that involve a pair of fields with odd fermion number. In particular, the time-ordered product of polynomials $B_1,\ldots,B_n$ which have even fermion number is invariant under permutations of its arguments.
\item\label{axiom6} Causality: If none of the points $x_1,\ldots,x_m$ is in the causal past of any of the points $x_{m+1},\ldots,x_n$ then
\begin{multline}\label{eq:T_causality}
 \T(B_1(x_1),\ldots,B_n(x_n)) 
 \\
 = \T(B_1(x_1),\ldots,B_m(x_m))\T(B_{m+1}(x_{m+1}),\ldots,B_n(x_n)).
\end{multline}
% on the subspace of $\mathcal{S}(\R^{4n})$ consisting of functions $f$ such that
% \begin{equation}
%  \supp f \subset \left\{ (x_1,\ldots,x_n)\,:\, \forall_{\substack{j=1,\ldots,m\\k=m+1,\ldots,n}} x_k \notin x_j + \overline{V}^+ \right\},
% \end{equation}
% where $\overline{V}^+$ is the closed future light cone.
\end{enumerate}

Causality is the most important property of the time-ordered products. In combination with graded-symmetry it implies, in particular, the identity
\begin{multline}\label{eq:T_noncoinciding}
 \T(B_1(x_1),\ldots,B_n(x_n))
 = \sum_{\pi\in\mathcal{P}_n} (-1)^{\mathbf{f}(\pi)}
 \theta(x_{\pi(1)}^0-x_{\pi(2)}^0)\ldots\theta(x_{\pi(n-1)}^0-x_{\pi(n)}^0) 
 \\
 \normord{B_{\pi(1)}(x_{\pi(1)})} \ldots \normord{B_{\pi(n)}(x_{\pi(n)})}
\end{multline}
for $n\geq 2$ and non-coinciding points $x_1,\ldots,x_n$ which justifies the name the \emph{time-ordered product}. By $\mathcal{P}_n$ we mean the set of permutations of $\{1,\ldots,n\}$; $\mathbf{f}(\pi)\in\Z/2\Z$ is the number of transpositions in $\pi$ that involve a pair of fields with odd fermion number. The construction of time-ordered products is thus reduced to finding an appropriate extension of the distribution on the RHS of \eqref{eq:T_noncoinciding}. The existence of an extension satisfying all axioms \ref{axiom1}-\ref{axiom6} will be shown in Section~\ref{sec:T_const}. The proof uses the advanced and retarded products defined in the next section. Taking into account the support properties of these distributions one proves that outside the main diagonal $x_1=\ldots=x_n$ the time-ordered product $\T(B_1(x),\ldots,B_n(x_n))$ is determined uniquely by the time-ordered products which have strictly less then $n$ arguments. As we will see the extension to the main diagonal is always possible but it is not unique.

The axioms \ref{axiom1}-\ref{axiom6} should be viewed as the minimal requirements which have to be imposed to make possible  this construction possible. We are then left with the problem of an appropriate selection from the class of extensions. The freedom in the definition of the time-ordered products will be restricted by imposing the so-called normalization conditions. In Section~\ref{sec:sd} we introduce the condition \ref{norm:sd} which bounds from above the Steinmann scaling degree of the VEVs of the time-ordered products. Other normalization conditions like e.g. the Lorentz covariance or unitarity are stated in Sections \ref{sec:norm_con} and \ref{sec:one}. In these sections we also prove their compatibility with the normalization condition \ref{norm:wAL} formulated in Section~\ref{sec:proof_scalar}, which is needed for the existence of the Wightman and Green functions. Note that only the axioms \ref{axiom1}-\ref{axiom6} and normalization conditions \ref{norm:sd} and \ref{norm:wAL} will be used in the proof of the existence of the weak adiabatic limit.

We stress that the time-ordered products, like any other $\Fa$ product, are operator-valued Schwartz distributions on $\mathcal{D}_0$ indexed by the symbolic fields $B\in\Fa$, and not the Wick polynomials $\normord{B(x)}\in\mathcal{S}'(\R^{4},L(\mathcal{D}_0))$. This is the feature of the so-called off-shell formalism \cite{stora2002pedagogical,dutsch2004causal,brouder2008relating}. Observe that, for example, the time-ordered product $\T(\square\varphi(x_1),\varphi(x_2))$ does not have to vanish even if $\varphi$ is a free field satisfying the wave equation. Its value depends on the chosen normalization conditions. For example, if the action Ward identities \cite{dutsch2004causal} is imposed it holds $\T(\square\varphi(x_1),\varphi(x_2))=(-\ri)\delta(x_1-x_2)$. Because this issue is not relevant for the proof of the existence of the Wightman and Green functions, it will not be further discussed in the thesis.

\section{Generating functional and \texorpdfstring{$\aT$, $\Adv$, $\Ret$, $\Dif$}{aT, A, R, D} products}\label{sec:generating}

In the EG construction of the time-ordered products, which will be presented in Section~\ref{sec:T_const}, one uses other $\Fa$ products besides the time-ordered product. They are most easily defined with the use of the generating functional of the time-ordered products, which is also known as the time-ordered exponential. If a model under consideration contains fermionic fields, then one is compelled to use the Grassmann-valued Schwartz functions. The set of such functions, which is denoted by $\mathcal{S}(\R^N,E)$, coincides with the algebraic tensor product of $\mathcal{S}(\R^N)$ and the Grassmann algebra $E$. The precise definitions of $E$ and $\mathcal{S}(\R^N,E)$ and the properties of these objects are given in Appendix~\ref{app:grassmann}.

Let $B_1,\ldots,B_m\in\Fa$ and $g_1,\ldots,g_m\in\mathcal{S}(\R^4,E)$. The time ordered exponential is a map that sends the elements of $\mathcal{S}(\R^4,E)\otimes \Fa$ into formal power series of elements of $L(\mathcal{D}_0)\otimes E$. By definition
\begin{multline}\label{eq:def_Texp}
 \sum_{j=1}^m g_j \otimes B_j ~~\mapsto~~ \Texp\left( \ri \int \rd^4 x \, \sum_{j=1}^m g_j(x) B_j(x)\right) := 
 \\
 \sum_{n=0}^\infty \frac{\ri^n}{n!} \sum_{j_1,\ldots,j_n=1}^m~ \int\rd^4 x_1\ldots\rd^4 x_n
 \,g_{j_1}(x_1)\ldots g_{j_n}(x_n)
 \T(B_{j_1}(x_1),\ldots,B_{j_n}(x_n)),
\end{multline}
where $g_1,\ldots,g_m\in\mathcal{S}(\R^4,E)$ and $B_1,\ldots,B_m\in\Fa$. Note that any element of $\mathcal{S}(\R^4,E)\otimes \Fa$ is of the form $\sum_{j=1}^m g_j \otimes B_j$. The argument of $\Texp$ should be treated symbolically. The RHS of the above equation is a formal power series in $g_1,\ldots,g_m$. 

In order to take advantage of the Grassmann-valued test functions we have to restrict attention to certain subspace of $\mathcal{S}(\R^4,E)\otimes \Fa$. An element of $\mathcal{S}(\R^4,E)\otimes \Fa$ is called admissible iff it is of the form $\sum_{j=1}^m g_j \otimes B_j$ for some $B_1,\ldots,B_m\in\Fh$ and $g_1,\ldots,g_m\in\mathcal{S}(\R^4,E)$ such that $g_j$ has even/odd Grassmann parity if $\mathbf{f}(B_j)$ is even/odd number. Note that we consider only homogenous symbolic fields $B\in\Fh$ which by their definition have definite fermion number $\mathbf{f}(B)\in\Z$. Actually, it is also possible to make a stronger assumption and suppose that the functions $g_j$ corresponding to $B_j$ with even fermion number are complex-valued (complex numbers can be viewed as the Grassmann numbers proportional to the identity in $E$). In what follows we consider only admissible elements of $\mathcal{S}(\R^4,E)\otimes \Fa$. For any admissible $\sum_{j=1}^m g_j \otimes B_j\in\mathcal{S}(\R^4,E)\otimes \Fa$ it holds
\begin{equation}\label{eq:commutativity_g}
 g_j(x)g_{j'}(y) = (-1)^{\mathbf{f}(B_j)\mathbf{f}(B_{j'})}  g_{j'}(y)g_j(x).
\end{equation}
If there are no fermion fields in the model under consideration then the there is no need to use the Grassmann-valued test functions. In this case we set $E=\C$ everywhere and identify $\mathcal{S}(\R^4,E)$ with $\mathcal{S}(\R^4)$.

Now let $\mathpzc{g}=\sum_{j=1}^m g_j \otimes B_j$ be an admissible element of $\mathcal{S}(\R^4,E)\otimes \Fa$. The generating functional of the time-ordered products is given by 
\begin{equation}
 S(\mathpzc{g}):= \Texp\left( \ri \int \rd^4 x \, \sum_{j=1}^m g_j(x) B_j(x)\right).
\end{equation}
Note that because of the identity \eqref{eq:commutativity_g} the integrand in the definition \eqref{eq:def_Texp} is invariant under permutations of its arguments $x_1,\ldots,x_n$ in the case of the above time-ordered exponential. As a result, the functional $S(\mathpzc{g})=S(\sum_{j=1}^m g_j \otimes B_j)$ generates the time-ordered products in the following sense
\begin{equation}\label{eq:time_ordered_derivative}
 \T(B_1(x_1),\ldots,B_m(x_m)) 
 = (-\ri)^m \frac{\delta}{\delta g_m(x_m)}\ldots\frac{\delta}{\delta g_1(x_1)} S(\mathpzc{g})\bigg|_{\mathpzc{g}=0}, 
\end{equation}
where we used the identities
\begin{equation}
\begin{aligned}
 \frac{\delta}{\delta g_j(x)} g_{j'}(y) &= \delta_{jj'} \delta(x-y),
 \\
 \frac{\delta}{\delta g_j(x)}\frac{\delta}{\delta g_{j'}(y)}&=(-1)^{\mathbf{f}(B_j)\mathbf{f}(B_{j'})}\frac{\delta}{\delta g_{j'}(y)}\frac{\delta}{\delta g_{j}(x)}.
\end{aligned}
\end{equation}

Now we are going to introduce the abbreviated notation which will be used throughout the thesis. We set
\begin{equation}\label{eq:I}
 I\equiv \left( B_1(x_1),\ldots,B_n(x_n)\right),
\end{equation}
\begin{equation}\label{eq:fun_der}
 \frac{\delta}{\delta \mathpzc{g}(I)} \equiv  \frac{\delta}{\delta g_n(x_n)}\ldots\frac{\delta}{\delta g_1(x_1)}
\end{equation}
and
\begin{equation}
F(I)\equiv F(B_1(x_1),\ldots,B_n(x_n)),
\end{equation}
where $B_1,\ldots,B_n\in\Fh$, $\mathpzc{g}=\sum_{j=1}^n g_j \otimes B_j$ is an admissible element of $\mathcal{S}(\R^4,E)\otimes \Fa$ and $F$ is some $\Fa$ product. We stress that $I$ is a sequence. We say that $I_1,\ldots,I_k$ is a partition of $I$ if $I_j$ are subsequences of $I$ and $\bigcup_{j=1}^k \hat{I}_j=\hat{I}$ and $\hat{I}_j\cap \hat{I}_k= \emptyset$ for all $j\neq k$, where $\hat{I}$ is the set of elements of $I$ (note that we consider only lists $I$ of distinct elements). The empty sequence is denoted by $\emptyset$. The number of elements of the sequence $I$ is denoted by $|I|$. The sequence which is obtained after the concatenation of the sequences $I_1$ and $I_2$ is denoted by $I_1+I_2$ (this operation is not Abelian). Note that for any partition $I_1,\ldots,I_k$ of the sequence~$I$ there exists always a unique permutation $\pi\in\mathcal{P}_n$ such that $I_1+\ldots+I_k=\pi(I)$, where $\pi(I)$ is the sequence permuted by $\pi$. In what follows we will frequently encounter sums of the form 
\begin{equation}\label{eq:sum}
 \sum^I_{\substack{I_1,\ldots,I_k\\I_1+\ldots+I_k=\pi(I)}} (-1)^{\mathbf{f}(\pi)}\, F_1(I_1)\ldots F_k(I_k),
\end{equation}
where $F_1,\ldots,F_k$ are some $\Fa$ products. The sum is carried out over all partitions $I_1,\ldots,I_k$ of the sequence~$I$. Given such partition, the permutation $\pi$ is determined uniquely by the the condition stated in the second line below the summation sign. Consequently, the factor $(-1)^{\mathbf{f}(\pi)}$ is well-defined for each term of the sum. We recall that $\mathbf{f}(\pi)\in\Z/2\Z$ is the number of transpositions in $\pi$ that involve a pair of arguments of $I$ with odd fermion number.

Let us note as an aside that if there are no fermions in the model under consideration then the notation $F(I)=F(\hat{I})$, where $\hat{I}$ is the set of elements of $I$, is meaningful for any product $F$ which is symmetric under permutations of its arguments (the time-ordered product is an example of such product). In this case the sum \eqref{eq:sum} is equivalent to the sum over disjoint subsets $\hat{I}_1,\ldots,\hat{I}_n$ of the set~$I$ of the form
\begin{equation}
 \sum^{\hat{I}}_{\hat{I}_1,\ldots,\hat{I}_k} F_1(\hat{I}_1)\ldots F_k(\hat{I}_k).
\end{equation}

Let us turn to the definition of the anti-time-ordered products. First, note that because $S(0)=\id$ the formal power series $S(\mathpzc{g})$ is invertible. The anti-time-ordered products are defined by
\begin{equation}
 \aT(I) 
 := (-\ri)^n \frac{\delta}{\delta \mathpzc{g}(I)} S(-\mathpzc{g})^{-1}\bigg|_{\mathpzc{g}=0},
\end{equation}
where $I$ and $ \frac{\delta}{\delta \mathpzc{g}(I)}$ were introduced in \eqref{eq:I} and \eqref{eq:fun_der}. By definition of the generating functional of the time-ordered products we have
\begin{equation}
 S(-\mathpzc{g})^{-1} = \left[\Texp\left( -\ri \int \rd^4 x \, \sum_{j=1}^n g_j(x) B_j(x)\right)\right]^{-1}.
\end{equation}
Since $S(0)=\id$ we can use the formal power series expansion of the function $K\mapsto(1-K)^{-1}=\sum_{n=0}^\infty K^n$ to obtain the formula which relates explicitly the anti-time-ordered products with the time-ordered products
\begin{equation}\label{eq:aT_standard}
 \aT(I) = \sum_{k=1}^{|I|} \sum^I_{\substack{I_1,\ldots,I_k\\I_1+\ldots+I_k=\pi(I)\\I_1,\ldots,I_k\neq \emptyset}} (-1)^{|I|+k+\mathbf{f}(\pi)}\T(I_1)\ldots \T(I_k).
\end{equation}
The anti-time-ordered products fulfill the condition \eqref{eq:T_noncoinciding} with the reverse ordering of operators in the second line.

A non-trivial but very important consequence of the axiom of causality \ref{axiom6} is the so-called {\it causal factorization property} 
\begin{equation}\label{eq:causal_fact}
 S(\mathpzc{g}'+\mathpzc{g}+\mathpzc{g}'') = S(\mathpzc{g}'+\mathpzc{g})S(\mathpzc{g})^{-1}S(\mathpzc{g}+\mathpzc{g}'')
\end{equation}
which holds whenever $\supp\,\mathpzc{g}' + \overline{V}^+  \cap \supp\,\mathpzc{g}''=\emptyset$, i.e. the support of $\mathpzc{g}''$ does not intersect the causal future of the support of $\mathpzc{g}'$. The support of $\mathpzc{g}$ may be arbitrary. The causal factorization property rewritten in the following way
\begin{equation}
  S(\mathpzc{g}'+\mathpzc{g}+\mathpzc{g}'')S(\mathpzc{g}+\mathpzc{g}'')^{-1} = S(\mathpzc{g}'+\mathpzc{g}) S(\mathpzc{g})^{-1}
\end{equation}
implies immediately that $S(\mathpzc{g}'+\mathpzc{g}) S(\mathpzc{g})^{-1}$ depends only on the behavior of $\mathpzc{g}$ in the causal future of $\supp\,\mathpzc{g}'$. Similarly, one can show that $S(\mathpzc{g})^{-1} S(\mathpzc{g}'+\mathpzc{g})$ depends only on the behavior of $\mathpzc{g}$ in the causal past of $\supp\,\mathpzc{g}'$.

In order to define the advanced and retarded products we consider two admissible elements of $\mathcal{S}(\R^4,E)\otimes\Fa$ of the form $\mathpzc{g}=\sum_{j=1}^n g_j \otimes B_j$ and $\mathpzc{h}=\sum_{j=1}^m h_j\otimes C_j$ and set $I=\left( B_1(y_1),\ldots,B_n(y_n)\right)$, $J=\left( C_1(x_1),\ldots,C_m(x_m)\right)$. The advanced product is given by
\begin{equation}\label{eq:def_adv}
 \Adv(I;J):=
 \\
 (-\ri)^{n+m} 
 \frac{\delta}{\delta \mathpzc{h}(J)}
 \frac{\delta}{\delta \mathpzc{g}(I)} 
 S(\mathpzc{g}+\mathpzc{h}) S(\mathpzc{g})^{-1} \bigg|_{\substack{\mathpzc{g}=0\\\mathpzc{h}=0}}.
\end{equation}
Similarly, the retarded product is defined in the following way
\begin{equation}\label{eq:def_ret}
 \Ret(I;J):=
 \\
 (-\ri)^{n+m} 
 \frac{\delta}{\delta \mathpzc{h}(J)}
 \frac{\delta}{\delta \mathpzc{g}(I)} 
 S(\mathpzc{g})^{-1} S(\mathpzc{g}+\mathpzc{h}) \bigg|_{\substack{\mathpzc{g}=0\\\mathpzc{h}=0}}.
\end{equation}
Moreover, we will also use the following products
\begin{align}\label{eq:def_dif}
 \Dif(I;J) &:=
 \Adv(I;J)
 -\Ret(I;J),
 \\
 \label{eq:def_adv_prime}
 \Adv'(I;J)
 &:=
 \Adv(I;J)
 -\T(I;J),
 \\
 \Ret'(I;J)
 &:=
 \Ret(I;J)
 -\T(I;J),
\end{align}
which do not have standard names. We will call them the $\Dif$, $\Adv'$ and $\Ret'$ products. The advanced and retarded products are given explicitly by
\begin{equation}\label{eq:adv_standard}
 \Adv(I;J) = \sum^I_{\substack{I_1,I_2\\I_1+J+I_2=\pi(I+J)}}(-1)^{|I_2|+\mathbf{f}(\pi)}\, \T(I_1,J)\aT(I_2)
\end{equation}
and
\begin{equation}\label{eq:ret_standard}
 \Ret(I;J) = \sum^I_{\substack{I_1,I_2\\I_2+I_1+J=\pi(I+J)}}(-1)^{|I_2|+\mathbf{f}(\pi)} \, \aT(I_2)T(I_1,J),
\end{equation}
respectively. Note that the condition $I_2+I_1+J=\pi(I+J)$ in the sum on the RHS of Equation \eqref{eq:ret_standard} can be replaced by the equivalent condition $I_2+I_1=\pi(I)$. Similar formulas hold also for $\Adv'$ and $\Ret'$ products but in these cases the sum is further restricted by the condition that $I_1\neq I$ or equivalently $I_2$ is not empty. Thus, $\Adv'$ and $\Ret'$ with $n$ arguments are expressed by the time-ordered products with at most $n-1$ arguments. The same is also true for the casual product $\Dif$. Moreover, we have
\begin{equation}\label{eq:dif_standard_com}
 \Dif(I;J) = \sum^I_{\substack{I_1,I_2\\I_1+J+I_2=\pi(I+J)}}(-1)^{|I_2|+\mathbf{f}(\pi)} \, [T(I_1,J),\aT(I_2)],
\end{equation}
where $[\cdot,\cdot]$ is the graded commutator \eqref{eq:com}. 

Let us list some of the properties of the products $\aT(I)$, $\Adv(I;J)$, $\Ret(I;J)$, $\Adv'(I;J)$ and $\Ret'(I;J)$ which follow from the fact that the time-ordered products satisfy the axioms \ref{axiom1}-\ref{axiom6}. It is evident that all these products are $\Fa$ products. The axioms \ref{axiom4}, \ref{axiom5} and an analog of the axiom \ref{axiom6} with the reverse order of factors on the RHS of \eqref{eq:T_causality} hold for the anti-time-ordered products. The products $\Adv(I;J)$, $\Ret(I;J)$, $\Adv'(I;J)$, $\Ret'(I;J)$ and $\Dif(I;J)$ are graded-symmetric separately in arguments collectively denoted by $I$ and $J$. Moreover, using the causal factorization property \eqref{eq:causal_fact} one proves that the advanced, retarded and causal products have the following support properties
\begin{equation}\label{eq:supp_adv_ret_dif}
\begin{aligned}
 \supp \Adv(I;J) &\subset \Gamma^+_{n,m},
 \\
 \supp \Ret(I;J) &\subset \Gamma^-_{n,m},
 \\
 \supp \Dif(I;J) &\subset \Gamma^+_{n,m}\cup \Gamma^-_{n,m},
\end{aligned}
\end{equation}
where
\begin{equation}\label{eq:def_gen_cones}
 \Gamma^\pm_{n,m}:=\{(y_1,\ldots,y_n;x_1,\ldots,x_m) ~:~\exists_{u} \forall_j ~y_j \in x_{u(j)} \pm \overline{V}^+ \}
\end{equation}
and $u:\,\{1,\ldots,n\}\to\{1,\ldots,m\}$. The condition in the definition of the sets $\Gamma^\pm_{n,m}$ means that each of the points $y_1,\ldots,y_n$ is in the causal future/past of some of the points $x_1,\ldots,x_m$. In particular, in the case when $J$ contains only one element, using translational invariance of the VEVs of the $\Fa$ products we obtain:
\begin{equation}\label{eq:supp_VEVs}
\begin{aligned}
 &\supp\,(\Omega|\Adv(B_1(x_1),\ldots,B_n(x_n);B_{n+1}(0))\Omega) \subset \Gamma_{n}^+,
 \\
 &\supp\,(\Omega|\Ret(B_1(x_1),\ldots,B_n(x_n);B_{n+1}(0))\Omega) \subset \Gamma_{n}^-,
 \\
 &\supp\,(\Omega|\Dif(B_1(x_1),\ldots,B_n(x_n);B_{n+1}(0))\Omega) \subset \Gamma_{n}^+ \cup \Gamma_{n}^-,
\end{aligned} 
\end{equation}
where
\begin{equation}\label{eq:def_cones}
 \Gamma_{n}^+ = -\Gamma_{n}^- := \{ (x_1,\ldots, x_{n}) \,:\, \forall_j \,x_j \in \overline{V}^+ \}.
\end{equation}
This implies that the equations
\begin{multline}\label{eq:adv_split}
 (\Omega|\Adv(B_1(x_1),\ldots,B_n(x_n);B_{n+1}(0))\Omega) 
 \\
 = 
 \theta(x_1^0)\ldots\theta(x_n^0)~ (\Omega|\Dif(B_1(x_1),\ldots,B_n(x_n);B_{n+1}(0))\Omega)
\end{multline}
and
\begin{multline}\label{eq:ret_split}
 (\Omega|\Ret(B_1(x_1),\ldots,B_n(x_n);B_{n+1}(0))\Omega) 
 \\
 = 
 -\theta(-x_1^0)\ldots\theta(-x_n^0)~ (\Omega|\Dif(B_1(x_1),\ldots,B_n(x_n);B_{n+1}(0))\Omega)
\end{multline}
hold outside the origin. Using the above equalities it is possible to define the distributions on the LHS for tests functions $g\in\mathcal{S}(\R^{4n})$ such that $0\notin\supp\,g$, given only the time-ordered products with at most $n$ arguments. In the next section we study the existence and uniqueness of the extension of the distributions of this type to the space of all Schwartz functions. Let us mention that the process of determining the Schwartz distributions on the LHS of \eqref{eq:adv_split}, \eqref{eq:ret_split} given their difference is known in the literature as the splitting procedure. 

\section{Extension of distribution}\label{sec:ext_dist}

Let $\mathcal{S}(\R^N\setminus\{0\})$ be the subset of $\mathcal{S}(\R^N)$ consisting of all Schwartz test
functions $g$ such that $0\notin\supp\,g$. We equip $\mathcal{S}(\R^N\setminus\{0\})$ with the topology
induced from $\mathcal{S}(\R^N)$. By definition $\mathcal{S}'(\R^N\setminus\{0\})$ is the dual
of $\mathcal{S}(\R^N\setminus\{0\})$. If $t\in\mathcal{S}'(\R^N)$ then $t|_{\mathcal{S}(\R^N\setminus\{0\})}\in\mathcal{S}'(\R^N\setminus\{0\})$. Let $t^0\in\mathcal{S}'(\R^N\setminus\{0\})$. Any distribution $t\in\mathcal{S}'(\R^N)$ such that $t|_{\mathcal{S}(\R^N\setminus\{0\})}=t^0$ is called an extension of $t^0$. 

\begin{dfn}\label{def:sd}
\emph{(Steinmann scaling degree \cite{steinmann1971perturbative}\footnote{The original definition by Steinmann was given only for $t\in\mathcal{S}'(\R^N)$ and differs by sign from the version usually found in the literature \cite{brunetti2000microlocal} which we quote.})}
Let $t\in\mathcal{S}'(\R^N\setminus\{0\})$. The scaling degree of $t$ with respect to the origin is given by
\begin{align}
 \label{eq:scaling_set}
 &\sd(t)
 :=
 \inf\left\{s\in \R~:~\forall_{g\in\mathcal{S}(\R^N\setminus\{0\})}
 \lim_{\lambda\searrow 0}\int\rd^N\!x\,\lambda^s t(\lambda x) g(x) =0 \right\} ,
 \\
 &\sd(t)\in \{-\infty\}\cup\R\cup\{+\infty\},
\end{align}
where by definition $\inf\emptyset=+\infty$. The scaling degree of $t\in\mathcal{S}'(\R^N)$ is defined analogously with the only difference that $g\in\mathcal{S}(\R^N)$ in \eqref{eq:scaling_set}. 
\end{dfn}
Let us list some of the properties of the Steinmann scaling degree:
\begin{enumerate}[leftmargin=*,label={(\arabic*)}]
\item if $t\in\mathcal{S}'(\R^N)$ then $\sd(t)<\infty$,
\item if $t\in\mathcal{S}'(\R^N)$ and $0\notin \supp\,t$, then $\sd(t)=-\infty$,
\item $\sd(\partial^\alpha t)=\sd(t)+|\alpha|$,
\item if $t\in\mathcal{S}'(\R^N)$ is homogeneous of degree $d$ we have $\sd(t)=-d$,
\item if $t\in\mathcal{S}'(\R^N)$ and $t'\in\mathcal{S}(\R^M)$ then $\sd(t\otimes t')\leq \sd(t)+\sd(t')$,
\item if $t\in\mathcal{S}'(\R^{4(n+1)})$ is translationally invariant then it holds $\sd(t)=\sd(\underline{t})$ (for definition of $\underline{t}$ see Remark \ref{rem:transl} in Section \sectionref{rem:transl}).
\end{enumerate}

\begin{thm}\label{thm:extension}
\emph{\cite{brunetti2000microlocal}}
Let $t^0\in\mathcal{S}'(\R^N\setminus\{0\})$ such that $\sd(t^0) <\infty$. There exists an extension $t\in\mathcal{S}'(\R^N)$ of $t^0$ such that $\sd(t)=\sd(t^0)$. 
\end{thm}

\section{Construction of the time-ordered product}\label{sec:T_const}

Now we are ready for the inductive construction of the time-ordered products. First, let us note that it is enough to define the time-ordered products for the monomials $A^r$ as they form a basis of $\Fa$. The starting point of the construction is the identity $\T(A^r(x))=~\normord{A^r(x)}$ which is part of the axiom \ref{axiom4}. Next, let $n\in\N_+$ and assume that the time-ordered products with at most $n$ arguments are defined such that they fulfill the axioms \ref{axiom1}-\ref{axiom6}. We are going to construct the time-ordered product with $n+1$ arguments
\begin{equation}\label{eq:time_ordered}
 \T(A^{r_1}(x_1),\ldots,A^{r_{n+1}}(x_{n+1})).
\end{equation}
By the use of axiom \ref{axiom3} this task is equivalent to the construction of the numerical distribution 
\begin{equation}\label{eq:vev_time_ordered}
 (\Omega|\T(A^{r_1}(x_1),\ldots,A^{r_{n+1}}(x_{n+1}))\Omega)
\end{equation}
in such a way that the time ordered products \eqref{eq:time_ordered} expressed in terms of \eqref{eq:vev_time_ordered} by the identity \eqref{eq:T_expansion} satisfy remaining axioms \ref{axiom1}, \ref{axiom2}, \ref{axiom4}, \ref{axiom5} and \ref{axiom6}.

As has been shown in Section~\ref{sec:generating} the products 
\begin{align}
 \Dif(A^{r_1}(x_1),\ldots,A^{r_n}(x_n);A^{r_{n+1}}(x_{n+1})), \label{eq:const_dif1}
 \\
 \Adv'(A^{r_1}(x_1),\ldots,A^{r_n}(x_n);A^{r_{n+1}}(x_{n+1})) \label{eq:const_adv_prime}
\end{align}
are expressed in terms of the time-ordered products with at most $n$ arguments and, consequently, are known. Thus, in particular the following distribution
\begin{equation}\label{eq:const_adv_split}
 \theta(x_1^0)\ldots\theta(x_n^0)~ (\Omega|\Dif(A^{r_1}(x_1),\ldots,A^{r_n}(x_n);A^{r_{n+1}}(0))\Omega) \in\mathcal{S}'(\R^{4n}\setminus\{0\})
\end{equation}
is given. By the support properties of the VEVs of $\Adv$ and $\Dif$ products \eqref{eq:supp_adv_ret_dif} the (yet unknown) distribution
\begin{equation}\label{eq:const_adv}
 (\Omega|\Adv(A^{r_1}(x_1),\ldots,A^{r_n}(x_n);A^{r_{n+1}}(0))\Omega) \in\mathcal{S}'(\R^{4n})
\end{equation}
coincides on $\mathcal{S}(\R^{4n}\setminus\{0\})$ with \eqref{eq:const_adv_split}. Using the support properties of the product $\Dif$ one can easily prove that the Steinmann scaling degree of the distribution \eqref{eq:const_adv_split} is not greater then the scaling degree of
\begin{equation}\label{eq:const_dif}
 (\Omega|\Dif(A^{r_1}(x_1),\ldots,A^{r_n}(x_n);A^{r_{n+1}}(0))\Omega) \in\mathcal{S}'(\R^{4n}).
\end{equation}
Since the latter is finite (the scaling degree of the Schwartz distribution cannot be infinite), by Theorem \ref{thm:extension} there exists an extension of \eqref{eq:const_adv_split} which we denote by 
\begin{equation}\label{eq:const_extension}
 \underline{\textrm{a}}_0^{r_1,\ldots,r_{n+1}}(x_1,\ldots,x_n) \in\mathcal{S}'(\R^{4n}).
\end{equation} 
If $\sum_{j=1}^{n+1}\mathbf{f}(A^{r_j})\neq 0$ then the distribution \eqref{eq:const_dif} vanishes. In this case we set $$\underline{\textrm{a}}_0^{r_1,\ldots,r_{n+1}}=0,$$ which is a particular extension of the distribution \eqref{eq:const_adv_split}. The distribution $\underline{\textrm{a}}_0^{r_1,\ldots,r_{n+1}}$ is used in the definition of the distribution \eqref{eq:const_adv} but need not coincide with it.

Using the extension $\underline{\textrm{a}}_0^{r_1,\ldots,r_{n+1}}$ of the distribution \eqref{eq:const_adv_split} we define
\begin{multline}\label{eq:const_T_sym}
 (\Omega|\T(A^{r_1}(x_1),\ldots,A^{r_{n+1}}(x_{n+1}))\Omega) 
 :=
 \\
 =\frac{1}{(n+1)!}\sum_{\pi\in \mathcal{P}_{n+1}}\, (-1)^{\mathbf{f}(\pi)}\, \textrm{t}_0^{r_{\pi(1)},\ldots,r_{\pi(n+1)}}(x_{\pi(1)},\ldots,x_{\pi(n+1)}).
\end{multline}
In the analogy to Equation \eqref{eq:def_adv_prime} defining $\Adv'(I;J)$ we set
\begin{multline}
 \textrm{t}_0^{r_1,\ldots,r_{n+1}}(x_1,\ldots,x_{n+1})
 := \textrm{a}_0^{r_1,\ldots,r_{n+1}}(x_1,\ldots,x_{n+1})
 \\
 -(\Omega|\Adv'(A^{r_1}(x_1),\ldots,A^{r_n}(x_n);A^{r_{n+1}}(x_{n+1}))\Omega),
\end{multline}
where
\begin{equation}
 \textrm{a}_0^{r_1,\ldots,r_{n+1}}(x_1,\ldots,x_{n+1}):=\underline{\textrm{a}}_0^{r_1,\ldots,r_{n+1}}(x_1-x_{n+1},\ldots,x_n-x_{n+1}).
\end{equation}

Any other possible definition of \eqref{eq:vev_time_ordered} compatible with the axioms may differ from the above one by a translationally-invariant distribution supported on the thin diagonal $x_1=\ldots=x_{n+1}$. Such distribution has to be of the from
\begin{equation}\label{eq:const_freedom}
 P(\partial)\, \delta(x_1-x_{n+1})\ldots\delta(x_n-x_{n+1}),
\end{equation}
where $P(p_1,\ldots,p_{n+1})$ is a polynomial with complex coefficients ($P$ cannot be completely arbitrary). The ambiguity in the definition of $(\Omega|\Adv(A^{r_1}(x_1),\ldots,A^{r_n}(x_n);A^{r_{n+1}}(x_{n+1}))\Omega) $ is the same as the ambiguity in the definition of the VEVs of the time-ordered products. In Section~\ref{sec:sd} the normalization condition which significantly reduces this freedom is introduced. The remaining ambiguity is characterized in Section~\ref{sec:freedom}.

If the extension $\underline{\textrm{a}}_0$ of the distribution \eqref{eq:const_adv_split} is chosen such that 
\begin{equation}
 \textrm{t}_0^{r_1,\ldots,r_{n+1}}(x_1,\ldots,x_{n+1})
\end{equation}
defined above is graded-symmetric, then the distribution \eqref{eq:const_adv} obtained in the construction equals $\underline{a}_0$. Otherwise, the distribution \eqref{eq:const_adv} is given by
\begin{multline}
 \underline{\textrm{a}}^{r_1,\ldots,r_{n+1}}(x_1,\ldots,x_n)
 := (\Omega|\T(A^{r_1}(x_1),\ldots,A^{r_n}(x_n),A^{r_{n+1}}(0))\Omega)
 \\
 -(\Omega|\Adv'(A^{r_1}(x_1),\ldots,A^{r_n}(x_n);A^{r_{n+1}}(0))\Omega),
\end{multline}
which differs from $\underline{\textrm{a}}_0$ by the distribution of the form \eqref{eq:const_freedom}.

Finally, the time-ordered products are defined with the use of the identity \eqref{eq:T_expansion}. One can show that fulfill all the axioms \ref{axiom1}-\ref{axiom6}. As an aside, we mention that there exist also an alternative proof of the existence of the time-ordered products \cite{popineau2016pedagogical,brunetti2000microlocal} which does not require using other products besides the time-ordered products.

\section{VEV of product of \texorpdfstring{$\Fa$}{F} products} \label{sec:aux}

Let $F$ and $F'$ be two $\Fa$ products of the form \eqref{eq:F_Fprime}. In this section we recall the formula \eqref{eq:vev_product_representation} which expresses the VEV of the product of $F$ and $F'$
\begin{equation}\label{eq:vev_prod_F_F}
 (\Omega| F(B_1(x_1),\ldots,B_k(x_k)) F'(B'_1(x'_1),\ldots,B'_{k'}(x'_{k'}))\Omega)
\end{equation}
in terms of the VEVs of $F$ and $F'$. First note that as a result of the causal Wick expansion (the condition \ref{axiom3} given in Section~\ref{sec:F_prod}) the expression \eqref{eq:vev_prod_F_F} can be represented by the sum over super-quadri-indices $s_1,\ldots,s_k,s'_1,\ldots,s'_{k'}$ of the terms which are up to a multiplicative constant of the form
\begin{multline}\label{eq:product_vev_formula}
 (\Omega|F(B_1^{(s_1)}(x_1),\ldots,B_k^{(s_k)}(x_k))\Omega)
 ~\times~
 (\Omega| F'(B_1^{\prime(s_1)}(x'_1),\ldots,B_{k'}^{\prime(s_{k'})}(x'_{k'}))\Omega)
 \\
 ~\times~(\Omega|\normord{A^{s_1}(x_1)\ldots A^{s_k}(x_k)}\,
 \normord{A^{s'_1}(x'_1)\ldots A^{s'_{k'}}(x'_{k'})}\Omega).
\end{multline}
Let us concentrate for a moment on the last factor
\begin{equation}\label{eq:product_two_kormal}
 (\Omega|\normord{A^{s_1}(x_1)\ldots A^{s_k}(x_k)}\,
 \normord{A^{s'_1}(x'_1)\ldots A^{s'_{k'}}(x'_{k'})}\Omega).
\end{equation}
As is well-known it can be expressed in terms of the Wightman two-point functions of generators (i.e. elements of the set $\mathcal{G}$ introduced in Section~\ref{sec:ff})
\begin{equation}\label{eq:two_point_function_gen}
(\Omega|\normord{\partial^\alpha\!A_{i}(x)}\,\normord{\partial^{\alpha'}\!A_{i'}(x')}\Omega).
\end{equation}
We will briefly remind the reader how this representation is obtained. First, we expand each Wick monomial appearing in the expression \eqref{eq:product_two_kormal} as a product of generators using Equation \eqref{eq:wick_monomial}. Next, we consider the complete pairing of all generators such that each generator belonging to
\begin{equation}\label{eq:normord_1}
\normord{A^{s_1}(x_1)\ldots A^{s_k}(x_k)}
\end{equation}
is paired with exactly one generator from 
\begin{equation}
\normord{A^{s'_1}(x'_1)\ldots A^{s'_{k'}}(x'_{k'})}.
\end{equation} 
To each pair of generators $\partial^{\alpha}\!A_{i}(x)$, $\partial^{\alpha'}\!\!A_{i'}(x')$ corresponds the factor \eqref{eq:two_point_function_gen}. The expression corresponding to the complete pairing is the product of factors corresponding to all pairs. It is of the form (up to a multiplicative constant)
\begin{equation}\label{eq:prod_two_point_functions}
 \prod_{j=1}^l  (\Omega|\normord{\partial^{\bar \alpha(j)}\!A_{\bar i(j)}(x_{\bar u(j)})}\,
 \normord{\partial^{\bar \alpha'(j)}\!A_{\bar i'(j)}(x'_{\bar u'(j)})}\Omega),
\end{equation}
where $\bar u,\bar u'$ are functions from $\{1,\ldots,l\}$ to $\{1,\ldots,k\}$ or $\{1,\ldots,k'\}$ respectively, $\bar i,\bar i'$ are functions from $\{1,\ldots,l\}$ to the set $\{1,\ldots,\mathrm{p}\}$ which is identified with the set $\mathcal{G}_0$ of the basic generators \eqref{eq:basic_gen}, $\bar \alpha,\bar \alpha'$ are functions from $\{1,\ldots,l\}$ to $\N_0^4$ (the set of multi-indices) and $l$ is the total number of pairs ($l$ is the same for every complete pairing). Finally, the VEV \eqref{eq:product_two_kormal} is a sum over all complete parings. The complete pairing is fully characterized by specification of the functions $\bar u$, $\bar u'$, $\bar i$, $\bar i'$, $\bar \alpha$, $\bar \alpha'$.

Given a list of functions $\bar u$, $\bar u'$, $\bar i$, $\bar i'$, $\bar \alpha$, $\bar \alpha'$ of the above type we shall define the following maps characterizing this list
\begin{equation}
 \bar\ext,\bar\der:\, \{1,\ldots,\mathrm{p}\}\simeq\mathcal{G}_0\to\N_0,
\end{equation}
where $\mathcal{G}_0$ is the set of the basic generators \eqref{eq:basic_gen}. We set
\begin{equation}\label{eq:ext_s}
 \bar\ext(i)\equiv \bar\ext(A_i):= |\{j\,: \bar i(j)=i\}|+ |\{j\,: \bar i'(j)=i\}|
\end{equation}
and
\begin{equation}\label{eq:der_s}
 \bar\der(i)\equiv \bar\der(A_i):= \sum_{\substack{j\in\{1,\ldots,l\}\\\bar i(j)=i}}|\bar \alpha(j)| + \sum_{\substack{j\in\{1,\ldots,l\}\\\bar i'(j)=i}}|\bar \alpha'(j)|.
\end{equation}
$\bar\ext(A_i)$ is the total number of times the field $A_i$ or its derivative appears in the expression \eqref{eq:prod_two_point_functions} after expanding each monomial $A^r$ into the product of the generators and $\bar\der(A_i)$ is the total number of derivatives acting on the field $A_i$ in this expression. Note that if $\bar\ext(A_i)=0$ then $\bar\der(A_i)=0$. 

Similarly, given a list of super-quadri indices $\mathbf{s}=(s_1,\ldots,s_k)$ we shall define functions 
\begin{equation}
 \ext_{\mathbf{s}}, \der_{\mathbf{s}}:\, \{1,\ldots,\mathrm{p}\}\simeq\mathcal{G}_0\to\N_0
\end{equation}
characterizing this list by setting
\begin{equation}\label{eq:ext}
 \ext_{\mathbf{s}}(i)\equiv \ext_{\mathbf{s}}(A_i):= \sum_{\alpha\in\N_0^4}  s(i,\alpha),
\end{equation}
and
\begin{equation}\label{eq:der}
 \der_{\mathbf{s}}(i)\equiv \der_{\mathbf{s}}(A_i):= \sum_{\alpha\in\N_0^4}  s(i,\alpha)~|\alpha|,
\end{equation}
where 
\begin{equation}\label{eq:total_index}
 s(i,\alpha):=\sum_{j=1}^n s_j(i,\alpha)
\end{equation}
is the total super-quadri-index of the list $\mathbf{s}$. For example, let us consider the list of fields $(A^{s_1}(x_1),\ldots,A^{s_k}(x_k))$. Expand all Wick monomials from this list into products of generators as in \eqref{eq:wick_monomial}. Now $\ext_{\mathbf{s}}(A_i)$ is the total number of times the field $A_i$ or its derivative appears in the list after the expansion and $\der_{\mathbf{s}}(A_i)$ is the total number of derivatives acting on the field $A_i$. Observe that if $\ext_{\mathbf{s}}(A_i)=0$ then $\der_{\mathbf{s}}(A_i)=0$. 

The VEV of the product of $F$ and $F'$ may be expressed as follows 
\begin{multline}\label{eq:vev_product_representation}
(\Omega|F(B_1(x_1),\ldots,B_k(x_k)) F'(B_k^{\prime}(x'_1),\ldots,B^{\prime}_{k'}(x'_{k'}))\Omega)
\\
 =\sum_{\bar u,\bar u',\bar i,\bar i',\bar \alpha,\bar \alpha'} ~\const~~
 \prod_{j=1}^l  (\Omega|\normord{\partial^{\bar \alpha(j)}\!A_{\bar i(j)}(x_{\bar u(j)})}\,
 \normord{\partial^{\bar \alpha'(j)}\!A_{\bar i'(j)}(x'_{\bar u'(j)})}\Omega)\times
 \\
 (\Omega|F(B^{(s_1)}_1(x_1),\ldots,B^{(s_k)}_k(x_k))\Omega) \, (\Omega| F'(B^{\prime(s'_1)}_1(x'_1),\ldots,B^{\prime(s'_{k'})}_{k'}(x'_{k'}))\Omega),
\end{multline}
where the sum is carried out over functions $\bar u,\bar u',\bar i,\bar i',\bar \alpha,\bar \alpha'$ of the form given above. The super-quadri-indices $s_1,\ldots,s_k$, $s'_1,\ldots,s'_k$ are determined uniquely by the conditions
\begin{equation}
 A^{s_n}\propto \prod_{\substack{j\in \{1,\ldots,l\} \\ \bar u(j)=n}} \partial^{\bar \alpha(j)} A_{\bar i(j)},~~~~~
 A^{s'_n}\propto \prod_{\substack{j\in \{1,\ldots,l\} \\ \bar u'(j)=n}} \partial^{\bar \alpha'(j)} A_{\bar i'(j)},
\end{equation}
where $\propto$ indicates the equality up to a factor $(-1)$. 
% and the sets $\mathcal{I}_j$, $\mathcal{I}'_j$ are given by
% \begin{equation}
%  \mathcal{I}_n =\{ j\in \{1,\ldots,l\} \,:\, \bar u(j)=n \},~~~~~
%  \mathcal{I}'_n =\{ j\in \{1,\ldots,l\} \,:\, \bar u'(j)=n \}.
% \end{equation}
In what follows it will be only important that the lists $\mathbf{s}=(s_1,\ldots,s_k)$, $\mathbf{s}'=(s'_1,\ldots,s'_k)$ and functions $\bar u,\bar u',\bar i,\bar i',\bar \alpha,\bar \alpha'$ satisfy the following constraints
\begin{equation}\label{eq:constraints}
 \ext_{\mathbf{s}}(A_i) + \ext_{\mathbf{s}'}(A_i) = \bar\ext(A_i),~~~~~\der_{\mathbf{s}}(A_i) + \der_{\mathbf{s}'}(A_i) = \bar\der(A_i).
\end{equation}
The term in the sum on the RHS of \eqref{eq:vev_product_representation} for which $l=0$ (i.e. $\mathbf{s}=\mathbf{s}'=0$ and $\bar\ext(A_i)=\ext_{\mathbf{s}}(A_i)=\ext_{\mathbf{s}'}(A_i)=0$ for $A_i\in\mathcal{G}_0$) is called the \emph{vacuum contribution}. The terms in the sum on the RHS of \eqref{eq:vev_product_representation} corresponding to functions $\bar u$, $\bar u'$, $\bar i$, $\bar i'$, $\bar \alpha$, $\bar \alpha'$, which satisfy the condition $\bar\ext(A_i)=\ext_{\mathbf{s}}(A_i)=\ext_{\mathbf{s}'}(A_i)=0$ for all massive fields $A_i\in\mathcal{G}_0$ (i.e. the super-quadri-indices from the lists $\mathbf{s}$, $\mathbf{s}'$ involve only massless fields) are called the \emph{massless contributions}. 

The above representation will play an important role in our proof of the existence of the Wightman and Green functions. In the next section we present its application by deriving a bound on the Steinmann scaling degree of the VEVs of the time-ordered products. 

We close this section with a lemma which further restricts the form of terms which may appear in the sum on the RHS of \eqref{eq:vev_product_representation} after integrating both sides with a Schwartz function with support in the momentum space in a sufficiently small neighborhood of the origin.

\begin{lem}\label{lem:aux_lemma}
\hspace{0em}
\begin{enumerate}[leftmargin=*,label={(\Alph*)}]
\item There exist neighborhoods $\mathcal{O}\subset\R^{4k}$ and $\mathcal{O}'\subset\R^{4k'}$ of the origins in $\R^{4k}$ and $\R^{4k'}$, respectively, such that the following holds:
Let $g\in\mathcal{S}(\R^{4k+4k'})$ be any test function with 
\begin{equation}
 \supp\,\F{g} \subset \R^{4k} \times \mathcal{O}' \cup \mathcal{O} \times \R^{4k'}.
\end{equation}
Then all contributions to the expansion \eqref{eq:vev_product_representation} vanish on $g$ with the exception of the vacuum and the massless contributions.

\item Representation \eqref{eq:vev_product_representation} of the VEV of the graded commutator $[F,F']$ does not contain the vacuum contribution.
\end{enumerate}
\begin{proof}
Part~(A) is a consequence of the conservation of the four-momentum. More precisely, the translational covariance of $F$ and $F'$ (the condition \ref{axiom1} given in Section~\ref{sec:F_prod}) implies that the Fourier transforms of the VEVs of $F$ and $F'$ contain Dirac deltas of the sum of the arguments of $F$ and $F'$, respectively. Moreover, it holds
\begin{equation}
 (\Omega|\normord{\partial^\alpha\!A_{i}(x)}\,\normord{\partial^{\alpha'}\!A_{i'}(x')}\Omega)
 =\partial^\alpha_x \partial^{\alpha'}_{x'} D^{(+)}_{ii'}(x-x').
\end{equation}
The distribution $D^{(+)}_{ii'}$ vanishes unless the fields $A_i$, $A_{i'}$ have the same mass $m\in[0,\infty)$ and the support of its Fourier transform is contained in the set 
\begin{equation}
 H_m=\{p\in\R^4\,:\,p^2=m^2,~p^0\geq 0\}.
\end{equation}
If $p_1\in H_{m_1}, \ldots, p_l\in H_{m_l}$ then
\begin{equation}
 p_1+\ldots+p_l \in \{p\in\R^4\,:\,p^2\geq m_{\min}^2,~p^0\geq 0\},~~~~~m_{\min}=\min(m_1,\ldots,m_l).
\end{equation}
As a result, if there are two-point functions of the massive fields in the product \eqref{eq:prod_two_point_functions}, then its Fourier transform does not intersect some neighborhood of the origin. The statement of Part~(A) of the lemma follows from the fact that the Fourier transform of the product of the distribution from the third line of \eqref{eq:vev_product_representation} and the distribution \eqref{eq:prod_two_point_functions} is the convolution of their Fourier transforms (because of the support properties of the above-mentioned distributions the convolution is well-defined).

Part~(B) follows from the definition of the graded-commutator of $\Fa$ products and the condition \ref{axiom2}, which are both given in Section~\ref{sec:F_prod}.
\end{proof}

\end{lem}

\section{Bound on the scaling degree}\label{sec:sd}

In order to restrict the freedom \eqref{eq:const_freedom} in the definition of the VEVs of the time-ordered products we impose the following normalization condition.
\begin{enumerate}[label=\bf{N.SD},leftmargin=*]
\item\label{norm:sd} Bound on the Steinmann scaling degree: For all $B_1,\ldots,B_k\in\Fh$ it holds
\begin{equation}\label{eq:sd_bound}
 \sd(\,(\Omega|\T(B_1(x_1),\ldots,B_k(x_k)) \Omega)\,) \leq \sum_{j=1}^k (\dim(B_j)+\CC),
\end{equation}
where $\sd(\cdot)$ is the scaling degree of a distribution and $\CC\in \{0,1\}$ is a fixed constant.
\end{enumerate}

Let us make a couple of remarks concerning the above condition. 
\begin{enumerate}[label=(\arabic*),leftmargin=*]
 \item The Steinmann scaling degree measures the UV behavior of the VEVs of distributions. The above normalization condition plays a crucial  role in the proof of the renormalizability of models of perturbative quantum field theories (cf. Section~\ref{sec:ren}).
 \item It follows from the translational-invariance of the VEVs of time-ordered products that
\begin{equation}
 \sd(\,(\Omega|\T(B_1(x_1),\ldots,B_k(x_k)) \Omega)\,) = \sd(\,(\Omega|\T(B_1(x_1),\ldots,B_{k-1}(x_{k-1}),B_k(0)) \Omega)\,).
\end{equation}
\item Because of the axiom \ref{axiom2} the RHS of the inequality \eqref{eq:sd_bound} is always an integer in the non-trivial cases.
\item By the axiom \ref{axiom4} if some of polynomials $B_1,\ldots,B_k$ are constants then the time-ordered product $\T(B_1(x_1),\ldots,B_k(x_k))$ is equal to the time ordered product with less arguments. The above bound on the scaling degree of the VEV of the latter product is not weaker than the bound on the VEV of the former (if $\CC=0$ they are both equivalent). Thus, imposing the bound \eqref{eq:sd_bound} only for $B_1,\ldots,B_k\in\Fh$ such that none of them is a constant results in an equivalent normalization condition.

\item Setting $\CC=0$ in the above condition yields the so-called minimal solution of the inductive problem which is considered to be standard \cite{epstein1973role,brunetti2000microlocal,fredenhagen2015perturbative}. The case $\CC>0$ corresponds in the BPHZ approach to the special form of the so-called over-subtractions \cite{zimmermann1969convergence}. We always assume that $\CC=0$ in the case of models with at least one interaction vertex of dimension $4$. However, in the case of models with interaction vertices of dimension 3 it is also possible to set $\CC=1$. In the latter case the scaling degree of the VEVs of time ordered products with $n+1$ arguments, constructed in Section~\ref{sec:T_const}, may be strictly larger than the scaling degree of \eqref{eq:const_dif} and \eqref{eq:const_adv_split}. The choice $\CC=1$ makes the UV behavior of the theory worse but at the same time allows more freedom in the construction of time ordered products. This freedom may be used to improve its IR properties. In Section~\ref{sec:the_model} we consider a model for which the proof of the existence of the Wightman and Green functions is possible only if one sets $\CC=1$ in the above condition. See also the discussion of the renormalizability of interacting models in Section~\ref{sec:ren}.
\end{enumerate}

Let us outline the inductive proof of the compatibility of the condition \ref{norm:sd} with the axioms \ref{axiom1}-\ref{axiom6}. First note that the bound \eqref{eq:sd_bound} holds for $k=1$ since $(\Omega|B(x)\Omega)=\const$ and in consequence, $\sd\left(\, (\Omega|B(x)\Omega)\,\right)=0 \leq \dim(B) + \CC$. Next, fix $n\in\N_+$ and assume that the bound \eqref{eq:sd_bound} holds for $k\leq n$. 

Using the notation introduced in the previous section we shall first show that the bound \eqref{eq:sd_bound} holds for VEVs of all $\Fa$ products which may be expressed as linear combinations of products of the time-ordered products with at most $n$ arguments. 
\begin{lem}\label{lem:sd}
Let $t\in\mathcal{S}'(\R^{4k})$, $t'\in\mathcal{S}'(\R^{4k'})$ be translationally invariant distributions. Then
\begin{equation}\label{eq:lem_product_sd}
 t(x_1,\ldots,x_k)~ t'(x'_1,\ldots,x'_{k'})~ 
 \prod_{j=1}^l  (\Omega|(\partial^{\bar \alpha(j)} A_{\bar i(j)}(x_{\bar u(j)}))
 (\partial^{\bar \alpha'(j)} A_{\bar i'(j)}(x'_{\bar u'(j)}))\Omega)
 %\prod_{k=1}^l \partial_x^{\bar\alpha(j)}\partial_{x'}^{\bar\alpha'(j)} 
 %D^{(+)}_{\bar i(j) \bar i'(j)}(x_{\bar u(j)}-x'_{\bar u'(j)})
\end{equation}
is well-defined  in the sense of H{\"o}rmander's product of distributions and its Steinmann scaling degree is bounded by 
\begin{equation}
 \sd(t)+\sd(t') + \sum_{i=1}^{\mathrm{p}} \left[ \bar\ext(A_i) \dim(A_i) + \bar\der(A_i)\right].
\end{equation}
The notation used in \eqref{eq:lem_product_sd} was introduced below Equation \eqref{eq:prod_two_point_functions}.
\begin{proof}
It follows from Lemma 6.6 of \cite{brunetti2000microlocal} that the expression \eqref{eq:lem_product_sd} is well-defined\footnote{Actually, the fact that \eqref{eq:lem_product_sd} is well-defined follows easily from the support properties of the Wightman two point functions in the momentum representation and the form of the Fourier transform of a translationally invariant distribution \eqref{eq:dist_trans_inv_F}.} and its scaling dimension is given by
\begin{equation}
 \sd(t)+\sd(t') + \sum_{j=1}^l\sd\left( 
 %\partial_x^{\alpha(j)}\partial_{x'}^{\alpha'(j)} D^{(+)}_{i(j) i'(j)}(x_{u(j)}-x'_{u'(j)}) 
 (\Omega|(\partial^{\bar \alpha(j)} A_{\bar i(j)}(x_{\bar u(j)}))
 (\partial^{\bar \alpha'(j)} A_{\bar i'(j)}(x'_{\bar u'(j)}))\Omega)
 \right).
\end{equation}
The thesis follows from the identity
\begin{equation}
\sd\left((\Omega|(\partial^\alpha\! A_{i}(x))(\partial^{\alpha'}\! A_{i'}(x'))\Omega) \right)
% \sd\left( \partial_x^\alpha\partial_{x'}^{\alpha'} D_{ii'}^{(+)}(x-x') \right) 
 = \dim(A_i) + \dim(A_{i'}) + |\alpha| + |\alpha'|.
\end{equation}
Actually, the identity of the above type is sometimes considered to be the definition of the canonical dimension of a field.
\end{proof}
\end{lem}

We will show that if the bound \eqref{eq:sd_bound} holds for the VEVs of the two $\Fa$~products then it also holds for the VEV of their product. To this end, we use the representation~\eqref{eq:vev_product_representation} and apply the above lemma. The statement follows from the equation
\begin{multline}\label{eq:sd_bound_other}
  \sum_{j=1}^k (\dim(B_j)+\CC) + \sum_{j=1}^{k'} (\dim(B'_j)+\CC) = 
  \sum_{j=1}^k (\dim(B_j^{(s_j)})+\CC) + \sum_{j=1}^{k'} (\dim(B_j^{\prime(s'_j)})+\CC) 
  \\
  + 
  \sum_{i=1}^{\mathrm{p}} \left[ \ext_{\mathbf{s}}(A_i) \dim(A_i) + \der_{\mathbf{s}}(A_i) \right]
  +
  \sum_{i=1}^{\mathrm{p}} \left[ \ext_{\mathbf{s}'}(A_i) \dim(A_i) + \der_{\mathbf{s}'}(A_i) \right]
\end{multline}
and the constraints \eqref{eq:constraints} which imply that the sum $\sum_{i=1}^{\mathrm{p}} \left[ \bar\ext(A_i) \dim(A_i) + \bar\der(A_i)\right]$ coincides with the expression from the last line of Equation \eqref{eq:sd_bound_other}. The functions 
$\ext_{\mathbf{s}}(A_i)$, $\der_{\mathbf{s}}(A_i)$ and $\ext_{\mathbf{s}'}(A_i)$, $\der_{\mathbf{s}'}(A_i)$ were defined in Section~\ref{sec:aux}.

By the result of the previous paragraph and the inductive assumption the bound \eqref{eq:sd_bound} holds for VEVs of all $\Fa$ products which may be expressed as linear combinations of products of the time-ordered products with at most $n$ arguments. In particular, it is valid for the VEVs of products \eqref{eq:const_dif1}, \eqref{eq:const_adv_prime}. To prove the inductive step it is enough to demand that in the construction presented in Section~\ref{sec:T_const} the scaling degree of the extension \eqref{eq:const_extension} of the distribution \eqref{eq:const_adv_split} is not bigger than the scaling degree of \eqref{eq:const_dif}. The extension with this property always exists according to Theorem \ref{thm:extension}.

\section{Normalization freedom}\label{sec:freedom}

Assume that all the time-ordered products with at most $n\in\N_+$ arguments are given and they fulfill the axioms \ref{axiom1}-\ref{axiom6} and the normalization condition \ref{norm:sd}. By the result of the previous section we know that it is possible to define the time-ordered product with $n+1$ arguments such that all the above-mentioned conditions are satisfied. In this section we will characterize the ambiguity in this definition. To this end, by the axiom~\ref{axiom3} it is enough to consider the family of numerical distributions
\begin{equation}\label{eq:const_freedom_time_ordered}
 \Fa^{n+1}\ni(B_1,\ldots,B_{n+1})\mapsto (\Omega|\T(B_1(x_1),\ldots,B_{n+1}(x_{n+1}))\Omega)\in\mathcal{S}'(\R^{4(n+1)}).
\end{equation}
We claim that any two possible definitions of the above family differ by the graded-symmetric family of distributions
\begin{equation}\label{eq:freedom_v}
  \Fa^{n+1}\ni(B_1,\ldots,B_{n+1})\mapsto v(B_1(x_1),\ldots,B_{n+1}(x_{n+1}))\in\mathcal{S}'(\R^{4(n+1)})
\end{equation}
such that for any $B_1,\ldots,B_{n+1}\in\Fh$ 
\begin{enumerate}[leftmargin=*,label={(\arabic*)}]
\item $v(B_1(x_1),\ldots,B_{n+1}(x_{n+1}))$ is of the form 
\begin{equation}\label{eq:freedom_form}
 \sum_{\substack{\gamma\\|\gamma|\leq\omega}} c_\gamma \partial^\gamma \delta(x_1-x_{n+1})\ldots \delta(x_{n}-x_{n+1})
\end{equation}
for some constants $c_\gamma\in\C$ indexed by multi-indices $\gamma$, $|\gamma|\leq\omega$, where
\begin{equation}\label{eq:omega}
 \omega = \sum_{j=1}^{n+1} (\dim(B_j)+\CC) - 4n,
\end{equation}
\item $v(B_1(x_1),\ldots,B_{n+1}(x_{n+1}))$ vanishes if $ \sum_{j=1}^{n+1} \mathbf{f}(B_j)\neq 0$.
\end{enumerate}
It follows that $v(B_1(x_1),\ldots,B_{n+1}(x_{n+1}))=0$ if $\omega<0$. In this case the distribution \eqref{eq:const_freedom_time_ordered} is determined uniquely by the time-ordered products with at most $n$ arguments. By definition the map of the form \eqref{eq:freedom_v} is graded-symmetric iff
\begin{equation}\label{eq:v_graded}
 v(B_1(x_1),\ldots,B_{n+1}(x_{n+1})) = (-1)^{\mathbf{f}(\pi)}v(B_{\pi(1)}(x_{\pi(1)}),\ldots,B_{\pi(n+1)}(x_{\pi(n+1)}))
\end{equation}
for all $B_1,\ldots,B_{n+1}\in\Fh$ and all permutations $\pi\in\mathcal{P}_{n+1}$, where $\mathbf{f}(\pi)\in\Z/2\Z$ is the number of transpositions in $\pi$ that involve a pair of fields with odd fermion number.

To prove that the distribution $v(B_1(x_1),\ldots,B_{n+1}(x_{n+1}))$ has to be indeed of the form \eqref{eq:freedom_form} we first recall from Section~\ref{sec:T_const} that any two possible definitions of $(\Omega|\T(B_1(x_1),\ldots,B_n(x_n),B_{n+1}(0)) \Omega)$ may only differ by
\begin{equation}
 P(\partial)\, \delta(x_1)\ldots\delta(x_n),
\end{equation}
where $P$ is some polynomial. It turns out that its degree $\deg(P)$ is bounded from above by~$\omega$. To show this we observe that the Steinmann scaling degree of the above distribution is equal $\deg(P)+4n$ and take advantage of the normalization condition \ref{norm:sd} which says that 
\begin{equation}\label{eq:sd_omega_bound}
 \sd(\,(\Omega|\T(B_1(x_1),\ldots,B_n(x_n),B_{n+1}(0)) \Omega)\,) \leq \omega + 4n.
\end{equation} 
Other properties of the family \eqref{eq:freedom_form} follow from the axioms \ref{axiom2} and \ref{axiom5}.

The same ambiguity exists in defining the VEVs of the advanced products
\begin{equation}\label{eq:const_freedom_advanced}
 \Fa^{n+1}\ni(B_1,\ldots,B_{n+1})\mapsto (\Omega|\Adv(B_1(x_1),\ldots,B_n(x_n);B_{n+1}(x_{n+1}))\Omega)\in\mathcal{S}'(\R^{4(n+1)}),
\end{equation}
since the difference between the distributions \eqref{eq:const_freedom_time_ordered} and \eqref{eq:const_freedom_advanced} is determined uniquely by the time-ordered products with at most $n$ arguments. 

We have characterized above the freedom in defining the time-ordered products in a~single step of the EG construction. Note that in general the VEVs of two time-ordered products satisfying the axioms \ref{axiom1}-\ref{axiom6} and the normalization condition \ref{norm:sd} may differ by distribution $v$ which is not of the form stated in this section. The relation between different definitions of the time-ordered products is fully described by the Main Theorem of Renormalization \cite{popineau2016pedagogical,pinter2001finite,dutsch2004causal,brunetti2009perturbative}.

%%%%%%%%%%%%%%%%%%%%%%%%%%%%%%%%%%%%%%%%%%%%%%%%%%%%%%%%%%%%%%%%%%%%%%%%%%%%%%%%%%%%%%%%%%%%%%%%%%%%
%%%%%%%%%%%%%%%%%%%%%%%%%%%%%%%%%%%%%%%%%%%%%%%%%%%%%%%%%%%%%%%%%%%%%%%%%%%%%%%%%%%%%%%%%%%%%%%%%%%%
%                                   MODELS
%%%%%%%%%%%%%%%%%%%%%%%%%%%%%%%%%%%%%%%%%%%%%%%%%%%%%%%%%%%%%%%%%%%%%%%%%%%%%%%%%%%%%%%%%%%%%%%%%%%%
%%%%%%%%%%%%%%%%%%%%%%%%%%%%%%%%%%%%%%%%%%%%%%%%%%%%%%%%%%%%%%%%%%%%%%%%%%%%%%%%%%%%%%%%%%%%%%%%%%%%
\chapter{Interacting models}\label{sec:models}

In order to define the model of the interacting quantum field theory one has to specify its interaction vertices 
\begin{equation}
 \mathcal{L}_1,\ldots,\mathcal{L}_\mathrm{q}\in\Fh
\end{equation}
which are distinguished homogeneous elements of the algebra of symbolic fields $\Fa$ satisfying the following conditions:
\begin{enumerate}[leftmargin=*,label={(\arabic*)}]
 \item $\mathcal{L}_1,\ldots,\mathcal{L}_\mathrm{q}$ are Lorentz scalars,
 \item $\mathcal{L}_l=\mathcal{L}_l^*$ for all $l\in\{1,\ldots,\mathrm{q}\}$,
 \item $\dim(\mathcal{L}_l)\leq 4$ for all $l\in\{1,\ldots,\mathrm{q}\}$,
 \item $\mathbf{f}(\mathcal{L}_l)=0$, $\mathbf{q}(\mathcal{L}_l)=0$ for all $l\in\{1,\ldots,\mathrm{q}\}$.
\end{enumerate}
The first condition is crucial for the Lorentz covariance of the model. The second one is needed for quantum-mechanical consistency of the model, e.g. for the unitarity of the scattering matrix. The third reflects the fact that we consider only renormalizable interactions. The last condition guarantees that the scattering matrix commutes with the fermion number operator and the electric charge. To prove the existence of the Wightman and Green function in the case of models with massless particles we impose some additional conditions on the interaction vertices which are formulated in Section~\ref{sec:proof_scalar} as Assumption~\ref{asm}. 

To each interaction vertex $\mathcal{L}_l$ we associate a parameter $e_l\in\R$ also called a coupling constant. The scattering matrix and the interacting fields are defined as formal power series in the independent parameters $e_1,\ldots,e_\mathrm{q}$. In some models the physical coupling constants accompanying different interaction terms are related. In this case we assume that there exists a set of independent parameters such that all  $e_1,\ldots,e_\mathrm{q}$ are polynomial functions of them (for example it may hold $e_1=e$, $e_2=e^2$ for some $e\in\R$). Note that some normalization conditions like, e.g. the Ward identities, which ensure the desired physical properties of the theory may depend crucially on the specific form of the above relations. However, in our proof of the existence of the weak adiabatic limit it is possible to assume that the parameters $e_1,\ldots,e_\mathrm{q}$ are independent. We stress that the relation between $e_1,\ldots,e_\mathrm{q}$ of the above-mentioned form may be always imposed in the final expression for the Wightman and Green functions by reorganizing the resulting formal power series. 

The S-matrix of the model under consideration is formally given by
\begin{multline}
 S = \Texp\left(\ri\, \int \rd^4 x \, \sum_{l=1}^\mathrm{q} e_l \mathcal{L}_l(x)\right)
 \\
 =
 \sum_{n=0}^\infty \frac{\ri^n}{n!} \sum_{l_1,\ldots,l_n=1}^\mathrm{q} e_{l_1}\ldots e_{l_n} 
 \int\rd^4 y_1\ldots\rd^4 y_n\,
 \T(\mathcal{L}_{l_1}(x_1),\ldots,\mathcal{L}_{l_n}(x_n)).
\end{multline}
Unfortunately, the above expression is ill-defined because $\T(\mathcal{L}_{l_1}(x_1),\ldots,\mathcal{L}_{l_n}(x_n))\in\mathcal{S}'(\R^{4n},L(\mathcal{D}_0))$ cannot be integrated with constant functions. In order to avoid this problem one replaces the coupling constants $e_1,\ldots,e_\mathrm{q}$ in the above expression by functions $e_1 g_1,\ldots,e_\mathrm{q}g_\mathrm{q}\in\mathcal{S}(\R^{4n})$, where $g_1,\ldots,g_\mathrm{q}\in\mathcal{S}(\R^{4n})$ are called the switching functions. They switch off the interaction as $|x|\to\infty$ and play the role of the IR regularization. The interacting model with the IR regularization is fully specified by 
\begin{equation}
 \mathpzc{g}=\sum_{l=1}^\mathrm{q} e_lg_l\otimes \mathcal{L}_l\in\mathcal{S}(\R^4)\otimes \Fa.
\end{equation}
After the modification which was outlined above we obtain the expression for the IR-regularized S-matrix
\begin{multline}\label{eq:S_matrix}
 S(\mathpzc{g}) = \Texp\left(\ri\, \int \rd^4 x \, \sum_{l=1}^\mathrm{q} e_l g_l(x) \mathcal{L}_l(x)\right)=
 \sum_{n=0}^\infty \frac{\ri^n}{n!} \sum_{l_1,\ldots,l_n} e_{l_1}\ldots e_{l_n} 
 \\
 \int\rd^4 y_1\ldots\rd^4 y_n\,
 g_{l_1}(y_1)\ldots g_{l_n}(y_n) \T(\mathcal{L}_{l_1}(x_1),\ldots,\mathcal{L}_{l_n}(x_n)).
\end{multline}
The RHS of the above equation is clearly well-defined as a formal power series in the coupling constants $e_1,\ldots,e_\mathrm{q}$. We note that because the interaction vertices have vanishing fermion numbers, $S(\mathpzc{g})$ is the result of evaluating the generating functional of the time-ordered products discussed in Section~\ref{sec:generating} at the argument $\mathpzc{g}$.

Obviously, $S(\mathpzc{g})$ cannot be identified with the physical S-matrix as it describes the scattering of particles in an unphysical theory with spacetime-dependent coupling constants. This is why we always refer to it as the IR-regularized S-matrix. The physical scattering operator is obtained by taking the limit $g_l(x)\to 1$ for all $l\in\{1,\ldots,\mathrm{q}\}$ of the operator $S(\mathpzc{g})$ in an appropriate sense to be discussed in Section~\ref{sec:adiabatic_limits} (if this limit exists). The interacting fields or the Wightman and Green functions are given by the analogous limits of the corresponding objects with the IR regularization.

In the next Section we discuss the notion of the renormalizability of the interacting models defined with the use of the time-ordered products satisfying our version of the normalization condition \ref{norm:sd}. In Sections \ref{sec:interacting} and \ref{sec:W_G_IR} we give the definitions of the interacting fields and the Wightman and Green functions with the IR regularization. We provide explicit expressions for the coefficients of their formal power series expansions in the coupling constants $e_1,\ldots,e_\mathrm{q}$. Moreover, in Section~\ref{sec:W_G_IR} we state a couple of formulas which will be needed in the proof of the existence of the physical Wightman and Green functions. 

\section{Renormalizability}\label{sec:ren}

The normalization condition \ref{norm:sd} was formulated in a form which is more general then the standard one. Let us discuss the implications of this generalization for renormalizability of interacting models. To this end, we consider the time-ordered products which appear in the expansion of $S(\mathpzc{g})$ in a formal power series in the coupling constants. They have the form
\begin{equation}
 \T(\mathcal{L}_{l_1}(x_1),\ldots,\mathcal{L}_{l_k}(x_k))
\end{equation}
for some $k\in\N_0$ and $l_1,\ldots,l_k \in \{1,\ldots,\mathrm{q}\}$. Note that by the axiom \ref{axiom3} the above time-ordered products are linear combinations of the terms of the form
\begin{equation}\label{eq:ren_time}
  (\Omega|\T(\mathcal{L}^{(s_1)}_{l_1}(x_1),\ldots,\mathcal{L}^{(s_k)}_{l_k}(x_k))\Omega) ~\normord{A^{s_1}(x_1)\ldots A^{s_k}(x_k)}
\end{equation}
for some super-quadri-indices $s_1,\ldots,s_k$. The time-ordered products under consideration are determined by the VEVs of the time-ordered products of sub-polynomials of the interaction vertices.

Our normalization condition \ref{norm:sd} implies that
\begin{equation}\label{eq:ren_bound}
 \sd(\,(\Omega|\T(\mathcal{L}^{(s_1)}_{l_1}(x_1),\ldots,\mathcal{L}^{(s_{n+1})}_{l_{n+1}}(x_{n+1}))\Omega)\,)
 \leq \omega  + 4n,
\end{equation}
where $\omega$, defined by Equation \eqref{eq:omega}, may be also rewritten in the form
\begin{equation}\label{eq:ren_omega}
 \omega = \left[ 4 - \sum_{j=1}^{n+1} (4-\CC-\dim(\mathcal{L}_{l_j})) \right] - \sum_{j=1}^{n+1} \dim(A^{s_l}).
\end{equation}
To obtain the above formula we used the identity $\dim(\mathcal{L}_l^{(s_l)})=\dim(\mathcal{L}_l)-\dim(A^{s_l})$. Note that in the thesis we will provide a few different expressions defining $\omega$. Their explicit form may be different but all of them refer to the number which appears on the RHS of the bound \eqref{eq:sd_omega_bound}.

Assume that the time-ordered products with at most $n$ arguments are given. As discussed in Section~\ref{sec:freedom}, the distribution
\begin{equation}\label{eq:ren_vev}
 (\Omega|\T(\mathcal{L}^{(s_1)}_{l_1}(x_1),\ldots,\mathcal{L}^{(s_{n+1})}_{l_{n+1}}(x_{n+1}))\Omega)
\end{equation}
is determined uniquely if the super-quadri-indices $s_1,\ldots,s_{n+1}$ are such that $\omega<0$. If $\omega\geq 0$ then there is an ambiguity in the definition of the distribution \eqref{eq:ren_vev} which may be removed by appropriate normalization. Let $N_n$ be the number of independent distributions \eqref{eq:ren_vev} with $n+1$ arguments which have to be normalized in order to be uniquely determined (the distributions which are related by some conditions implied by the axioms \ref{axiom1}-\ref{axiom6} do not count as independent). If $N_n$ is bounded by a constant independent of $n$ then the model is said to be renormalizable, otherwise it is unrenormalizable. If $\sum_{n=1}^\infty N_n$ is finite then the model is said to be super-renormalizable. A closer inspection of Equation \eqref{eq:ren_omega} reveals that the model is renormalizable in view of the above definition iff 
\begin{equation}
 4 - \sum_{j=1}^{n+1} (4-\CC-\dim(\mathcal{L}_{l_j}))
\end{equation}
is bounded from above by a constant independent of $n$. It holds iff $\dim(\mathcal{L}_l)+\CC\leq 4$ for all $l\in\{1,\ldots,\mathrm{q}\}$. Since we restrict attention to renormalizable models in this thesis we admit $\CC=1$ in the normalization condition \ref{norm:sd} only if $\dim(\mathcal{L}_l)\leq 3$ for all $l\in\{1,\ldots,\mathrm{q}\}$. If $\dim(\mathcal{L}_l)=4$ for at least one vertex then we always assume that $\CC=0$. In the three examples of the interacting models we will present in Section~\ref{sec:examples} it holds $\dim(\mathcal{L}_l)+\CC=4$ for all $l\in\{1,\ldots,\mathrm{q}\}$. In fact, our proof of existence of the Wightman and Green functions is valid only under this assumptions. If the above-mentioned condition is satisfied then the formula \eqref{eq:ren_omega} for $\omega$ acquires a simpler form
\begin{equation}\label{eq:ren_omega2}
 \omega = 4 - \sum_{j=1}^{n+1} \dim(A^{s_j}) 
 = 4 - \sum_{i=1}^{\mathrm{p}} [\dim(A_i) \ext_{\mathbf{s}}(A_i) + \der_{\mathbf{s}}(A_i)],
\end{equation}
where $\mathrm{p}$ is the number of basic generators, the functions $\ext_{\mathbf{s}}(\cdot)$, $\der_{\mathbf{s}}(\cdot)$ were defined in Section~\ref{sec:aux} and $\mathbf{s}=(s_1,\ldots,s_{n+1})$.

\section{Interacting fields with IR regularization}\label{sec:interacting}

In order to define interacting fields we introduce the extended scattering operator with the IR regularization:
\begin{equation}\label{eq:S_extended}
 S(\mathpzc{g}+\mathpzc{h}) = \Texp\left( \ri\, \int \rd^4 x \, \sum_{l=1}^\mathrm{q} e_l g_l(x) \mathcal{L}_l(x) + \ri \int \rd^4 x \, h(x)C(x)\right),
\end{equation}
where 
\begin{equation}
 \mathpzc{g}=\sum_{l=1}^\mathrm{q} e_lg_l\otimes \mathcal{L}_l
 ~~~~\textrm{and}~~~~
 \mathpzc{h}=h\otimes C,~~~h\in\mathcal{S}(\R^4),~C\in\Fh.
\end{equation}
The extended S-matrix with the IR regularization is a formal power series in $e_1,\ldots,e_\mathrm{q}$ and $h$.

Using the so-called Bogoliubov formula \cite{bogoliubov1959introduction} one defines the interacting fields with the IR regularization:  the advanced field
\begin{multline}\label{eq:bogoliubov_adv}
 C_\adv(\mathpzc{g};x) 
 :=(-\ri) \frac{\delta}{\delta h(x)}
  S(\mathpzc{g}+\mathpzc{h})S(\mathpzc{g})^{-1}\bigg|_{\mathpzc{h}=0} 
 =
 \sum_{n=0}^\infty \frac{\ri^n}{n!} \sum_{l_1,\ldots,l_n=1}^\mathrm{q}e_{l_1}\ldots e_{l_n} \\\times
 \int\rd^4 y_1\ldots\rd^4 y_n\,
 g_{l_1}(y_1)\ldots g_{l_n}(y_n)
 ~\Adv(\mathcal{L}_{l_1}(y_1),\ldots,\mathcal{L}_{l_n}(y_n);C(x)) 
\end{multline}
and the retarded field
\begin{multline}\label{eq:bogoliubov_ret}
 C_\ret(\mathpzc{g};x) 
 :=(-\ri) \frac{\delta}{\delta h(x)}
 S(\mathpzc{g})^{-1} S(\mathpzc{g}+\mathpzc{h})\bigg|_{\mathpzc{h}=0} 
 =
 \sum_{n=0}^\infty \frac{\ri^n}{n!} \sum_{l_1,\ldots,l_n=1}^\mathrm{q}e_{l_1}\ldots e_{l_n}\\ \times
 \int\rd^4 y_1\ldots\rd^4 y_n\,
 g_{l_1}(y_1)\ldots g_{l_n}(y_n)
 ~\Ret(\mathcal{L}_{l_1}(y_1),\ldots,\mathcal{L}_{l_n}(y_n);C(x)). 
\end{multline}
Note that if the function $\mathpzc{g}$ has compact support then the advanced and retarded fields coincide with the free field $\normord{C(x)}$ outside the past and future of $\supp\,\mathpzc{g}$, respectively. Moreover, for any $C,C'\in\Fh$ it holds \cite{epstein1973role,dutsch1999local}
\begin{equation}
 [C_{\adv/\ret}(\mathpzc{g};x),C'_{\adv/\ret}(\mathpzc{g};x')] = 0
\end{equation}
if $x$ and $x'$ are spatially separated. In models with a single interaction vertex $\mathcal{L}$ we often write $C_{\adv/\ret}(g;x)$ instead of $C_{\adv/\ret}(\mathpzc{g};x)$, where $\mathpzc{g}=eg\otimes \mathcal{L}$.

% \begin{equation}
%  C_\adv(\mathpzc{g};x) =(\ri) \frac{\delta}{\delta h(x)}
%   S(\mathpzc{g})S(\mathpzc{g}+\mathpzc{h})^{-1}\bigg|_{\mathpzc{h}=0}  
% \end{equation}
% 
% \begin{equation}
%  \aT_\adv(\mathpzc{g};C_1(x_1),\ldots,C_{m}(x_m))  =\ri^m \frac{\delta}{\delta h_m(x_m)}\ldots\frac{\delta}{\delta h_1(x_1)}
%   S(\mathpzc{g})S(\mathpzc{g}+\mathpzc{h})^{-1}\bigg|_{\mathpzc{h}=0}  
% \end{equation}

Now, let $\mathpzc{h}=\sum_{j=1}^m h_j\otimes C_j$ be an admissible element of $\mathcal{S}(\R^4,E)\otimes\Fa$ (cf. Section~\ref{sec:generating}) and set
\begin{equation}\label{eq:S_extended2}
 S(\mathpzc{g}+\mathpzc{h}) = \Texp\left( \ri\, \int \rd^4 x \, \sum_{l=1}^\mathrm{q} e_l g_l(x) \mathcal{L}_l(x) + \ri \int \rd^4 x \, \sum_{j=1}^m h_j(x)C_j(x)\right).
\end{equation}
The time ordered-product of the following advanced fields with the IR regularization 
\begin{equation}
 C_{1,\adv}(\mathpzc{g};x_1), \ldots,  C_{m,\adv}(\mathpzc{g};x_m)
\end{equation}
is given by \cite{epstein1973role}
\begin{multline}\label{eq:time_ordered_adv}
 \T_\adv(\mathpzc{g};C_1(x_1),\ldots,C_{m}(x_m))  
 := (-\ri)^{m} \frac{\delta}{\delta h_m(x_m)}\ldots\frac{\delta}{\delta h_1(x_1)} 
 S(\mathpzc{g}+\mathpzc{h})S(\mathpzc{g})^{-1}\bigg|_{\mathpzc{h}=0}
 =
 \\
 \sum_{n=0}^\infty  \frac{\ri^n}{n!} \sum_{l_1,\ldots,l_n=1}^\mathrm{q}e_{l_1}\ldots e_{l_n} \int\rd^4  y_1\ldots\rd^4 y_n\, g_{l_1}(y_1)\ldots g_{l_n}(y_n)
 \\
 \Adv(\mathcal{L}_{l_1}(y_1),\ldots,\mathcal{L}_{l_n}(y_n);C_1(x_1),\ldots,C_n(x_m)). 
\end{multline}
Using an analogous formula one defines the time-ordered products of the retarded fields
\begin{multline}\label{eq:time_ordered_ret}
 \T_\ret(\mathpzc{g};C_1(x_1),\ldots,C_{m}(x_m))   
 := (-\ri)^{m} \frac{\delta}{\delta h_m(x_m)}\ldots\frac{\delta}{\delta h_1(x_1)} 
 S(\mathpzc{g})^{-1}S(\mathpzc{g}+\mathpzc{h})\bigg|_{\mathpzc{h}=0}
 =
 \\
 \sum_{n=0}^\infty  \frac{\ri^n}{n!} \sum_{l_1,\ldots,l_n=1}^\mathrm{q}e_{l_1}\ldots e_{l_n} \int\rd^4  y_1\ldots\rd^4 y_n\, g_{l_1}(y_1)\ldots g_{l_n}(y_n)
 \\
 \Ret(\mathcal{L}_{l_1}(y_1),\ldots,\mathcal{L}_{l_n}(y_n);C_1(x_1),\ldots,C_n(x_m)).
\end{multline}
The advanced and retarded fields with the IR regularization as well as their time-ordered products are formal power series in the coupling constants $e_1,\ldots,e_\mathrm{q}$. The formulas \eqref{eq:time_ordered_adv} and \eqref{eq:time_ordered_ret} hold for arbitrary $C_1,\ldots,C_m\in\Fh$ and thus determine uniquely the multi-linear map
\begin{equation}
 \Fa^n \ni (C_1,\ldots,C_m)\mapsto  \T_{\adv/\ret}(\mathpzc{g};C_1(x_1),\ldots,C_{n}(x_n)) \in \mathcal{S}'(\R^4,L(\mathcal{D}_0)) \llbracket e\rrbracket,
\end{equation}
where $\mathcal{S}'(\R^4,L(\mathcal{D}_0)) \llbracket e\rrbracket$ is the space of formal power series in the coupling constants $e_1,\ldots,e_\mathrm{q}$ with coefficients in $\mathcal{S}'(\R^4,L(\mathcal{D}_0))$. The time-ordered products of advanced and retarded fields are not $\Fa$ products in the sense of the definition from Section~\ref{sec:F_prod}. However, they satisfy the following generalizations of the axioms \ref{axiom4}, \ref{axiom5}, \ref{axiom6} and \ref{axiom2} which hold for all $C_1,\ldots,C_n\in\Fh$:
\begin{enumerate}[label=\bf{A'\!.\arabic*},leftmargin=*]
\setcounter{enumi}{3}
\item$\T_{\adv/\ret}(\mathpzc{g};\emptyset)=\id$, $\T_{\adv/\ret}(\mathpzc{g};C(x))=\,C_{\adv/\ret}(\mathpzc{g};x)$,
\begin{equation}
\T_{\adv/\ret}(\mathpzc{g};C_1(x_1),\ldots,C_{n}(x_n),1(x_{n+1})) = \T_{\adv/\ret}(\mathpzc{g};C_1(x_1),\ldots,C_{n}(x_n)),
\end{equation}
where $1$ on the LHS of the above equality is the unity in $\Fa$.
\item Graded symmetry:
\begin{equation}
 \T_{\adv/\ret}(\mathpzc{g};C_1(x_1),\ldots,C_{n}(x_n)) = (-1)^{\mathbf{f}(\pi)}\T_{\adv/\ret}(\mathpzc{g};C_{\pi(1)}(x_{\pi(1)}),\ldots,C_{\pi(n)}(x_{\pi(n)})),
\end{equation}
where $\pi$ is any permutation of the set $\{1,\ldots,n\}$ and $\mathbf{f}(\pi)\in\Z/2\Z$ is the number of transpositions in $\pi$ that involve a pair of fields with odd fermion number.
\item\label{axiomPrime:causality} Causality:
\begin{multline}
 \T_{\adv/\ret}(\mathpzc{g};C_1(x_1),\ldots,C_{n}(x_n))
 \\
 = \T_{\adv/\ret}(\mathpzc{g};C_1(x_1),\ldots,C_{m}(x_m))\T_{\adv/\ret}(\mathpzc{g};C_{m+1}(x_{m+1}),\ldots,C_{n}(x_n))
\end{multline}
if none of the points $x_1,\ldots,x_m$ is in the causal past of any of the points $x_{m+1},\ldots,x_n$.
\setcounter{enumi}{1}
\item If $\mathbf{f}(C_1)+\ldots+\mathbf{f}(C_n)\neq0$ then
\begin{equation}
 (\Omega|\T_{\adv/\ret}(\mathpzc{g};C_1(x_1),\ldots,C_{n}(x_n)) \Omega) = 0.
\end{equation}
\end{enumerate}
For example, condition \ref{axiomPrime:causality} follows immediately from the definitions \eqref{eq:time_ordered_adv}, \eqref{eq:time_ordered_ret} of the time-ordered products of the advanced and retarded fields and the causal factorization property \eqref{eq:causal_fact}. Note that because of the presence of the switching functions the time-ordered products of the advanced and retarded fields are not translationally covariant. 

\section{Wightman and Green functions with IR regularization}\label{sec:W_G_IR}

The Wightman function of the advanced/retarded fields with the IR regularization are by definition the VEVs of the product of the advanced/retarded fields
\begin{equation}\label{eq:wightman_IR}
 (\Omega| C_{1,\adv/\ret}(\mathpzc{g};x_1),\ldots,C_{m,\adv/\ret}(\mathpzc{g};x_m) \Omega),
\end{equation}
whereas the Green functions of the advanced/retarded fields with the IR regularization are defined by
\begin{equation}\label{eq:green_IR}
 (\Omega| \T_{\adv/\ret}(\mathpzc{g};C_1(x_1),\ldots,C_{m}(x_m)) \Omega),
\end{equation}
where $\mathpzc{g}=\sum_{l=1}^\mathrm{q} e_lg_l\otimes\mathcal{L}_l$. In order to obtain the physical Wightman and Green functions one has to remove the IR regularization by taking the limit $g_l(x)\to 1$ in an appropriate sense. To be more specific, we consider the so-called adiabatic limit \cite{epstein1973role}, which is discussed in Section~\ref{sec:weak_adiabatic_general}, of each term in the expansion  of \eqref{eq:wightman_IR} and \eqref{eq:green_IR} in powers of the coupling constants $e_1,\ldots,e_\mathrm{q}$. In what follows, we list a number of formulas which will be needed in Chapter~\ref{Sec:existence_wAL}, where the proof of existence of this limit is presented. 

Let us start with the definition of the generalized advanced and retarded products with a partition. To this end, we introduce the notation
\begin{equation}
\begin{aligned}
 I &= (B_1(y_1),\ldots,B_n(y_n)),
 \\
 J &= (C_1(x_1),\ldots,C_m(x_m)),
\end{aligned}
\end{equation}
fix a strictly increasing sequence of natural numbers
\begin{equation}
 P = (p_0,\ldots,p_k),
\end{equation}
such that $p_0=0$ and $p_k=m$ and define a partition $J_1,\ldots,J_k$ of $J$ such that $J_j$ is a contiguous subsequence of $J$ starting at the position $p_j+1$ and ending at $p_{j+1}$; in particular it holds $J=J_1+\ldots+J_k$. The generalized advanced product with the partition $P$ is defined by
\begin{multline}\label{eq:def_gen_adv}
 \Adv(I;J;P):=
 \\
 (-\ri)^{n+m} 
 \frac{\delta}{\delta \mathpzc{h}_k(J_k)}\ldots\frac{\delta}{\delta \mathpzc{h}_1(J_1)}\frac{\delta}{\delta \mathpzc{g}(I)}~
  S(\mathpzc{g}+\mathpzc{h}_1)S(\mathpzc{g})^{-1} \ldots  S(\mathpzc{g}+\mathpzc{h}_k)S(\mathpzc{g})^{-1}\bigg|_{\substack{\mathpzc{g}=0\\\mathpzc{h}=0}},
\end{multline}
where $\mathpzc{g}=\sum_{j=1}^n g_j\otimes B_j$ and $\mathpzc{h}_k=\sum_{j=p(k-1)}^{p(k)} h_j\otimes C_j$. The generating functional $S(\mathpzc{g})$ and the notation used in the above equation was introduced in Section~\ref{sec:generating}. Similarly, the generalized retarded product with the partition $P$ is given by
\begin{multline}\label{eq:def_gen_ret}
 \Ret(I;J;P):=
 \\
 (-\ri)^{n+m} 
 \frac{\delta}{\delta \mathpzc{h}_k(J_k)}\ldots\frac{\delta}{\delta \mathpzc{h}_1(J_1)}\frac{\delta}{\delta \mathpzc{g}(I)}~
 S(\mathpzc{g})^{-1}S(\mathpzc{g}+\mathpzc{h}_1) \ldots S(\mathpzc{g})^{-1} S(\mathpzc{g}+\mathpzc{h}_k)\bigg|_{\substack{\mathpzc{g}=0\\\mathpzc{h}=0}}.
\end{multline}
The generalized $\Dif$ product with the partition $P$ is by definition
\begin{equation}\label{eq:def_dif_gen}
 \Dif(I;J;P):= \Adv(I;J;P) - \Ret(I;J;P).
\end{equation}
Let us consider a few special cases. If $k=1$, i.e. $P=(0,m)$ we have
\begin{equation}
\begin{aligned}
 \Adv(I;J;P) = \Adv(I;J),
 \\
 \Ret(I;J;P) = \Ret(I;J),
 \\
 \Dif(I;J;P) = \Dif(I;J),
\end{aligned}
\end{equation}
where the products on the RHS are the standard $\Adv$, $\Ret$, $\Dif$ products defined in Section~\ref{sec:generating}. If $n=0$, i.e. $I=\emptyset$ we have
\begin{equation}\label{eq:def_gen_T}
 \Adv(\emptyset;J;P) = \Ret(\emptyset;J;P) =\T(J_1)\T(J_2)\ldots \T(J_k) =:\T(J;P).
\end{equation}
It follows from the very definition that the products $\Adv(I;J;P)$, $\Ret(I;J;P)$, $\Dif(I;J;P)$ are $\Fa$ products and they are graded-symmetric in variables denoted collectively by $I$. Moreover, if $I$ is the list of the interaction vertices
\begin{equation}\label{eq:I_lagrangians}
  I = (\mathcal{L}_{l_1}(y_1),\ldots,\mathcal{L}_{l_n}(y_n)),
\end{equation}
then we have
\begin{equation}\label{eq:gen_adv_ret_dif_std}
\begin{aligned}
 \Adv(I;J;P) &= (-\ri)^{n} \frac{\delta}{\delta \mathpzc{g}(I)}\T_\adv(\mathpzc{g};J_1) \ldots \T_\adv(\mathpzc{g};J_k) ,
 \\
 \Ret(I;J;P) &=  (-\ri)^{n} \frac{\delta}{\delta \mathpzc{g}(I)}\T_\ret(\mathpzc{g};J_1) \ldots \T_\ret(\mathpzc{g};J_k) ,
\end{aligned} 
\end{equation}
since by assumption the interaction vertices have vanishing fermion number. Because $\T_{\adv/\ret}(\mathpzc{g};C(x))=\,C_{\adv/\ret}(\mathpzc{g};x)$, the terms of order $e_{l_1}\ldots e_{l_n}$ in the formal expansions of the Wightman and Green functions may be expressed by the VEVs of the products $\Adv(I;J;P)$ and $\Ret(I;J;P)$. More precisely, the terms of order $e_{l_1}\ldots e_{l_n}$ in the expansion of the expressions \eqref{eq:wightman_IR} and \eqref{eq:green_IR} in powers of the coupling constants have the form
\begin{multline}\label{eq:w_g_adv}
 \int\rd^4 y_1\ldots\rd^4 y_n\,g_{l_1}(y_1)\ldots g_{l_n}(y_n)  
 \\
 (\Omega|\Adv(\mathcal{L}_{l_1}(y_1),\ldots,\mathcal{L}_{l_n}(y_n);C_1(x_1),\ldots,C_m(x_m);P)\Omega),
\end{multline}
where $P=(0,1,2,\ldots,m)$ in the case of Wightman functions and $P=(0,m)$ in the case of Green functions (in what follows we consider only these two types of sequences~$P$). Note that $J_1=(C_1(x_1)),\ldots,J_k=(C_{k}(x_k))$ and $k=m$ if $P=(0,1,2,\ldots,m)$ and $J_1=J=(C_1(x_1),\ldots,C_m(x_m))$ and $k=1$ if $P=(0,m)$. The formula \eqref{eq:w_g_adv} holds for the Wightman and Green functions with the IR regularization defined in terms of the advanced fields. In the case of the Wightman and Green functions with the IR regularization defined in terms of the retarded fields the product $\Adv(I;J;P)$ has to be replaced by $\Ret(I;J;P)$.

Note that the Green functions are expressed by the ordinary advanced product $\Adv(I;J)$ introduced in Section \ref{sec:generating}. This is not true for the Wightman functions. In the formula \eqref{eq:w_g_adv} the arguments of $\Adv(I;J;P)$ collectively denoted by $I$ are always of the form \eqref{eq:I_lagrangians} and the identity \eqref{eq:gen_adv_ret_dif_std} apply. However, in the inductive proof of the existence of the limit $g_l(x)\to1$ of \eqref{eq:w_g_adv} presented in Section~\ref{Sec:existence_wAL} we will have to consider also the case when $I$ is the list of sub-polynomials of the interaction vertices
\begin{equation}
  I = (\mathcal{L}^{(s_1)}(y_1),\ldots,\mathcal{L}^{(s_n)}(y_n)).
\end{equation}

We are now going to derive the formulas which play a key role in the proof of the existence of the limit $g_l(x)\to1$. Using the identity
\begin{multline}
 S(\mathpzc{g})^{-1} S(\mathpzc{g}+\mathpzc{h}_1)~ \ldots~ S(\mathpzc{g})^{-1} S(\mathpzc{g}+\mathpzc{h}_k)
 \\
 = S(\mathpzc{g})^{-1}\left(S(\mathpzc{g}+\mathpzc{h}_1)~ \ldots~ S(\mathpzc{g})^{-1} S(\mathpzc{g}+\mathpzc{h}_k)S(\mathpzc{g})^{-1}\right)S(\mathpzc{g})
\end{multline}
and the definitions of $\T(I)$, $\aT(I)$, $\Adv(I;J;P)$, $\Ret(I;J;P)$ we obtain:
\begin{equation}
 \Ret(I;J;P)
 =
 \sum^I_{\substack{I_1,I_2,I_3\\I_1+I_2+J+I_3=\pi(I+J)}} (-1)^{|I_1|+\mathbf{f}(\pi)}~ \aT(I_1) \Adv(I_2;J;P) \T(I_3)
\end{equation}
and
\begin{equation}\label{eq:dif_T_Adv_aT}
 \Dif(I;J;P)
 =
 -\sum^I_{\substack{I_1,I_2,I_3\\I_1+I_2+J+I_3=\pi(I+J)\\I_2 \neq I }} (-1)^{|I_1|+\mathbf{f}(\pi)} \aT(I_1) \Adv(I_2;J;P) \T(I_3).
\end{equation}
It follows from the identities
\begin{equation}
 S(\mathpzc{g})^{-1} S(\mathpzc{g}) = \id = S(\mathpzc{g}) S(\mathpzc{g})^{-1} 
\end{equation}
and the definitions of $\T(I)$, $\aT(I)$ that for $I\neq \emptyset$ it holds
\begin{equation}
 \sum^I_{\substack{I_1,I_2\\I_1+I_2=\pi(I)}} (-1)^{|I_1|+\mathbf{f}(\pi)} \aT(I_1) \T(I_2) = 0
\end{equation}
and 
\begin{equation}
 \sum^I_{\substack{I_1,I_2\\I_1+I_2=\pi(I)}} (-1)^{|I_2|+\mathbf{f}(\pi)} \T(I_1) \aT(I_2) = 0.
\end{equation}
Using the above equations and the formula \eqref{eq:dif_T_Adv_aT} we obtain two important identities
\begin{equation}\label{eq:dif_com}
 \Dif(I;J;P)
 =
 -\sum^I_{\substack{I_1,I_2,I_3\\I_1+I_2+J+I_3=\pi(I+J)\\I_2 \neq I }} (-1)^{|I_1|+\mathbf{f}(\pi)} [\aT(I_1) \Adv(I_2;J;P), \T(I_3)]
\end{equation}
and
\begin{equation}\label{eq:dif_com2}
 \Dif(I;J;P)
 =
 -\sum^I_{\substack{I_1,I_2,I_3\\I_1+I_2+J+I_3=\pi(I+J)\\I_2 \neq I }} (-1)^{|I_1|+\mathbf{f}(\pi)} [\aT(I_1), \Adv(I_2;J;P)] \T(I_3),
\end{equation}
where $[\cdot,\cdot]$ is the graded commutator \eqref{eq:com}. Note that using any of the above formulas \eqref{eq:dif_T_Adv_aT}, \eqref{eq:dif_com}, \eqref{eq:dif_com2} we can express the product $\Dif(I;J;P)$ with $|I|=n$ by the anti-time-ordered products $T(I_1)$, the time-ordered products $\T(I_3)$ and the generalized advanced products $\Adv(I_2;J;P)$ with $|I_1|,|I_2|,|I_3|<n$. 

Finally, note that as a consequence of the causal factorization property \eqref{eq:causal_fact} the products $\Adv(I;J;P)$, $\Ret(I;J;P)$, $\Dif(I;J;P)$ have the same support properties \eqref{eq:supp_adv_ret_dif} as the standard advanced, retarded and $\Dif$ products introduced in Section~\ref{sec:generating}. The proof of the existence of the weak adiabatic limit given in Section~\ref{Sec:existence_wAL} is based on the above-mentioned support properties of the products $\Adv(I;J;P)$, $\Ret(I;J;P)$, $\Dif(I;J;P)$ and Equations \eqref{eq:dif_com} and \eqref{eq:dif_com2}. The presence of commutator in these formulas is very useful because it allows to take advantage of Part~(B) of Lemma \ref{lem:aux_lemma}. 

Let us remark that in the proof due to Blanchard and Seneor a different, more complicated, formula for $\Dif(I;J;P)$ was used. It allows to express the product $\Dif(I;J;P)$ with $|I|=n$ as a linear combination of terms which are obtained by taking iterated graded-commutators of the standard advanced products and $\Adv(I';J;P)$ with $|I'|<n$. This formula is most easily derived with the use of the Steinmann's arrow calculus \cite{epstein1973role,steinmann1960wightman}. We do not present it here since it will not be needed in our proof.  

\section{Examples of interacting models}\label{sec:examples}

\subsection{Spinor QED}\label{sec:spinor_qed}

The quantum electrodynamics (QED) is the theory which describes the interactions between photons and electrons. In its perturbative formulation photons are the quanta of the free massless electromagnetic field and electrons -- the quanta of the massive Dirac spinor field. In EG approach the quantized electromagnetic field has to be defined using the Gupta-Bleuler formalism \cite{bleuler1950neue,gupta1950theory,scharf2016gauge,strocchi2013introduction}\footnote{An alternative could be the use of the string-local vector potential introduced in \cite{mund2004string}. The generalization of the EG approach to string-local fields is under investigations \cite{mund2016string,gracia2017chirality}.}. In this formalism the electromagnetic potential $A_\mu$ is well-defined local field of dimension $1$ which is an operator-valued distribution acting on the Fock space with indefinite Poincar{\'e} invariant inner product (cf. Section~\ref{sec:Wick}). Its components are solutions of the wave equation. The Maxwell equations are satisfied only if an additional constraint $\partial_\mu A^\mu =0$ is imposed. The spinor field is defined in the standard way. Besides QED (which we also call the massive spinor QED) we also consider the model with massless electrons which is called the massless spinor QED.  

The basic generators of the massive and massless spinor QED are the components of the massless vector field $A_\mu$ and the massive or massless spinor field $\psi_a$ and its Hermitian conjugate  $\psi_a^*$. As a result the set of the basic generators
\begin{equation}
 \mathcal{G}_0 = \{ A_\mu,~\mu\in\{0,1,2,3\},~~\psi_a,~\psi^*_a,~a\in\{1,2,3,4\}\}
\end{equation}
contains $12$ elements. The components of the massless vector field are denoted by $A_\mu$, whereas the symbols $A_i\in\mathcal{G}_0$ and $A^r\in\Fa$ denote any of the basic generators and the monomial \eqref{eq:wick_def}, respectively. We have $\dim(A_\mu)=1$, $\dim(\psi_a)=3/2$, $\mathbf{f}(A_\mu)=0$, $\mathbf{f}(\psi_a)=-\mathbf{f}(\psi_a^*)=1$, $\mathbf{q}(\psi)=-\mathbf{q}(\psi^*)=-1$ (the spinor field $\normord{\psi(f)}$ acting on the vacuum creates a state with one negatively-charged electron).

There is one coupling constant $e$ and one interaction vertex
\begin{equation}
 \mathcal{L} = A_\mu j^\mu,
\end{equation}
where 
\begin{equation}\label{eq:electric_spinor}
 j^\mu = \overline{\psi}\gamma^\mu\psi
\end{equation}
is the free electric current. We have $\dim(\mathcal{L})=4$ and set $\CC=0$ in \ref{norm:sd}. The list of all sub-polynomials of the interaction vertex was given in \eqref{eq:qed_sub_pol}

Finally, because the interaction vertex does not contain fields with derivatives, the distribution
\begin{equation}\label{eq:qed_sub}
 (\Omega|\T(\mathcal{L}^{(s_1)}(x_1),\ldots,\mathcal{L}^{(s_{n+1})}(x_{n+1}))\Omega)
\end{equation}
vanishes if any of the super-quadri-indices $s_1,\ldots,s_{n+1}$ involves fields with derivatives. As a result, in the case of the present model we can restrict attention to super-quadri-indices $s_1,\ldots,s_{n+1}$ which do not involve fields with derivatives. It holds
\begin{equation}
 \sd(\,(\Omega|\T(\mathcal{L}^{(s_1)}(x_1),\ldots,\mathcal{L}^{(s_{n+1})}(x_{n+1}))\Omega)\,) \leq \omega + 4n,
\end{equation}
where
\begin{equation}\label{eq:omega_spinor_QED}
 \omega = 4 - \sum_{\mu=0}^3\ext_{\mathbf{s}}(A_\mu) -\frac{3}{2}\sum_{a=1}^4\ext_{\mathbf{s}}(\psi_a) - \frac{3}{2}\sum_{a=1}^4\ext_{\mathbf{s}}(\overline{\psi}_a).
\end{equation}
The above equation for $\omega$ coincides with the identity \eqref{eq:ren_omega2} for lists $\mathbf{s}=(s_1,\ldots,s_{n+1})$ of super-quadri-indices which do not involve fields with derivatives. Moreover, because of the axiom \ref{axiom2} the VEV of the distribution \eqref{eq:qed_sub} vanishes unless $\sum_{a=1}^4\ext_{\mathbf{s}}(\psi_a)=\sum_{a=1}^4\ext_{\mathbf{s}}(\overline{\psi}_a)$. As a result in non-trivial cases $\omega\in\Z$.

\subsection{Scalar QED}\label{sec:scalar_qed}

The scalar QED is a model very similar to the spinor QED. Its basic generators are the components of the massless vector field and the massive or massless charged scalar field. The massless vector field is defined with the use of the Gupta-Bleuler formalism as in the case of the spinor QED. The model with massive scalar field is called the massive scalar QED. The model with massless scalar field -- the massless scalar QED. The set of the basic generators
\begin{equation}
 \mathcal{G}_0=\{A_\mu,~\mu\in\{0,1,2,3\},~~\phi,~~\phi^*\},
\end{equation}
contains $6$ elements. There are no fermionic fields in the model. We have $\dim(A_\mu)=\dim(\phi)=\dim(\phi^*)=1$, $\mathbf{q}(A_\mu)=0$, $\mathbf{q}(\phi)=-\mathbf{q}(\phi^*)=-1$. 

By definition there is one coupling constant $e$ and one interaction vertex
\begin{equation}\label{eq:scalar_qed_std_vertex}
 \mathcal{L} = A_\mu j^\mu,
\end{equation}
where
\begin{equation}\label{eq:electric_scalar}
 j^\mu = \ri [\varphi^*(\partial^\mu\varphi) -(\partial^\mu\varphi^*)\varphi]
  =: \ri \varphi^*\overset{\leftrightarrow}{\partial}{}^\mu\varphi
\end{equation}
is the free electric current. We have $\dim(\mathcal{L})=4$ and set $\CC=0$ in \ref{norm:sd}. Note that the vertex
\begin{equation}\label{eq:scalar_qed_vertex}
 A^\mu A_\mu\varphi^* \varphi
\end{equation}
which appears in the standard Lagrangian of the scalar QED is not included in the list of the interaction vertices. This assumption is standard in the context of the causal perturbation theory \cite{scharf2014,dutsch1993scalar}. The term of the form \eqref{eq:scalar_qed_std_vertex} appears in the expansion of the scattering matrix and the interacting fields in powers of the coupling constant as a consequence of the gauge invariance. In particular, in the case of the second order correction to the scattering matrix the above statement follows form the definition \eqref{eq:scalar_qed_derivatives} of the time-ordered product of the fields $\phi$ and $\phi^*$, which is a consequence of the Ward identities.

Finally, we note that in the case of the scalar QED $\omega$ given by \eqref{eq:ren_omega2} acquires the form
\begin{equation}\label{eq:omega_scalar_QED}
 \omega = 4 - \sum_{\mu=0}^3\ext_{\mathbf{s}}(A_\mu) -\ext_{\mathbf{s}}(\phi) - \ext_{\mathbf{s}}(\phi^*) - \der_{\mathbf{s}}(\phi) - \der_{\mathbf{s}}(\phi^*).
\end{equation}

\subsection{Scalar model}\label{sec:the_model}

We close the list of examples with a simple model based on two scalar, charge-free fields: a massive field $\psi$ and a massless field $\varphi$, forming the set of basic generators
\begin{equation}
 \mathcal{G}_0=\{\varphi,~\psi\}.
\end{equation}
Interaction is supplied by one interaction vertex in the form of monomial
\begin{equation}
 \mathcal{L}=\frac{1}{2}\,\varphi\psi^2,
\end{equation}
with the corresponding coupling constant $e$. Since $\dim(\psi)=\dim(\varphi)=1$ it holds $\dim(\mathcal{L})=3$ and the model is super-renormalizable according to the standard classification. The interaction vertex is a monomial. Let us list all of its sub-monomials
\begin{equation}
 \mathcal{L},~ \psi^2, ~\psi\varphi,~ \varphi,~ \psi,~1.
\end{equation}

In what follows, we shall refer to this example as the scalar model. To our best knowledge, this model has not been considered in the literature before. Our motivation for investigating it is twofold. First, it may be treated as a toy model for considering IR problem in the massive spinor QED (this issue will not be pursued in the thesis). Second, the model will serve as an illustration of usefulness of our generalized normalization condition \ref{norm:sd}. As mentioned in Sections \ref{sec:sd} and \ref{sec:ren} there are, a priori, two possible choices of the constant $\CC$ in the this condition. In the case of the standard choice $\CC=0$ the model is super-renormalizable. It turns out, however, that in this case the condition \ref{norm:sd} is too restrictive and it is impossible to define the time-ordered products such that the normalization condition needed for the existence of the Wightman and Green functions is satisfied. In contrast, the choice c=1 puts the model in the class of renormalizable (not super-renormalizable) theories, but enables the definition of physical Wightman and Green functions. Moreover we have
\begin{equation}\label{eq:omega_scalar_model}
  \omega = 4 -\ext_{\mathbf{s}}(\varphi) - \ext_{\mathbf{s}}(\psi). 
\end{equation}
The above equation is a special case of \eqref{eq:ren_omega2} valid if the super-quadri-indices from the list $\mathbf{s}$ do not involve fields with derivatives. Since the interaction vertex does not contain such fields this assumption does not not imply any loss generality.

%%%%%%%%%%%%%%%%%%%%%%%%%%%%%%%%%%%%%%%%%%%%%%%%%%%%%%%%%%%%%%%%%%%%%%%%%%%%%%%%%%%%%%%%%%%%%%%%%%%%
%%%%%%%%%%%%%%%%%%%%%%%%%%%%%%%%%%%%%%%%%%%%%%%%%%%%%%%%%%%%%%%%%%%%%%%%%%%%%%%%%%%%%%%%%%%%%%%%%%%%
%                                    ADIABATIC
%%%%%%%%%%%%%%%%%%%%%%%%%%%%%%%%%%%%%%%%%%%%%%%%%%%%%%%%%%%%%%%%%%%%%%%%%%%%%%%%%%%%%%%%%%%%%%%%%%%%
%%%%%%%%%%%%%%%%%%%%%%%%%%%%%%%%%%%%%%%%%%%%%%%%%%%%%%%%%%%%%%%%%%%%%%%%%%%%%%%%%%%%%%%%%%%%%%%%%%%%

\chapter{Adiabatic limits}\label{sec:adiabatic_limits}

The S-matrix, the interacting fields and the Wightman and Green functions considered in Chapter~\ref{sec:models} were constructed assuming that all the coupling constants are replaced by spacetime-dependent functions vanishing rapidly at infinity. This assumption is clearly unphysical but it makes the scattering theory easy -- the asymptotic states of the interacting theory are simply the multiparticle Fock states and the free fields used in the construction are not only auxiliary objects needed to built the interaction vertices but are also the asymptotic fields of the theory. In order to make physical predictions one should define the theory with switching function $g_l\equiv1$. However, even in the case of purely massive theories the expressions for the S-matrix \eqref{eq:S_matrix}, the interacting operators \eqref{eq:bogoliubov_adv}, \eqref{eq:bogoliubov_ret} or Wightman and Green functions \eqref{eq:wightman_IR}, \eqref{eq:green_IR} in general do not make sense for $g_l\equiv 1$. To circumvent this problem the so-called adiabatic limit is carried out. 

For any $g\in\mathcal{S}(\R^4)$ such that $g(0)=1$ one defines the one parameter family of Schwartz functions
\begin{equation}
 g_\epsilon(x)= g(\epsilon x),\quad \epsilon>0,
\end{equation} 
or equivalently in the momentum space
\begin{equation}\label{eq:adiabatic_momentum}
 \F{g}_\epsilon(q) = \frac{1}{\epsilon^4} \F{g}(q/\epsilon),~~~~\int \mP{q}\, \F{g}(q) = 1.
\end{equation}
Note that
\begin{equation}\label{eq:adiabatic_g_limit}
 \lim_{\epsilon\searrow0}g_\epsilon(x)=1,~~~~~ \lim_{\epsilon\searrow0}\F{g}_\epsilon(q)=(2\pi)^4\delta(q)
\end{equation}
in the sense of distributions.

Assume that each switching function $g_l$ is replaced by the one-parameter family $g_{l,\epsilon}$ defined above. In order to define the physical S-matrix, the physical interacting fields and their time-ordered products or the physical Wightman and Green functions one replaces $\mathpzc{g}=\sum_{l=1}^\mathrm{q} e_lg_l\otimes \mathcal{L}_l$ which appear in the IR-regularized definitions of these objects given in the previous chapter by $\mathpzc{g}_\epsilon=\sum_{l=1}^\mathrm{q} e_lg_{l,\epsilon}\otimes \mathcal{L}_l$. In this way a one-parameter family of objects with the IR regularization parametrized by $\epsilon>0$ is obtained. The corresponding physical objects are obtained by taking the limit $\epsilon\searrow 0$, if the limit exists. The procedure described above is called the adiabatic limit. One can distinguish three more particular types of this procedure:
\begin{enumerate}[leftmargin=*,label={(\arabic*)}]
 \item The algebraic adiabatic limit: The existence of the net of local abstract algebras of interacting fields (the word limit is used here in a figurative sense). 
 \item The weak adiabatic limit: the existence of the Wightman and Green functions and the existence of the vacuum state on the net of local algebras. 
\item The strong adiabatic limit: the existence of the physical S-matrix.  
\end{enumerate}
The thesis is devoted to the study of the existence of the weak adiabatic limit in models containing massless particles. Let us, however, briefly describe all three of the above types of the adiabatic limit.

\section{Algebraic adiabatic limit}\label{sec:algebraic_adiabatic}

The existence of the local abstract algebras of interacting operators is based on the following observation made in \cite{brunetti2000microlocal}. For simplicity, we assume that there is only one interaction vertex $\mathcal{L}$. The switching function is denoted by $g$, the coupling constant -- by $e$ and the interacting operators by $C_\adv(g;h)=C_\adv(\mathpzc{g};h)$, where $\mathpzc{g}=eg\otimes \mathcal{L}$. Let $\mathcal{O}$ be an open bounded subset of $\R^4$ and $g,g'\in\mathcal{D}(\R^4)$ coincide in a neighborhood of $J^+(\mathcal{O})\cap J^-(\mathcal{O})$, where $J^\pm(\mathcal{O}):=\mathcal{O}+\overline{V}^\pm$ is the causal future/past of the set $\mathcal{O}$. Then, as a result of the causal factorization formula \eqref{eq:causal_fact}, there exists a unitary operator $V(g',g)$ such that
\begin{equation}
 V(g',g) C_\adv(g;h) V(g',g)^{-1} = C_\adv(g';h)
\end{equation}
for any $C\in\Fa$ and $h\in\mathcal{D}(\R^4)$, $\supp\,h\subset\mathcal{O}$. Let us introduce the following denotation
\begin{equation}
 \mathcal{D}_\mathcal{O} := \{ g\in \mathcal{D}(\R^4) \,:\, g \equiv 1 \textrm{ on a neighborhood of } J^+(\mathcal{O})\cap J^-(\mathcal{O})\}
\end{equation}
for bounded open sets $\mathcal{O}\subset\R^4$ and define for $h\in\mathcal{D}(\R^4)$ such that $\supp\, h\subset \mathcal{O}$ the function
\begin{equation}\label{eq:algebraic_adiabatic_int_field}
 C_\adv(\cdot;h): \mathcal{D}_\mathcal{O} \ni g \mapsto C_\adv(g;h)\in L(\mathcal{D}_0)\llbracket e\rrbracket,
\end{equation}
where $L(\mathcal{D}_0)\llbracket e\rrbracket$ is the space of formal power series in $e$ with coefficients in $L(\mathcal{D}_0)$. We define the addition, multiplication and conjugation of these maps by the pointwise operations. The local algebra of interacting fields localized in $\mathcal{O}$ denoted by $\mathfrak{F}(\mathcal{O})$ is by definition the ${}^*$-algebra over $\C\llbracket e\rrbracket$ with unity $\id$ generated by $C_\adv(\cdot;h)$ with $h\in\mathcal{D}(\R^4)$, $\supp \,h\subset \mathcal{O}$ and $C\in\Fa$. For $\mathcal{O}'\subset\mathcal{O}$ we define the embedding
\begin{equation}
 \iota_{\mathcal{O}\mathcal{O}'}:~\mathfrak{F}(\mathcal{O}')\rightarrow \mathfrak{F}(\mathcal{O})
\end{equation}
by the restriction of the function \eqref{eq:algebraic_adiabatic_int_field} to $\mathcal{D}_{\mathcal{O}}\subset\mathcal{D}_{\mathcal{O}'}$. We have 
\begin{equation}
 \iota_{\mathcal{O}\mathcal{O}'}\circ\iota_{\mathcal{O}'\mathcal{O}''}= \iota_{\mathcal{O}\mathcal{O}''}.
\end{equation}
The global algebra of interacting fields $\mathfrak{F}$ is the inductive limit of the net $\mathcal{O}\to\mathfrak{F}(\mathcal{O})$ with canonical embeddings
\begin{equation}
 \iota_\mathcal{O}: \mathfrak{F}(\mathcal{O}) \to \mathfrak{F}.
\end{equation}
The action of the Poincar{\'e} transformations on $\mathfrak{F}$ is given by the automorphisms defined in the following way on the generators of $\mathfrak{F}$
\begin{equation}
 \alpha_{a,\Lambda}(C_\adv(\cdot;h)) = (\rho(\Lambda^{-1})C)_\adv(\cdot;h_{a,\Lambda}),
\end{equation}
where $h_{a,\Lambda}(x) = h(\Lambda^{-1}(x-a))$ and $\rho$ is the representation of the Lorentz group acting on $\Fa$ introduced in Section~\ref{sec:ff}.

As shown in \cite{dutsch2001algebraic,fredenhagen2015perturbative} the net $\mathfrak{F}(\mathcal{O})$ constructed above fulfills the axioms: (1)~isotony, (2) Poincar{\'e} covariance and (3) Einstein causality (see \cite{haag2012local} for their definition) in the sense of formal power series. The construction of the algebra $\mathfrak{F}$ and the net $\mathcal{O}\to\mathfrak{F}(\mathcal{O})$ outlined above is called in the literature the algebraic adiabatic limit.

In the case of theories with local symmetries the algebra $\mathfrak{F}$ has the interpretation of the algebra of interacting fields. From physical point of view more important is the algebra of observables, which is certain quotient algebra consisting of equivalence classes of gauge-invariant fields. The construction of the algebra of observables was carried out by D{\"u}tsch and Fredenhagen in \cite{dutsch1999local} in the case of QED and by Hollands in \cite{hollands2008renormalized} in the case of non-abelian Yang-Mills theories without matter.

\section{Weak adiabatic limit}\label{sec:weak_adiabatic_general}

The weak adiabatic limit allows to define the physical Wightman and Green functions. Let $\mathpzc{g}_\epsilon=\sum_{l=1}^\mathrm{q} e_lg_{l,\epsilon}\otimes \mathcal{L}_l$ and $g_{l,\epsilon}(x)=g_l(\epsilon x)$ and $C_1,\ldots,C_m\in \Fa$. Consider the following distributional limits, calculated separately for each term of the formal expansion in powers of the coupling constants: 
\begin{equation}\label{eq:wig_limit}
 \Wig(C_1(x_1),\ldots,C_m(x_m)) :=
 \lim_{\epsilon\searrow 0}\, (\Omega|C_{1,\adv/\ret}(\mathpzc{g}_\epsilon;x_1)\ldots C_{m,\adv/\ret}(\mathpzc{g}_\epsilon;x_m)\Omega)
\end{equation}
and
\begin{equation}\label{eq:gre_limit}
 \Gre(C_1(x_1),\ldots,C_m(x_m)) :=
 \lim_{\epsilon\searrow 0}\, (\Omega|\T_{\adv/\ret}(\mathpzc{g}_\epsilon;C_1(x_1),\ldots,C_{m}(x_m)\Omega).
\end{equation}
We say that a model has the weak adiabatic limit if the above limits exist and their values are 
\begin{enumerate}[label=(\arabic*),leftmargin=*]
 \item independent of the choice of $g_1,\ldots,g_\mathrm{q}\in\mathcal{S}(\R^4)$ such that $g_1(0)=\ldots=g_\mathrm{q}(0)=1$,
 \item the same in the case when all the fields in \eqref{eq:wig_limit} and \eqref{eq:gre_limit} are advanced and retarded.
\end{enumerate}
If the adiabatic limit exists then the formulas \eqref{eq:wig_limit} and \eqref{eq:gre_limit} determine the Schwartz distributions $\Wig(C_1(x_1),\ldots,C_m(x_m))\in\mathcal{S}'(\R^{4m})$ and $\Gre(C_1(x_1),\ldots,C_m(x_m))\in\mathcal{S}'(\R^{4m})$, which are called the Wightman and Green functions, respectively. In Chapter~\ref{sec:proof_scalar} we prove the existence of the weak adiabatic limit in a large class of models. Subsequently, in Section~\ref{sec:properties} we show that the Wightman and Green functions defined in this way have a number of properties (if the time-ordered products satisfy the normalization conditions listed in Sections \ref{sec:norm_con} and \ref{sec:one}). At this place let us only note that they are Poincar{\'e} covariant and satisfy the relativistic spectrum condition. 

Suppose that the weak adiabatic limit exists and the covariant inner product on $\mathcal{D}_0$ is positive-definite. Since the latter assumption does not hold in models with vector fields the result below do not apply to them. Moreover, for simplicity let us suppose that there is only one interaction vertex $\mathcal{L}$ and one coupling constant $e$. It turns out that the existence of the weak adiabatic limit may be used to define the Poincar{\'e} invariant vacuum state on the algebra $\mathfrak{F}$ introduced in the last section.  
\begin{dfn}\label{def:positivity}
The linear functional $\sigma:\,\mathfrak{F}\to\C\llbracket e\rrbracket$ is called a state if it is normalized $\sigma(\id)=1$, real $\sigma(\mathbf{B}^*)=\overline{\sigma(\mathbf{B})}$ and positive $\sigma(\mathbf{B}^* \mathbf{B})\geq 0$ for any $\mathbf{B}\in\mathfrak{F}$.
\end{dfn}
\begin{dfn}[\cite{dutsch1999local}]\label{dfn:positive_formal}
A formal power series $a\in\C\llbracket e\rrbracket$ is positive iff there exists $b\in\C\llbracket e\rrbracket$ such that $a=\overline{b} b$ and $b\neq 0$. A formal power series $a\in\C\llbracket e\rrbracket$ is non-negative iff $a$ is positive or $a=0$. 
\end{dfn}

\begin{thm}
The unique linear functional $\sigma:\mathfrak{F}\to\C\llbracket e\rrbracket$ given by
\begin{equation}\label{eq:state}
 \sigma(C_\adv(\cdot;h_1)\ldots C_\adv(\cdot;h_n)) = \lim_{\epsilon\searrow0} \, (\Omega|C_\adv(g_\epsilon;h_1)\ldots C_\adv(g_\epsilon;h_n) \Omega)
\end{equation}
for any $n\in\N_0$, $h_1,\ldots,h_n\in\mathcal{D}(\R^4)$ is a Poincar{\'e} invariant vacuum state.
\begin{proof}
Let $\mathcal{O}\subset\R^4$ be bounded open set such that $\supp\,h_1,\ldots,\supp\,h_n\subset\mathcal{O}$.  For any $\mathcal{O}$ and $g\in\mathcal{D}(\mathcal{O})$ we have $g_\epsilon\in\mathcal{D}(\mathcal{O})$ for all $\epsilon\in(0,1)$. It follows from the existence of the weak adiabatic limit that the limit on the RHS of Equation \eqref{eq:state} exists and is independent of $g$. As a result, the functional $\sigma$ is well-defined. It is evident that $\sigma$ is normalized and real. The positivity follows from the lemma below. The Poincar{\'e} invariance of the state $\sigma$,
\begin{equation}
 \sigma(\alpha_{a,\Lambda}(\mathbf{B})) = \sigma(\mathbf{B})~~~\textrm{for all}~~ \mathbf{B}\in\mathfrak{F},
\end{equation}
is a consequence of the Poincar{\'e} covariance of the Wightman functions.
\end{proof}
\end{thm}

\begin{lem}
If $\mathbf{B} \in L(\mathcal{D}_0)\llbracket e\rrbracket$ then $(\Omega|\mathbf{B}^* \mathbf{B}\Omega)\in\C\llbracket e\rrbracket$ is non-negative. 
\begin{proof}
We have $\mathbf{B}=\sum_{n=0}^\infty e^n \mathbf{B}_n$ for some $\mathbf{B}_n\in L(\mathcal{D}_0)$. Let $m\in\N_0\cup\{\infty\}$ be the largest number such that $(\Omega|\mathbf{B}_n^* \mathbf{B}_n^{\phantom{*}}\Omega)=0$ for all $n < m$. Then 
\begin{equation}
 |(\Omega|\mathbf{B}_{n}^* \mathbf{B}_{n'}^{\phantom{*}}\Omega)|^2\leq(\Omega|\mathbf{B}_{n}^* \mathbf{B}_{n}\Omega) (\Omega|\mathbf{B}_{n'}^* \mathbf{B}_{n'}^{\phantom{*}}\Omega)=0
\end{equation} 
if $n < m$ or $n'<m$. In the case $m=\infty$ we have $(\Omega|\mathbf{B}^* \mathbf{B}\Omega)=0$. Otherwise, the first non-vanishing term of the expansion of $(\Omega|\mathbf{B}^* \mathbf{B}\Omega)$ in powers of the coupling constant is $e^{2m}(\Omega|\mathbf{B}_m^* \mathbf{B}_m^{\phantom{*}}\Omega)> 0$. This implies that the formal power series $(\Omega|\mathbf{B}^* \mathbf{B}\Omega)$ is non-negative.
\end{proof}
\end{lem}

In the case of models with vector fields such as QED the covariant inner product on $\mathcal{D}_0$ is indefinite and the functional $\sigma$ given by \eqref{eq:state} is not positive. Nevertheless, it is expected that it can be used to define the Poincar\'e invariant state on the algebra of observables.

\section{Strong adiabatic limit}\label{sec:strong_adiabatic}

The strong adiabatic limit allows to define the physical S-matrix of the given model as the limit of the form $\lim_{\epsilon\searrow 0}S(\mathpzc{g}_\epsilon)$, where the S-matrix with the IR regularization $S(\mathpzc{g})$ was defined in \eqref{eq:S_matrix}. So far the existence of the strong adiabatic limit has been shown only in purely massive models containing only massive scalar fields \cite{epstein1976adiabatic}. We do not deal with the strong adiabatic limit in the thesis. Let us, however, present the results of \cite{epstein1976adiabatic} in some detail.

To this end, we first introduce the following subspace in the Fock space
\begin{multline}\label{eq:dom1}
 \mathcal{D}_1 :=\Span\bigg\{ \int\rd^4 x_1\ldots\rd^4 x_n\,f(x_1,\ldots,x_n)\,A_{i_1}(x_1)\ldots A_{i_n}(x_n)\Omega ~:
 \\
 ~n\in\N_0, ~i_1,\ldots,i_n\in\{1,\ldots,\mathrm{p}\},~\F{f}\in C_\textrm{c}^\textrm{H}(\R^{4n})\bigg\},
\end{multline}
where $C_\textrm{c}^\textrm{H}(\R^N)$ is the space of H{\"o}lder continuous functions $\R^N\to\C$ of compact support. The elements of $\mathcal{D}_1$ are vectors in the Fock space with finite number of non-vanishing multiparticle wave functions which are H{\"o}lder continuous and have compact support in the momentum representation. The subspace $\mathcal{D}_1$ is dense in $\mathcal{H}$. As was shown in \cite{epstein1976adiabatic} the operator $S(\mathpzc{g})$ maps $\mathcal{D}_1$ into $\mathcal{D}_1$ and the limit 
\begin{equation}\label{eq:strong_limit}
 \lim_{\epsilon\searrow0}S(\mathpzc{g}_\epsilon) \Psi =: S\Psi
\end{equation}
exists as a formal power series in coupling constants for any $\Psi\in\mathcal{D}_1$ and is again an element of $\mathcal{D}_1$. The operator $S:\mathcal{D}_1 \to\mathcal{D}_1$ defined above is unitary and has the interpretation of the physical $S$-matrix of the model. The proof of the existence of the limit \eqref{eq:strong_limit} relies on the fact that the vacuum and the one-particle hyperboloid are isolated from the rest of the energy-momentum spectrum.

Because of the infrared problem the above limit is not expected to exist in most interacting theories containing massless particles. In particular the term of order $e$ of 
\begin{equation}\label{eq:strong_limit_D_0}
 \lim_{\epsilon\searrow0}S(\mathpzc{g}_\epsilon) \Psi
\end{equation}
is divergent for any $\Psi \in\mathcal{D}_0$ in the models introduced in Section~\ref{sec:examples}. It is expected that the limit exists after suitable modification of the definition of $S(\mathpzc{g})$ which takes into account the long-range character of interactions in the models from Section~\ref{sec:examples}. In fact, with the use of non-trivial asymptotic dynamics, based on the ideas due to Dollard \cite{dollard1964asymptotic} and Kulish and Faddeev \cite{kulish1970asymptotic}, we are going to propose a modified definition of the S-matrix $S(\mathpzc{g})$ with the IR regularization $g\in\mathcal{S}(\R^4)$ in the scalar model. Our preliminary results show that with this modification the strong adiabatic limit exists in the first and the second order in the coupling constant $e$, provided $\CC=1$ in the normalization condition \ref{norm:sd}. These results will be presented elsewhere.

%%%%%%%%%%%%%%%%%%%%%%%%%%%%%%%%%%%%%%%%%%%%%%%%%%%%%%%%%%%%%%%%%%%%%%%%%%%%%%%%%%%%%%%%%%%%%%%%%%%%
%%%%%%%%%%%%%%%%%%%%%%%%%%%%%%%%%%%%%%%%%%%%%%%%%%%%%%%%%%%%%%%%%%%%%%%%%%%%%%%%%%%%%%%%%%%%%%%%%%%%
%                                    WEAK ADIABATIC LIMIT
%%%%%%%%%%%%%%%%%%%%%%%%%%%%%%%%%%%%%%%%%%%%%%%%%%%%%%%%%%%%%%%%%%%%%%%%%%%%%%%%%%%%%%%%%%%%%%%%%%%%
%%%%%%%%%%%%%%%%%%%%%%%%%%%%%%%%%%%%%%%%%%%%%%%%%%%%%%%%%%%%%%%%%%%%%%%%%%%%%%%%%%%%%%%%%%%%%%%%%%%%
\part{Weak adiabatic limit}

\chapter{Wightman and Green functions}\label{Sec:existence_wAL}

In this chapter we present our main results. The objective is to prove the existence of the weak adiabatic limit (wAL) in a large class of models including all models with interaction vertices of dimension four. To this end it is enough to show that the limit
\begin{multline}\label{eq:wAL}\tag{wAL}
 \lim_{\epsilon\searrow 0}\int\rd^4 y_1\ldots\rd^4 y_n\,\rd^4 x_1\ldots\rd^4 x_m\, g_\epsilon(y_1,\ldots,y_n) \, f(x_1,\ldots,x_m)
 \\
 (\Omega|\Adv(\mathcal{L}_{l_1}(y_1),\ldots,\mathcal{L}_{l_n}(y_n);C_1(x_1),\ldots,C_m(x_m);P)\Omega)
\end{multline}
and the analogous limit with the generalized advanced product $\Adv(I;J;P)$ replaced by the generalized retarded product $\Ret(I;J;P)$
\begin{enumerate}[leftmargin=*,label={(\arabic*)}]
 \item  exist for any $n,m\in\N_0$, $l_1,\ldots,l_n\in\{1,\ldots,\mathrm{q}\}$, $C_1,\ldots,C_m\in\Fa$, $f\in\mathcal{S}(\R^{4m})$ and arbitrary sequence $P$ of the form which was considered in Section~\ref{sec:W_G_IR},
 \item are independent of $g\in\mathcal{S}(\R^{4n})$ such that $g(0,\ldots,0)=1$ and
 \item have the same values in the retarded and advanced case.
\end{enumerate} 
By definition $g_\epsilon(y_1,\ldots,y_n):=g(\epsilon y_1,\ldots,\epsilon y_n)$. The previous statement follows directly from the results of Section~\ref{sec:W_G_IR} and the definition of the weak adiabatic limit given in Section~\ref{sec:weak_adiabatic_general}.

We consider first theories with only massive particles. In this case the existence of the limit \eqref{eq:wAL} was shown by Epstein and Glaser in \cite{epstein1973role}. In Section~\ref{sec:weak_massive_proof} we present a modified version of their proof which shall be regarded as the preparation for the more involved proof for theories with massless particles outlined in Section~\ref{sec:idea} and presented in full detail in Section~\ref{sec:proof_scalar}. Sections \ref{sec:math}, \ref{sec:prod} and \ref{sec:split} contain intermediate results needed in Section~\ref{sec:proof_scalar}. The proof of the existence of the Wightman and Green functions for theories with massless particles requires the introduction of certain normalization condition of the time-ordered products. This condition is stated in Section~\ref{sec:proof_scalar} as the normalization condition \ref{norm:wAL}. In Chapter~\ref{ch:comp} it is shown that this normalization condition is compatible with other normalization conditions which are usually imposed on the time-ordered products.

In what follows we frequently refer to the identities and notation introduced in Sections \ref{sec:W_G_IR} and \ref{sec:aux}.

\section{Proof for massive theories}\label{sec:weak_massive_proof}

The proof of the existence of the weak adiabatic limit for theories with only massive particles relies on the presence of the mass gap in the energy-momentum spectrum of these theories. It means that the vacuum state is separated from the rest of the spectrum. This property, which holds only in the case of purely massive models, greatly simplifies the proof of the existence of the weak adiabatic limit. It turns out that the presence of the mass gap implies that there is a neighborhood $\mathcal{O}$ of $0$ in $\R^{4n}$ such that the distribution
\begin{multline}\label{eq:weak_massive_proof}
 \F{\mathrm{a}}(q_1,\ldots,q_n)= \int \rd^4 y_1\ldots\rd^4 y_n\rd^4 x_1\ldots\rd^4 x_m 
 \exp(\ri q_1\cdot y_1+\ldots+\ri q_n\cdot y_n) f(x_1,\ldots,x_m)
 \\
 (\Omega|\Adv(\mathcal{L}_{l_1}(y_1),\ldots,\mathcal{L}_{l_n}(y_n);C_1(x_1),\ldots,C_m(x_m);P)\Omega)
\end{multline}
restricted to test functions supported in $\mathcal{O}$ is a smooth function. Before proving this let us observe that with the use of the above distribution the limit \eqref{eq:wAL} can be rewritten in the form
\begin{equation}\label{eq:weak_massive_adiabatic}
 \lim_{\epsilon\searrow0}\int\mP{q_1}\ldots\mP{q_n}\,\F{\mathrm{a}}(-q_1,\ldots,-q_n)\, \F{g}_\epsilon(q_1,\ldots,q_n)
\end{equation}
Assuming that $\F{a}$ is smooth in $\mathcal{O}$, the existence of the above limit is an immediate consequence the following obvious lemma. 

\begin{lem}\label{lem:weak_massive_simple}

\hspace{0mm}

\begin{enumerate}[leftmargin=*,label={(\Alph*)}]
\item Let $t\in C(\R^N)$ For every $g\in\mathcal{S}(\R^N)$ we have
\begin{equation}
 \lim_{\epsilon\searrow0} \int \frac{\rd^N\!q}{(2\pi)^N}\, t(-q)\,
 \frac{1}{\epsilon^N}g(q/\epsilon)
 = t(0) \int \frac{\rd^N\!q}{(2\pi)^N} g(q).
\end{equation}
\item For any $g\in\mathcal{S}(\R^N)$ and $\chi\in C^\infty(\R^N)$ such that $0\notin\supp\,\chi$ we have 
\begin{equation}
 \lim_{\epsilon\searrow0}\frac{1}{\epsilon^N}g(q/\epsilon)\,\chi(q)=0
\end{equation}
in the topology of $\mathcal{S}(\R^N)$. As a consequence if $t\in\mathcal{S}'(\R^N)$, $0\notin\supp\,t$ then for every $g\in\mathcal{S}(\R^N)$ it holds
\begin{equation}
\lim_{\epsilon\searrow0} \int \frac{\rd^N\!q}{(2\pi)^N}\, t(-q)\,
 \frac{1}{\epsilon^N}g(q/\epsilon) = 0.
\end{equation}
\end{enumerate}
\end{lem}

Let us show that \eqref{eq:weak_massive_proof} is indeed a smooth function in some neighborhood of the origin. To this end, for each $n\in\N_+$ we define the function $\Theta_n\in C^\infty(\R^{4n})$, called the regularized splitting function, such that
\begin{equation}\label{eq:def_example_splitting_fun}
 \Theta_n(y_1,\ldots,y_n)
 :=\rho\left(\frac{3n(y_1^0+\ldots+y_n^0)}{|(y_1,\ldots,y_n)|}\right) ~~~~\textrm{for}~~~|(y_1,\ldots,y_n)|\geq \ell,
\end{equation}
where $\ell$ is arbitrary positive constant of the dimension of length, $$|(y_1,\ldots,y_n)|=(|y_1|^2+\ldots+|y_n|^2)^{1/2}$$ and $\rho\in C^\infty(\R)$ is a real function having the following properties:
\begin{enumerate}[leftmargin=*,label={(\arabic*)}]
\item $\forall_{s\in\R}\,0\leq\rho(s)\leq1$,
\item $\forall_{s\in\R}\,\rho(s)=1-\rho(-s)$,
\item $\forall_{s>1}\,\rho(s)=1$, $\forall_{s<-1}\,\rho(s)=0$.
\end{enumerate}
The precise form of the functions $\Theta_n$ for $|(y_1,\ldots,y_n)| < \ell$ is irrelevant for our purposes. For $|(y_1,\ldots,y_n)|\geq \ell$ it holds:
\begin{equation}
 (1-\Theta_n(y_1,\ldots,y_n))=\Theta_n(-y_1,\ldots,-y_n)
\end{equation}
and
\begin{equation}\label{eq:supp_theta}
 \Theta_n(\pm y_1,\ldots,\pm y_n) = 0~~~~\textrm{if}~~~\mp(y_1^0+\ldots+y_n^0)\geq\frac{1}{3n}|(y_1,\ldots,y_n)|.
\end{equation}
Moreover, $\F{\Theta}_n$ -- the Fourier transform of $\Theta_n$ -- is a smooth function outside the origin and $\F{\Theta}_n(k)$ vanishes at infinity faster than any power of $|k|$ (the proof of this statement is postponed to Section~\ref{sec:math}; it is a simple consequence of Lemma \ref{lem:splitting_theta_alpha}).

By the definition \eqref{eq:def_dif_gen} of the generalized product $\Dif$ the following identity holds
\begin{equation}\label{eq:decomposition_weak_massive}
\begin{split}
 (\Omega|\Adv(\mathcal{L}_{l_1}(y_1),\ldots,\mathcal{L}_{l_n}&(y_n);C_1(x_1),\ldots,C_m(x_m);P)\Omega)=
 \\
 (1-\Theta_n(y_1,\ldots,y_n))\,&(\Omega|\Adv(\mathcal{L}_{l_1}(y_1),\ldots,\mathcal{L}_{l_n}(y_n);C_1(x_1),\ldots,C_m(x_m);P)\Omega)
 \\
 +\Theta_n(y_1,\ldots,y_n)\,&(\Omega|\Ret(\mathcal{L}_{l_1}(y_1),\ldots,\mathcal{L}_{l_n}(y_n);C_1(x_1),\ldots,C_m(x_m);P)\Omega)
 \\
 +\Theta_n(y_1,\ldots,y_n)\,&(\Omega|\Dif(\mathcal{L}_{l_1}(y_1),\ldots,\mathcal{L}_{l_n}(y_n);C_1(x_1),\ldots,C_m(x_m);P)\Omega).
\end{split} 
\end{equation}
We will show that the contribution to \eqref{eq:weak_massive_proof} from each of the three terms of the RHS of Equation \eqref{eq:decomposition_weak_massive} is a smooth function in some neighborhood of zero. For the first two terms this follows from the lemma below with $q'_1=\ldots=q'_m=0$, $B_1=\mathcal{L}_{l_1},\ldots,B_{n}=\mathcal{L}_{l_n}$. The lemma will be also used in the proof of the existence of the weak adiabatic limit for theories with massless particles. Its proof uses only the support properties of the functions $\Theta_n$, $(1-\Theta_n)$ and of the generalized advanced and retarded products.

\begin{lem}\label{lem:splitting_ind_two_smooth}
Let $n,m\in\N_0$ and $B_1,\ldots,B_n,C_1,\ldots,C_m\in\Fa$. Moreover, let $P$ be any sequence of the form considered in Section~\ref{sec:W_G_IR}. For every $f\in\mathcal{S}(\R^{4m})$ the distribution
\begin{multline}\label{eq:lem_splitting_ind_two_smooth}
 (q_1,\ldots,q_n,q'_1,\ldots,q'_m)\mapsto \int\rd^4 y_1\ldots\rd^4 y_n\rd^4 x_1\ldots\rd^4 x_m 
 \\[2pt]
 \exp(\ri q_1\cdot y_1+\ldots+\ri q_n\cdot y_n + \ri q'_1\cdot x_1+\ldots+\ri q'_m\cdot x_m) ~f(x_1,\ldots,x_m)
 \\[6pt]
 ~\Theta_n(y_1,\ldots,y_n)\,
 (\Omega|\Ret(B_1(y_1),\ldots,B_n(y_n);C_1(x_1),\ldots,C_m(x_m);P)\Omega)
\end{multline} 
is a smooth function. The same holds for 
\begin{equation}
 (1-\Theta_n(y_1,\ldots,y_n))\,
 (\Omega|\Adv(B_1(y_1),\ldots,B_n(y_n);C_1(x_1),\ldots,C_m(x_m);P)\Omega).
\end{equation}
\begin{proof}
Because of the support property of the retarded distribution (see Section~\ref{sec:W_G_IR}) and the presence of the function $\Theta_n$ the integrand in \eqref{eq:lem_splitting_ind_two_smooth} for $|(y_1,\ldots,y_n)|\geq \ell$ may be non-zero only in the region
\begin{multline}\label{eq:proof_inclusion}
 \Gamma^-_{n,m} \cap \{ (y_1,\ldots,y_n):\, 
 (y_1^0+\ldots+y_n^0)+\tfrac{1}{3n}|(y_1,\ldots,y_n)|\geq0\} \times \R^{4m}
 \\
 \subset \{ (y_1,\ldots,y_n;x_1,\ldots,x_m): |(y_1,\ldots,y_n)| \leq \const \, |(x_1,\ldots,x_m)| \},
\end{multline} 
where the cone $\Gamma^-_{n,m}$ is given by \eqref{eq:def_gen_cones}. To prove the above inclusion we first note that for any $(y_1,\ldots,y_n;x_1,\ldots,x_m)\in\Gamma^-_{n,m}$ it holds
\begin{equation} \label{eq:proof_inclusion_bound}
 y^0_j \leq \,|(x_1,\ldots,x_m)|
 ~~~~\textrm{and}~~~~
 |\vec{y}_j| \leq \,2|(x_1,\ldots,x_m)| - y^0_j.
\end{equation}
Since $(y_1^0+\ldots+y_n^0) + \frac{1}{3n}|(y_1,\ldots,y_n)|\geq 0$ we have
\begin{equation}
 -y^0_j \leq (y^0_1 + \ldots + y^0_n) - y^0_j + \frac{1}{3n}(|\vec{y}_1|+\ldots+|\vec{y}_n|) + \frac{1}{3n}(|y^0_1|+\ldots+|y^0_n|).
\end{equation}
Combining the above inequality with the first bound in \eqref{eq:proof_inclusion_bound} we get
\begin{equation}\label{eq:proof_inclusion_third_bound} 
 |y^0_j| \leq (n-1) \,|(x_1,\ldots,x_m)| + \frac{1}{3n}(|\vec{y}_1|+\ldots+|\vec{y}_n|) + \frac{1}{3n}(|y^0_1|+\ldots+|y^0_n|).
\end{equation}
After summing both sides of the second bound in \eqref{eq:proof_inclusion_bound} and of the bound \eqref{eq:proof_inclusion_third_bound} over $j$ from $1$ to $n$ we obtain
\begin{align} 
 &|\vec{y}_1|+\ldots+|\vec{y}_n| \leq 2n \,|(x_1,\ldots,x_m)| + |y^0_1|+\ldots+|y^0_n|,
 \\
 &\frac{2}{3}(|y^0_1|+\ldots+|y^0_n|) \leq n(n-1) \,|(x_1,\ldots,x_m)| + \frac{1}{3}(|\vec{y}_1|+\ldots+|\vec{y}_n|),
\end{align}
respectively. The inclusion \eqref{eq:proof_inclusion} follows from the above bounds.

As a result there exists a function $\chi\in C^\infty(\R^{4n+4m})$ such that $\chi\equiv 1$ in some neighborhood of the support of the integrand in \eqref{eq:lem_splitting_ind_two_smooth} and 
\begin{equation}
 (y_1,\ldots,y_n;x_1,\ldots,x_m)\mapsto \chi(y_1,\ldots,y_n;x_1,\ldots,x_m) f(x_1,\ldots,x_m)
\end{equation}
is a Schwartz function. To show the statement of the lemma we replace $f$ in \eqref{eq:lem_splitting_ind_two_smooth} by the above function and use the following fact. For any $t\in\mathcal{S}'(\R^N)$ and $h\in\mathcal{S}(\R^N)$, the Fourier transform of $h(x)t(x)$ is a smooth function.
\end{proof}
\end{lem}

Let us investigate the third term on the RHS of Equation \eqref{eq:decomposition_weak_massive} involving the product $\Dif$. We will first show that the distribution
\begin{multline} \label{eq:weak_massive_f_dif}
 \F{\mathrm{d}}(q_1,\ldots,q_n):=\int \rd^4 x_1\ldots\rd^4 x_m 
 \exp(\ri q_1\cdot y_1+\ldots+\ri q_n\cdot y_n) f(x_1,\ldots,x_m)
 \\
 (\Omega|\Dif(\mathcal{L}_{l_1}(y_1),\ldots,\mathcal{L}_{l_n}(y_n);C_1(x_1),\ldots,C_m(x_m);P)\Omega)
\end{multline}
vanishes in some neighborhood of zero. To this end, using the identity \eqref{eq:dif_com} we represent $\F{d}$ as a linear combination of the following distributions
\begin{multline} 
 (q_1,\ldots,q_n)\mapsto \int \rd^4 y_1\ldots\rd^4 y_n\rd^4 x_1\ldots\rd^4 x_m 
 \exp(\ri q_1\cdot y_1+\ldots+\ri q_n\cdot y_n) f(x_1,\ldots,x_m)
 \\
 (\Omega|[\aT(I_1)\Adv(I_2;B_1(x_1),\ldots,B_m(x_m)),\T(I_3)]\Omega),
\end{multline}
where $I_1+I_2+I_3$ is some permutation of a sequence $(\mathcal{L}_{l_1}(y_1),\ldots,\mathcal{L}_{l_n}(y_n))$. The above distributions vanish in some neighborhood of the origin as a consequence of both parts of Lemma \ref{lem:aux_lemma} and the assumption that there are no massless fields. Note that the validity of the last statement follows ultimately from the presence of the mas gap in the energy-momentum spectrum. This is the only place in the proof where we use the assumption that all fields are massive.

The contribution to \eqref{eq:weak_massive_proof} from the last term on the RHS of Equation \eqref{eq:decomposition_weak_massive} is of the form
\begin{equation}
 (q_1,\ldots,q_n)\mapsto\int\mP{k_1}\ldots\mP{k_n}\, \F{\mathrm{d}}(k_1,\ldots,k_n) \,\F{\Theta}_n(q_1-k_1,\ldots,q_n-k_n).
\end{equation}
Because of support properties of \eqref{eq:weak_massive_f_dif} and the smoothness of $\F{\Theta}_n$ outside the origin the above distribution is indeed a smooth function in some neighborhood of the origin. The above result implies the existence of the limit \eqref{eq:wAL}. Since $\Dif(I;J;P)=\Adv(I;J;P)-\Ret(I;J;P)$ and the distribution $\F{d}$ vanishes in some neighborhood of zero the limit \eqref{eq:wAL} with the $\Adv$ product replaced by $\Ret$ also exists and has the same value as \eqref{eq:wAL}. This shows the existence of the weak adiabatic limit in massive theories.

\section{Idea of proof for theories with massless particles}\label{sec:idea}

The proof of the existence of the Wightman and Green functions in massive models was based on the fact that the distribution \eqref{eq:weak_massive_f_dif} vanishes in some neighborhood of the origin. This in turn follows from the existence of the mass gap in the energy-momentum spectrum of massive theories and is no longer true when massless particles are present. In fact, calculations in low orders of the perturbation theory show that in theories with massless particles the distribution \eqref{eq:weak_massive_proof} is usually not a continuous function in any neighborhood of zero. The condition of smoothness of \eqref{eq:weak_massive_proof} in some neighborhood of zero proved in the previous section for massive theories has to be weakened. 

First note that by Part~(B) of Lemma \ref{lem:weak_massive_simple} it is enough to control the behavior of the distribution \eqref{eq:weak_massive_proof} in a vicinity of the origin. To quantify the regularity of distributions $t\in\mathcal{S}'(\R^N)$ near $0\in\R^N$ we shall introduce a distributional condition  $t(q)=O^{\textrm{dist}}(|q|^\delta)$, generalizing the condition $f(q)=O(|q|^\delta)$ expressed in terms of the standard big O notation, which applies when $f$ is a function. In particular, if a~distribution $t\in\mathcal{S}'(\R^N)$ is given in some neighborhood of zero by a continuous function such that $t(q)=O(|q|^\delta)$, or equivalently $|t(q)|\leq \const\, |q|^\delta$ for $q$ in some neighborhood of zero, then $t(q)=O^{\textrm{dist}}(|q|^\delta)$. We will prove that in a large class of models with massless particles (see Assumption \ref{asm} for the precise specification of this class) it is possible to normalize the time-ordered products such that the distribution \eqref{eq:weak_massive_proof} is of the form
\begin{equation}\label{eq:idea_a}
 \F{\mathrm{a}}(q_1,\ldots,q_n) = c + O^{\textrm{dist}}(|q_1,\ldots,q_n|^\delta)
\end{equation}
for some $c\in\C$ and $\delta>0$ and the distribution \eqref{eq:weak_massive_f_dif} is of the form
\begin{equation}\label{eq:idea_d}
 \F{\mathrm{d}}(q_1,\ldots,q_n) = O^{\textrm{dist}}(|q_1,\ldots,q_n|^\delta)
\end{equation}
for some $\delta>0$. As we will see the existence of the weak adiabatic limit follows immediately from the above form of distributions $\F{\mathrm{a}}$ and $\F{\mathrm{d}}$. It was shown in the previous section that the above conditions are satisfied in particular in purely massive theories. Indeed, it was established that $\F{\mathrm{a}}$ is a smooth function in some neighborhood of the origin and $\F{\mathrm{d}}$ vanishes in some neighborhood of the origin. 

The conditions \eqref{eq:idea_a} and \eqref{eq:idea_d} are satisfied if the time-ordered products fulfill the normalization condition \ref{norm:wAL}, which will be formulated in Section~\ref{sec:proof_scalar}. It says that for all lists $\mathbf{u}=(u_1,\ldots,u_{k})$ of super-quadri-indices which involve only massless fields\footnote{We recall that a super-quadri-index $u$ involves only massless fields if $u(i,\alpha)=0$ for all $i$ such that $A_i$ is a massive field. If a super-quadri-index $u$ involves only massless fields then the monomial $A^u$ is a product of massless generators.} it holds
\begin{equation}
 (\Omega|\T(\F{\mathcal{L}}_{l_1}^{(u_1)}(q_1), \ldots, \F{\mathcal{L}}_{l_{k}}^{(u_{k})}(q_{k}))\Omega)
 =(2\pi)^4 \delta(q_1+\ldots+q_{k})\,\F{\underline{t}}(q_1,\ldots,q_{k-1}), 
\end{equation}
where $\partial_q^\gamma \F{\underline{t}}(0) = 0$ for all multi-indices $\gamma$ such that $|\gamma|<\omega$ and 
\begin{equation}
 \omega =  4 - \sum_{i=1}^{\mathrm{p}} [\dim(A_i) \ext_{\mathbf{u}}(A_i) + \der_{\mathbf{u}}(A_i)]
\end{equation}
(the above formula is equivalent to \eqref{eq:ren_omega2}). The value of the distribution $\partial_q^\gamma\F{\underline{t}}$ at zero is defined in the sense of {\L}ojasiewicz (cf. Definition \ref{def:lojasiewicz} in the next section). The condition \ref{norm:wAL} involves only the time-ordered products of some of the sub-polynomials of the interaction vertices, namely, those which are of the form $\mathcal{L}_l^{(u)}$ for some $l\in\{1,\ldots\mathrm{q}\}$  and super-quadri index $u$ involving only massless fields. For example, in the case of the massive spinor QED the only such sub-polynomials are the vertex $\mathcal{L}$ and the components of the electric current $j^\mu$.  The condition \ref{norm:wAL} implies in particular that the photon self-energy corrections (called the vacuum polarization corrections) have zero of order $2$ at vanishing external momentum in the sense of {\L}ojasiewicz.  

Besides \ref{norm:wAL} we introduce a seemingly stronger normalization condition \ref{norm:wAL2} which says that for any $\Fa$ product $F$ which is a product of the time-ordered products and for all super-quadri-indices  $u_1,\ldots,u_{k}$ involving only massless fields
\begin{equation}\label{eq:idea_wAL}
\begin{aligned}
&(\Omega|F(\F{\mathcal{L}}_{l_1}^{(u_1)}(q_1), \ldots, \F{\mathcal{L}}_{l_{k}}^{(u_{k})}(q_{k}))\Omega)
 =(2\pi)^4 \delta(q_1+\ldots+q_{k})\,\F{\underline{t}}(q_1,\ldots,q_{k-1}),
\\
&\textrm{~~where~~}\F{\underline{t}}(q_1,\ldots,q_{k-1})=O^{\textrm{dist}}(|q_1,\ldots,q_{k-1}|^{\omega-\varepsilon}) \textrm{~~for every~~} \varepsilon>0.
\end{aligned} 
\end{equation}
Theorem \ref{thm:main1} shows that it is possible to define the time-ordered products such that the conditions \ref{norm:sd} and \ref{norm:wAL2} are satisfied simultaneously. Moreover, it shows that \ref{norm:wAL} is equivalent to the normalization condition \ref{norm:wAL2}. The condition \ref{norm:wAL} has more transparent interpretation; however, it is the condition \ref{norm:wAL2} that is used in the proof of the existence of the Wightman and Green functions.

The proof of the compatibility of \ref{norm:sd} and \ref{norm:wAL2} (stated in Theorem \ref{thm:main1}) is by induction with respect to $k$. We assume that the time ordered products with at most $n$ arguments satisfy the conditions \ref{norm:sd} and \ref{norm:wAL2} and prove that it is possible to define time-ordered products with $n+1$ arguments such that these conditions hold. The proof of the inductive step is divided into two parts. The first part is based on the following fact which is an immediate consequence of the statement (1') of Theorem \ref{thm:product_families} stated in Section~\ref{sec:prod} entitled \emph{Product}. Let $F$ and $F'$ be two $\Fa$ products. Assume that their VEVs satisfy the condition \eqref{eq:idea_wAL}. Then the VEV of their product \eqref{eq:F_product} also satisfies this condition. Consequently, the VEVs of the $\Dif$ and $\Adv'$ products with $n+1$ arguments fulfill the condition \eqref{eq:idea_wAL}. The second part of the proof of the inductive step is based on Theorem \ref{thm:split} from Section~\ref{sec:split} entitled \emph{Splitting}. Using this theorem we show that if the VEV of the $\Dif$ product with $n+1$ arguments satisfies the condition \eqref{eq:idea_wAL} then it is possible to define the advanced product with $n+1$ arguments such that its VEV also satisfies the condition \eqref{eq:idea_wAL} and \ref{norm:sd} is not violated.

We now turn to Theorem \ref{thm:main2} which states the existence of the weak adiabatic limit in the class of models satisfying our assumptions. Let us explain the intuitive content of this theorem. Consider the distribution
\begin{multline}\label{eq:idea_adv}
 \F{\textrm{a}}^{u_1,\ldots,u_{k+m}}(q_1,\ldots,q_k;q'_1,\ldots,q'_m):=\int\rd^4 y_1\ldots\rd^4 y_k\rd^4 x_1\ldots\rd^4 x_m 
 \\
 \exp(\ri q_1\cdot y_1+\ldots+\ri q_k\cdot y_k + \ri q'_1\cdot x_1+\ldots+\ri q'_m\cdot x_m) ~f(x_1,\ldots,x_m)
 \\
 (\Omega|\Adv(\mathcal{L}_{l_1}^{(u_1)}(y_1),\ldots,\mathcal{L}_{l_k}^{(u_k)}(y_k);C_1^{(u_{k+1})}(x_1),\ldots,C^{(u_{k+m})}_m(x_m);P)\Omega).
\end{multline} 
Because of the presence of a Schwartz function $f$, it is a continuous function of $q'_1,\ldots,q'_m$. It holds 
\begin{equation}
 \F{\textrm{a}}^{0,\ldots,0}(q_1,\ldots,q_k;0,\ldots,0) = \F{\textrm{a}}(q_1,\ldots,q_k),
\end{equation}
where the distribution $\F{\textrm{a}}$ is given by \eqref{eq:weak_massive_proof}. We will prove that:
\begin{enumerate}[leftmargin=*,label={(\Alph*)}]
 \item For all super-quadri-indices $u_1,\ldots,u_{k+m}$ which involve only massless fields and at least one of them is non-zero it holds 
\begin{equation}\label{eq:idea_2}
 \F{\textrm{a}}^{u_1,\ldots,u_{k+m}}(q_1,\ldots,q_k;q'_1,\ldots,q'_m) =O^\textrm{dist}(|q_1,\ldots,q_k|^{d-\varepsilon})
\end{equation}
for any $\varepsilon>0$, where
\begin{equation}
 d:= 1 - \sum_{i=1}^{\mathrm{p}} [\dim(A_i) \ext_{\mathbf{u}}(A_i)+ \der_{\mathbf{u}}(A_i)] 
\end{equation}
and $\mathbf{u}=(u_1,\ldots,u_{k+m})$.
\item There exists $c\in\C$ such that 
\begin{equation}\label{eq:idea_3}
 \F{\textrm{a}}(q_1,\ldots,q_k) =c+ O^\textrm{dist}(|q_1,\ldots,q_k|^{1-\varepsilon})
\end{equation}
for any $\varepsilon>0$.
\end{enumerate}
The first of the above conditions is needed in the inductive proof of the second one which implies the existence of the weak adiabatic limit.

The proof of the theorem is conducted by induction on $k$. It relies on Equation \eqref{eq:dif_com2} which allows to express $\Dif(I;J;P)$ with $|I|=n$ in terms of the anti-time-ordered products $\aT(I_1)$, the generalized advanced products $\Adv(I_2;J;P)$ and time-ordered products $\T(I_3)$ with $|I_1|,|I_2|,|I_3|<n$. The VEV of $\Adv(I_2;J;P)$ satisfies the condition \eqref{eq:idea_2} by the induction assumption and the VEVs of $\aT(I_1)$ and $\T(I_3)$ fulfill the condition \eqref{eq:idea_wAL}. Note that we actually use the condition \ref{norm:wAL2} only with $F$ being the time-ordered or anti-time-ordered product.  Using the above-mentioned properties and the statement (2') and (3') of Theorem \ref{thm:product_families} in Section~\ref{sec:prod} entitled \emph{Product} one shows that for all super-quadri-indices $u_1,\ldots,u_{n+m}$ which involve only massless fields it holds 
\begin{equation}\label{eq:idea_2d}
 \F{\textrm{d}}^{u_1,\ldots,u_{n+m}}(q_1,\ldots,q_n;q'_1,\ldots,q'_m) =O^\textrm{dist}(|q_1,\ldots,q_n|^{d-\varepsilon})
\end{equation}
for any $\varepsilon>0$, where the distribution $\F{\textrm{d}}^{u_1,\ldots,u_{n+m}}(q_1,\ldots,q_k;q'_1,\ldots,q'_m)$ is defined by \eqref{eq:idea_adv} with the $\Adv$ product replaced by the $\Dif$ product. Because of the presence of the graded-commutator in the formula \eqref{eq:dif_com2} it is possible to prove \eqref{eq:idea_2d} for all $u_1,\ldots,u_{n+m}$ which involve only massless fields using the validity of \eqref{eq:idea_2} for all $u_1,\ldots,u_{n+m}$ which involve only massless fields such that at least of of them is non-zero.  The above condition implies that
\begin{equation}\label{eq:idea_3d}
 \F{\textrm{d}}(q_1,\ldots,q_n) =O^\textrm{dist}(|q_1,\ldots,q_n|^{1-\varepsilon})
\end{equation}
for any $\varepsilon>0$, where the distribution $\F{\textrm{d}}$ is given by \eqref{eq:weak_massive_f_dif}. Using the above results about the distributions $\F{\textrm{d}}^{u_1,\ldots,u_{n+m}}(q_1,\ldots,q_k;q'_1,\ldots,q'_m)$ and $\F{\textrm{d}}(q_1,\ldots,q_n)$ as well as Theorem \ref{thm:split_gen} from Section~\ref{sec:split} entitled \emph{Splitting} we obtain that the distribution $\F{\textrm{a}}^{u_1,\ldots,u_{n+m}}(q_1,\ldots,q_k;q'_1,\ldots,q'_m)$ satisfies the condition \eqref{eq:idea_2} with $k=n$ and the distribution $\F{\textrm{a}}(q_1,\ldots,q_k)$ satisfies the condition \eqref{eq:idea_3}  with $k=n$.

\subsubsection{Summary of sections with preliminary results}

Let us describe in more detail the content of the next three sections in which we gather intermediate results needed in the proof of the existence of the weak adiabatic limit outlined above. In Section~\ref{sec:math} entitled {\it Mathematical preliminaries} we introduce the regularity condition for Schwartz distributions $t\in\mathcal{S}'(\R^N\times\R^M)$ denoted by $t(q,q')=O^\mathrm{dist}(|q|^\delta)$, gather a number of results regarding the distributions $t$ of that type, recall the definition of the value of a distribution at a point in the sense of {\L}ojasiewicz and present in the abstract terms the UV regularized splitting procedure. In order to quantify the regularity near the origin in the momentum space of translationally-invariant distributions (frequently appearing in our considerations) we introduce the notion of the \underline{IR}-index and IR-index of a distribution. These indices are defined with the use of the notation $t(q,q')=O^\mathrm{dist}(|q|^\delta)$. Their definitions are given in Section~\ref{sec:prod} entitled {\it Product}. This section is devoted to the investigation of the \underline{IR}- and IR-index of the VEV of the product of two $\Fa$ products whose VEVs have specific \underline{IR}- or IR-indices. In Section  \ref{sec:split}, entitled  {\it Splitting}, we prove results which allow to determine the \underline{IR}- or IR-index of the VEVs of the advanced products $\Adv$ given the \underline{IR}- or IR-index of the VEVs of the products $\Dif$. To this end, we use the decomposition \eqref{eq:decomposition_weak_massive} and Lemma \ref{lem:splitting_ind_two_smooth}, stated in the previous section, and Theorem \ref{thm:math_splitting}, which will be proved in the next section.

%%%%%%%%%%%%%%%%%%%%%%%%%%%%%%%%%%%%%%%%%%%%%%%%%%%%%%%%%%%%%%%%%%%%%%%%%%%%%%%%%%%%%%%%%%%%%%%%%%%%
%%%%%%%%%%%%%%%%%%%%%%%%%%%%%%%%%%%%%%%%%%%%%%%%%%%%%%%%%%%%%%%%%%%%%%%%%%%%%%%%%%%%%%%%%%%%%%%%%%%%
%                                    MATH
%%%%%%%%%%%%%%%%%%%%%%%%%%%%%%%%%%%%%%%%%%%%%%%%%%%%%%%%%%%%%%%%%%%%%%%%%%%%%%%%%%%%%%%%%%%%%%%%%%%%
%%%%%%%%%%%%%%%%%%%%%%%%%%%%%%%%%%%%%%%%%%%%%%%%%%%%%%%%%%%%%%%%%%%%%%%%%%%%%%%%%%%%%%%%%%%%%%%%%%%%

\section{Mathematical preliminaries}\label{sec:math}

\begin{dfn}\label{def:O_not}
Let $t\in\mathcal{S}'(\R^N\times \R^M)$. For $\delta\in\R$ we write 
\begin{equation}
 t(q,q') = O^{\mathrm{dist}}(|q|^\delta),
\end{equation}
where $q\in\R^N$ and $q'\in\R^M$ iff there exist a neighborhood $\mathcal{O}$ of the origin in $\R^N\times\R^M$ and a family of functions $t_\alpha\in C(\mathcal{O})$ indexed by multi-indices $\alpha$ such that 
\begin{enumerate}[label=(\arabic*),leftmargin=*]
 \item $t_\alpha \equiv 0$ for all but finite number of multi-indices $\alpha$,
 \item $|t_\alpha(q,q')|\leq \const\,|q|^{\delta+|\alpha|}$ for $(q,q')\in\mathcal{O}$,
 \item $t(q,q') = \sum_{\alpha}
 \partial_q^\alpha t_{\alpha}(q,q')$ for $(q,q')\in\mathcal{O}$.
\end{enumerate}
Note that the differential operator $\partial^\alpha_q$ and the factor $|q|^{\delta+|\alpha|}$ above involve only the variable $q\in\R^N$. If $N=0$ we write $t(q')=O^{\mathrm{dist}}(|\cdot|^\delta)$. By definition for any $\delta\leq 0$ we have $t(q')=O^{\mathrm{dist}}(|\cdot|^\delta)$ iff $t\in C(\R^M)$ and for $\delta>0$ we have $t(q')=O^\mathrm{dist}(|\cdot|^\delta)$ iff $t=0$. 
\end{dfn}

\noindent Let us make a couple of remarks about the above definition:
\begin{enumerate}[label=(\arabic*),leftmargin=*]
 \item In our applications the exponent $\delta$ which appears in Definition \ref{def:O_not} will never be an integer. Usually we set $\delta = d-\varepsilon$, where $d\in\Z$ and $\varepsilon\in(0,1)$. 

 \item The condition $t(q,q')=O^{\mathrm{dist}}(|q|^\delta)$ controls the behavior of the distribution $t$~only near the origin. In particular if a distribution $t\in\mathcal{S}'(\R^N\times\R^M)$ vanishes in a neighborhood of $0$ then $t(q,q')=O^{\mathrm{dist}}(|q|^{\delta})$ for arbitrarily large $\delta$. Moreover, if $t(q,q')=O^{\mathrm{dist}}(|q|^\delta)$ then $t(q,q')=O^{\mathrm{dist}}(|q|^{\delta'})$ for all $\delta'\leq \delta$. 

 \item If $t\in\mathcal{S}'(\R^N\times\R^M)$, $t(q,q')=O^{\mathrm{dist}}(|q|^\delta)$ then there exist neighborhoods $\mathcal{O}_1$ and $\mathcal{O}_2$ of the origin in $\R^N$ and $\R^M$, respectively, such that   for every $g\in\mathcal{S}(\R^N)$, $\supp\,g\subset\mathcal{O}_1$ the distribution
 \begin{equation}
  \int \frac{\rd^N q}{(2\pi)^N} \, t(q,q') g(q) 
 \end{equation}
 is a continuous function for $q'\in\mathcal{O}_2$. In particular, the distribution $t(q,0)$ is well defined and it holds $t(q,0)=O^{\mathrm{dist}}(|q|^\delta)$.
 
 \item If $t\in C(\R^N\times\R^M)$ is such that $|t(q,q')|\leq \const\,|q|^\delta$ in some neighborhood of $0$ then $t(q,q')=O^{\mathrm{dist}}(|q|^\delta)$. There are, however, $t\in C(\R^N\times\R^M)$ such that $t(q,q')=O^{\mathrm{dist}}(|q|^\delta)$ and the bound $|t(q,q')|\leq \const\,|q|^\delta$ is violated in every neighborhood of $0$. An example of such function in the case $N=1$, $M=0$ and $\delta=2$ is $t(q)=q \sin(1/q)+3q^2\cos(1/q) = \partial_q [q^3 \cos(1/q)]$. 

 \item Finally, let us remark that a very similar characterization of the regularity of distributions near the origin in one dimension was introduced by Estrada in \cite{estrada1998regularization} for the investigation of the existence of an extension $t\in\mathcal{D}'(\R)$ of a distribution $t^0\in\mathcal{D}'((0,\infty))$. 

\end{enumerate}

\begin{dfn}\label{def:lojasiewicz}
We say that a distribution $t\in\mathcal{S}'(\R^N)$ has a value $t(0)\in\C$ at zero in the sense of {\L}ojasiewicz \cite{lojasiewicz1957valeur} iff the limit below
\begin{equation}
 t(0):=\lim_{\epsilon\searrow0}\int \frac{\rd^N q}{(2\pi)^N}\, t(q) g_\epsilon(q)
\end{equation}
exists for any $g\in\mathcal{S}(\R^N)$ such that $\int \frac{\rd^N q}{(2\pi)^N} g(q)=1$, $g_\epsilon(q)=\epsilon^{-N} g(q/\epsilon)$. The value $t(0)$ is usually called in the physical literature the adiabatic limit of the distribution $t$ at $0$.

We say that the distribution $t\in\mathcal{S}'(\R^N)$ has zero of order $\omega\in\N_+$ at the origin in the sense of {\L}ojasiewicz iff $\partial^\gamma_q t(q)\big|_{q=0}=0$ for all multi-indices $\gamma$ such that $|\gamma|<\omega$, where $\partial^\gamma_q t(q)\big|_{q=0}$ is defined in the sense of {\L}ojasiewicz.
\end{dfn}

\begin{thm}\label{thm:math_adiabatic_limit}

\hspace{0mm}

\begin{enumerate}[label=(\Alph*),leftmargin=*]
 \item 
Let $t\in\mathcal{S}'(\R^N)$ such that
\begin{equation}
 t(q)=c+O^\mathrm{dist}(|q|^\delta),
\end{equation}
where $\delta > 0$ and $c\in\C$. We have $t(0)=c$ in the sense of {\L}ojasiewicz. 

\item If
\begin{equation}
 t(q)=O^\mathrm{dist}(|q|^\delta),
\end{equation}
where $\delta+1>\omega\in\N_+$, then $t$ has zero of order $\omega$ at the origin in the sense of {\L}ojasiewicz.
\end{enumerate}
\begin{proof}
Let us begin with the proof of part (A). It follows from Definition \ref{def:O_not} that there exists $\chi\in\mathcal{D}(\R^N)$, $\chi\equiv 1$ on some neighborhood of $0$, such that
\begin{equation}
 \chi(q) t(q) = \chi(q) c+ \chi(q)\sum_{\alpha}
 \partial_q^\alpha t_{\alpha}(q)
\end{equation}
for all $q\in\R^N$. Using the notation from Definition \ref{def:lojasiewicz} we obtain 
\begin{equation}
 \lim_{\epsilon\searrow0} (1-\chi(q))g_\epsilon(q)= 0~~~\textrm{in}~~\mathcal{S}(\R^N)
\end{equation}
and
\begin{multline}
 \lim_{\epsilon\searrow0} \sum_{\alpha} 
 \int\frac{\rd^N q}{(2\pi)^N}\,
 \partial_q^\alpha t_{\alpha}(q)\, \chi(q)\,g_\epsilon(q)
 \\
 =\lim_{\epsilon\searrow0} \epsilon^{-|\alpha|}  \sum_{\alpha} (-1)^{|\alpha|}
 \int\frac{\rd^N q}{(2\pi)^N}\,t_{\alpha}(\epsilon q)\, \partial_q^\alpha(\chi(\epsilon q)\,g(q)) = 0,
\end{multline}
since $|t_{\alpha}(q)|\leq\const\,|q|^{|\alpha|+\delta}$ and by assumption $\delta>0$. Thus,
\begin{equation}
 t(0)=\lim_{\epsilon\searrow0}\int\frac{\rd^N q}{(2\pi)^N}\,\chi(q)c(q) \,g_\epsilon(q) = c.
\end{equation}
This finishes the proof of Part~(A). Part~(B) follows from Part~(A) and the fact that $\partial_q^\gamma t(q)=O^\mathrm{dist}(|q|^{\delta-|\gamma|})$ if $t(q)=O^\mathrm{dist}(|q|^\delta)$.
\end{proof}
\end{thm}

\begin{dfn}\label{def:splitting_function_theta}
The function $\Theta:\R^N\to\R$ is called a UV regular splitting function in $\R^N$ iff
\begin{enumerate}[label=(\arabic*),leftmargin=*]
 \item $\Theta$ is smooth,
 \item $0\leq\Theta(y)\leq1$,
 \item $\forall_{\lambda>1}\Theta(\lambda y)=\Theta(y)$ for $|y|>\ell$,
\end{enumerate}
where $\ell$ is some positive constant of dimension of length.
\end{dfn}
An example of a UV regular splitting function is the function $\Theta_n$ defined in Section~\ref{sec:weak_massive_proof}. The UV regularized splitting of a Schwartz distribution $t\in\mathcal{S}'(\R^N\times\R^M)$ is defined in the position space by
\begin{equation}
 t_\Theta(y,x):=\Theta(y)t(y,x).
\end{equation}
The result of the splitting $t_\Theta$ is again a Schwartz distribution. Equivalently, in the momentum space the splitting of a distribution $t\in\mathcal{S}'(\R^N\times\R^M)$ is given by
\begin{equation}\label{eq:splitting_general_momentum}
 \int \frac{\rd^N q}{(2\pi)^N}\,\FF{t_\Theta}(q,q') g(q) = \int \frac{\rd^N k}{(2\pi)^N}\, \F{t}(k,q')\int \frac{\rd^N q}{(2\pi)^N}\,\F{\Theta}(q-k)g(q)
\end{equation} 
for any $g\in\mathcal{S}(\R^N)$. In the rest of this section we consider only $t\in\mathcal{S}'(\R^N\times\R^M)$ in the momentum space and omit tilde over $t(q,q')$ and $t_\Theta(q,q')$.

\begin{lem}\label{lem:splitting_function_derivative}
Let $N\geq 2$ and $\Theta$ be a UV regular splitting function in $\R^N$. There exist functions $\F{\Theta}_\beta,\F{\Theta}^\mathrm{hom}_\beta,\F{\Theta}^\mathrm{rest}_\beta:\,\R^N\to\C$ for each multi-index $\beta$, $|\beta|=1$ such that
\begin{equation}\label{eq:lem_theta}
 \F{\Theta}(k) = \sum_{|\beta|=1}  \partial^\beta_k \F{\Theta}_\beta(k),
 ~~~~\F{\Theta}_\beta(k) = \F{\Theta}^\mathrm{hom}_\beta(k) + \F{\Theta}^\mathrm{rest}_\beta(k),
\end{equation}
where 
\begin{enumerate}[label=(\arabic*),leftmargin=*]
 \item $\F{\Theta}^\mathrm{hom}_\beta$ is smooth on $\R^N\setminus\{0\}$ and homogeneous of degree $-n+1$,
 \item $\F{\Theta}^\mathrm{rest}_\beta \in C^\infty(\R^N)$,
 \item $\F{\Theta}_\beta(k)$ vanishes at infinity faster than any power of $|k|$.
\end{enumerate}
It follows that $\F{\Theta}_\beta$ is absolutely integrable.
\begin{proof}
Let $\Theta_\beta(y):=(-\ri)\Theta(y) y^\beta/|y|^2$, $\Theta^\mathrm{hom}_\beta(y)$ be the unique homogeneous distribution of degree $-1$ such that $\Theta^\mathrm{hom}_\beta(y)=\Theta_\beta(y)$ for $|y|>\ell$ and $\Theta^\mathrm{rest}_\beta(y)=\Theta_\beta(y)-\Theta^\mathrm{hom}_\beta(y)$. The first equality in \eqref{eq:lem_theta} follows from
\begin{equation}
 \Theta(y) = \sum_{|\beta|=1}  \ri y^\beta \Theta_\beta(y).
\end{equation}
It is evident that $\F{\Theta}^\mathrm{rest}_\beta \in C^\infty(\R^N)$ as $\Theta^\mathrm{rest}_\beta(y)$ is of compact support. We have
\begin{equation}
 \int\frac{\rd^N k}{(2\pi)^N}\,\F{\Theta}_\beta(k)\, \tilde{g}(k)
 =\int\rd^N y\, \left[(-\Delta_y)^{m}\Theta_\beta(y)\right] 
  \int \frac{\rd^N k}{(2\pi)^N}\, \frac{1}{|k|^{2m}} \tilde{g}(k)\exp(-\ri k y)
\end{equation}
for every $g\in\mathcal{S}(\R^N)$ such that $0\notin\supp\,g$. It follows that the function $\F{\Theta}_\beta$ vanishes at infinity faster than any power of $|k|$ and is a smooth function outside the origin because 
 \begin{equation}
  |(-\Delta_y)^{m}\Theta_\beta(y)|\leq \const \, (1+|y|)^{-2m-1},
 \end{equation}
and we can choose $m$ at will. Since $\Theta^\mathrm{hom}_\beta$ is homogeneous of degree $-1$,  $\F{\Theta}^\mathrm{hom}_\beta$ is homogeneous of degree $-n+1$. 
\end{proof}
\end{lem}

\begin{thm}\label{thm:math_splitting}
Let $N\geq 2$ and $t\in\mathcal{S}'(\R^N\times\R^M)$, $t(q,q') = O^\mathrm{dist}(|q|^{d-\varepsilon})$, $d\in\Z$, $\varepsilon\in(0,1)$. There exist $c_\gamma\in C(\R^M)$ for multi-indices $\gamma$, $|\gamma|<d$ such that
\begin{equation}
 t_\Theta(q,q') = O^\mathrm{dist}(|q|^{d-\varepsilon}) + \sum_{|\gamma|< d} c_\gamma(q') q^\gamma.  
\end{equation}
\begin{proof}
Suppose that $0\notin \supp \, t$. It follows from \eqref{eq:splitting_general_momentum} that $t_\Theta(q,q')$ is a smooth function of $q$ continuous in $q'$ in some neighborhood of the origin. This concludes the proof in this case. Thus, because of the assumption $t(q,q') = O^\mathrm{dist}(|q|^{d-\varepsilon})$ it is enough to consider distributions $t$ which are globally of the form
\begin{equation}
 t(q,q') = \sum_{\alpha}
 \partial_q^\alpha t_{\alpha}(q,q'),
\end{equation}
where the functions $t_\alpha\in C(\R^N\times\R^M)$ are of compact support and satisfy the conditions listed in Definition \ref{def:O_not} with $\mathcal{O}=\R^N\times\R^M$ and $\delta=d-\varepsilon$. By the definition \eqref{eq:splitting_general_momentum} of $t_\Theta$ we get
\begin{equation}
 \int \frac{\rd^N q}{(2\pi)^N}\,t_\Theta(q,q') g(q) 
 =
 \sum_{\alpha} (-1)^{|\alpha|}
  \int \frac{\rd^N k}{(2\pi)^N}\, t_\alpha(k,q')
 \int \frac{\rd^N q}{(2\pi)^N}\,\F{\Theta}(q-k)\, \partial_q^\alpha g(q)
\end{equation}
for any $g\in\mathcal{S}(\R^N)$. Using Lemma \ref{lem:splitting_function_derivative} we obtain
\begin{equation}\label{eq:splitting_thm_terms}
 t_\Theta(q,q')
 =
 \sum_{\substack{\alpha,\beta\\|\beta|=1}} \partial^{\alpha+\beta}_q  
 \int \frac{\rd^N k}{(2\pi)^N}\,t_\alpha(k,q') \,\F{\Theta}_\beta(q-k),
\end{equation}
where the integrand on the RHS of the above equation is absolutely integrable. The theorem follows from Lemma \ref{lem:splitting_theta_alpha} with $D=d+|\alpha|\geq1$.
\end{proof}
\end{thm}

\begin{lem}\label{lem:splitting_theta_alpha}
Let $N\geq2$, $t\in C(\R^N\times\R^M)$ be of compact support, $|t(q,q')|\leq \const\, |q|^{D-\varepsilon}$ for all $(q,q')\in\R^M\times\R^N$ and fixed $D\in\N_+$ and $\varepsilon\in(0,1)$. There exist $t'\in C(\R^N\times\R^M)$ and $c_\gamma\in C(\R^M)$ for multi-indices $\gamma$, $|\gamma|\leq D$ such that
\begin{equation}\label{eq:lem_splitting_identity}
 \int\frac{\rd^N k}{(2\pi)^N}\,t(k,q') \F{\Theta}_\beta(q-k) 
 = t'(q,q')
 + \sum_{|\gamma|\leq D} c_\gamma(q') q^\gamma
\end{equation}
and $|t'(q,q')|\leq \const\, |q|^{D+1-\varepsilon}$ for all $(q,q')\in\R^M\times\R^N$.
\begin{proof}
We have
\begin{equation}\label{eq:lem_splitting_second_term}
 c_\gamma(q') = \frac{1}{\gamma!} \int\frac{\rd^N k}{(2\pi)^N}\,t(k,q') \left.\left[\partial_q^\gamma \F{\Theta}_\beta(q-k)\right]\right|_{q=0}
\end{equation}
and
\begin{equation}\label{eq:lem_splitting_first_term}
 t'(q,q')=\int\frac{\rd^N k}{(2\pi)^N}\, t(k,q') 
 \left[\F{\Theta}_\beta(q-k)- \F{\Theta}_\beta^D(q,k)\right], 
\end{equation} 
where
\begin{equation}\label{eq:lem_splitting_Taylor_def}
 \F{\Theta}_\beta^D(q,k):= \sum_{|\gamma|\leq D} \frac{1}{\gamma!}\, q^\gamma \left.\left[\partial_q^\gamma \F{\Theta}_\beta(q-k)\right]\right|_{q=0}
\end{equation}
is the Taylor approximation of degree $D$ of the function $q\mapsto \F{\Theta}_\beta(q-k)$. We will show below that expressions  \eqref{eq:lem_splitting_second_term} and \eqref{eq:lem_splitting_first_term} are well-defined and have the required properties. Using Lemma \ref{lem:splitting_theta_alpha} we get
\begin{equation}\label{eq:lem_splitting_Taylor_sum_bound}
 \left.\left[\partial_q^\gamma \F{\Theta}_\beta(q-k)\right]\right|_{q=0}  \leq \const\, \frac{1}{|k|^{N-1+|\gamma|}}
\end{equation}
and
\begin{equation}\label{eq:lem_splitting_Taylor_rest_bound}
 \left|\F{\Theta}_\beta(q-k)- \F{\Theta}_\beta^D(q,k)\right| \leq
 \const\,\sup_{\lambda\in[0,1]} \frac{|q|^{D+1}}{|\lambda q-k|^{N+D}}.
\end{equation}
The second bound is a consequence of the Taylor theorem. Because of the assumptions about $t$ the functions $c_\gamma$ given by \eqref{eq:lem_splitting_second_term} are well-defined and continuous.

Now let us turn to the function $t'$ defined by \eqref{eq:lem_splitting_first_term}. Upon introducing new integration variables $u=k/|q|$, we find 
\begin{multline}\label{eq:lem_splitting_def_t_prime}
 t'(q,q')=|q|^{D+1-\varepsilon}  
 \\
 \times\int\frac{\rd^N u}{(2\pi)^N}~ |q|^{-(D-\varepsilon)}
  t(|q|u,q') ~~|q|^{N-1}\left[\F{\Theta}_\beta(q-|q|u)- \F{\Theta}_\beta^D(q,|q|u)\right].
\end{multline}
Since $|\lambda q-|q|u|\geq \frac{1}{2}|q| |u|$ for $|u|\geq 2$ and $\lambda\in[0,1]$, it follows from \eqref{eq:lem_splitting_Taylor_rest_bound} and $|t(q,q')|\leq \const \,|q|^{D-\varepsilon}$ that for $|u|\geq2$ the integrand above is bounded by $\const\,|u|^{-N-\varepsilon}$. For $|u|<2$ we use the inequalities:
\begin{equation}
\begin{split}
 |q|^{N-1}|\F{\Theta}_\beta(q-|q|u)|\leq \const \, |q/|q|-u|^{-N+1},
 \\[4pt]
 |u|^{D-\varepsilon}|q|^{N-1}|\F{\Theta}_\beta^D(q,|q|u)| \leq \const\,  |u|^{-N+1-\varepsilon}.
\end{split}
\end{equation}
Consequently, the integral in \eqref{eq:lem_splitting_def_t_prime} exists and is a bounded function of $q$ and $q'$. The continuity of $t'$ is evident because the LHS of \eqref{eq:lem_splitting_identity} is a continuous function of $q$ and $q'$ and $c_\gamma\in C(\R^M)$.
\end{proof}
\end{lem}

%===================================================================================================
%===================================================================================================
%===================================================================================================
\section{Product}\label{sec:prod}
%===================================================================================================
%===================================================================================================
%===================================================================================================

\begin{dfn}[\underline{IR}-index]\label{def:IR1}
The translationally invariant distribution 
\begin{equation}
 t(x_1,\ldots,x_{n+1})\in\mathcal{S}'(\R^{4(n+1)})
\end{equation}
has \underline{IR}-index $d\in\Z$ iff for any $\varepsilon > 0$ it holds
\begin{equation}
 \F{\underline{t}}(q_1,\ldots,q_n) = O^\mathrm{dist}(|q_1,\ldots,q_n|^{d-\varepsilon}),
\end{equation}
where $\underline{t}(x_1,\ldots,x_{n})=t(x_1,\ldots,x_{n},0)$ (cf. Remark \ref{rem:transl} in Section \sectionref{rem:transl}).  If $n=0$ and $d\leq 0$ then by definition the distribution $t$ is an arbitrary constant. If $n=0$ and $d>0$ then $t=0$. Observe that if $t\in\mathcal{S}'(\R^{4(n+1)})$ vanishes in some neighborhood of $0$ in the momentum space then the \underline{IR}-index of $t$ is arbitrary large. If $t$ has \underline{IR}-index $d\in\Z$ then it also has \underline{IR}-index $d'\in\Z$ for any $d'\leq d$. 
\end{dfn}

The following lemma is an immediate consequence of Part~(B) of Theorem \ref{thm:math_adiabatic_limit}.
\begin{lem}\label{lem:IR_der}
Let $t\in\mathcal{S}'(\R^{4(n+1)})$ be translationally invariant distribution which has \underline{IR}-index $d$. Then the distribution $\F{\underline{t}}$ has zero of order $d$ at the origin in the sense of {\L}ojasiewicz (cf. Definition \ref{def:lojasiewicz}). 
\end{lem}

\begin{dfn}[IR-index]\label{def:IR2}
The distribution 
\begin{equation}
 t(x_1,\ldots,x_n;x'_1,\ldots,x'_m)\in\mathcal{S}'(\R^{4n}\times\R^{4m})
\end{equation}
has IR-index $d\in\Z$, with respect to the variables $x_1,\ldots,x_n$ iff for any $\varepsilon > 0$ and every $f\in\mathcal{S}(\R^{4m})$ it holds
\begin{multline}\label{eq:IR_index_integration}
 \int\mP{p_1}\ldots\mP{p_m}\,\F{t}(q_1,\ldots,q_n;q'_1-p_1,\ldots,q'_m-p_m) \F{f}(p_1,\ldots,p_m) 
 \\
 = O^\mathrm{dist}(|q_1,\ldots,q_n|^{d-\varepsilon}).
\end{multline}
The LHS of the above equation is the Fourier transform of the distribution 
\begin{equation}
 f(x'_1,\ldots,x'_{m'})\, t(x_1,\ldots,x_n;x'_1,\ldots,x'_{m'}).
\end{equation}
By definition if $n=0$ and $d\leq 0$ then the above condition is equivalent to saying that the distribution $t$ is a continuous function. If $n=0$ and $d>0$ then $t=0$. Observe that if $t\in\mathcal{S}'(\R^{4n}\times\R^{4m})$ vanishes in some neighborhood of $0\times \R^{4m}$ in the momentum space then the IR-index of $t$ with respect to the first $n$ variables is arbitrary large.  If $t$ has IR-index $d\in\Z$ then it also has IR-index $d'\in\Z$ for any $d'\leq d$. 
\end{dfn}

The rationale for introducing the \underline{IR}- and IR-index instead of using the notation $\tilde{t}(q,q')=O^\mathrm{dist}(|q|^\delta)$ is twofold. First, the formulations of Theorems \ref{thm:product} and \ref{thm:product_families} presented in this section are more transparent and concise in the position representation (albeit the proof of Theorem \ref{thm:product} is performed entirely in the momentum representation). Secondly, it will be important in the proof of the existence of Wightman and Green functions that for some distributions $t$ it hold $\tilde{t}(q,q')=O^\mathrm{dist}(|q|^\delta)$ with the exponent $\delta$ arbitrarily close to an integer. Note that $\delta$ cannot be an integer because Theorem \ref{thm:math_splitting} is formulated only for distributions $\tilde{t}(q,q')=O^\mathrm{dist}(|q|^\delta)$ with $\delta\notin\Z$. The presence of $\varepsilon$ in the above definitions is essential in the investigation of the splitting of distributions carried out in the next section. In contrast, in this section it does not play any role.

\begin{thm}\label{thm:product}
Let $t\in\mathcal{S}'(\R^{4n})$, $t'\in\mathcal{S}'(\R^{4n'})$ be translationally invariant distributions. Consider the translationally invariant distribution
\begin{equation}\label{eq:product_thm_dist}
 t(x_1,\ldots,x_n)~ t'(x'_1,\ldots,x'_{n'})~ 
  \prod_{k=1}^l  (\Omega|\normord{\partial^{\bar \alpha(k)} A_{\bar i(k)}(x_{\bar u(k)})}\,
 \normord{\partial^{\bar \alpha'(k)} A_{\bar i'(k)}(x'_{\bar u'(k)})}\Omega)
\end{equation}
such that for all $k\in\{1,\ldots,l\}$ the fields $A_{\bar i(k)}$, $A_{\bar i'(k)}$ are massless. In the above equation we used the notation introduced in Section~\ref{sec:aux}. Note that the distribution \eqref{eq:product_thm_dist} is well-defined as a result of Lemma 6.6 of \cite{brunetti2000microlocal} (cf. Lemma \ref{lem:sd}). Assume that one of the following conditions is satisfied:
\begin{enumerate}[leftmargin=*,label={(\arabic*)}]
 \item $t$ and $t'$ have \underline{IR}-indices $d$ and $d'$, respectively, 
 \item $t$ has \underline{IR}-index $d$ and $t'$ has IR-index $d'$ with respect to the first $k'\leq n'$ variables,
 \item $t$ has IR-index $d$ with respect to the first $k\leq n$ variables and $t'$ has \underline{IR}-index $d'$
\end{enumerate}
and set
\begin{equation}
 d''=d+d'+ \sum_{i=1}^{\mathrm{p}} [\dim(A_i) \bar\ext(A_i)+ \bar\der(A_i)] - 4.
\end{equation}
For the definition of $\bar\ext(A_i)$ and $\bar\der(A_{i})$ see Equations \eqref{eq:ext_s} and \eqref{eq:der_s}. Observe that, by assumption, $\bar\ext(A_i)=0$, $\bar\der(A_i)=0$ unless $A_i$ is a massless field. In respective cases, the distribution \eqref{eq:product_thm_dist} has
\begin{enumerate}[leftmargin=*,label={(\arabic*')}]
 \item \underline{IR}-index $d''$,
 \item IR-index $d''$ with respect to all variables but $x'_{k'+1},\ldots,x'_{n'}$,
 \item IR-index $d''$ with respect to all variables but $x_{k+1},\ldots,x_{n}$.
\end{enumerate}
The case (1') of this theorem is an analog of Lemma \ref{lem:sd}. The Steinmann scaling degree considered in that lemma controls the UV behavior of the distribution \eqref{eq:product_thm_dist}, whereas in the present theorem we study the IR properties of this distribution.
\begin{proof}
First, let us observe that for massless fields $A_i$ and $A_{i'}$ it holds
\begin{multline}
 (\Omega|\normord{\partial^{\alpha} A_{i}(x)}\,\normord{\partial^{\alpha'} A_{i'}(x')}\Omega)
 =
 \partial^{\alpha}_x \partial^{\alpha'}_{x'} (\Omega|\normord{A_{i}(x)}\,\normord{A_{i'}(x')}\Omega)
 \\
 =
 \sum_{\gamma\in\N_0^4}c_{ii'}^\gamma  \int \mH{0}{k} \, k^{\alpha+\alpha'+\gamma} \exp(-\ri k\cdot (x-x')) ,
\end{multline}
where $\rd\mu_0(k) = \frac{1}{(2\pi)^3} \rd^4 k\,\theta(k^0) \delta(k^2)$ and $c_{ii'}^\gamma   \in\C$ are constants such that if $c_{ii'}^\gamma\neq 0$ then $\dim(A_i)=\dim(A_{i'})$ and $|\gamma|=\dim(A_i)+\dim(A_{i'})-2$. The above statement follows immediately from the form of the explicit expressions for the (non-vanishing) two-point functions of the massless basic generators \eqref{eq:basic_gen} considered in the thesis:
\begin{enumerate}[leftmargin=*,label={(\alph*)}]
 \item the two-point function of the scalar, possibly charged, fields ($\dim(\phi)\!=\!\dim(\phi^*)\!=\!1$)
\begin{equation}
 (\Omega|\normord{\phi^*(x)}\,\normord{\phi(y)}\Omega) = -\ri D^{(+)}_0(x-y),
\end{equation}
\item the two-point function of the vector fields ($\dim(A_\mu)=3/2$)
\begin{equation}
 (\Omega|\normord{A_{\mu}(x)}\,\normord{A_{\nu}(y)}\Omega) = \ri g_{\mu\nu} D^{(+)}_0(x-y),
\end{equation}
\item the two-point function of the Dirac spinor fields ($\dim(\psi)=\dim(\psi^*)=1$)
\begin{equation}
\begin{aligned}
 (\Omega|\normord{\psi_a(x)}\,\normord{\psi^*_b(y)}\Omega) &= -\ri (\ri \slashed{\partial}_x \gamma^0)_{ab} D^{(+)}_0(x-y),
 \\
 (\Omega|\normord{\psi^*_b(y)}\,\normord{\psi_a(x)}\Omega) &= \ri(\ri \slashed{\partial}_x\gamma^0)_{ab} D^{(+)}_0(y-x),
\end{aligned} 
\end{equation}
\item the two-point function of the ghost fields ($\dim(u)=\dim(\tilde{u})=1$)
\begin{equation}
\begin{aligned}
 (\Omega|\normord{u(x)}\,\normord{\tilde{u}(y)}\Omega) = \ri D^{(+)}_0(x-y),
 \\
 (\Omega|\normord{\tilde{u}(x)}\,\normord{u(y)}\Omega) = \ri D^{(+)}_0(x-y), 
\end{aligned} 
\end{equation}
\end{enumerate}
where
\begin{equation}
 D^{(+)}_0(x):=\ri \int\mH{0}{k}\,\exp(-\ri k\cdot x).
\end{equation}
As a consequence the Fourier transform of the expression \eqref{eq:product_thm_dist} is a sum of terms which can be represented (up to a multiplicative constant) in each of the following three forms:
\leqnomode
\begin{multline}\label{eq:Fourier_product_T_Tprime_one}
\tag{A} (2\pi)^4 \delta(q_1+\ldots+q_n+q'_1+\ldots+q'_{n'}) 
 \\
 \int\mH{0}{k_1}\ldots\mH{0}{k_l}
 \,(2\pi)^4 \delta(q_1+\ldots+q_n-k_1-\ldots-k_l) 
 \\
 k^\beta\,\underline{\F{t}}(q_1-k(I_1),\ldots,q_{n-1}-k(I_{n-1})) 
 \,\underline{\F{t}}'(q'_1+k(I'_1),\ldots,q'_{n'-1}+k(I'_{n'-1})),
\end{multline}
\begin{multline}\label{eq:Fourier_product_T_Tprime_two}
\tag{B} \int\mH{0}{k_1}\ldots\mH{0}{k_l}
 \,(2\pi)^4 \delta(q_1+\ldots+q_n-k_1-\ldots-k_l) 
 \\
 k^\beta\,\underline{\F{t}}(q_1-k(I_1),\ldots,q_{n-1}-k(I_{n-1})) 
 \,\F{t}'(q'_1+k(I'_1),\ldots,q'_{n'}+k(I'_{n'})),
\end{multline}
\begin{multline}\label{eq:Fourier_product_T_Tprime_three}
\tag{C} \int\mH{0}{k_1}\ldots\mH{0}{k_l}
 \,(2\pi)^4 \delta(q'_1+\ldots+q'_{n'}+k_1+\ldots+k_l) 
 \\
 k^\beta\,\F{t}(q_1-k(I_1),\ldots,q_{n}-k(I_{n})) 
 \,\underline{\F{t}}'(q'_1+k(I'_1),\ldots,q'_{n'-1}+k(I'_{n'-1})), 
\end{multline}
\reqnomode
where $\beta$ is a multi-index such that
\begin{equation}
 |\beta| = \sum_{i=1}^{\mathrm{p}} [\dim(A_i) \bar\ext(A_i)+ \bar\der(A_i)] - 2l,
\end{equation}
the sets $I_1,\ldots,I_n$, $I'_1,\ldots,I'_{n'}$ are subsets of $\{1,\ldots,l\}$ such that
\begin{equation}
 \bigcup_{i=1}^{n} I_i = \bigcup_{i=1}^{n'} I'_i = \{1,\ldots,l\}
 ~~\textrm{and}~~I_i\cap I_j = \emptyset, 
 ~~I'_i\cap I'_j = \emptyset~~\textrm{for}~~i\neq j
\end{equation}
and 
\begin{equation}
 k(I) := \sum_{i\in I} k_i
\end{equation}
for any subset $I$ of $\{1,\ldots,l\}$. We remind the reader that the distribution $\underline{t}$ and $\underline{t}'$ were defined in terms of translationally-invariant distributions $t$ and $t'$ in Remark \ref{rem:transl} in Section \sectionref{rem:transl}.

The representations \eqref{eq:Fourier_product_T_Tprime_one}, \eqref{eq:Fourier_product_T_Tprime_two} and \eqref{eq:Fourier_product_T_Tprime_three} will be used to prove the statements (1'), (2'), (3') of the theorem, respectively. Let us denote the distribution \eqref{eq:product_thm_dist} by $t''$. The statements (1')-(3') of the theorem follow from Lemma \ref{lem:IR_index_product_general} and the fact that for $t\in\mathcal{S}'(\R^N\times\R^M)$ the condition $t(q,q')=O^\mathrm{dist}(|q|^\delta)$ is invariant under the linear changes of variables collectively denoted by $q\in\R^M$. Note that the Fourier transform of $\underline{t}''$ may be easily read off from the representation \eqref{eq:Fourier_product_T_Tprime_one} which is used in the proof of the statement (1') of the theorem.
\end{proof}
\end{thm}

The following fact will be needed in the proof of Lemma \ref{lem:IR_index_product_general}.
\begin{lem}\label{lem:IR_index_product_general_aux}
The Riesz distribution
\begin{equation}\label{eq:lem_IR_index_product_general_s}
 \R^4\ni k \mapsto s(k) = \frac{\pi^3}{4}\, k^2\theta(k^0)\theta(k^2) \in\R
\end{equation}
has the following properties:
\begin{enumerate}[leftmargin=*,label={(\arabic*)}]
 \item $s\in C(\R^4)$,
 \item $\supp \,s\subset \{k\in\R^4:~k^2\geq 0,\, k^0\geq 0\}$,
 \item $|s(k)|\leq \const \,|k|^2$ for all $k\in\R^4$,
 \item $(2\pi)^4\delta(k) = \square^3 s(k)$ where $\square$ is d'Alembertian.
\end{enumerate}
\end{lem}

\begin{lem}\label{lem:IR_index_product_general}
Let $t \in\mathcal{S}'(\R^{4n})$, $t'\in\mathcal{S}'(\R^{4n'})$ be such that
\begin{equation}
\begin{gathered}
 t(q_1,\ldots,q_n)=O^\mathrm{dist}(|q_1,\ldots,q_k|^{\delta}),
 \\
 t'(q'_1,\ldots,q'_{n'})=O^\mathrm{dist}(|q'_1,\ldots,q'_{k'}|^{\delta'}),
\end{gathered}
\end{equation}
where $1\leq k\leq n$, $1\leq k'\leq n'$, $\delta,\delta'\in\R$. Then for any linear functionals
\begin{equation}\label{eq:functionals_K}
 K_1,\ldots,K_n,K'_1,\ldots,K'_{n'}: \R^{4l} \rightarrow \R^4
\end{equation}
it holds
\begin{multline}\label{eq:thm_prod_form}
 t''(Q,q_1,\ldots,q_n,q'_1,\ldots,q'_{n'}):=
 \int\mH{0}{k_1}\ldots\mH{0}{k_l}\,
 (2\pi)^4\delta(Q-k_1-\ldots-k_l)
 \\
 k^\beta\,t(q_1-K_1(k),\ldots,q_n-K_n(k)) 
 \,t'(q'_1+K'_1(k),\ldots,q'_{n'}+K'_{n'}(k)) 
 \\
 = O^\mathrm{dist}(|Q,q_1,\ldots,q_k,q'_1,\ldots,q'_{k'}|^{\delta''}),
\end{multline}
where $k=(k_1,\ldots,k_l)$ and $\delta'' = \delta + \delta'+2l+|\beta|-4$.
\begin{proof}
The variables $q_1,\ldots,q_n$ will be denoted collectively by $q$, the variables $q_1,\ldots,q_k$ -- by $\mathbf{q}$ and similarly for the primed variables. By Definition \ref{def:O_not} the distributions $t$ and $t'$ admit (in some neighborhood of the origin) the following representations: 
\begin{equation}
 t(q)=\sum_\alpha\partial_\mathbf{q}^\alpha t_\alpha(q)~~~~\textrm{and}
 ~~~~ t'(q')=\sum_{\alpha'} \partial_{\mathbf{q}'}^{\alpha'} t'_{\alpha'}(q'),
\end{equation}
where the function $t_\alpha$ and $t'_{\alpha'}$ are continuous and such that
\begin{equation}\label{eq:lem_product_bounds}
 |t_\alpha(q)|\leq \const\, |\mathbf{q}|^{|\alpha|+\delta}~~~~\textrm{and}
 ~~~~|t'_{\alpha'}(q')|\leq\const\, |\mathbf{q}'|^{|\alpha'|+\delta'}.
\end{equation}
%The functions $t_\alpha$ and $t'_{\alpha'}$ vanish unless $|\alpha|+\delta\geq0$ and $|\alpha'|+\delta'\geq0$, respectively. 

The distribution $t''$ defined by \eqref{eq:thm_prod_form} is expressed in some neighborhood of $0$ by the following sum
\begin{equation}\label{eq:IR_index_product_general_derivatives}
 t''(Q,q,q')=\sum_{\alpha,\alpha'} \partial_{\mathbf{q}}^{\alpha}\partial_{\mathbf{q}'}^{\alpha'} \square_{Q}^3 t''_{\alpha\alpha'}(Q,q,q'),
\end{equation}
where
\begin{multline}\label{eq:thm_IR_index_product_general_T}
 t''_{\alpha\alpha'}(Q,q,q') := \int\mH{0}{k_1}\ldots\mH{0}{k_l}\,k^\beta
 s(Q-k_1-\ldots-k_l)
 \\
 t_\alpha(q_1-K_1(k),\ldots,q_n-K_n(k)) 
 ~t'_{\alpha'}(q'_1+K'_1(k),\ldots,q'_{n'}+K'_{n'}(k))
\end{multline}
and $s$ is a continuous function given in Lemma \ref{lem:IR_index_product_general_aux}. To prove the above statement we observe that as a result of the support properties of the function $s$ and of the measure $\rd\mu_0$ the integration region in \eqref{eq:thm_IR_index_product_general_T} may be restricted to $|k_1|,\ldots,|k_l|\leq |Q|$. Thus, it follows from the linearity of the maps \eqref{eq:functionals_K} that \eqref{eq:IR_index_product_general_derivatives} holds for $Q,q,q'$ in sufficiently small neighborhood of $0$ in $\R^{4n+4n'+4}$.

The continuity of the functions $t''_{\alpha\alpha'}$ is evident. Thus, to prove the statement of the lemma it is enough to show that 
\begin{equation}\label{eq:lem_product_bound}
 |t''_{\alpha\alpha'}(Q,q,q')|\leq \const \,|Q,q,q'|^{|\alpha|+\delta+|\alpha'|+\delta'+2l+|\beta|+2}.
\end{equation}
Upon changing the variables of integration $u_i=k_i/|Q|$ in \eqref{eq:thm_IR_index_product_general_T} we obtain
\begin{multline}
 t''_{\alpha\alpha'}(Q,q,q') = |Q|^{2l+|\beta|}
 \int_{|u_j|\leq 1}\mH{0}{u_1}\ldots\mH{0}{u_l}\,u^\beta
 s(Q-|Q| u_1-\ldots-|Q| u_l)
 \\ 
 t_\alpha(q_1-|Q| K_1(u),\ldots,q_n-|Q| K_n(u)) 
 t'_{\alpha'}(q'_1+|Q| K'_1(u),\ldots,q'_{n'}+|Q| K'_{n'}(u)).
\end{multline}
The bound \eqref{eq:lem_product_bound} follows from \eqref{eq:lem_product_bounds} and the inequalities
\begin{equation}
 |s(Q-|Q| u_1-\ldots-|Q| u_l)|\leq \const\, |Q|^2
\end{equation}
and
\begin{equation}
 |\mathbf{q}-|Q|\mathbf{K}(u)|^\eta,|\mathbf{q}'-|Q|\mathbf{K}'(u)|^\eta \leq \const\, |Q,q,q'|^{\eta}
\end{equation}
valid for $|u_1|,\ldots,|u_l|\leq 1$ and $\eta\geq 0$, where $\mathbf{K}(u)=(K_1(u),\ldots,K_k(u))$ and similarly for $\mathbf{K}'(u)$. This ends the proof.
\end{proof}
\end{lem}

Finally, we arrive at the main result of this section.

\begin{thm}\label{thm:product_families}
Let $F$ and $F'$ be $\Fa$ products with $n$ and $n'$ arguments, respectively. Fix $B_1,\ldots,B_n$, $B'_1,\ldots,B'_{n'}\in\Fh$ and set 
\begin{equation}\label{eq:thm_product_d}
 d = d_0 - \sum_{i=1}^{\mathrm{p}} [\dim(A_i)\ext_\mathbf{s}(A_i)+ \der_\mathbf{s}(A_i)],
 ~~~~
 d'= d'_0 - \sum_{i=1}^{\mathrm{p}} [\dim(A_i)\ext_{\mathbf{s}'}(A_i)+ \der_{\mathbf{s}'}(A_i)]
\end{equation}
for any lists of super-quadri-indices $\mathbf{s}=(s_1,\ldots,s_n)$ and $\mathbf{s}'=(s'_1,\ldots,s_{n'})$, where $d_0,d'_0\in\Z$ are some fixed constants. For definitions of $\ext_{\mathbf{s}}(A_i)$ and $\der_\mathbf{s}(A_i)$ see \eqref{eq:ext} and \eqref{eq:der}. Assume that the VEVs of the operator-valued distributions
\begin{equation}\label{eq:product_families}
 F(B_1^{(s_1)}(x_1),\ldots,B_n^{(s_n)}(x_n)),
 ~~~~~
 F'(B_1^{\prime(s'_1)}(x'_1),\ldots,B_{n'}^{\prime(s'_{n'})}(x'_{n'}))
\end{equation}
satisfy one of the following conditions. They have
\begin{enumerate}[leftmargin=*,label={(\arabic*)}]
 \item \underline{IR}-indices $d$ and $d'$, respectively,
 \item \underline{IR}-index $d$ and IR-index $d'$ with respect to the first $k'\leq n'$ variables, respectively,
 \item IR-index $d$ with respect to the first $k\leq n$ variables and \underline{IR}-index $d'$, respectively,
\end{enumerate}
for all super-quadri-indices $s_1, \ldots, s_n$, $s'_1,\ldots, s'_{n'}$ which involve only massless fields. By the axiom \ref{axiom2} it holds $d,d'\in\Z$ unless the VEVs of the products \eqref{eq:product_families} vanish. 

Then the VEV of the product of $F$ and $F'$ 
\begin{equation}\label{eq:product_T_bis}
 F(B_1(x_1),\ldots,B_n(x_n))F'(B_1^{\prime}(x'_1),\ldots,B_{n'}^{\prime}(x'_{n'}))
\end{equation}
and the graded commutator of $F$ and $F'$
\begin{equation}\label{eq:product_com}
 [F(B_1(x_1),\ldots,B_n(x_n)),F'(B_1^{\prime}(x'_1),\ldots,B_{n'}^{\prime}(x'_{n'}))]
\end{equation}
have 
\begin{enumerate}[leftmargin=*,label={(\arabic*')}]
 \item \underline{IR}-index $d_0+d'_0-4$,
 \item IR-index $d_0+d'_0-4$ with respect to all variables but $x'_{k'+1},\ldots,x'_{n'}$,
 \item IR-index $d_0+d'_0-4$ with respect to all variables but $x_{k+1},\ldots,x_{n}$,
\end{enumerate} 
in the respective cases. In the case of the graded commutator it is enough that the assumptions (1)-(3) are satisfied for all super-quadri-indices $s_1, \ldots, s_n$, $s'_1, \ldots, s'_{n'}$ which involve only massless fields such that at least one of them is non-zero.

\begin{proof}
The proof is based on the formula \eqref{eq:vev_product_representation} which we reproduce below for the convenience of the reader
\begin{multline}\label{eq:vev_product_representation2}
(\Omega|F(B_1(x_1),\ldots,B_n(x_n)) F'(B_n^{\prime}(x'_1),\ldots,B^{\prime}_{n'}(x'_{n'}))\Omega)
\\
 =\sum_{\bar u,\bar u',\bar i,\bar i',\bar \alpha,\bar \alpha'} ~\const~~
 \prod_{k=1}^l  (\Omega|\normord{\partial^{\bar \alpha(k)}\!A_{\bar i(k)}(x_{\bar u(k)})}\,
 \normord{\partial^{\bar \alpha'(k)}\!A_{\bar i'(k)}(x'_{\bar u'(k)})}\Omega)\times
 \\
 (\Omega|F(B^{(s_1)}_1(x_1),\ldots,B^{(s_n)}_n(x_n))\Omega) \, (\Omega| F'(B^{\prime(s'_1)}_1(x'_1),\ldots,B^{\prime(s'_{n'})}_{n'}(x'_{n'}))\Omega).
\end{multline}
The notation used in the above equation was introduced in Section~\ref{sec:aux}. We will show that the statements (1')-(3') hold independently for each term of the above sum. First note that as a result of Definitions \ref{def:IR1} and \ref{def:IR2} of the \underline{IR}- and IR-index and Part~(A) of Lemma \ref{lem:aux_lemma} this is true for all the terms of the sum on the RHS of the formula \eqref{eq:vev_product_representation2} for which at least one of the super-quadri-indices $s_1,\ldots,s_n$, $s'_1,\ldots,s'_{n'}$ involves a massive field. The statements in the cases (1')-(3') follow now directly from Theorem \ref{thm:product} and the constraints \eqref{eq:constraints}. In order to prove the claim about the graded commutator we use Part~(B) of Lemma  \ref{lem:aux_lemma} and the previous result. 
\end{proof}
\end{thm}

%===================================================================================================
%===================================================================================================
%===================================================================================================
\section{Splitting}\label{sec:split}
%===================================================================================================
%===================================================================================================
%===================================================================================================

\begin{thm}\label{thm:split}
Assume that the distribution
\begin{equation}
  (\Omega|\Dif(B_1(y_1),\ldots,B_n(y_n);B_{n+1}(y_{n+1}))\Omega)
\end{equation}
has \underline{IR}-index $d\in\Z$. Then there exist constants $c_\alpha\in\C$ for multi-indices $\alpha$, $|\alpha|<d$ such that the distribution
\begin{equation}
  (\Omega|\Adv(B_1(y_1),\ldots,B_n(y_n);B_{n+1}(y_{n+1}))\Omega) - \sum_{\substack{\alpha\\|\alpha|<d}} c_\alpha \partial^\alpha_y \delta(y_1-y_{n+1})\ldots \delta(y_n-y_{n+1})
\end{equation}
has \underline{IR}-index $d$.
\begin{proof}
Let $I=(B_1(y_1),\ldots,B_n(y_n))$. Using the definition \eqref{eq:def_dif} of the $\Dif$ product we get
\begin{equation}\label{eq:split1_decomposition}
\begin{split}
 (\Omega|\Adv(I;B_{n+1}(0))\Omega)
 =(1-\Theta_n(y_1,\ldots,y_n))\,&(\Omega|\Adv(I;B_{n+1}(0))\Omega)
 \\
 +\Theta_n(y_1,\ldots,y_n)\,&(\Omega|\Ret(I;B_{n+1}(0))\Omega)
 \\
 +\Theta_n(y_1,\ldots,y_n)\,&(\Omega|\Dif(I;B_{n+1}(0))\Omega),
\end{split} 
\end{equation}
where the functions $\Theta_n$ were introduced in Section~\ref{sec:weak_massive_proof}. It is enough to prove that the Fourier transform of each term of the RHS of the above equation is of the form 
\begin{equation}\label{eq:split1_form}
 \tilde{t}(q_1,\ldots,q_n) = \sum_{|\gamma|<d} c_\gamma (-\ri)^{|\gamma|} q^\gamma + O^\mathrm{dist}(|q_1,\ldots,q_n|^{d-\varepsilon}) 
\end{equation}
for any $\varepsilon\in(0,1)$ and some constants $c_\gamma\in\C$, where $\gamma$ is a multi-index. 

For the last term it is an immediate consequence of Theorem \ref{thm:math_splitting} with $M=0$ and Definition \ref{def:IR1} of the \underline{IR}-index. Because of the support property of the retarded distribution (see Section~\ref{sec:W_G_IR}) and of the function $\Theta_n$  (see Section~\ref{sec:weak_massive_proof}) the support of the second term on the RHS of \eqref{eq:split1_decomposition} is contained in the union of the regions: $$\{(y_1,\ldots,y_n)\in\R^{4n}\,:\,|(y_1,\ldots,y_n)|\leq \ell\}$$ and 
\begin{equation}
 \Gamma^-_{n} \cap \{ (y_1,\ldots,y_n):\, 
 (y_1^0+\ldots+y_n^0)+\tfrac{1}{3n}|(y_1,\ldots,y_n)|\geq0\}.
\end{equation} 
It follows from the definition \eqref{eq:def_cones} of the cone $\Gamma^-_{n}$ that the second region is empty. Thus, the support of the second term on the RHS of Equation \eqref{eq:split1_decomposition} vanishes for $|(y_1,\ldots,y_n)|> \ell$. The same is true for the first term on the RHS of this equation. Consequently, the Fourier transforms of these terms are smooth functions. To show that they are of the form \eqref{eq:split1_form} it is enough to use the Taylor theorem. 
\end{proof}
\end{thm}

\begin{thm}\label{thm:split_gen}
Assume that the distribution
\begin{equation}\label{eq:thm_split_gen_D}
 (\Omega|\Dif(B_1(y_1),\ldots,B_n(y_n);C_1(x_1),\ldots,C_m(x_m);P)\Omega)
\end{equation}
has IR-index $d\in\Z$ with respect to $y_1,\ldots,y_n$.
\begin{enumerate}[leftmargin=*,label={(\Alph*)}]
\item The distribution
\begin{equation}\label{eq:splitting2_proof_index}
 (\Omega|\Adv(B_1(y_1),\ldots,B_n(y_n);C_1(x_1),\ldots,C_m(x_m);P)\Omega)
\end{equation}
has IR-index $\min(d,0)$ with respect to $y_1,\ldots,y_n$. 
\item If $d=1$ then for every $f\in\mathcal{S}(\R^{4m})$ and $\varepsilon>0$ there exists $c\in\C$ such 
\begin{multline}
 \int\rd^4 x_1\ldots\rd^4 x_m\,(\Omega|\Adv(\F{B}_1(q_1),\ldots,\F{B}_n(q_n);C_1(x_1),\ldots,C_m(x_m))\Omega) \,f(x_1,\ldots,x_m) 
 \\
 = O^\mathrm{dist}(|q_1,\ldots,q_n|^{1-\varepsilon})+ c.
\end{multline}
\end{enumerate}
\begin{proof}
Let $I= (B_1(y_1),\ldots,B_n(y_n))$, $J= (C_1(x_1),\ldots,C_m(x_m))$. Using the definition \eqref{eq:def_dif_gen} of the generalized $\Dif$ product we obtain 
\begin{equation}\label{eq:split_gen_decomposition}
\begin{split}
 (\Omega|\Adv(I;J;P)\Omega)
 =(1-\Theta_n(y_1,\ldots,y_n))\,&(\Omega|\Adv(I;J;P)\Omega)
 \\
 +\Theta_n(y_1,\ldots,y_n)\,&(\Omega|\Ret(I;J;P)\Omega)
 \\
 +\Theta_n(y_1,\ldots,y_n)\,&(\Omega|\Dif(I;J;P)\Omega),
\end{split} 
\end{equation}
where the functions $\Theta_n$ were introduced in Section~\ref{sec:weak_massive_proof}. By Definition of the IR-index \ref{def:IR2} it holds
\begin{multline}\label{eq:thm_split_gen_D_int}
 \int\mP{p_1}\ldots\mP{p_m}\,(\Omega|\Dif(\F{B}_1(q_1),\ldots,\F{B}_n(q_n);\F{C}_1(q'_1-p_1),\ldots,\F{C}_m(q'_m-p_m))\Omega)
 \\ \times\F{f}(p_1,\ldots,p_m) 
 = O^\mathrm{dist}(|q_1,\ldots,q_n|^{d-\varepsilon})
\end{multline}
for every $\varepsilon>0$ and $f\in\mathcal{S}(\R^{4m})$. We will prove that for $t(y_1,\ldots,y_n;x_1,\ldots,x_m)$ being any of the three terms on the RHS of Equation \eqref{eq:split_gen_decomposition} there exist $c_\gamma\in C(\R^{4m})$ for multi-indices $\gamma$, $|\gamma|<d$ such that
\begin{multline}\label{eq:splitting2_proof}
 \int\mP{p_1}\ldots\mP{p_m}\,\tilde{t}(q_1,\ldots,q_n;q'_1-p_1,\ldots,q'_m-p_m) \F{f}(p_1,\ldots,p_m) 
 \\
 = O^\mathrm{dist}(|q_1,\ldots,q_n|^{d-\varepsilon}) + \sum_{|\gamma|< d} c_\gamma(q') q^\gamma.
\end{multline}
For the first and the second terms this follows form Lemma \ref{lem:splitting_ind_two_smooth} from Section~\ref{sec:weak_massive_proof} and the Taylor theorem. For the last term it is an immediate consequence of Theorem \ref{thm:math_splitting}. Thus, Equation \eqref{eq:splitting2_proof} is valid also for $t(y_1,\ldots,y_m;x_1,\ldots,x_m)=(\Omega|\Adv(I;J;P)\Omega)$. Part~(A) follows now from Definition of the IR-index since for $d\leq 0$ the last term on the RHS of \eqref{eq:splitting2_proof} is absent. 

Next, we observe that by the above result and the third remark below Definition \ref{def:O_not} at the beginning of Section~\ref{sec:math} we have
\begin{multline}
 \int\mP{p_1}\ldots\mP{p_m}\,(\Omega|\Adv(\F{B}_1(q_1),\ldots,\F{B}_n(q_n);\F{C}_1(p_1),\ldots,\F{C}_m(p_m))\Omega)
 \\ \times\F{f}(-p_1,\ldots,-p_m) 
 = O^\mathrm{dist}(|q_1,\ldots,q_n|^{d-\varepsilon})+ \sum_{|\gamma|< d} c_\gamma(0) q^\gamma.
\end{multline}
To show Part~(B) we note that the last term on the RHS of the above equation is a constant for $d=1$.
\end{proof}
\end{thm}

%===================================================================================================
%===================================================================================================
\section{Proof for theories with massless particles}\label{sec:proof_scalar}
%===================================================================================================
%===================================================================================================

In this section we prove the existence of the weak adiabatic limit in theories containing massless particles. The idea of the proof was described in Section~\ref{sec:idea}. Our proof is valid only if some of the time-ordered products of the sub-polynomials of the interaction vertices satisfy certain condition which is formulated below as the normalization condition \ref{norm:wAL}. In Theorem \ref{thm:main1}, we prove that this condition may be imposed in all models satisfying Assumption \ref{asm} below and is compatible with the normalization condition \ref{norm:sd}. We also show that \ref{norm:wAL} may be rewritten as the normalization condition \ref{norm:wAL2}. The latter form is used in the proof of Theorem \ref{thm:main2} which states the existence of the Wightman and Green functions. The proof of the compatibility of \ref{norm:wAL} with other standard normalization conditions usually imposed on the time-ordered products like e.g. the unitarity or Poincar{\'e} covariance is postponed to Chapter~\ref{ch:comp}. 

The existence of the weak adiabatic limit in theories with massless particles needs some specific normalization of time-ordered products. In Appendix~\ref{sec:mass} we show that the correct mass normalization of all massless fields is necessary for the existence of the weak adiabatic limit. Note that this condition is a part of the normalization condition \ref{norm:wAL}. The Wightman and Green functions cannot be defined by means of the weak adiabatic limit in models in which the correct mass normalization of massless fields is not possible. An example of such model is the massless $\varphi^3$ theory.

Our proof of the existence of the Wightman and Green functions applies to the following class of models.
\begin{asm}\label{asm}
We assume that all interaction vertices $\mathcal{L}_1,\ldots,\mathcal{L}_\mathrm{q}$ of the models under consideration have vanishing fermion number and satisfy one of the following conditions:
\begin{enumerate}[leftmargin=*,label={(\arabic*)}]
\item $\dim(\mathcal{L}_l)=3$ and $\mathcal{L}_l$ contains at least one massive field\hspace{0.3mm}\footnote{A polynomial $B\in\Fa$ contains at least one massive field if it is a combination of products of generators, each with a massive field factor.} for all $l\in\{1,\ldots,\mathrm{q}\}$,
\item $\dim(\mathcal{L}_l)=4$ for all $l\in\{1,\ldots,\mathrm{q}\}$.
\end{enumerate}
Moreover, we assume throughout that the normalization condition \ref{norm:sd} holds with $\CC=1$ in the case (1) and $\CC=0$ in the case (2) or, stated differently,
\begin{equation}\label{eq:proof_dim}
 \CC = 4 - \dim(\mathcal{L}_l).
\end{equation}
\end{asm}
Other conditions which have to be fulfilled by the interaction vertices are listed in the introduction to Chapter~\ref{sec:models}. They ensure the correct physical properties of the model but do not play any role in the proof of the existence of the weak adiabatic limit presented in this section. We will, however, use them afterward to prove that the Wightman and Green functions have all the standard properties. Note that in the proof of the existence of the Wightman and Green function in purely massive theories we only use the fact that the interaction vertices have vanishing fermion number.

For any list of super-quadri indices $\mathbf{u}=(u_1,\ldots,u_k)$ we set
\begin{equation}\label{eq:proof_omega}
 \omega = 4 - \sum_{i=1}^{\mathrm{p}} [\dim(A_i) \ext_{\mathbf{u}}(A_i)+ \der_{\mathbf{u}}(A_i)] = 4 - \sum_{j=1}^k \dim(A^u).
\end{equation}
The functions $\ext_{\mathbf{u}}(\cdot)$, $\der_{\mathbf{u}}(\cdot)$ are given by \eqref{eq:ext} and \eqref{eq:der} and $\mathrm{p}$ is the number of basic generators. The above definition of $\omega$ coincides with the one introduced in Chapter~\ref{sec:models}. In what follows it will be important that $\omega\in\Z$ (the \underline{IR}- and IR-index are by definitions integers). By the axiom \ref{axiom2} this is certainly true unless the distribution
\begin{equation}
 (\Omega|\T(\mathcal{L}^{(u_1)}_{l_1}(x_1),\ldots,\mathcal{L}^{(u_{k})}_{l_k}(x_k))\Omega)
\end{equation}
vanishes.

\begin{enumerate}[label=\bf{N.wAL},leftmargin=*]
\item\label{norm:wAL} 
Let $k\in\N_+$, $l_1,\ldots,l_k\in\{1,\ldots,\mathrm{q}\}$ and $\mathbf{u}=(u_1,\ldots,u_k)$ be a list of super-quadri-indices such that every $u_j$ involves only massless fields (i.e. $\ext_{\mathbf{u}}(A_i)=0$ unless $A_i$ is a massless field). The Fourier transform of the distribution
\begin{equation}\label{eq:wAL_vev}
 (\Omega|\T(\mathcal{L}^{(u_1)}_{l_1}(x_1),\ldots,\mathcal{L}^{(u_{k})}_{l_k}(x_k))\Omega)
\end{equation}
is of the form
\begin{equation}
 (2\pi)^4 \delta(q_1+\ldots+q_k)\,\F{\underline{t}}_{l_1,\ldots,l_k}^{u_1,\ldots,u_k}(q_1,\ldots,q_{k-1}), 
\end{equation}
where $\F{\underline{t}}_{l_1,\ldots,l_k}^{u_1,\ldots,u_k}$ has zero of order $\omega$ at $q_1=\ldots=q_{k-1}=0$ in the sense of {\L}ojasiewicz, i.e. for all multi-indices $\gamma$ such that $|\gamma|<\omega$ it holds
\begin{equation}
  \partial^\gamma_q \F{\underline{t}}_{l_1,\ldots,l_k}^{u_1,\ldots,u_k}(q_1,\ldots,q_n)\big|_{q=0} = 0
\end{equation}
in the sense of {\L}ojasiewicz (cf. Definition \ref{def:lojasiewicz}). If $u_1,\ldots,u_k$ are such that $\omega\leq 0$ then there is no condition on $\F{\underline{t}}_{l_1,\ldots,l_k}^{u_1,\ldots,u_k}$. 
\end{enumerate}

\begin{enumerate}[label=\bf{N.wAL'},leftmargin=*]
\item\label{norm:wAL2} 
Let $F$  be any $\Fa$ product which is a combination of products of the time-ordered products. The distribution 
\begin{equation}\label{eq:wAL_dist2}
 (\Omega|F(\mathcal{L}^{(u_1)}_{l_1}(x_1),\ldots,\mathcal{L}^{(u_{k})}_{l_k}(x_k))\Omega)
\end{equation}
has \underline{IR}-index $\omega$ given by \eqref{eq:proof_omega} for all $k\in\N_+$, $l_1,\ldots,l_k\in\{1,\ldots,\mathrm{q}\}$ and any list of super-quadri-indices $\mathbf{u}=(u_1,\ldots,u_k)$ such that each $u_j$ involves only massless fields (i.e. $\ext_{\mathbf{u}}(A_i)=0$ unless $A_i$ is a massless field). In particular $F$ may be the time-ordered product $\T$ or the anti-time-ordered product $\aT$.
\end{enumerate}

Before we show that the above normalization conditions are equivalent and are compatible with \ref{norm:sd} and the axioms \ref{axiom1}-\ref{axiom6} let us make a few observations.
\begin{enumerate}[label=(\arabic*),leftmargin=*]
\item The condition \ref{norm:wAL2} implies \ref{norm:wAL} by Lemma \ref{lem:IR_der}. 

\item The condition \ref{norm:wAL} says that the time-ordered products corresponding to the scattering processes involving only massless particles in the incoming and outgoing states, in particular the self-energy corrections, have to be properly normalized.

\item The condition \ref{norm:wAL2} which is seemingly stronger than \ref{norm:wAL} will be needed in the proof of Theorem \ref{thm:main2} which guarantees the existence of the Wightman and Green functions. 

\item Let $\Fa_\mathcal{L}$ be any maximal linearly independent subset of
\begin{equation}\label{eq:proof_set}
 \{\mathcal{L}^{(u)}_{l} \in\Fa\,:\, l\in\{1,\ldots,\mathrm{q}\}, ~u~~ \textrm{involves only massless fields} \}.
\end{equation}
The normalization condition \ref{norm:wAL} is equivalent to its modified version involving only the time-ordered products \eqref{eq:wAL_vev} of elements of $\Fa_\mathcal{L}$. An analogous statement is also true for \ref{norm:wAL2}.

The above statement is not trivial because elements of the set \eqref{eq:proof_set} may be linearly dependent. Actually, it is true only because we do not consider models with interaction vertices of different dimensions. Its proof is based on the equality \eqref{eq:proof_dim} which implies that $\dim(\mathcal{L}^{(u)}_l)=\dim(\mathcal{L}_l)-\dim(A^u) = 4-\CC -\dim(A^u)$.  As a result, $\omega$ given by \eqref{eq:proof_omega} which depends explicitly on $u_1,\ldots,u_k$ may be expressed in terms of the interaction vertices sub-polynomials $B_1=\mathcal{L}^{(u_1)}_{l_1},\ldots,B_k=\mathcal{L}^{(u_{k})}_{l_k}$ which appear in \eqref{eq:wAL_vev} or \eqref{eq:wAL_dist2}. In fact, $\omega=\omega'$, where by definition
\begin{equation}\label{eq:proof_omega_prime}
 \omega':=4-\sum_{j=1}^{k} (4-\CC- \dim(B_j))
\end{equation}
for any $B_1,\ldots,B_k\in\Fh$.

\item Assume that the time-ordered products with at most $n$ arguments are given. We will characterize the ambiguity in defining the time-ordered product with $n+1$ arguments assuming that it satisfies one of the normalization conditions: \ref{norm:wAL} or \ref{norm:wAL2} (the normalization freedom is the same for each of these conditions; in fact, these conditions are equivalent). Any two possible definitions of
\begin{equation}
 \Fa^{n+1}\ni(B_1,\ldots,B_{n+1})\mapsto (\Omega|\T(B_1(x_1),\ldots,B_{n+1}(x_{n+1}))\Omega)\in\mathcal{S}'(\R^{4(n+1)})
\end{equation}
differ by the graded-symmetric map
\begin{equation}
  \Fa^{n+1}\ni(B_1,\ldots,B_{n+1})\mapsto v(B_1(x_1),\ldots,B_{n+1}(x_{n+1}))\in\mathcal{S}'(\R^{4(n+1)})
\end{equation}
such that for all $B_1,\ldots,B_{n+1}\in\Span_\C\Fa_\mathcal{L}$ the distribution $v(B_1(x_1),\ldots,B_{n+1}(x_{n+1}))$ is of the form
\begin{equation}\label{eq:proof_freedom}
 \sum_{\substack{\gamma\\|\gamma|\geq\omega'}} c_\gamma \partial^\gamma \delta(x_1-x_{n+1})\ldots \delta(x_{n}-x_{n+1})
\end{equation}
for some constants $c_\gamma\in\C$ indexed by multi-indices $\gamma$, $|\gamma|\geq\omega'$ such that only finite number of $c_\gamma$ is non-zero. The above statement may be proved with the use of the method from Section~\ref{sec:freedom}. Note that the distribution \eqref{eq:proof_freedom} has \underline{IR}-index $\omega'$. Moreover, its Fourier transform is of the form $(2\pi)^4\delta(q_1+\ldots+q_{n+1}) P(q_1,\ldots,q_n)$, where the polynomial $P(q_1,\ldots,q_n)$ has zero of order $\omega'$ at $q_1=\ldots=q_n=0$.

\item We always impose the condition \ref{norm:wAL} (or equivalently \ref{norm:wAL2}) simultaneously with \ref{norm:sd}. In this case by result of Section~\ref{sec:freedom} the ambiguity in defining the time-ordered product with $n+1$ arguments may be described as above with the additional stipulation that $c_\gamma$ in the expression \eqref{eq:proof_freedom} may be non-zero only for $|\gamma|=\omega$.
\end{enumerate}

\begin{thm}\label{thm:main1} Suppose that Assumption \ref{asm} holds.
\begin{enumerate}[leftmargin=*,label={(\Alph*)}]
\item It is possible to define the time-ordered products in such a way that both normalization conditions \ref{norm:sd} and \ref{norm:wAL2} are satisfied. This is achieved as follows. 

Assume that the time-ordered products with at most $n$ arguments are given and fulfill the conditions \ref{norm:sd} and \ref{norm:wAL2}. Define the time-ordered products with $n+1$ arguments arbitrarily such that the condition \ref{norm:sd} holds. It is possible to modify this definition such that the conditions \ref{norm:sd} and \ref{norm:wAL2} hold simultaneously. To this end, it is enough to redefine the VEV of time-ordered product with $n+1$ arguments 
\begin{equation}\label{eq:thm_main1_time}
 \Fa^{n+1}\ni(B_1,\ldots,B_{n+1})\mapsto (\Omega|\T(B_1(x_1),\ldots,B_{n+1}(x_{n+1}))\Omega)\in\mathcal{S}'(\R^{4(n+1)})
\end{equation}
by adding to it the graded-symmetric map
\begin{equation}\label{eq:comp_delta_form}
 \Fa^{n+1}\ni(B_1,\ldots,B_{n+1})\mapsto v(B_1(x_1),\ldots,B_{n+1}(x_{n+1}))\in\mathcal{S}'(\R^{4(n+1)})
\end{equation}
such that 
\begin{enumerate}[label=(\roman*),leftmargin=*]
\item $v(B_1(x_1),\ldots,B_{n+1}(x_{n+1}))$ is non-zero only for $B_1,\ldots,B_{n+1}\in\Span_\C\Fa_\mathcal{L}$,
\item $v(B_1(x_1),\ldots,B_{n+1}(x_{n+1}))=0$ for all $B_1,\ldots,B_{n+1}\in\Fh$ such that $\mathbf{f}(B_1)+\ldots+\mathbf{f}(B_{n+1})\neq 0$,
\item for all $B_1,\ldots,B_{n+1}\in\Fa$ the distribution $v(B_1(x_1),\ldots,B_{n+1}(x_{n+1}))$ is of the form
\begin{equation}
 \sum_{\substack{\gamma\\|\gamma|<\omega'}} c_\gamma \partial^\gamma \delta(x_1-x_{n+1})\ldots \delta(x_{n}-x_{n+1})
\end{equation}
for some constants $c_\gamma\in\C$ indexed by multi-indices $\gamma$, $|\gamma|<\omega'$, where $\omega'$ is given by \eqref{eq:proof_omega_prime}.
\end{enumerate}
The above redefinition is compatible with the normalization condition \ref{norm:sd}. 

%If $\mathbf{f}(B_1)+\ldots+\mathbf{f}(B_{n+1})\neq 0$ then then the redefinition of \eqref{eq:thm_main1_time} is neither needed nor possible by the axiom \ref{axiom2}.

\item The normalization condition \ref{norm:wAL} and \ref{norm:wAL2} are equivalent.
\end{enumerate}

\begin{proof}
First, let us prove inductively Part~(A) of the theorem. The induction is with respect to the number of arguments of the time-ordered products which we denote by $k$. For $k=1$ we have to consider the distributions of the form
\begin{equation}\label{eq:thm_main_one}
 (\Omega|\T(\mathcal{L}^{(u)}_{l}(x))\Omega) = (\Omega|\mathcal{L}^{(u)}_{l}(x)\Omega),
\end{equation}
where the super-quadri-index $u$ involves only massless fields and $l\in\{1,\ldots,\mathrm{q}\}$. By assumption $\mathcal{L}_{l}$ contains at least one massive field or $\dim(\mathcal{L}_l)=4$. First note that \eqref{eq:thm_main_one} is always a constant, which is non-zero iff $\mathcal{L}^{(u)}_{l}$ contains a non-zero constant term in the decomposition into monomials. The later condition is never satisfied if $\mathcal{L}_l$ contains a massive field. Thus, \eqref{eq:thm_main_one} is zero in this case. If $\dim(\mathcal{L}_l)=4$ then $\mathcal{L}_l^{(u)}$ may contain a non-zero constant term in the decomposition into monomials only if $\dim(A^u)=4$ which implies that $\omega=4-\dim(A^u)=0$ (set $k=1$ in Equation \eqref{eq:proof_omega}). As a result, the normalization condition \ref{norm:wAL} is satisfied in both cases. It is also evident that the distribution \eqref{eq:thm_main_one} has the correct \underline{IR}-index $\omega$ (see Definition \ref{def:IR1} of the \underline{IR}-index), thus, the condition \ref{norm:wAL2} holds as well.

In the inductive step we will use the following fact which is an immediate consequence of the statement (1') of Theorem \ref{thm:product_families}. Let $F$ and $F'$ be two $\Fa$ products with $n$ and $n'$ arguments, respectively. Assume that for all $l_1,\ldots,l_n,l'_1,\ldots,l'_{n'}\in\{1,\ldots,\mathrm{q}\}$ and all lists $\mathbf{u}=(u_1,\ldots,u_{n})$ and $\mathbf{u}'=(u_1,\ldots,u_{n'})$ of super-quadri-indices involving only massless fields the VEVs of the operator-valued distributions
\begin{equation}
 F(\mathcal{L}_{l_1}^{(u_1)}(x_1),\ldots,\mathcal{L}_{l_n}^{(u_n)}(x_n)),
 ~~~~~
 F'(\mathcal{L}_{l'_1}^{(u'_1)}(x'_1),\ldots,\mathcal{L}_{l'_{n'}}^{(u'_{n'})}(x'_{n'}))
\end{equation}
have \underline{IR}-indices
\begin{equation}
 d = 4 - \sum_{i=1}^{\mathrm{p}} [\dim(A_i) \ext_{\mathbf{u}}(A_i)+ \der_{\mathbf{u}}(A_i)] ,
 ~~~~~
 d' = 4 - \sum_{i=1}^{\mathrm{p}} [\dim(A_i) \ext_{\mathbf{u}'}(A_i)+ \der_{\mathbf{u}'}(A_i)],
\end{equation}
respectively. Then the \underline{IR}-index of 
\begin{equation}
 F(\mathcal{L}_{l_1}^{(u_1)}(x_1),\ldots,\mathcal{L}_{l_n}^{(u_n)}(x_n))
 F'(\mathcal{L}_{l'_1}^{(u'_1)}(x'_1),\ldots,\mathcal{L}_{l'_{n'}}^{(u'_{n'})}(x'_{n'}))
\end{equation}
equals
\begin{equation}
 d'' = 4 - \sum_{i=1}^{\mathrm{p}} [\dim(A_i) (\ext_{\mathbf{u}+\mathbf{u}'}(A_i)+ \der_{\mathbf{u}+\mathbf{u}'}(A_i)],
\end{equation}
where $\mathbf{u}+\mathbf{u}'$ is the concatenation of the lists $\mathbf{u}$ and $\mathbf{u}'$.

We proceed to the proof of the inductive step. We remind the reader that we denoted by $k$ the number of arguments of the time-ordered product. Fix $n\in\N_+$ and assume that Part~(A) holds for all $k\leq n$ (it is enough to assume that the normalization condition \ref{norm:wAL2} holds only for $F$ being the time-ordered product). We will demonstrate that it is true for $k=n+1$. Using the inductive hypothesis and the result from the previous paragraph we obtain that for all $m\in\N_+$, super-quadri-indices $u_1,\ldots,u_{m}$, which involve only massless fields and $\Fa$ products $F$ which may be expressed as a linear combinations of products of the time-ordered products with at most $n$ arguments the distribution
\begin{equation}\label{eq:thm_main1_F}
 (\Omega|F(\mathcal{L}_{l_1}^{(u_1)}(x_1),\ldots,\mathcal{L}_{l_m}^{(u_{m})}(x_m))\Omega) 
\end{equation}
has \underline{IR}-index $\omega$ given by \eqref{eq:proof_omega} with $k=m$. In particular, for all $B_1,\ldots,B_{n+1}\in \Fa_\mathcal{L}$ the distributions
\begin{gather}
 (\Omega|\Dif(B_1(x_1),\ldots,B_n(x_n);B_{n+1}(x_{n+1}))\Omega),
 \\
 (\Omega|\Adv'(B_1(x_1),\ldots,B_n(x_n);B_{n+1}(x_{n+1}))\Omega) 
\end{gather}
have \underline{IR}-indices $\omega'$ given by \eqref{eq:proof_omega_prime} with $k=n+1$.

With the use of the above results, Theorem \ref{thm:split} and the relation \eqref{eq:def_adv_prime} between the products $\Adv$ and $\Adv'$ we get that for all $B_1,\ldots,B_{n+1}\in \Fa_\mathcal{L}$ there exist constants $c_\gamma\in\C$ indexed by multi-indices $\gamma$, $|\gamma|<\omega'$ such that 
\begin{equation}
 (\Omega|\T(B_1(x_1),\ldots,B_{n+1}(x_{n+1}))\Omega)
 - \sum_{\substack{\gamma\\|\gamma|<\omega'}} c_\gamma \partial^\gamma \delta(x_1-x_{n+1})\ldots \delta(x_n-x_{n+1})
\end{equation}
has \underline{IR}-index $\omega'$ given by \eqref{eq:proof_omega_prime} with $k=n+1$. If $B_1,\ldots,B_{n+1}$ are such that $\omega'\leq 0$ then the second term is absent. This proves that the condition \ref{norm:wAL2} restricted to $\omega'\leq 0$ is satisfied for $k=n+1$ (this is evident for $F=\T$; the generalization for arbitrary $F$ considered in \ref{norm:wAL2} follows from the statement about the distribution \eqref{eq:thm_main1_F} made above). If $\omega'\geq 1$ then the time-ordered products \eqref{eq:thm_main1_time} may be redefined, as explained in the formulation of Part~(A) of the theorem, such that the condition \ref{norm:wAL2} is fulfilled. The above claim is evident if we drop the assumption that the map $v$ is graded-symmetric (cf. Section~\ref{sec:freedom}). However, if map $v$ satisfies all the conditions stated in Part~(A) of the theorem but is not graded-symmetric then the map
\begin{multline}
 v_{\textrm{sym}}(B_1(x_1),\ldots,B_{n+1}(x_{n+1})) :=
 \\
 \frac{1}{(n+1)!}\sum_{\pi\in \mathcal{P}_{n+1}}\, (-1)^{\mathbf{f}(\pi)} v(B_{\sigma(1)}(x_{\sigma(1)}),\ldots,B_{\sigma(n+1)}(x_{\sigma(n+1)}))
\end{multline}
has all the needed properties.

We now turn to the proof of the second part of the theorem. First, as we already observed the condition \ref{norm:wAL2} implies \ref{norm:wAL} by Lemma \ref{lem:IR_der}. Thus, if the time-ordered products are defined as discussed above then both conditions \ref{norm:wAL2} and \ref{norm:wAL} are fulfilled. As a result, Part~(B) of the theorem follows from the observation (5) below \ref{norm:wAL2}.
\end{proof}
\end{thm}

Now we state our main theorem according to which the weak adiabatic limit exits in models satisfying Assumption \ref{asm}.

\begin{thm}\label{thm:main2}
Let us assume that the time ordered products satisfy the normalization condition \ref{norm:wAL}. Fix $m\in\N_0$, $C_1,\ldots,C_m\in\Fa$ and a sequence $P$ of the form considered in Section~\ref{sec:W_G_IR}.
\begin{enumerate}[leftmargin=*,label={(\Alph*)}]
\item
Suppose that $\sum_{j=1}^{m}\mathbf{f}(C_j)=0$. For any $k\in\N_0$ and any list $\mathbf{u}=(u_1,\ldots,u_{k+m})$ of super-quadri-indices such that every $u_j$ involves only massless fields (i.e. $\ext_{\mathbf{u}}(A_i)=0$, $\der_{\mathbf{u}}(A_i)=0$ unless $A_i$ is a massless field) and at least one super-quadri-index from $\mathbf{u}$ is non-zero (i.e. $\sum_{i=1}^\mathrm{p}\ext_{\mathbf{u}}(A_i)\geq 1$) the distribution
\begin{equation}\label{eq:main_distribution_adv}
 (\Omega|\Adv(\mathcal{L}_{l_1}^{(u_1)}(y_1),\ldots,\mathcal{L}_{l_k}^{(u_k)}(y_k);C_1^{(u_{k+1})}(x_1),\ldots,C_m^{(u_{k+m})}(x_m);P)\Omega)
\end{equation}
has IR-index 
\begin{equation}\label{eq:main2_IR}
 d= 1 - \sum_{i=1}^{\mathrm{p}} [\dim(A_i) \ext_{\mathbf{u}}(A_i)+ \der_{\mathbf{u}}(A_i)]
\end{equation}
with respect to the variables $y_1,\ldots,y_k$ for all $l_1,\ldots,l_k\in\{1,\ldots,\mathrm{q}\}$. Note that $d\in\Z$ if the distribution \eqref{eq:main_distribution_adv} is non-zero. 
\item 
For all $k\in\N_0$, $g\in\mathcal{S}(\R^{4n})$, $f\in\mathcal{S}(\R^{4m})$ and $l_1,\ldots,l_k\in\{1,\ldots,\mathrm{q}\}$ the limits:
\begin{multline}\label{eq:thm_wAL}
 \lim_{\epsilon\searrow 0}\int\rd^4 y_1\ldots\rd^4 y_n\,\rd^4 x_1\ldots\rd^4 x_m\, g(\epsilon y_1,\ldots,\epsilon y_n) \, f(x_1,\ldots,x_m)
 \\
 (\Omega|\Adv(\mathcal{L}_{l_1}(y_1),\ldots,\mathcal{L}_{l_k}(y_k);C_1(x_1),\ldots,C_m(x_m);P)\Omega)
\end{multline}
and the limit of the analogous expression with the product $\Adv(I;J;P)$ replaced by $\Ret(I;J;P)$ exist and coincide. This implies the existence of the weak adiabatic limit (for the definition of the weak adiabatic limit see Section~\ref{sec:weak_adiabatic_general}).
\end{enumerate}
\begin{proof}
We will prove Part~(A) of the theorem by induction with respect to $k$. In the case $k=0$ by Equation \eqref{eq:def_gen_T} we have $\Adv(\emptyset;J;P) = \T(J;P)$. It follows that for any $f\in\mathcal{S}(\R^{4m})$
\begin{equation}
 \int\mP{p_1}\ldots\mP{p_m}\,\tilde{f}(p_1,\ldots,p_m) \,
 (\Omega|\T(\F{C}_1^{(u_{1})}(q'_1-p_1),\ldots,\F{C}_m^{(u_{m})}(q'_m-p_m);P)\Omega)
\end{equation}
is a smooth function of $q'_1,\ldots,q'_m$. This implies Part~(A) for $k=0$ as a result of Definition \ref{def:IR2} of the IR-index. Part~(B) is trivially true in this case. 

Now, let us assume that Part~(A) holds for $k\leq n-1$, $n\in\N_+$. We shall prove that both Part~(A) and (B) hold for $k=n$. We will first show that for $k=n$ and any list $\mathbf{u}=(u_1,\ldots,u_{k+m})$ of super-quadri-indices such that every $u_j$ involves only massless fields, including the case when all $u_j$ vanish, the distribution
\begin{equation}\label{eq:thm_main2_Dif}
 (\Omega|\Dif(\mathcal{L}^{(u_1)}(y_1),\ldots,\mathcal{L}^{(u_k)}(y_k);C_1^{(u_{k+1})}(x_1),\ldots,C_m^{(u_{k+m})}(x_m);P)\Omega)
\end{equation}
has IR-index $d$ given by \eqref{eq:main2_IR} with respect to the variables $y_1,\ldots,y_k$. The proof of this fact is based on Equation \eqref{eq:dif_com2} which is quoted below for the convenience of the reader
\begin{equation}\label{eq:thm_main2_dif_com}
 \Dif(I;J;P)
 =
 -\sum_{\substack{I_1,I_2,I_3\\I_1+I_2+J+I_3=\pi(I+J)\\I_2 \neq I }} (-1)^{|I_1|+\mathbf{f}(\pi)} [\aT(I_1), \Adv(I_2;J;P)] \T(I_3),
\end{equation}
where 
\begin{equation}\label{eq:main_thm_I_J}
 I = (\mathcal{L}^{(u_1)}(y_1),\ldots,\mathcal{L}^{(u_k)}(y_k)),~~~~~J=(C_1^{(u_{k+1})}(x_1),\ldots,C_m^{(u_{k+m})}(x_m)).
\end{equation}
Using the statement (2') of Theorem \ref{thm:product_families} we obtain that for all $k\leq n$, partitions $I_1,I_2$ of $I$ and super-quadri-indices $u_1,\ldots,u_{k+m}$ which involve only massless fields the VEV of the product
\begin{equation}\label{eq:thm_main2_T_bis_1}
 F(I_1,I_2;J):= [\aT(I_1), \Adv(I_2;J;P)]
\end{equation}
has IR-index $d$ with respect to the variables $y_1,\ldots,y_k$. Next, using the above result and the statement (3') of Theorem \ref{thm:product_families} we get that for $k=n$, all partitions $I_1,I_2,I_3$ of $I$ and super-quadri-indices $u_1,\ldots,u_{k+m}$ which involve only massless fields the VEV of the product
\begin{equation}\label{eq:thm_main2_T_bis_2}
 F(I_1,I_2;J) \T(I_3) 
\end{equation}
has IR-index $d$ with respect to the variables $y_1,\ldots,y_k$. In each of the two above cases the $\Fa$ products $F$, $F'$ and the polynomials $B_j$, $B'_j$ which appear in the statement of Theorem \ref{thm:product_families} are chosen in such a way that the distributions \eqref{eq:thm_main2_T_bis_1} and \eqref{eq:thm_main2_T_bis_2} coincide with \eqref{eq:product_com} and \eqref{eq:product_T_bis}, respectively. It follows from  Theorem \ref{thm:main1} that the normalization condition \ref{norm:wAL2} is satisfied for all $k\in\N_+$. Thus, by the inductive hypothesis the assumptions of Theorem \ref{thm:product_families} are fulfilled for the above choices of $F$, $F'$ and $B_j$, $B'_j$ with 
\begin{equation}
 d_0=4-\sum_{i=1}^{\mathrm{p}} [\dim(A_i) \ext_{\mathbf{u}_1}(A_i)+ \der_{\mathbf{u}_1}(A_i)]~~ \textrm{and}~~ 
 d'_0=1-\sum_{i=1}^{\mathrm{p}} [\dim(A_i) \ext_{\mathbf{u}_2}(A_i)+ \der_{\mathbf{u}_2}(A_i)]
\end{equation}
in Equations \eqref{eq:thm_product_d} in the case of the commutator \eqref{eq:thm_main2_T_bis_1} and 
\begin{equation}
 d_0=1-\sum_{i=1}^{\mathrm{p}} [\dim(A_i) \ext_{\mathbf{u}_{12}}(A_i)+ \der_{\mathbf{u}_{12}}(A_i)]~~ \textrm{and}~~ 
 d'_0=4-\sum_{i=1}^{\mathrm{p}} [\dim(A_i) \ext_{\mathbf{u}_3}(A_i)+ \der_{\mathbf{u}_3}(A_i)]
\end{equation}
in Equations \eqref{eq:thm_product_d} in the case of the product \eqref{eq:thm_main2_T_bis_2}, where $\mathbf{u}_1$, $\mathbf{u}_2$, $\mathbf{u}_3$, $\mathbf{u}_{12}$ are the lists of those super-quadri-indices $u_1,\ldots,u_{k+m}$ which appear in the lists $I_1$, $I_2+J$, $I_3$, $I_1+I_2+J$, respectively. This shows that the distribution \eqref{eq:thm_main2_Dif} has IR-index $d$ with respect to the variables $y_1,\ldots,y_k$.

Part~(A) for $k=n$ follows now from Part~(A) of Theorem \ref{thm:split_gen} since by assumption in this case the IR-index $d$ is non-positive. To show Part~(B) for $k=n$, we first observe that the distribution
\begin{equation}\label{eq:thm_main2_Dif2}
 (\Omega|\Dif(\mathcal{L}_{l_1}(y_1),\ldots,\mathcal{L}_{l_k}(y_k);C_1(x_1),\ldots,C_m(x_m);P)\Omega)
\end{equation}
has IR-index $d=1$. Indeed, this is the distribution \eqref{eq:thm_main2_Dif} with $u_1=\ldots=u_{k+m}=0$. Thus, by Part~(B) of Theorem \ref{thm:split_gen} for every $f\in\mathcal{S}(\R^{4m})$ and $\varepsilon>0$ there exists $c\in\C$ such that
\begin{multline}\label{eq:thm_main2_adv}
 \int\rd^4 x_1\ldots\rd^4 x_m\,(\Omega|\Adv(\F{\mathcal{L}}_{l_1}(q_1),\ldots,\F{\mathcal{L}}_{l_n}(q_n);C_1(x_1),\ldots,C_m(x_m))\Omega) \,f(x_1,\ldots,x_m) 
 \\
 = O^\mathrm{dist}(|q_1,\ldots,q_n|^{1-\varepsilon})+ c.
\end{multline}
As result of Theorem \ref{thm:math_adiabatic_limit} the limit \eqref{eq:thm_wAL} exists and is equal $c$ for any $g\in\mathcal{S}(\R^{4n})$. Since the distribution \eqref{eq:thm_main2_Dif2} has IR-index $d=1$, it fulfills the conditions analogous to \eqref{eq:thm_main2_adv} with $c=0$. It follows that the limit \eqref{eq:thm_wAL} with the product $\Adv(I;J;P)$ replaced by $\Dif(I;J;P)$ exist and is equal $0$. As a result the limit \eqref{eq:thm_wAL} with the product $\Adv(I;J;P)$ replaced by $\Ret(I;J;P)$ coincides with \eqref{eq:thm_wAL}. This finishes the proof of Part~(B).
\end{proof}
\end{thm}

Let us make a remark about a possible generalization of the result of this section.
\begin{rem}\label{rem:main_vacuum}
The condition \ref{norm:wAL} implies in particular that the distribution \eqref{eq:wAL_dist2} with $u_1=\ldots=u_k=0$,
\begin{equation}
 (\Omega|\T(\F{\mathcal{L}}_{l_1}(q_1),\ldots,\F{\mathcal{L}}_{l_k}(q_k))\Omega) = (2\pi)^4 \delta(q_1+\ldots+q_k)\F{\underline{t}}(q_1,\ldots,q_{k-1})
\end{equation}
is normalized in such a way that $\F{\underline{t}}(q_1,\ldots,q_{k-1})$ has zero of order $4$ at $q_1=\ldots=q_{k-1}$ in the sense of {\L}ojasiewicz. However, it turns out that this normalization of the above distribution is not necessary for the existence of the weak adiabatic limit -- it is enough that the normalization condition \ref{norm:wAL3} stated below holds. The condition \ref{norm:wAL3}, which is imposed on the VEVs of the advanced products, can also be rewritten in terms of the truncated VEVs of the time-ordered products (they are defined e.g. in \cite{epstein1973role}). Part~(A) of Theorem \ref{thm:main1} remains valid if one replaces everywhere \ref{norm:wAL2} by \ref{norm:wAL3}. To show that Theorem \ref{thm:main2} is valid under the assumption that the condition \ref{norm:wAL3} is satisfied (instead of \ref{norm:wAL})  the formula \eqref{eq:thm_main2_dif_com} has to be replaced by more complicated one. It may be obtained with the use of the Steinmann's arrow calculus \cite{epstein1973role,steinmann1960wightman} and was used in the proof of the existence of the Wightman and Green functions due to Blanchard and Seneor \cite{blanchard1975green}. This formula allows to express the distribution \eqref{eq:thm_main2_Dif} with $k=n$ in terms of the distributions \eqref{eq:main_distribution_adv} with $k<n$ and the distributions \eqref{eq:wAL_dist3} with at least one non-zero  super-quadri-index $u_1,\ldots,u_k$. 
\end{rem}

\begin{enumerate}[label=\bf{N.wAL${}_{\adv}$},leftmargin=*]
\item\label{norm:wAL3} 
Let $k\in\N_+$, $l_1,\ldots,l_k\in\{1,\ldots,\mathrm{q}\}$ and $\mathbf{u}=(u_1,\ldots,u_k)$ be a list of super-quadri-indices such that every $u_j$ involve only massless fields (i.e. $\ext_{\mathbf{u}}(A_i)=0$ unless $A_i$ is a massless field). If at least one super-quadri-index $u_1,\ldots,u_k$ is non-zero then the Fourier transform of the distribution
\begin{equation}\label{eq:wAL_dist3}
 (\Omega|\Adv(\mathcal{L}^{(u_1)}_{l_1}(x_1),\ldots,\mathcal{L}^{(u_{k-1})}_{l_{k-1}}(x_{k-1});\mathcal{L}^{(u_{k})}_{l_k}(x_k))\Omega)
\end{equation}
is of the form
\begin{equation}
 (2\pi)^4 \delta(q_1+\ldots+q_k)\,\underline{a}_{l_1,\ldots,l_k}^{u_1,\ldots,u_k}(q_1,\ldots,q_{k-1}), 
\end{equation}
where $\underline{a}_{l_1,\ldots,l_k}^{u_1,\ldots,u_k}$ has zero of order $\omega$ at $q_1=\ldots=q_{k-1}=0$ in the sense of {\L}ojasiewicz. If $u_1,\ldots,u_k$ are such that $\omega\leq 0$ then there is no condition on $\F{\underline{a}}_{l_1,\ldots,l_k}^{u_1,\ldots,u_k}$. 
\end{enumerate}

Now we are going to discuss in more detail the models introduced in Section~\ref{sec:examples}. For each model we give the explicit formulas for the index $\omega$ appearing in the normalization condition \ref{norm:wAL}. We also indicate the time-ordered products which have to be normalized correctly for the condition \ref{norm:wAL} to hold. 

\begin{enumerate}[leftmargin=*,label={(\Alph*)}]
\item The massive spinor and scalar QED. 

The only massless fields in the model are the components of the vector potential $A_\mu$. There is only one interaction vertex $\mathcal{L}=j^\mu A_\mu$ and it does not contain derivatives of $A_\mu$. As a result, without loss of generality, we restrict attention in \ref{norm:wAL} and Theorem \ref{thm:main2} to lists $\mathbf{u}=(u_1,\ldots,u_k)$ of super-quadri-indices which involve only the fields $A_\mu$ without derivatives, i.e. $u_j(\partial^\alpha\!A_i)=0$ unless $\alpha=0$ and $A_i$ is one of the components of the vector potential $A_\mu$. Thus, $\der_\mathbf{u}(A_i)=0$ for all $A_i$ and
\begin{equation}
 \omega= 4 - \sum_{\mu=0}^3\ext_{\mathbf{u}}(A_\mu).
\end{equation}
The normalization condition \ref{norm:wAL} involves distributions of the form
\begin{equation}\label{eq:qed_dist}
 (\Omega|\T(j^{\mu_1}(x_1),\ldots,j^{\mu_{n}}(x_{n}),\mathcal{L}(x_{n+1}),\ldots,\mathcal{L}(x_k))\Omega)
\end{equation}
with $n\in\{0,1,2,3\}$ and says that their Fourier transforms are of the form
\begin{equation}
 (2\pi)^4 \delta(q_1+\ldots+q_k)\,\F{\underline{t}}^{\mu_1,\ldots,\mu_n}_{n,k}(q_1,\ldots,q_{k-1}), 
\end{equation}
where $\F{\underline{t}}^{\mu_1,\ldots,\mu_n}_{n,k}(q_1,\ldots,q_{k-1})$ has zero of order $\omega=4-n$ at $q_1=\ldots=q_{k-1}=0$ in the sense of {\L}ojasiewicz. By Remark \ref{rem:main_vacuum} we can drop the case $n=0$ as it is not really needed for the existence of the Wightman and Green functions. For $n\in\{1,2,3\}$ the correct normalization of \eqref{eq:qed_dist} is, in fact, a consequence of the Ward identities \ref{norm:ward} formulated in Section~\ref{sec:norm_con} (it may be proved using the technique introduced in that section). Thus, the existence of the weak adiabatic limit in QED already follows from the standard normalization conditions and no further constraints on the time-ordered products are necessary.

\item The scalar model. 

The only massless field is the scalar field $\varphi$. There is one interaction vertex $\mathcal{L}=\frac{1}{2} \varphi \psi^2$ and it does not contain derivatives of $\varphi$. Without loss of generality, we restrict attention in \ref{norm:wAL} and Theorem \ref{thm:main2} to lists $\mathbf{u}=(u_1,\ldots,u_k)$ of super-quadri-indices which involve only the field $\varphi$ without derivatives. We have
\begin{equation}
 \omega= 4 - \ext_{\mathbf{u}}(\varphi)
\end{equation}
with $n\in\{0,1,2,3\}$ (again, by Remark \ref{rem:main_vacuum} the case $n=0$ may be omitted) and says that their Fourier transform is of the form
\begin{equation}
 (2\pi)^4 \delta(q_1+\ldots+q_k)\,\F{\underline{t}}_{n,k}(q_1,\ldots,q_{k-1}), 
\end{equation}
where $\F{\underline{t}}_{n,k}(q_1,\ldots,q_{k-1})$ has zero of order $\omega=4-n$ at $q_1=\ldots=q_{k-1}=0$ in the sense of {\L}ojasiewicz. The above-mentioned distributions contribute to the scattering processes involving up to three particles in the incoming or outgoing states all of which are massless. The corresponding Feynman diagrams have the following names: the vacuum bubble for $n=0$, the tadpole for $n=1$ and the self-energy for $n=2$ (the diagram with $n=3$ external lines does not have a special name).

In the case of the scalar model \ref{norm:wAL} does not follow from any other normalization condition. By Theorem \ref{thm:main1} the condition \ref{norm:wAL} can always be imposed simultaneously with \ref{norm:sd} (we recall that $\CC=1$ in \ref{norm:sd} as follows from the Assumption \ref{asm}). The compatibility of these conditions with standard normalization conditions which guarantee the unitarity, the Poincar{\'e} covariance and the validity of the field equations in the model follows from results of Sections \ref{sec:norm_con} and \ref{sec:one} of the next chapter.

Let us comment on the assumption $\CC=1$ in the normalization condition \ref{norm:sd} and its necessity for the existence of the Wightman and Green functions. If we suppose that $\CC=0$, then as a result of formula \eqref{eq:omega_scalar_model}, the distributions $\F{\underline{t}}_{2,k}$ with $k\geq 3$ are determined uniquely in terms of the time-ordered products with a lower number of arguments and cannot be redefined. It is unlikely that they satisfy the condition \ref{norm:wAL} (for a generic value of the mass of the field $\psi$). Even the normalization of the mass\footnote{By the normalization of mass of the massless field $\varphi$ we mean the vanishing of the distributions $\F{\underline{t}}_{2,k}(q_1,\ldots,q_{k-1})$ at $q_1=\ldots=q_{k-1}=0$ in the sense of {\L}ojasiewicz. Note that according to the condition \ref{norm:wAL} these distributions have zero of order $2$ at $q_1=\ldots=q_{k-1}=0$.} of the field $\varphi$ which is part of \ref{norm:wAL} is not possible if $\CC=0$. Since, as shown in Appendix \ref{sec:mass}, the correct mass normalization is required for the existence of the Wightman and Green functions (in the second order of the perturbation theory), setting $\CC=1$ in \ref{norm:sd} seems to be necessary. 

Observe that, the distributions $\F{\underline{t}}_{n,k}$ with odd $n$, which correspond to Feynman diagrams with odd number of external massless lines, do not have to vanish in the scalar model. Let us remark that the vanishing of the diagrams with odd number of massless external lines and no massive external lines is known in QED as Furry’s theorem. It is a consequence of the normalization condition \ref{norm:pct} formulated in Section~\ref{sec:norm_con}, the charge-conjugation-invariance of the interaction vertex $\mathcal{L}$ of QED and the following behavior of the vector potential $A_\mu$ under the charge conjugation transformation $A_\mu \to -A_\mu$ (for the proof of the Furry's theorem in the present setting see e.g. \cite{scharf2014}). The generalization of the Furry's theorem to the case of the scalar model is not possible because in this model both the interaction vertex $\mathcal{L}$ and the massless field $\varphi$ are charge-conjugation-invariant (it is not possible to define the charge conjugation such that $\varphi\to-\varphi$ under this transformation). Consequently, in particular, the diagrams with $n=1$ external massless line (called tadpoles) are present. This, however, does not pose any problems as the correct normalization of these diagrams dictated by the normalization condition \ref{norm:wAL} can always be achieved (according to \ref{norm:wAL} the distribution $\F{\underline{t}}_{n,k}$ with $n=1$ must have zero of order three at the origin in the sense of {\L}ojasiewicz). In contrast to the method of Blanchard and Seneor \cite{blanchard1975green} our proof of the existence of the Wightman and Green functions applies also to models with tadpole diagrams. 
\end{enumerate}

Next let us consider examples of models which contain only massless particles. For such models the normalization condition \ref{norm:wAL} involves all the distributions of the form
\begin{equation}
 (\Omega|\T(\mathcal{L}_{l_1}^{(u_1)}(x_1),\ldots,\mathcal{L}_{l_k}^{(u_{k})}(x_k))\Omega)
\end{equation}
where $\mathcal{L}_{l_1},\ldots,\mathcal{L}_{l_k}$ are interaction vertices and $u_1,\ldots,u_k$ are arbitrary super-quadri indices such that $\omega\geq 1$ (for the explicit form of $\omega$ in different models see the examples below). As we will show in Section~\ref{sec:ahs} in the case of purely massless models the condition \ref{norm:wAL} is a consequence of the normalization condition \ref{norm:sc} which says that the VEVs of the time-ordered products of polynomials of massless fields scale almost homogeneously in the sense of Definition \ref{def:aHS} stated in the next chapter.

\begin{enumerate}[leftmargin=*,label={(\Alph*)}]
\setcounter{enumi}{2}

\item The massless spinor and scalar QED. 

All basic generators are massless. In the case of the spinor QED without loss of generality, we restrict attention in \ref{norm:wAL} and Theorem \ref{thm:main2} to lists $\mathbf{u}=(u_1,\ldots,u_k)$ of super-quadri-indices which does not involve derivatives (i.e. $\der_\mathbf{u}(A_i)=0$ for all basic generators $A_i$). This is, however, not possible in the scalar QED since its interaction vertex contains derivatives of the fields $\varphi$ and $\varphi^*$. We have 
\begin{equation}
 \omega=4-\sum_{\mu=0}^3 \ext_\mathbf{u}(A_\mu)-\frac{3}{2}\sum_{a=1}^4 (\ext_\mathbf{u}(\psi_a)+\ext_\mathbf{u}(\overline{\psi}_a))
\end{equation}
in the case of the spinor QED and
\begin{equation}
 \omega=4-\sum_{\mu=0}^3 \ext_\mathbf{u}(A_\mu)-\ext_\mathbf{u}(\phi)-\ext_\mathbf{u}(\phi^*)-\der_{\mathbf{u}}(\phi)-\der_{\mathbf{u}}(\phi^*)
\end{equation}
in the case of the scalar QED.

\item Yang-Mills theory without matter. 

Our results apply also to models containing anti-commuting scalar fields such as ghosts. In particular, our method can be used to show the existence of the Wightman and Green functions in the non-abelian Yang-Mills theories without matter. The normalization condition \ref{norm:sc} is a part of the definition of these theories due to Hollands \cite{hollands2008renormalized}. Consequently, in this formulation the condition \ref{norm:wAL} is automatically satisfied. Hollands considered the model defined on an arbitrary globally hyperbolic spacetime -- here we restrict attention to the case of the flat Minkowski spacetime. The existence of the Wightman and Green functions of the interacting fields follows form Theorem \ref{thm:main2}. Before taking the adiabatic limit all background fields such as the anti-fields have to be set equal to zero.
\end{enumerate}

In the case of models in which \ref{norm:wAL} is not a consequence of any other normalization condition (an example of such model is the scalar model) one has to check that it is compatible with the standard normalization conditions which guarantee e.g. the unitarity and Poincar{\'e} covariance of the model. This is done in Sections \ref{sec:norm_con} and \ref{sec:one} of the next chapter.

\subsubsection{Comparison with the method due to Blanchard and Seneor}

Let us compare our method of the proof of the existence of the weak adiabatic limit with the one used by Blanchard and Seneor \cite{blanchard1975green}. First, let us point the similarities. The general structure of our proof resembles that of \cite{blanchard1975green}. Theorems \ref{thm:main1} and \ref{thm:main2} have their counterparts in the paper by Blanchard and Seneor. In both cases the theorems are proved by induction with respect to the order of the perturbation expansion and the proofs of the inductive steps are divided into two parts. In the first part the statement about the product $\Dif$ is established. In the second -- the splitting of the product $\Dif$ is investigated in order to demonstrate that the product $\Adv$ satisfies appropriate regularity condition.

Our method differs in a few important respects, which allow us to obtain a significantly more general result. One of the most important features of our proof is the use of the notion of regularity of a distribution expressed with the use of the notation $\F{t}(q,q')=O^\textrm{dist}(|q|^\delta)$. The regularity condition was formulated in such a way that the proofs of the theorems stated in Section~\ref{sec:prod} entitled \emph{Product} are relatively easy. The proof that the regularity condition is preserved under the splitting of distributions is perhaps not as transparent as the proof of the regularity of the products of distributions, but its only technical part is the proof of Lemma~\ref{lem:splitting_theta_alpha}. 

In broad terms, our condition on $t\in\mathcal{S}'(\R^N)$
\begin{equation}\label{eq:our_cond}
 \F{t}(q) = O^\textrm{dist}(|q|^{d-\varepsilon}) ~~\textrm{for all}~~\epsilon>0
\end{equation}
corresponds to the condition
\begin{equation}\label{eq:BS_semi}
 \left|\int\rd^N x\, t(x) g(x)\right|\leq \sum_{\substack{\alpha\\[1mm] \max(0,d)\leq|\alpha|\leq P}} 
 \sup_{x\in\R^N} (1+|x|)^{|\alpha|-d+1-\varepsilon} |\partial^\alpha g(x)|
\end{equation}
for all $g\in\mathcal{S}(\R^N)$ and some $\varepsilon>0$, where $d,P\in\Z$, in \cite{blanchard1975green}. In fact, these authors use a more complicated seminorm than the one on the RHS of \eqref{eq:BS_semi}, which allows them to control IR and UV behavior simultaneously.  In contrast, in the thesis we take advantage of the Steinmann scaling degree to solve the UV problem. Our regularity condition \eqref{eq:our_cond} controls only IR properties of distributions.

There are also other minor differences between the two methods. For example, we use a simpler formula for $\Dif(I;J;P)$ in the inductive step of Theorem \ref{thm:main2} since the absence of the vacuum contributions is not important in our method (see the last paragraph of Section~\ref{sec:W_G_IR}). The appropriate normalization of the time-ordered products is crucial in the case of both methods. The formulation of the condition \ref{norm:wAL} with the use of the value of a distribution at zero in the sense of {\L}ojasiewicz is particularly transparent and facilitates the proof of its compatibility with all the standard normalization conditions which is presented in the next chapter.

\chapter{Compatibility of normalization conditions}\label{ch:comp}

In this chapter we consider standard normalization conditions usually imposed on the time-ordered products. We show that these conditions are compatible with the normalization condition \ref{norm:wAL} introduced in Section~\ref{sec:proof_scalar} which is needed for the existence of the weak adiabatic limit. First we show that in purely massless models the condition \ref{norm:wAL} is implied by almost homogeneous scaling of VEVs of time-ordered products. In models with both massless and massive particles the condition \ref{norm:wAL} need not follow from any other normalization conditions. Its compatibility with the standard normalization conditions is established in Sections \ref{sec:norm_con} and \ref{sec:one}. In Section~\ref{sec:properties} we list the properties of the Wightman and Green functions which follows from the normalization conditions discussed in this chapter.

\section{Almost homogeneous scaling}\label{sec:ahs}

In this section we assume that all fields under consideration are massless and set $\CC=0$ in the normalization condition \ref{norm:sd}. Following \cite{hollands2001local,hollands2002existence} we introduce the definition of the almost homogeneous scaling of a distribution.
\begin{dfn}\label{def:aHS}
A distribution $t\in \mathcal{S}'(\R^N)$ or $\mathcal{S}'(\R^N\setminus \{0\})$ scales almost homogeneously with degree $D\in\R$ and power $P\in\N_0$ iff
\begin{equation}\label{eq:a_h_d}
(\sum_{j=1}^N x_j\partial_{x_j}+D)^{P+1}t(x)=0
\end{equation}
and $P$ is the minimal natural number with this property. If $P=0$ the above condition states that the distribution $t$ is homogeneous of degree $-D$.
\end{dfn}

The condition \eqref{eq:a_h_d} is equivalent to
\begin{equation}
 (\rho\partial_\rho)^{P+1}\left(\rho^D t(\rho x)\right)=0.
\end{equation}
Thus, $t$ scales almost homogeneously with degree $D$ and power $P$ iff $\rho^D t(\rho x)$ is a polynomial of $\log\rho$ with degree $P$. This implies that in particular $\sd(t)=D$, where $\sd(\cdot)$ is the Steinmann scaling degree. Moreover,
\begin{equation}
 (\rho\partial_\rho)^{P+1}\left(\rho^{D-N} \F{t}(q/\rho)\right)=0.
\end{equation}
Hence, it holds $\sd(\F{t})=N-D$.

As shown in \cite{dutsch2004causal}, if the normalization condition \ref{norm:sd} holds with $\CC=0$ then the time-ordered products of polynomials of massless fields can be normalized such that the following condition holds.
\begin{enumerate}[label=\bf{N.aHS},leftmargin=*]
\item\label{norm:sc} Almost homogeneous scaling:
For all polynomials $B_1,\ldots,B_k\in\Fh$ of massless fields and their derivatives the distribution
\begin{equation}
 (\Omega|\T(B_1(x_1),\ldots,B_k(x_k))\Omega)
\end{equation}
scales almost homogeneously with degree 
\begin{equation}
 D = \omega + 4(k-1) = \sum_{j=1}^{k}\dim(B_j).
\end{equation}
\end{enumerate}
The above normalization condition is a special case of the condition called {\it Scaling} in \cite{dutsch2004causal} which was imposed also for the time-ordered products of massive fields as a substitute of \ref{norm:sd}. It is clear that in the case of purely massless models \ref{norm:sc} is stronger than \ref{norm:sd} since for any $t\in\mathcal{S}'(\R^N)$ which scales almost homogeneously with degree $D$ it holds $\sd(t)=D$. The condition similar to \ref{norm:sc} was introduced for the first time in \cite{hollands2001local,hollands2002existence} in the context of the quantum field theory in curved spacetime. For massless fields this method of normalization was used e.g. in \cite{grigore2001scale,gracia2003improved,lazzarini2003improved}.

\begin{thm}
If all fields under consideration are massless and $\CC=0$ in \ref{norm:sd} then the normalization condition \ref{norm:sc} implies \ref{norm:wAL}.
\begin{proof}
By the comment below Definition \ref{def:aHS} the following equality
\begin{equation}
 \sd\left( (\Omega|\T(\F{B}_1(q_1),\ldots,\F{B}_k(q_k))\Omega) \right) = \sum_{j=1}^{k} (4 -\dim(B_j))
\end{equation}
follows from the normalization condition \ref{norm:sc}. As a result, we obtain
\begin{equation}
 \sd\left( \F{\underline{t}}(q_1,\ldots,q_{k-1}) \right)  = \sum_{j=1}^{k} (4 -\dim(B_j))-4 = -\omega,
\end{equation}
where $\F{\underline{t}}\in\mathcal{S}'(\R^{4(k-1)})$ is such that
\begin{equation}
 (\Omega|\T(\F{B}_1(q_1),\ldots,\F{B}_k(q_k))\Omega) = (2\pi)^4 \delta(q_1+\ldots+q_k)\,\F{\underline{t}}(q_1,\ldots,q_{k-1}).
\end{equation}
Thus, for any $g\in\mathcal{S}(\R^{4(k-1)})$ and multi-index $\gamma$, $|\gamma|<\omega$ it holds
\begin{equation}
 \lim_{\epsilon\searrow 0} \int\mP{q_1}\ldots\mP{q_{k-1}} \, \left(\partial^\gamma_q\F{\underline{t}}(q_1,\ldots,q_{k-1})\right)\, g_\epsilon(q_1,\ldots,q_{k-1}) = 0, 
\end{equation}
where $g_\epsilon$ is defined in terms $g$ as in Definition \ref{def:lojasiewicz} of the value of a distribution at a point in the sense of {\L}ojasiewicz. This implies the normalization condition \ref{norm:wAL}.
\end{proof}
\end{thm}

\section{Standard normalization conditions}\label{sec:norm_con}

In this section we list the standard normalization conditions which are usually imposed on the time-ordered products and show their compatibility with the normalization condition \ref{norm:wAL}, which is needed for the existence of the Wightman and Green functions.

\begin{enumerate}[label=\bf{N.U},leftmargin=*]
\item\label{norm:u} Unitarity:
\begin{equation}\label{eq:n_U}
 \T(B_1(x_1),\ldots,B_k(x_k))^* = \aT(B_k^*(x_k),\ldots,B_1^*(x_1))
\end{equation}
for all $B_1,\ldots,B_k\in\Fa$. The definitions of the adjoints in $\Fa$ and $L(\mathcal{D}_0)$ are given in Section~\ref{sec:ff} and \ref{sec:Wick}, respectively. The condition under consideration implies that the generating functional $S(\sum_{j=1}^n g_j\otimes B_j)$ introduced in Section~\ref{sec:generating} is unitary if $B_j=B_j^*$ have even fermion number and $g_j\in\mathcal{S}(\R^4)$ are real-valued for all $j\in\{1,\ldots,n\}$.
\end{enumerate}
\begin{enumerate}[label=\bf{N.P},leftmargin=*]
\item\label{norm:p} Poincar\'e covariance: 
\begin{multline}\label{eq:n_P}
 U(a,\Lambda)\T(B_1(x_1),\ldots,B_k(x_k))U(a,\Lambda)^{-1} 
 \\
 = \T((\rho(\Lambda)B_1)(\Lambda x_1+a),\ldots,(\rho(\Lambda)B_k)(\Lambda x_k+a)),
\end{multline}
where $B_1,\ldots,B_k\in\Fa$, $\rho$ is the representation of $SL(2,\C)$ acting on $\mathcal{F}$ and $U$ is the unitary representation of Poincar{\'e} group on $\mathcal{D}_0$.
\end{enumerate}
\begin{enumerate}[label=\bf{N.CPT},leftmargin=*]
\item\label{norm:pct} Covariance with respect to the discrete group of CPT transformations (the charge conjugation, the spatial-inversion and the time-reversal): 
\begin{multline}\label{eq:n_CPT}
 U(g)\T(B_1(x_1),\ldots,B_k(x_k))U(g)^{-1} 
 \\
 =\T((\rho(g)B_1)(\Lambda(g)x_1),\ldots,(\rho(g)B_k)(\Lambda(g)x_k)),
\end{multline}
where $g$ is element of the CPT group, $U(g)$ is its representation on $\mathcal{D}_0$ and $\rho(g)$ -- its representation on $\Fa$.
\end{enumerate}

\begin{enumerate}[label=\bf{N.W},leftmargin=*]
\item\label{norm:ward} Ward identities for the electric current $j^\mu$ in the spinor and scalar QED: For any $B_1,\ldots,B_{k-1}\in\Fh$ which are sub-polynomials of the interaction vertex $\mathcal{L}$ it holds
\begin{multline}\label{eq:n_W}
 \partial^x_{\mu}\T(j^\mu(x),B_1(x_1),\ldots,B_{k-1}(x_{k-1}))
 \\
 = \ri \sum_{j=1}^{k-1} \mathbf{q}(B_j) \,\delta(x_j-x)\,\T(B_1(x_1),\ldots,B_{k-1}(x_{k-1})),
\end{multline}
where $\mathbf{q}(B)$ is the charge number. The interaction vertex $\mathcal{L}$, the electric current $j^\mu$ and the charge number were defined in in Sections \ref{sec:spinor_qed} and \ref{sec:scalar_qed}. Note that in the EG approach the Ward identities are imposed on the time-ordered products of the free fields.
\end{enumerate}

The outline of the proof of the compatibility of the axioms \ref{axiom1}-\ref{axiom6} and all the normalization conditions \ref{norm:sd}, \ref{norm:u}, \ref{norm:p}, \ref{norm:pct}, \ref{norm:one} may be found e.g. in \cite{boas2000gauge}. The proof is based on results from several papers. The normalization conditions \ref{norm:u} and \ref{norm:p} was investigated by Epstein and Glaser in \cite{epstein1973role}. However, their proof of the Poincar{\'e} covariance has restricted applicability. The general proof that is possible to define the time-ordered products in such a way that the condition \ref{norm:p} holds was given for the first time by Popineau and Stora in \cite{popineau2016pedagogical} using  the vanishing of the cohomologies of the $SL(2,\C)$ group (see also \cite{scharf2014,prange1999lorentz,bresser1999lorentz,lazzarini2003improved}). The CPT covariance (the condition \ref{norm:pct}) was studied in QED e.g. in \cite{scharf2014}. In the case of QED the compatibility of the Ward identities with other normalization conditions was shown e.g. in Appendix B of \cite{dutsch1999local} (see also \cite{scharf2014,grigore2001gauge,dutsch1994gauge,dutsch1993scalar,dutsch1990gauge}).

We will prove that it is possible to carry out the construction of the time-ordered products such that the normalization conditions \ref{norm:sd}, \ref{norm:u}-\ref{norm:ward} and \ref{norm:wAL2} (which is equivalent to \ref{norm:wAL}) are satisfied. Let assume that the time-ordered products with at most $n$ arguments were constructed such that all these conditions hold. It follows from references cited above that we can define the time-ordered products with $n+1$ arguments such that \ref{norm:sd}, \ref{norm:u}-\ref{norm:ward} are fulfilled. As a result of Theorem \ref{thm:main1} it is possible to modify this definition, possibly violating \ref{norm:u}-\ref{norm:ward}, such that the conditions \ref{norm:sd} and \ref{norm:wAL2} hold. To this end, one redefines
\begin{equation}
 \Fa^{n+1}\ni(B_1,\ldots,B_{n+1})\mapsto (\Omega|\T(B_1(x_1),\ldots,B_{n+1}(x_{n+1}))\Omega)\in\mathcal{S}'(\R^{4(n+1)})
\end{equation}
by adding to it the graded-symmetric map
\begin{equation}
 \Fa^{n+1}\ni(B_1,\ldots,B_{n+1})\mapsto v(B_1(x_1),\ldots,B_{n+1}(x_{n+1}))\in\mathcal{S}'(\R^{4(n+1)})
\end{equation}
such that $v(B_1(x_1),\ldots,B_{n+1}(x_{n+1}))$ is non-zero only for $B_1,\ldots,B_n\in\Span_\C\Fa_\mathcal{L}$ (for the definition of the set $\Fa_\mathcal{L}$ see \eqref{eq:proof_set}) and for all $B_1,\ldots,B_{n+1}\in\Fa$ the distribution $v(B_1(x_1),\ldots,B_{n+1}(x_{n+1}))$ is of the form
\begin{equation}\label{eq:comp_delta_form3}
 \sum_{\substack{\gamma\\|\gamma|<\omega'}} c_\gamma \partial^\gamma \delta(x_1-x_{n+1})\ldots \delta(x_{n}-x_{n+1})
\end{equation}
for some constants $c_\gamma\in\C$ indexed by multi-indices $\gamma$, $|\gamma|<\omega'$, where
\begin{equation}\label{eq:comp_omega}
 \omega':=4-\sum_{j=1}^{n+1} (4-\CC- \dim(B_j)).
\end{equation}
The new time-ordered products are defined with the use of the modified VEVs introduced above and the axiom \ref{axiom3}. 

After this modification the difference between the VEVs of the LHS and RHS of Equations \eqref{eq:n_U}, \eqref{eq:n_P}, \eqref{eq:n_CPT} and \eqref{eq:n_W} with $k=n+1$ need not vanish if all their arguments belong to $\Span_\C\Fa_\mathcal{L}$ and $\omega'\geq 1$. Moreover, the difference, if non-zero, has to be of the form 
\begin{equation}\label{eq:comp_delta_form2}
 \sum_{\substack{\gamma\\|\gamma|<d}} c_\gamma \partial^\gamma \delta(x_1-x_{n+1})\ldots \delta(x_{n}-x_{n+1})
\end{equation}
with $d=\omega'$ in the case of the first four equations and $d=\omega'+1$ in the case of the last equation. 

On the other hand, the condition \ref{norm:wAL2} implies that the difference between the VEVs of the LHS and RHS of Equations \eqref{eq:n_U}, \eqref{eq:n_P} and \eqref{eq:n_CPT} with $k=n+1$ has \underline{IR}-index $\omega'$. Since the \underline{IR}-index of the distribution of the form \eqref{eq:comp_delta_form2} is strictly bounded from above by $\omega$, the VEVs of the LHS and RHS of the above-mentioned equations are equal. By the axiom \ref{axiom3} the result implies that \ref{norm:u}, \ref{norm:p} and \ref{norm:pct} hold after the above redefinition of the time-ordered products.

It remains to investigate the VEVs of both sides of Equation \eqref{eq:n_W} with $k=n+1$ and $B_1,\ldots,B_{n}\in\Span_\C\Fa_\mathcal{L}$. Observe that $j^\mu\in\Fa_\mathcal{L}$. It follows from Definition \ref{def:IR1} of the \underline{IR}-index that each derivative increases the \underline{IR}-index by one. Thus, the \underline{IR}-index of the VEV of the LHS of \eqref{eq:n_W} equals $d_{\textrm{L}}=\omega'+1$, where 
\begin{equation}
 \omega'=4-\sum_{j=1}^{n} (4- \dim(B_j)) - (4 - \dim(j^\mu)) = 3 -\sum_{j=1}^{n} (4- \dim(B_j)),
\end{equation}
since $\dim(j^\mu)=3$ and $\CC=0$ in the case of the spinor and scalar QED. The \underline{IR}-index of the VEV of the RHS is the same as the \underline{IR}-index of 
\begin{equation}
 (\Omega|\T(B_1(x_1),\ldots,B_{n}(x_n))\Omega),
\end{equation}
which equals 
\begin{equation}
 d_{\textrm{R}} = 4-\sum_{j=1}^{n} (4 - \dim(B_j)) = d_{\textrm{L}}.
\end{equation}
Consequently, by the argument given in the previous paragraph, the VEVs of the LHS and RHS of \eqref{eq:n_W} coincide. Using the axiom \ref{axiom3}, the normalization condition \ref{norm:one} introduced in the next section and the technique from Appendix B of \cite{dutsch1999local} one shows that \ref{norm:ward} is indeed fulfilled. This finishes the proof of the compatibility of the normalization conditions. 

It turns out that the normalization condition \ref{norm:wAL} restricts the form of the possible violation of the Ward identities in QED. In Chapter~\ref{sec:generalized_central} we introduce a stronger normalization condition which can be imposed in the massive spinor QED. The ward identities follow from this normalization condition.

\section{Field equations}\label{sec:one}

In this section we consider the normalization condition which, in particular, implies that the interacting fields satisfy the field equations. We study in some detail models introduced in Section~\ref{sec:examples}. At the end we argue that the normalization condition \ref{norm:wAL} and \ref{norm:one} are compatible.

\begin{enumerate}[label=\bf{N.FE},leftmargin=*]
\item\label{norm:one} Let $B,B_1,\ldots,B_k\in\Fh$ be the sub-polynomials of the interaction vertex such that $B\in\mathcal{G}$ ($\mathcal{G}$ is the set of generators; a generator is an elements of $\Fa$ of the form $\partial^{\alpha}\!A_{i}$, where $A_i$ is a basic generator and $\alpha$ is a multi-index). It holds: 
\begin{multline}\label{eq:one}
 (\Omega|\T(B(x),B_1(x_1),\ldots,B_k(x_k))\Omega)
 = 
 \sum_{j=1}^k\! \sum_{C\in\mathcal{G}} (-1)^{\mathbf{f}_j(C)} ~
 (\Omega|\T(B(x),C(x_j))\Omega)\times
 \\ (\Omega|\T\big(B_1(x_1),\ldots,\frac{\partial B_j}{\partial C}(x_j),\ldots,B_k(x_k)\big)\Omega),
\end{multline}
where the factor $(-1)^{\mathbf{f}_{j}(C)}$ can be read off from the equation
\begin{equation}
 \normord{B_1(x_1)\ldots B_{j-1}(x_{j-1}) C(x)}\, 
 =\,(-1)^{\mathbf{f}_j(C)}\, \normord{C(x) B_1(x_1)\ldots B_{j-1}(x_{j-1})}.
\end{equation}
\end{enumerate}

The above condition says that the time-ordered product of sub-polynomials of the interaction vertex with a generator among its arguments is uniquely determined by the time-order products with less arguments. Without loss of generality the second sum on the RHS of Equation \eqref{eq:one} may be carried out over the generators $C$ which are contained in the interaction vertex. For convenience of the reader, we list these generators for each of the models introduced in Section~\ref{sec:examples}: 
\begin{enumerate}[label=(\arabic*),leftmargin=*]
\item the spinor QED: $A_\mu$, $\psi_a$, $\overline{\psi}_a$,
\item the scalar QED: $A_\mu$, $\phi$, $\phi^*$, $\partial_\mu \phi$, $\partial_\mu \phi^*$,
\item the scalar model: $\varphi$, $\psi$.
\end{enumerate}
If the normalization condition \ref{norm:sd} holds with $\CC=0$ (the standard case) then for $|\alpha|+|\alpha'|\leq1$ the distributions
\begin{equation}
 (\Omega|\T(\partial^\alpha\!A_i(x),\partial^{\alpha'}\!A_{i'}(x'))\Omega)
\end{equation}
are determined uniquely in terms of the standard Feynman propagators and its derivatives. Moreover, in the case of the scalar QED we have
\begin{equation}\label{eq:scalar_qed_derivatives}
 (\Omega|\T(\partial_\mu \phi^*(x),\partial_\nu \phi(x'))\Omega) := \partial^x_\nu\partial_\mu^{x'} (\Omega|\T(\phi^*(x) \phi(x'))\Omega)-\ri g_{\mu\nu} \delta(x-x')
\end{equation}
and
\begin{equation}
 (\Omega|\T(\partial_\mu \phi(x),\partial_\nu \phi(x'))\Omega) := 0,~~~(\Omega|\T(\partial_\mu \phi^*(x),\partial_\nu \phi^*(x'))\Omega):=0.
\end{equation}
This is the only possible definition of the above distributions which has correct scaling degree and is compatible with the charge-conjugation invariance and the Ward identities
\begin{equation}
\begin{aligned}
 \partial_\mu^{x'}\T(\phi(x)j^\mu(x')) &= -\ri \delta(x-x') \phi(x'),
 \\
 \partial_\mu^{x'}\T(\phi^*(x)j^\mu(x')) &= \ri \delta(x-x') \phi^*(x').
\end{aligned} 
\end{equation}
Thus, in the case of the spinor and scalar QED the factor $(\Omega|\T(B_1(x_1),C(x_j))\Omega)$ which appears in the identity \eqref{eq:one} is fixed uniquely. This is however not true for the scalar model. Because in the case of this model we assume that $\CC=1$ in the normalization condition \ref{norm:sd} there is some freedom in the definition of $(\Omega|\T(A_i(x)A_{i'}(x'))\Omega)$. We remove this ambiguity by adding the condition that this expression coincides with the standard Feynman propagator.

The normalization condition \ref{norm:one} may be rewritten in several equivalent forms, which are given e.g. in \cite{dutsch1999local,boas2000gauge}. It allows to determine uniquely all tree-level corrections to the S-matrix and, as we already mentioned, implies that the interacting fields satisfy the field equations. In order to substantiate the latter statement let us discuss as an example the scalar model. The condition \ref{norm:one} implies that
\begin{multline}
 \Adv(B_1(x_1),\ldots,B_k(x_k);B(x))
 = 
 \sum_{j=1}^k\! \sum_{C\in\{\varphi,\psi\}}~
 (\Omega|\Adv(C(x_j);B(x))\Omega)\times
 \\ \Adv\big(B_1(x_1),\ldots,B_{j-1}(x_{j-1}),B_{j+1}(x_{j+1}),\ldots,B_k(x_k);\frac{\partial B_j}{\partial C}(x_j)\big),
\end{multline}
where $\Adv$ is the advanced product, $B\in\{\varphi,\psi\}$ and $B_1,\ldots,B_k$ are the interaction vertex sub-monomials. By the assumption made above we have
\begin{equation}
\begin{aligned}
 (\Omega|\Adv(\varphi(y);\varphi(x))\Omega) &= D^\adv_0(y-x),\phantom{0}~~~~(\Omega|\Adv(\varphi(y);\psi(x))\Omega) = 0,
 \\
 (\Omega|\Adv(\psi(y);\varphi(x))\Omega) &=0,\phantom{D^\adv_0(y-x)}~~~~(\Omega|\Adv(\psi(y);\psi(x))\Omega) = D_m^\adv(y-x),
\end{aligned} 
\end{equation}
where $D^\adv_0(x)$ and $D^\adv_m(x)$ are the standard advanced Green functions for the wave equation and the Klein-Gordon equation with mass $m$, respectively. As a result, in the case $B_1=\ldots=B_k=\mathcal{L}$ using the definition \eqref{eq:bogoliubov_adv} of the advanced field with the IR regularization we obtain
\begin{equation}
  \square_x \varphi_\adv(g;x) = \tfrac{1}{2} g(x)\, (\psi^2)_\adv(g;x)
\end{equation}
and similarly for the massive field
\begin{equation}
  (\square_x+ m^2) \psi_\adv(g;x) =  g(x)\, (\varphi\psi)_\adv(g;x).
\end{equation}
After taking the algebraic adiabatic limit discussed in Section~\ref{sec:algebraic_adiabatic} we get
\begin{equation}
  \square_x \varphi_\adv(\cdot;x) = \tfrac{1}{2}\, (\psi^2)_\adv(\cdot;x), ~~~~
  (\square_x+ m^2) \psi_\adv(\cdot;x) = (\varphi\psi)_\adv(\cdot;x),
\end{equation}
which are the interacting field equations. %The analogous equations in the case of the spinor QED are given in \eqref{eq:eqs_motion}.

It is known that the condition \ref{norm:one} can be simultaneously satisfied with \ref{norm:sd} and all the normalization conditions which were considered in the previous section. It remains to show that \ref{norm:one} and \ref{norm:wAL} are compatible. This is not trivial only if the interaction vertex contains no more then one massive field. To see this recall that \ref{norm:wAL} is imposed only on the time-ordered products of polynomials $\mathcal{L}_l^{(u)}$, where $u$ is a super-quadri-index involving only massless fields. However, such polynomial $\mathcal{L}_l^{(u)}$ may be a generator (denoted by $B$ in \ref{norm:one}) only if $\mathcal{L}_l$ contains at most one massive field. As a result there is nothing to check in the massive spinor or scalar QED and the scalar model. In the case of the massless spinor or scalar QED the normalization \ref{norm:wAL} follows from \ref{norm:sc}, and thus, is automatically compatible with all the standard normalization conditions.

\section{Properties of Wightman and Green functions}\label{sec:properties}

If the time-ordered products satisfy the normalization conditions formulated in the previous section, then the Wightman and Green functions have a number of important properties. In fact, the Wightman functions fulfill almost all the standard axioms which are listed in \cite{streater2000pct} in the sense of formal power series in the coupling constants. It is only not clear whether the cluster decomposition property holds. Let $C_1,\ldots,C_m\in\Fa$ be arbitrary polynomials. The following conditions are satisfied.
\begin{enumerate}[leftmargin=*,label={(\arabic*)}]
\item Poincar{\'e} covariance:
\begin{equation}
 \Wig(C_1(x_1),\ldots,C_m(x_m))=\Wig((\rho(\Lambda)C_1)(\Lambda x_1+a),\ldots,(\rho(\Lambda)C_m)(\Lambda x_k+a)),
\end{equation}
where $\rho$ is the representation of $SL(2,\C)$ acting on $\Fa$. 
\item Spectrum condition: The Fourier transform of the Wightman function
\begin{equation}
 \Wig(\F{C}_1(p_1),\ldots,\F{C}_m(p_m))
\end{equation}
has the support contained in 
\begin{equation}
 \left\{(p_1,\ldots,p_m)\in\R^{4m}\,:\, \sum_{j=1}^m p_j =0, ~~
 \forall_{k}~\sum_{j=1}^k p_j \in \overline{V}^+ \right\}.
\end{equation}
For the proof of this property see \cite{epstein1973role} (p. 267).
\item Hermiticity:
\begin{equation}
 \overline{\Wig(C_1(x_1),\ldots,C_m(x_m))}=\Wig(C^*_m(x_m),\ldots,C^*_1(x_1)).
\end{equation}
The definition of the adjoint in $\Fa$ is given in Section~\ref{sec:ff}.
\item Local (anti)commutativity: For $C_k,C_{k+1}\in \Fh$ it holds
\begin{equation}
 \Wig(\ldots,C_k(x_k),C_{k+1}(x_{k+1}),\ldots)
 = (-1)^{\mathbf{f}(C_k)\mathbf{f}(C_{k+1})} \Wig(\ldots,C_{k+1}(x_{k+1}),C_k(x_k),\ldots)
\end{equation}
if $x_k$ and $x_{k+1}$ are spatially separated, i.e. $(x_k-x_{k+1})^2\leq 0$.
\item Positive definiteness condition: Let $j_0\in\N_+$, $\{1,\ldots,j_0\} \ni j\mapsto f_j \in S(\R^{4n_j})$ and $C_{j,1},\ldots,C_{j,n_j}\in\mathcal{F}$, where $n_j\in\N_+$. Then the formal power series  
\begin{multline}
 \sum_{j,k\in\N_0}\int\rd^4 x_1,\ldots\rd^4 x_{n_j}\rd^4 y_1,\ldots\rd^4 y_{n_k}\, \overline{f_j(x_1,\ldots,x_{n_j})}f_k(y_1,\ldots,y_{n_k})\,
 \\
 \Wig(C^*_{j,1}(x_1),\ldots,C^*_{j,n_j}(x_{n_j}),C_{k,1}(y_1),\ldots,C_{k,n_k}(y_{n_k}))
\end{multline}
is non-negative in the sense of Definition \ref{dfn:positive_formal}. The above condition is satisfied only in models which can be defined on the Fock space with a positive-definite inner product and follows from the positivity of the vacuum state introduced in Section~\ref{sec:weak_adiabatic_general}. 

The above condition is violated in every model with vector fields, in particular, in QED. Nevertheless, it is expected that the Wightman functions of observables can be consistently defined in these theories and fulfill the positive definiteness condition.
\item Field equations: For example in the massless $\varphi^4$ theory it holds
\begin{equation}
 \square_x \Wig(C_1(x_1),\ldots,\varphi(x),\ldots,C_m(x_m)) 
 + \frac{\lambda}{3!} \Wig(C_1(x_1),\ldots,\varphi^3(x),\ldots,C_m(x_m)) = 0.
\end{equation}
Similar equations are satisfied in other models. Note that this condition is not included in the Wightman axioms.
\end{enumerate}

The Green function are of less fundamental character then the Wightman functions. Because of the LSZ reduction formula they play an important role in the computation of the elements of the S-matrix.  However, in theories with massless particles the singularities of their Fourier transforms near the mass shells usually do not have the required form \cite{kibble1968coherent2} and the application of the LSZ reduction formula is not possible. When studying scattering, one usually considers the Green functions with the IR regularization. The regularization may be removed only in the expressions for the inclusive cross-sections \cite{yennie1961infrared}. 

Let us list the properties of the Green functions which can be easily proved in the EG approach to perturbative QFT. 

\begin{enumerate}[leftmargin=*,label={(\arabic*)}]
\item Poincar{\'e} covariance:
\begin{equation}
 \Gre(C_1(x_1),\ldots,C_m(x_m))=\Gre((\rho(\Lambda)C_1)(\Lambda x_1+a),\ldots,(\rho(\Lambda)C_m)(\Lambda x_k+a)),
\end{equation}
where $\rho$ is the representation of $SL(2,\C)$ acting on $\Fa$.

\item Graded-symmetry:
\begin{equation}
 \Gre(C_1(x_1),\ldots,C_m(x_m)) = (-1)^{\mathbf{f}(\pi)}\Gre(C_{\pi(1)}(x_{\pi(1)}),\ldots,C_{\pi(m)}(x_{\pi(m)})).
\end{equation}
for all $C_1,\ldots,C_k\in\Fh$, where $\mathbf{f}(\pi)\in\Z/2\Z$ is the number of transpositions in the permutation $\pi\in\mathcal{P}_m$ that involve a pair of fields with odd fermion number.
\item Causality:
\begin{equation}
 \Gre(C_1(x_1),\ldots,C_m(x_m)) = \Wig(C_1(x_1),\ldots,C_m(x_m))
\end{equation}
if for all $j\in\{1,\ldots,m-1\}$ the point $x_{j}$ is not in the causal past of any of the points $x_{j+1},\ldots,x_m$.

\end{enumerate}
For non-coinciding points the Green function are determined uniquely by the Wightman functions with the use of the following formula
\begin{multline}
  \Gre(C_1(x_1),\ldots,C_m(x_m)) = \sum_{\pi\in\mathcal{P}_m} (-1)^{\mathbf{f}(\pi)}
 \theta(x_{\pi(1)}^0-x_{\pi(2)}^0)\ldots\theta(x_{\pi(m-1)}^0-x_{\pi(m)}^0)
 \\
 \Wig(C_{\sigma(1)}(x_{\sigma(1)}),\ldots,C_{\sigma(m)}(x_{\sigma(m)})).
\end{multline}

%===================================================================================================
%===================================================================================================
\chapter{Central normalization condition}\label{sec:generalized_central}
%===================================================================================================
%===================================================================================================

In this chapter we formulate the normalization condition for the time-ordered products which is a generalization of the condition introduced by Epstein and Glaser in \cite{epstein1973role}. Let us describe the content of the latter condition. It involves the VEVs of the advanced products of the form 
\begin{equation}
 (\Omega|\Adv(B_1(x_1),\ldots,B_n(x_n);B_{n+1}(0))\Omega)
\end{equation}
and says that their Fourier transforms have zero of order $\omega+1$ at the origin, where 
\begin{equation}
 \omega = \sum_{j=1}^{n+1} (\dim(B_j)+\CC) - 4n.
\end{equation}
The advanced product which fulfills the above condition is known in the literature as the central or symmetrical solution of the splitting problem, or sometimes the central or symmetrical extension\footnote{Finding a solution of {\it the splitting problem}, which is {\it an extension} of the distribution \eqref{eq:const_adv_split}, is part of the EG construction described in Section~\ref{sec:T_const}.}. This condition fixes uniquely all the time-ordered products and is compatible with the standard normalization conditions. So far the existence of the central splitting solution has been proved in theories without massless fields \cite{epstein1973role}. However, an expectation that a condition of a similar type can also be imposed in the massive spinor QED has been expressed by some authors \cite{scharf2014,dutsch1990gauge,dutsch1996finite,hurth1995nonabelian}. 

The normalization condition introduced in this chapter, which we call the central normalization condition, is a natural generalization of the above condition due to Epstein and Glaser to models with massless particles. In the case of the massive spinor QED it fixes uniquely all the time-ordered products of the sub-polynomials of the interaction vertex, and in particular implies the Ward identities.

In the first section of this chapter we remind the reader the proof of the existence of the central splitting solution in theories without massless fields. In Section~\ref{sec:central_general} we formulate the central normalization and show that it is possible to define the time-ordered products such that this condition is satisfied. Finally, in Section~\ref{sec:application_qed} we consider the applications of the central normalization condition in the massive spinor QED and make a comment about the case of the massive scalar QED.

\section{Central splitting solution in massive theories} \label{sec:cetral_splitting_massive}

In this section all fields under consideration are massive. For details we refer the reader to Sections 5.2 and 6.2 of \cite{epstein1973role}. First, let us note that as a result of the formula \eqref{eq:dif_standard_com} and both parts of Lemma \ref{lem:aux_lemma} the Fourier transforms of the distributions 
\begin{equation}\label{eq:central_massive_dif}
 (\Omega|\Dif(A^{r_1}(x_1),\ldots,A^{r_n}(x_n);A^{r_{n+1}}(0))\Omega)
\end{equation}
vanish in some neighborhood $\mathcal{O}\subset\R^{4n}$ of $0$. Moreover, because of the support properties in the position space \eqref{eq:supp_VEVs} of the advanced and retarded products, the Fourier transforms of the distributions
\begin{gather}
 \label{eq:central_massive_adv}
 (\Omega|\Adv(A^{r_1}(x_1),\ldots,A^{r_n}(x_n);A^{r_{n+1}}(0))\Omega),
 \\
 \label{eq:central_massive_ret}
 (\Omega|\Ret(A^{r_1}(x_1),\ldots,A^{r_n}(x_n);A^{r_{n+1}}(0))\Omega)
\end{gather} 
are boundary values of functions $F^\pm$ which are holomorphic in the tubes
\begin{equation}
 \R^{4n} + i\Gamma^\pm,
\end{equation}
respectively. Since the distribution \eqref{eq:central_massive_dif} is the difference of \eqref{eq:central_massive_adv} and \eqref{eq:central_massive_ret}, these boundary values coincide in $\mathcal{O}$.
%(see Theorem 2-9 of \cite{streater2000pct})
Thus, by the edge of the wedge theorem (Theorem 2-16 of \cite{streater2000pct}), the holomorphic functions $F^\pm$ are restrictions of a single function holomorphic in some complex neighborhood of $\mathcal{O}$. As a result, the Fourier transforms of the distributions \eqref{eq:central_massive_adv} and \eqref{eq:central_massive_ret} are analytic functions in some neighborhood of zero. 

Next, we prove that the Fourier transform of the distribution
\begin{equation}\label{eq:central_A}
 (\Omega|\Adv(A^{r_1}(x_1),\ldots,A^{r_n}(x_n);A^{r_{n+1}}(x_{n+1}))\Omega)
\end{equation}
is graded-symmetric in the sense of Equation \eqref{eq:v_graded} in some neighborhood of zero. To this end, it is enough to notice that for test functions $g\in\mathcal{S}(\R^{4(n+1)})$ such that $\supp\,\F{g}$ is contained in sufficiently small neighborhood of the origin it holds
\begin{multline}\label{eq:adv_sym}
 (\Omega|\Adv(I;A^{r_{n+1}}(x_{n+1}))\Omega)
 \\
%  = \sum_{k=1}^{|I|}\sum_{\substack{I_1,\ldots,I_k\\I'_1+I_2\ldots+I_k = \pi(I')}} (-1)^{k-1+\mathbf{fg}(\pi)}\, (\Omega|\T(I'_1)\Omega)(\Omega|\T(I_2)\Omega)\ldots(\Omega|\T(I_k)\Omega)
%  \\
  = \sum_{k=1}^{|I'|}\frac{1}{k}\sum_{\substack{I_1,\ldots,I_k\\I_1+\ldots+I_k = \pi(I')\\I_1,\ldots,I_k\neq \emptyset}} (-1)^{k-1+\mathbf{f}(\pi)}\, (\Omega|\T(I_1)\Omega)(\Omega|\T(I_2)\Omega)\ldots(\Omega|\T(I_k)\Omega),
\end{multline}
where 
\begin{equation}
 I=(A^{r_1}(x_1),\ldots,A^{r_n}(x_n)),~~I'=I+(A^{r_{n+1}}(x_{n+1})).%,~~I_1'=I_1+(A^{r_{n+1}}(x_{n+1})).
\end{equation}
To obtain the above identity we first express the advanced product as a product of the time-ordered products. Using the formulas \eqref{eq:adv_standard} and \eqref{eq:aT_standard} we get
\begin{equation}
 \Adv(I;A^{r_{n+1}}(x_{n+1})) = \sum_{\substack{I_1,\ldots,I_k\\I_1+I_0+I_2+\ldots+I_k = \pi(I')\\I_1,\ldots,I_k\neq \emptyset}} (-1)^{k-1+\mathbf{f}(\pi)}\,
 \T(I_1,A^{r_{n+1}}(x_{n+1}))\T(I_2)\ldots\T(I_k),
\end{equation}
where $I_0=(A^{r_{n+1}}(x_{n+1}))$.  Next, we apply Part~(A) of Lemma \ref{lem:aux_lemma} to the VEV of the RHS of the above equation.  This lemma says that only the vacuum contributions are present in the representation \eqref{eq:vev_product_representation} of this VEV if we restrict attention to Schwartz functions $g\in\mathcal{S}(\R^{4(n+1)})$ such that $\supp\,\F{g}$ is contained in sufficiently small neighborhood of the origin. The resulting expression may be written in a form \eqref{eq:adv_sym}, which is explicitly graded-symmetric under the permutation of the elements of the sequence~$I'$.

Let us recall that the distribution \eqref{eq:central_A} can be redefined by adding to it the distribution $v$ given by \eqref{eq:freedom_v} which satisfies the conditions stated in Section~\ref{sec:freedom}. Using the fact that the Fourier transform of \eqref{eq:central_A} is graded-symmetric we show that the following normalization conditions can always be satisfied.
\begin{enumerate}[label=\bf{N.C${}_m$},leftmargin=*]
\item\label{norm:central} 
For all $n\in\N_+$, super-quadri-indices $r_1,\ldots,r_{n+1}$ which involve only massive fields and multi-indices $\gamma$ such that $|\gamma|\leq \omega$ it holds
\begin{equation}\label{eq:central_extension_condition}
 \partial^\gamma_q \F{\underline{a}}^{r_1,\ldots,r_{n+1}}(q_1,\ldots,q_n)\big|_{q=0} = 0,
\end{equation}
where 
\begin{equation}\label{eq:central_omega}
 \omega = \sum_{j=1}^{n+1} (\dim(A^{r_j})+\CC) - 4n 
\end{equation}
and $\F{\underline{a}}^{r_1,\ldots,r_{n+1}}(q_1,\ldots,q_n)$ is the Fourier transform of the distribution
\begin{equation}\label{eq:norm_cent_adv}
 (\Omega|\Adv(A^{r_1}(x_1),\ldots,A^{r_n}(x_n);A^{r_{n+1}}(0))\Omega).
\end{equation}
\end{enumerate}
The advanced  products which fulfill the above condition are usually called in the literature the central solution of the splitting procedure. Note that this normalization condition fixes uniquely the time-ordered products of all polynomials of massive fields (we assume throughout that the normalization condition \ref{norm:sd} holds). To see this, we use the results of Section~\ref{sec:freedom}.  It is enough to note that the Fourier transform of the distribution \eqref{eq:freedom_form} is of the form $(2\pi)^4\delta(q_1+\ldots+q_{n+1}) P(q_1,\ldots,q_n)$, where $P$ is a polynomial of degree bounded by $\omega$ and the condition $ \partial^\gamma_q P(q_1,\ldots,q_n)|_{q=0} = 0$ is satisfied for all multi-indices $\gamma$, $|\gamma|\leq \omega$ iff $P=0$.

\section{Central normalization condition}\label{sec:central_general}

The normalization condition \ref{norm:central} was formulated only for massive theories. In the~form in which it is stated above it makes sense only if the distribution $\F{\underline{a}}^{r_1,\ldots,r_{n+1}}$ is a function at least $\omega$ times differentiable in some neighborhood of zero. This is, however, unlikely to be true if arguments of the time-ordered products contain massless fields\footnote{The condition \ref{norm:central} certainly cannot be imposed for the time-ordered products of polynomials of only massless fields, because of the logarithmic singularities of the Fourier transforms of their VEVs.}. For this reason, we first rewrite the condition \ref{norm:central} in an equivalent form which is more suitable for subsequent generalization to the case of models with massless particles.

Because the function $\F{\underline{a}}^{r_1,\ldots,r_{n+1}}$ in \ref{norm:central} is analytic in some neighborhood of zero, the condition \eqref{eq:central_extension_condition} is clearly equivalent to 
\begin{equation}
 \F{\underline{a}}^{r_1,\ldots,r_{n+1}}(q_1,\ldots,q_n) = O^{\mathrm{dist}}(|q_1,\ldots,q_n|^{\omega+1-\varepsilon}),~~~~\forall_{\varepsilon\in(0,1)},
\end{equation}
where the notation $O^{\mathrm{dist}}(|q|^{\delta})$ was introduced in Definition \ref{def:O_not}. By Definition \ref{def:IR1} of the \underline{IR}-index the above condition may be also expressed
by saying that for all super-quadri-indices $r_1,\ldots,r_n$ such that $\omega\geq 0$ the distribution 
\begin{equation}
 (\Omega|\Adv(A^{r_1}(x_1),\ldots,A^{r_n}(x_n);A^{r_{n+1}}(x_{n+1}))\Omega)
\end{equation}
has \underline{IR}-index equal $\omega+1$. Using the identity \eqref{eq:adv_sym} and Part~(A) of Theorem \ref{thm:product} with $l=0$ one can prove inductively on $n\in\N_0$ that the condition \ref{norm:central} can be written in yet another equivalent form.

\begin{enumerate}[label=\bf{N.C${}'_m$},leftmargin=*]
\item\label{norm:central2}
For all $n\in\N_+$, super-quadri-indices $r_1,\ldots,r_{n+1}$ which involve only massive fields and such that $\omega\geq 0$ the distribution 
\begin{equation}
 (\Omega|\T(A^{r_1}(x_1),\ldots,A^{r_{n+1}}(x_{n+1}))\Omega)
\end{equation}
has \underline{IR}-index equal $\omega+1$ ($\omega$ is given by \eqref{eq:central_omega}).
\end{enumerate}
Note that because the VEVs of the time-ordered and advanced product are related by Equation \eqref{eq:adv_sym}, the Fourier transform of
\begin{equation}
 \underline{t}^{r_1,\ldots,r_{n+1}}(x_1,\ldots,x_n):=
 (\Omega|\T(A^{r_1}(x_1),\ldots,A^{r_n}(x_n),A^{r_{n+1}}(0))\Omega)
\end{equation}
is usually not a smooth function in any neighborhood of zero, even if all $r_1,\ldots,r_{n+1}$ involve only massive fields. Indeed, the difference between $\F{\underline{a}}^{r_1,\ldots,r_{n+1}}$ and $\F{\underline{t}}^{r_1,\ldots,r_{n+1}}$ is a combination of terms which contain products of the Dirac deltas. Nevertheless, it follows from \ref{norm:central2} and Lemma \ref{lem:IR_der} that for all multi-indices $\gamma$ such that $|\gamma|\leq \omega$ it holds
\begin{equation}
 \partial^\gamma_q \F{\underline{t}}^{r_1,\ldots,r_{n+1}}(q_1,\ldots,q_n)\big|_{q=0} = 0,
\end{equation}
where the value at $q_1=\ldots=q_n=0$ of the distribution $\partial^\gamma_q \F{\underline{t}}^{r_1,\ldots,r_{n+1}}$ is defined in the sense of {\L}ojasiewicz.

The following normalization condition is a generalization of the condition \ref{norm:central2} to the case of the time-ordered products of both massive and massless fields.
\begin{enumerate}[label=\bf{N.C},leftmargin=*]
\item\label{norm:gc}
Let $k\in\N_+$ and $A^{r_1},\ldots,A^{r_k}\in\Fa$ be arbitrary monomials. Consider the following distribution
\begin{equation}\label{eq:cent_dist}
 (\Omega|\T(A^{r_1}(x_1),\ldots,A^{r_k}(x_k))\Omega).
\end{equation}
\begin{enumerate}[label=(\arabic*),leftmargin=*]
\item The distribution \eqref{eq:cent_dist} has \underline{IR}-index  
\begin{equation}\label{eq:gc_index1}
  d_1 = 4 + \sum_{j=1}^{k} \dim(A^{r_j}) - 4k.
\end{equation}
\item If all of the super-quadri-indices $r_1,\ldots,r_k$ involve at least one massive field, then the distribution \eqref{eq:cent_dist} has \underline{IR}-index
\begin{equation}\label{eq:gc_index2}
 d_2=5+\sum_{j=1}^{k} (\dim(A^{r_j})+\CC) - 4k.
\end{equation}
\item %If $k\geq k_0$ and 
If all but one of the super-quadri-indices $r_1,\ldots,r_k$ involve at least one massive field, then the distribution \eqref{eq:cent_dist} has \underline{IR}-index
\begin{equation}\label{eq:gc_index3}
 d_3=5+\sum_{j=1}^{k} \dim(A^{r_j}) -4k.
\end{equation}
%where $k_0\geq 2$ is some constant.
\end{enumerate}
\end{enumerate}
Let us make a few remarks before we show that the above normalization condition may be imposed on the time-ordered products simultaneously with the condition \ref{norm:sd}.
\begin{enumerate}[label=(\arabic*),leftmargin=*]
 \item Parts (1) and (2) of \ref{norm:gc} imply the normalization condition \ref{norm:wAL2} (and its equivalent version \ref{norm:wAL}).
 \item Part~(2) of \ref{norm:gc} fixes uniquely the time-ordered products to which it refers. This part implies the condition \ref{norm:central2} (and its equivalent version \ref{norm:central}).
 \item If all monomials $A^{r_1},\ldots,A^{r_k}$ are products of massless fields, then only Part~(1) of \ref{norm:gc} applies. Using the technique from Section~\ref{sec:ahs} one shows that this part of \ref{norm:gc} is equivalent to the normalization condition \ref{norm:sc}. Note that the time-ordered products of monomials of massless fields are not fixed uniquely by \ref{norm:gc} or \ref{norm:sc}.
 \item Part~(3) of \ref{norm:gc} is a generalization of Part~(2). If the normalization condition \ref{norm:sd} holds in the standard form (i.e. with $\CC=0$) it fixes uniquely all the time-ordered products with at least two arguments being monomials such that all of them but one contains at least one factor being a massive field.
 \item A practical method of constructing the time-ordered products satisfying Part~(1) of the central normalization condition has been given in \cite{gracia2003improved,lazzarini2003improved}.
\end{enumerate}

\begin{thm}\label{thm:cent}
It is possible to define the time-ordered products such that they satisfy simultaneously the normalization conditions \ref{norm:sd} and \ref{norm:gc}.
\begin{proof}
The proof goes along the lines of the proof of Theorem \ref{thm:main1}. We first check that Parts (1) and (2) of the normalization condition \ref{norm:gc} holds for $k=1$. Next we assume that \ref{norm:sd} and Parts (1) and (2) of \ref{norm:gc} hold for all $k\leq n$, $n\in\N_+$ and prove that it is possible to define the time-ordered products with $n+1$ arguments such that they satisfy simultaneously the normalization conditions \ref{norm:sd} and \ref{norm:gc} including Part~(3). 

To this end, we first show that for super-quadri-indices $r_1,\ldots,r_{n+1}$ of the form stated in each part of \ref{norm:gc} the distributions 
\begin{gather}\label{eq:gc_D_A}
 (\Omega|\Dif(A^{r_1}(x_1),\ldots,A^{r_n}(x_n);A^{r_{n+1}}(x_{n+1}))\Omega),
 \\
 (\Omega|\Adv'(A^{r_1}(x_1),\ldots,A^{r_n}(x_n);A^{r_{n+1}}(x_{n+1}))\Omega) 
\end{gather}
have \underline{IR}-indices $d_1$, $d_2$ or $d_3$ given by Equations \eqref{eq:gc_index1}, \eqref{eq:gc_index2} or \eqref{eq:gc_index3} with $k=n+1$. The above statements for Parts (1), (2), (3) of \ref{norm:gc} follow from the inductive assumption, the fact that the products $D$ and $A'$ with $n+1$ arguments are expressed by the products of the time-ordered products with at most $n$ arguments and Parts (A), (B), (C) of the lemma below.

Next we apply the procedure of UV regularized splitting as described in the proof of Theorem \ref{thm:main1}. Because $d_1,d_2,d_3\leq \omega +1$, where $\omega$ is defined by \eqref{eq:omega}, using the ambiguity in defining the time-ordered products it is possible to satisfy both the conditions \ref{norm:sd} and \ref{norm:gc}.
\end{proof}
\end{thm}

\begin{lem}
Let $F$ and $F'$ be $\Fa$ products with $n$ and $n'$ arguments, respectively, where $n,n'\in\N_+$. Consider the operator-valued distributions
\begin{equation}\label{eq:gc_proof_FF}
 F(A^{r_1}(x_1),\ldots,A^{r_n}(x_n)), ~~~~~F'(A^{r'_1}(x'_1),\ldots,A^{r'_{n'}}(x'_{n'})) 
\end{equation}
and their product
\begin{equation}\label{eq:gc_proof_FF_prod}
 F(A^{r_1}(x_1),\ldots,A^{r_n}(x_n))F'(A^{r'_1}(x'_1),\ldots,A^{r'_{n'}}(x'_{n'})). 
\end{equation}
\begin{enumerate}[label=(\Alph*),leftmargin=*]
 \item  If for all super-quadri-indices $r_1,\ldots,r_n$, $r'_1,\ldots,r'_{n'}$ the VEVs of distributions \eqref{eq:gc_proof_FF} have \underline{IR}-indices given by
\begin{equation}
  d_1 = 4 + \sum_{j=1}^{n} \dim(A^{r_j}) - 4n,~~~~~d'_1 = 4 + \sum_{j=1}^{n'} \dim(A^{r'_j}) - 4n',
\end{equation}
respectively, then for all super-quadri-indices $r_1,\ldots,r_n$, $r'_1,\ldots,r'_{n'}$ the VEV of their product \eqref{eq:gc_proof_FF_prod} has \underline{IR}-index
\begin{equation}
  d''_1 = 4 + \sum_{j=1}^{n} \dim(A^{r_j}) + \sum_{j=1}^{n'} \dim(A^{r'_j}) - 4(n+n').
\end{equation}

\item  If for all super-quadri-indices $r_1,\ldots,r_n$, $r'_1,\ldots,r'_{n'}$ such that each of them involves at least one massive field the VEVs of the distributions \eqref{eq:gc_proof_FF} have \underline{IR}-indices given by
\begin{equation}
  d_2 = 5 + \sum_{j=1}^{n} (\dim(A^{r_j})+\CC) - 4n,~~~~~d'_2 = 5 + \sum_{j=1}^{n'} (\dim(A^{r'_j})+\CC) - 4n',
\end{equation}
respectively, then for all super-quadri-indices $r_1,\ldots,r_n$, $r'_1,\ldots,r'_{n'}$ such that each of them involves at least one massive field the VEV of their product \eqref{eq:gc_proof_FF_prod} has \underline{IR}-index
\begin{equation}
  d''_2 = 5 + \sum_{j=1}^{n} (\dim(A^{r_j})+\CC) + \sum_{j=1}^{n'} (\dim(A^{r'_j})+\CC) - 4(n+n').
\end{equation}

\item Under the assumptions of Parts (A) and (B) above the VEV of the product \eqref{eq:gc_proof_FF_prod} has \underline{IR}-index
\begin{equation}
  d''_3 = 5 + \sum_{j=1}^{n} \dim(A^{r_j}) + \sum_{j=1}^{n'} \dim(A^{r'_j}) - 4(n+n')
\end{equation}
for all super-quadri-indices $r_1,\ldots,r_n$, $r'_1,\ldots,r'_{n'}$ such that all but one involve at least one massive field.
\end{enumerate}
\begin{proof}
The lemma follows almost immediately from the statement (1') of Theorem \ref{thm:product_families}. Note that that if $B=A^r$ then $B^{(s)}$ is up to a constant equal $A^{r-s}$. Moreover, we have $\dim(A^{r-s})=\dim(A^{r})-\dim(A^{s})$. Thus,
\begin{equation}
 \sum_{j=1}^n \dim(A^{r_j-s_j})= \sum_{j=1}^n \dim(A^{r_j}) -  \sum_{i=1}^{\mathrm{p}} [\dim(A_i) \ext_\mathbf{s}(A_i)+ \der_\mathbf{s}(A_i)]
\end{equation}
for any lists of super-quadri-indices $\mathbf{r}=(r_1,\ldots,r_n)$ and $\mathbf{s}=(s_1,\ldots,s_n)$. 

Let us note that it is crucial for the validity of Part~(B) that the super-quadri-indices $s_j$ and $s'_j$ which appear in the formulation of Theorem \ref{thm:product_families} involve only massless fields. Observe also that the VEV of the product \eqref{eq:gc_proof_FF_prod} in the case of Part~(B) has even higher \underline{IR}-index equal $d''_1+1$.  In Theorem \ref{thm:cent} we use only the weaker statement given in Part~(B) because the \underline{IR}-index equal $d''_1+1$  cannot be preserved by the splitting procedure unless \ref{norm:sd} is violated.
\end{proof}
\end{lem}

\section{Application in massive QED}\label{sec:application_qed}

The Part~(3) of the condition \ref{norm:gc} implies the following normalization condition which holds in the massive spinor QED and combined with the condition \ref{norm:sd} fixes uniquely all the time-ordered products of the sub-polynomials of the interaction vertex.
\begin{enumerate}[label=\bf{N.C${}_{\textrm{QED}}^{\textrm{spinor}}$},leftmargin=*]
 \item\label{norm:qed_spinor} The distribution
\begin{equation}\label{eq:norm_c_qed_dist}
 (\Omega|\T(\mathcal{L}^{(u_1)}(y_1),\ldots,\mathcal{L}^{(u_{k})}(y_k))\Omega) 
\end{equation} 
has IR-index 
\begin{equation}\label{eq:norm_qed_d}
 d = \omega+1 = 5 - \ext_\mathbf{u}(A) -\frac{3}{2}\ext_{\mathbf{u}}(\psi) - \frac{3}{2}\ext_{\mathbf{u}}(\overline{\psi})
\end{equation}
for any $k\in\N_+$ and any list of super-quadri-indices $\mathbf{u}=(u_1,\ldots,u_k)$  such that $\ext_\mathbf{u}(\psi)+\ext_\mathbf{u}(\overline{\psi})\leq 2$.% and $\mathcal{L}^{(u_j)}$ is not a constant for any $j\in\{1,\ldots,k\}$. 
\end{enumerate}
%Note that if $\mathcal{L}^{(u_j)}$ is a constant for some $j$ then by the axiom \ref{axiom4} the time-ordered product $\T(\mathcal{L}^{(u_1)}(y_1),\ldots,\mathcal{L}^{(u_{k})}(y_k))$ coincides with the time-ordered product with less arguments.
The above condition implies all the standard normalization conditions \ref{norm:u}, \ref{norm:p}, \ref{norm:pct}, \ref{norm:one}, \ref{norm:ward}, which were formulated in Sections \ref{sec:norm_con} and \ref{sec:one}, for the time-ordered products of the sub-polynomials of the interaction vertex.

Let us consider now the analog of the condition \ref{norm:qed_spinor} in the case of the massive scalar QED.

\begin{enumerate}[label=\bf{N.C${}_{\textrm{QED}}^{\textrm{scalar}}$},leftmargin=*]
 \item\label{norm:qed_scalar} The distribution
\begin{equation}\label{eq:qed_scalar_dist}
 (\Omega|\T(\mathcal{L}^{(u_1)}(y_1),\ldots,\mathcal{L}^{(u_{k})}(y_k))\Omega) 
\end{equation} 
has IR-index 
\begin{equation}
 d = \omega+1 = 5 - \ext_\mathbf{u}(A) -\ext_{\mathbf{u}}(\phi) - \ext_{\mathbf{u}}(\phi^*) -\der_\mathbf{u}(\phi) - \der_\mathbf{u}(\phi^*),
\end{equation}
for any $k\in\N_+$ and any list of super-quadri-indices $\mathbf{u}=(u_1,\ldots,u_k)$  such that $\ext_\mathbf{u}(\phi)+\ext_\mathbf{u}(\phi^*)\leq 2$.% and $\mathcal{L}^{(u_j)}$ is not a constant for any $j\in\{1,\ldots,k\}$. 
\end{enumerate}
It turns out that the above condition is not compatible with the Ward identities. It is violated by the following distribution
\begin{equation}\label{eq:norm_qed_scalar}
 (\Omega|\T(\partial_\mu\phi^*(x_1),\partial_\nu\phi(x_2))\Omega),
\end{equation}
which, as follows from Ward identities, has to be normalized such that the equality \eqref{eq:scalar_qed_derivatives} holds. On the other hand, the distribution \eqref{eq:norm_qed_scalar} has \underline{IR}-index $d$ specified in the normalization condition \ref{norm:qed_scalar} only if
\begin{equation}
 (\Omega|\T(\partial_\mu\phi(x),\partial_\nu\phi(x'))\Omega)= \partial^x_\nu\partial_\mu^{x'} (\Omega|\T(\phi^*(x) \phi(x'))\Omega).
\end{equation}
The condition \ref{norm:qed_scalar} may be compatible with the Ward identities in the scalar QED with two interaction vertices \eqref{eq:scalar_qed_std_vertex} and \eqref{eq:scalar_qed_vertex}. Since these vertices have dimension four the proof of the existence of the weak adiabatic limit given in Section~\ref{sec:proof_scalar} is also valid for this model. The relation between the standard formulation of the scalar QED,  presented in Section~\ref{sec:scalar_qed}, and the one with two interaction vertices will be studied elsewhere.

%===================================================================================================
%===================================================================================================
\chapter{Summary and outlook}
%===================================================================================================
%===================================================================================================

We proved the existence of the weak adiabatic limit in the Epstein-Glaser approach to the perturbative quantum field theory under the assumption that the time-ordered products fulfill certain normalization condition. We shown that it is possible to define the time-ordered products such that this condition is satisfied in all models with interaction vertices of the canonical dimension equal to four and all models with interaction vertices of the canonical dimension equal to three provided each of them contains at least one massive field. The result implies the existence of the Wightman and Green functions in the above models. The Wightman and Green functions have most of the standard properties following from the Wightman axioms. Our result can also be used to define a real and Poincar\'e invariant functional on the algebra of interacting fields obtained by means of the algebraic adiabatic limit. In the case of models which do not contain vector fields this functional is positive. Because it satisfies the relativistic spectrum condition it can be interpreted as the vacuum state. As expected, in the quantum electrodynamics and other models with vector fields the obtained Wightman functions do not fulfill the positive definiteness condition and the above functional defined on the algebra of interacting fields is not a state since it is not positive.

In the Epstein-Glaser approach the models containing vector fields have to be defined by using the Gupta-Bleuler method. Hence, some of the fields which appear in the construction of these models actually do not correspond to any physical degrees of freedom. This issue is closely related to the fact that the inner product on the Fock space used in the definition of these models is not positive-definite. Moreover, the interacting fields obtained by means of the algebraic adiabatic limit do not satisfy the correct equations of motion unless some constraints are imposed. For example, in the case of the quantum electrodynamics the Maxwell equations are violated unless the Lorenz condition $\partial_\mu A^\mu(\cdot,x)=0$ is enforced. For this reason, besides the algebra of interacting fields one defines the algebra of observables which is a quotient algebra consisting of equivalence classes of gauge-invariant fields. By definition the above-mentioned constraints are satisfied in the algebra of observables. In the case of the quantum electrodynamics the algebra of observables was constructed by D{\"u}tsch and Fredenhagen in \cite{dutsch1999local} and in the case of the non-abelian Yang-Mills theories without matter fields -- by Hollands in \cite{hollands2008renormalized}. Moreover, it was shown that each of the subalgebras of these algebras consisting of observables localized in a bounded open subset of the Minkowski spacetime can be represented on the pre-Hilbert space with a positive-definite inner product (in fact, in \cite{hollands2008renormalized} this was proved even in the case of curved spacetime). The definition of a state on the global algebra of observables, in particular the definition of the vacuum state, is an open problem. An interesting generalization of our results would be the proof that the previously-mentioned Poincar\'e invariant functional on the algebra of interacting fields can be used to define the functional on the algebra of observables in the case of the quantum electrodynamics and non-abelian Yang-Mills theories without matter, and the resulting functional is positive. The proof of positivity is non-trivial because the inner product on the space on which these models are defined is indefinite. 

An interesting possible alternative to the Gupta-Bleuler method  could be the utilization of the string-local fields which were introduced in \cite{mund2004string}. In this approach all fields can be defined on the linear space with positive-definite inner product, but the vector fields are localized not at single points but on semi-infinite strings extending to space-like or null infinity. It was argued in \cite{mund2016string} that this type of locality is sufficient for a perturbative construction of interacting models. The crucial condition in this construction is the principle of string independence which replaces the principle of gauge invariance. The generalization of the Epstein-Glaser approach to string-local fields is currently under investigation.

The techniques developed in the proof of the existence of the weak adiabatic limit were also used in the thesis to establish that it is possible to define the time-ordered products in such a way that they satisfy the condition called the central normalization condition. This result implies, in particular, the existence of the central splitting solution in the quantum electrodynamics with a massive spinor field. The time-ordered products of sub-polynomials of the interaction vertex of the quantum electrodynamics normalized in this way are fixed uniquely and satisfy the standard normalization conditions including the Ward identities. The central normalization condition implies in particular that the self-energies of the photon and the electron vanish three times at zero in the sense of {\L}ojasiewicz. The above normalization of the photon self-energy is compatible with the on-shell normalization. This is not true for the electron. The self-energy of the electron may be redefined in such a way that its mass is normalized correctly. However, because of the infrared problem \cite{kibble1968coherent2} it is not possible to satisfy the on-shell normalization condition for electron in the standard form which is stated in \cite{epstein1976adiabatic}. The correct normalization of the electron wave function requires a formulation of a new normalization condition which takes into account the fact that it is an infraparticle~\cite{schroer1963infrateilchen}.

The Wightman and Green functions characterize the local properties of the theory and are well-defined even in models with long-range interactions such as the quantum electrodynamics and the Yang-Mills theories. These theories suffer from the infrared problem which emerges when one investigates the scattering of particles. For example, in the case of the quantum electrodynamics the strong adiabatic limit does not exist even for the first order correction to the scattering matrix. Another manifestation of the infrared problem is a non-standard form of the mass-shell singularities of the Green functions \cite{kibble1968coherent2} which makes impossible the application of the LSZ reduction formula in the computation of the S-matrix elements. For this reason, in theories with long-range interactions one usually restricts attention to the so-called inclusive cross sections which, however, characterize the scattering event only partially. The correct definition of the scattering matrix in models with long-range interactions is still an unsolved problem. 

%===================================================================================================
%===================================================================================================
\appendix
%===================================================================================================
%===================================================================================================

\numberwithin{equation}{chapter}
\numberwithin{thm}{chapter}
\numberwithin{dfn}{chapter}
\numberwithin{rem}{chapter}
\numberwithin{asm}{chapter}

\chapter{Grassmann algebra}\label{app:grassmann}

The Grassmann algebra of a vector space $V$ over $\C$ denoted by $\Lambda(V)$ is the quotient algebra of the tensor algebra of $V$ by the two-sided ideal generated by all elements of the form $x \otimes x$ for $x\in V$. The Grassmann algebra is also known as the exterior or alternating algebra. The elements of $\Lambda(V)$ are called Grassmann numbers. The product of two Grassmann numbers is called the exterior or wedge product and will be denoted by $\Lambda(V)\times \Lambda(V)\ni (v_1,v_2)\mapsto v_1v_2\in \Lambda(V)$. It holds
\begin{equation}
 \Lambda(V)=\bigoplus_{n\in\N_0} \Lambda^n(V) = \Lambda^\textrm{even}(V) \oplus \Lambda^\textrm{odd}(V).
\end{equation}
By definition $\Lambda^n(V)$ is the linear span of $v_1\ldots v_n \in \Lambda(V)$, where $v_1,\ldots,v_n\in \Lambda^1(V)\simeq V$ and 
\begin{equation}
 \Lambda^\textrm{even}(V):=\bigoplus_{n\in\N_0} \Lambda^{2n}(V),~~~~
 \Lambda^\textrm{odd}(V):=\bigoplus_{n\in\N_0} \Lambda^{2n+1}(V).
\end{equation}
A Grassmann number belonging to $\Lambda^\textrm{even}(V)$ or $\Lambda^\textrm{odd}(V)$ have even, respectively, odd parity. Other elements of $\Lambda(V)$ do not have definite parity. Grassmann numbers $v_1,v_2$ with definite parity anti-commute if $v_1,v_2\in \Lambda^\textrm{odd}(V)$ and commute otherwise. Note that if $n=\dim(V)$ then $\Lambda^m(V)=\{0\}$ for all $m>n$.

Let $V$ be an infinite-dimensional vector space and set $E=\Lambda(V)$. The space of Schwartz functions valued in the Grassmann algebra is by definition 
\begin{equation}
 \mathcal{S}(\R^N,E):=\mathcal{S}(\R^N)\otimes E,
\end{equation}
where $\otimes$ is the algebraic tensor product. Any element of $\mathcal{S}(\R^N,E)$ is of the form
\begin{equation}
 g= \sum_{n=1}^m  g_n \otimes e_n,
\end{equation}
where $g_1,\ldots,g_m\in\mathcal{S}(\R^N)$ and $e_1,\ldots,e_m\in E$. The value of $g$ at the point $x\in\R^N$ is equal
\begin{equation}
 g(x)= \sum_{n=1}^m  g_n(x) e_n \in E.
\end{equation}
We say that the Grassmann-valued Schwartz function $g\in \mathcal{S}(\R^N,E)$ has even/odd parity if $g(x)$ has even/odd parity for all $x\in\R^N$. If $g\in \mathcal{S}(\R^N,E)$ and $h\in\mathcal{S}(\R^M,E)$ then there exists a unique $f\in\mathcal{S}(\R^{N+M},E)$ such that
\begin{equation}
 f(x,y) = g(x) h(y) 
\end{equation}
for any $x\in\R^N$ and $y\in\R^M$. The above statement remains true if $g\in \mathcal{S}(\R^N)$ or $h\in\mathcal{S}(\R^M)$. Let $g\in \mathcal{S}(\R^N,E)$ and $h\in\mathcal{S}(\R^M,E)$ such that $g$ and $h$ have definite parity. If $g$ and $h$ have odd parity then $g(x) h(y) = - h(y) g(x)$, otherwise $g(x) h(y) = h(y) g(x)$.
 
Let $T\in\mathcal{S}'(\R^N,L(\mathcal{D}))$ be an operator-valued Schwartz distribution on $\mathcal{D}$ and $g= \sum_{n=1}^m  g_n \otimes e_n\in \mathcal{S}(\R^N,E)$ be arbitrary Grassmann-valued Schwartz function. By definition
\begin{equation}
 \int\rd^Nx\,T(x) g(x) := \sum_{n=1}^m \left[\int\rd^Nx\,T(x) g_n(x)\right] \otimes e_n \in L(\mathcal{D})\otimes E.
\end{equation}

\chapter{Gell-Mann and Low formula}\label{sec:magic_formula}

In the standard approach to the quantum field theory the Green functions are defined by the following formula due to Gell-Mann and Low \cite{gell1951bound}
\begin{equation}
 \Gre(C_1(x_1),\ldots,C_m(x_m)) =
 \frac{\sum_{n=0}^\infty \frac{\ri^ne^n}{n!} \int\rd^4 y_1\ldots\rd^4 y_n\,(\Omega|\T(I_n,C_1(x_1),\ldots,C_m(x_m))\Omega)}{\sum_{n=0}^\infty \frac{\ri^ne^n}{n!} \int\rd^4 y_1\ldots\rd^4 y_n\,(\Omega|\T(I_n)\Omega)},
\end{equation}
where $I_n=(\mathcal{L}(y_1),\ldots,\mathcal{L}(y_n))$. For simplicity, we assumed that there is only one interaction vertex. The Gell-Mann and Low formula is not well-defined in its standard from stated above because of the presence of the formal integrals. For this reason, we consider instead the IR-improved Gell-Mann and Low formula, already introduced in \cite{dutsch1997slavnov}, which states that
\begin{equation}\label{eq:app_G_GL_lim}
 \Gre^{\textrm{GL}}(C_1(x_1),\ldots,C_m(x_m)) := \lim_{\epsilon\searrow0} \Gre^{\textrm{GL}}_\epsilon(C_1(x_1),\ldots,C_m(x_m)),
\end{equation}
where
\begin{multline}\label{eq:app_G_GL_def}
 \Gre^{\textrm{GL}}_\epsilon(C_1(x_1),\ldots,C_m(x_m)) :=
 \\
 \frac{\sum_{n=0}^\infty \frac{\ri^ne^n}{n!} \int\rd^4 y_1\ldots\rd^4 y_n\,g_\epsilon(y_1)\ldots g_\epsilon(y_n)\,(\Omega|\T(I_n,C_1(x_1),\ldots,C_m(x_m))\Omega)}{\sum_{n=0}^\infty \frac{\ri^n}{n!} \int\rd^4 y_1\ldots\rd^4 y_n\,g_\epsilon(y_1)\ldots g_\epsilon(y_n)\,(\Omega|\T(I_n)\Omega)}.
\end{multline}
The above distribution is well-defined for any $\epsilon>0$, but the existence of the limit on the RHS of \eqref{eq:app_G_GL_lim} is non-trivial. 

We will show that for models satisfying Assumption \ref{asm} the limit \eqref{eq:app_G_GL_lim} exists and coincides with the EG definition of the corresponding Green function. This is a generalization of the results of  \cite{dutsch1997slavnov}, where a similar statement was proved for purely massive models. Let us first note that
\begin{equation}\label{eq:app_G_GL}
 \Gre^{\textrm{GL}}_\epsilon(C_1(x_1),\ldots,C_m(x_m))=
 (-\ri)^{m} \frac{\delta}{\delta h_m(x_m)}\ldots\frac{\delta}{\delta h_1(x_1)} 
  \frac{(\Omega|S(\mathpzc{g}_\epsilon+\mathpzc{h})\Omega)}{(\Omega|S(\mathpzc{g}_\epsilon)\Omega)}  \bigg|_{\mathpzc{h}=0}, 
\end{equation}
where $S(\mathpzc{g}+\mathpzc{h})$ and $S(\mathpzc{g})$ are given by \eqref{eq:S_extended2} and \eqref{eq:S_matrix}, respectively. On the other hand, the Green functions in the EG approach are given by
\begin{equation}
 \Gre^\rEG(C_1(x_1),\ldots,C_m(x_m)) := \lim_{\epsilon\searrow0} \Gre^\rEG_\epsilon(C_1(x_1),\ldots,C_m(x_m)),
\end{equation}
where
\begin{multline}\label{eq:app_G_EG}
 \Gre^\rEG_\epsilon(C_1(x_1),\ldots,C_m(x_m)):=
 \\
 (-\ri)^{m} \frac{\delta}{\delta h_m(x_m)}\ldots\frac{\delta}{\delta h_1(x_1)} 
 (\Omega|S(\mathpzc{g}_\epsilon+\mathpzc{h}) S(\mathpzc{g}_\epsilon)^{-1}\Omega)  \bigg|_{\mathpzc{h}=0}. 
\end{multline}

\begin{thm}
Suppose that a model under consideration satisfies Assumption \ref{asm} formulated in Section~\ref{sec:proof_scalar}. If the time-ordered products fulfill the normalization condition~\ref{norm:wAL} then it holds
\begin{equation}\label{eq:app_thm}
 \lim_{\epsilon\searrow0} \Gre^{\mathrm{GL}}_\epsilon(C_1(x_1),\ldots,C_m(x_m))
 =
 \lim_{\epsilon\searrow0}\Gre^\rEG_\epsilon(C_1(x_1),\ldots,C_m(x_m)) ,
\end{equation}
where the limit on the RHS of the above equation exists by Theorem \ref{thm:main2} and defines the Green function in the EG approach.
\begin{proof}
Let $J=(C_1(x_1),\ldots,C_m(x_m))$. Since 
\begin{equation}
 (\Omega|S(\mathpzc{g}_\epsilon+\mathpzc{h})\Omega) = (\Omega|S(\mathpzc{g}_\epsilon+\mathpzc{h})S(\mathpzc{g}_\epsilon)^{-1}S(\mathpzc{g}_\epsilon)\Omega),
\end{equation}
it follows from \eqref{eq:time_ordered_adv} and \eqref{eq:S_matrix} that
\begin{multline}
 (-\ri)^{m} \frac{\delta}{\delta h_m(x_m)}\ldots\frac{\delta}{\delta h_1(x_1)}  (\Omega|S(\mathpzc{g}_\epsilon+\mathpzc{h})\Omega) \bigg|_{\mathpzc{h}=0}
 =
 \\
 \sum_{n=0}^\infty \frac{\ri^n e^n}{n!} 
 \int\rd^4 y_1\ldots\rd^4 y_n\,g_\epsilon(y_1)\ldots g_\epsilon(y_n)
 \,\sum_{\substack{I'_n,I''_n\\I'_n+I''_n=\pi(I_n)}} (-1)^{\mathbf{f}(\pi)} (\Omega|\Adv(I'_n;J)\T(I''_n)\Omega).
\end{multline}
Note that the factor $(-1)^{\mathbf{f}(\pi)}$ equals $1$ because the fields from $I$ have vanishing fermion number. 

Now we are going to show that 
\begin{equation}\label{eq:GL_dist}
 (\Omega|\Adv(I'_n;J)\T(I''_n)\Omega) -
 (\Omega|\Adv(I'_n;J)\Omega)\,(\Omega|\T(I''_n)\Omega)
\end{equation}
has IR-index $d=1$ with respect to all variables from the lists $I'_n$ and $I''_n$. To this end, we use the following variant of Theorem \ref{thm:product_families}: Assume that the $\mathcal{F}$ products \eqref{eq:product_families} satisfy the assumptions (1)-(3) of Theorem \ref{thm:product_families} for all super-quadri-indices $s_1, \ldots, s_n$, $s'_1, \ldots, s'_{n'}$ which involve only massless fields such that at least one of them is non-zero, then the distribution
\begin{multline}
 (\Omega|F(B_1(x_1),\ldots,B_n(x_n))F'(B_1^{\prime}(x'_1),\ldots,B_{n'}^{\prime}(x'_{n'}))\Omega) 
 \\
 - (\Omega|F(B_1(x_1),\ldots,B_n(x_n))\Omega)\,(\Omega|F'(B_1^{\prime}(x'_1),\ldots,B_{n'}^{\prime}(x'_{n'}))\Omega)
\end{multline}
fulfills the conditions (1')-(3') stated in this theorem. The above statement about \eqref{eq:GL_dist} follows from Part~(A) of Theorem \ref{thm:main2}, the normalization condition \ref{norm:wAL2} with $F$ being the time-ordered product and the above reformulation of Theorem \ref{thm:product_families} (more precisely we use the statement (3')).

As a result, we obtain that all coefficients in the formal expansion in powers of the coupling constant $e$ of 
\begin{equation}
  (-\ri)^{m} \frac{\delta}{\delta h_m(x_m)}\ldots\frac{\delta}{\delta h_1(x_1)} \left[ (\Omega|S(\mathpzc{g}_\epsilon+\mathpzc{h})\Omega)  - (\Omega|S(\mathpzc{g}_\epsilon+\mathpzc{h})S(\mathpzc{g}_\epsilon)^{-1}\Omega)\,(\Omega|S(\mathpzc{g}_\epsilon)\Omega) \right] \bigg|_{\mathpzc{h}=0}
\end{equation} 
after integrating with arbitrary Schwartz function $f(x_1,\ldots,x_m)$ are of order $O(\epsilon^{1-\varepsilon})$ for any $\varepsilon>0$. It follows from the normalization condition \ref{norm:wAL} that
\begin{equation}
 (\Omega|S(\mathpzc{g}_\epsilon)\Omega) =  O(\epsilon^{-\varepsilon})
\end{equation}
for any $\varepsilon>0$. The above equation means that all coefficients in the formal expansion in powers of the coupling constant $e$ of the LHS are of order $O(\epsilon^{-\varepsilon})$ for any $\varepsilon>0$. Since the first term of this expansion is equal $1$ we have
\begin{equation}
 (\Omega|S(\mathpzc{g}_\epsilon)\Omega)^{-1} =  O(\epsilon^{-\varepsilon})
\end{equation}
for any $\varepsilon>0$. Consequently, each term of the $e$ expansion of the difference between the RHS of Equations \eqref{eq:app_G_GL} and \eqref{eq:app_G_EG} integrated with arbitrary Schwartz function $f(x_1,\ldots,x_m)$ is of the order $O(\epsilon^{1-2\varepsilon})$ for any $\varepsilon>0$. This finishes the proof.
\end{proof}
\end{thm}

If Assumption \ref{asm} is satisfied and the interaction vertex contains at least one massive field, then according to Part~(2) of the normalization condition \ref{norm:gc} in Section~\ref{sec:central_general} the distribution
\begin{equation}\label{eq:app_T}
 (\Omega|\T(\mathcal{L}(x_1),\ldots,\mathcal{L}(x_n))\Omega)
\end{equation}
can be normalized in such a way that its \underline{IR}-index equals $d=5$. As a result 
\begin{equation}
 (\Omega|S(\mathpzc{g}_\epsilon)\Omega) -1 =  O(\epsilon^{1-\varepsilon})
\end{equation}
for any $\varepsilon >0$ and both sides of \eqref{eq:app_thm} are equal
\begin{equation}
 \lim_{\epsilon\searrow0} 
 ~(-\ri)^{m} \frac{\delta}{\delta h_m(x_m)}\ldots\frac{\delta}{\delta h_1(x_1)} 
 (\Omega|S(\mathpzc{g}_\epsilon+\mathpzc{h})\Omega)  \bigg|_{h=0}.
\end{equation}
This shows that for the above-mentioned normalization of \eqref{eq:app_T} the denominators in \eqref{eq:app_G_GL_def} and \eqref{eq:app_G_GL} may be omitted.

\chapter{Weak adiabatic limit in second order of perturbation theory}\label{sec:mass}

In this this appendix we show that in models with massless particles, in contrast to purely massive models, the weak adiabatic limit exists only if some normalization conditions are imposed on the time-ordered products. We prove that the correct mass normalization of all massless fields which are sub-polynomials of the interaction vertices is necessary for the existence of the weak adiabatic limit in the second order of the perturbation theory. As a result, in the case of theories with massless particles one cannot define the Wightman and Green functions for every normalization of the time-ordered products. Note that the correct mass normalization is a part of the normalization condition \ref{norm:wAL}, which is needed in our proof of the existence of the weak adiabatic limit.

For simplicity we consider the scalar model defined in Section~\ref{sec:the_model}. Similar results hold also in other models, in particular all models introduced in Section~\ref{sec:examples}. Let us recall that this model is defined in terms of two scalar fields: a massive field $\psi$ and a massless field $\varphi$. We will show that the correction of order $e^2$ to the Green function 
\begin{equation}\label{eq:app_green}
 \int\rd^4 x_1\rd^4 x_2 \,f(x_1,x_2)\,\Gre(\varphi(x_1),\varphi(x_2))
\end{equation}
in the scalar model is well-defined for any $f\in\mathcal{S}(\R^8)$ if and only if the Fourier transform of the distribution
\begin{equation}\label{eq:app1_vev}
 (\Omega|\T(\psi^2(x_1),\psi^2(x_2))\Omega),
\end{equation}
which is the second order correction to the self-energy of the field $\varphi$, is of the form 
\begin{equation}\label{eq:app1_sigma}
 (2\pi)^4\delta(q_1+q_2) \Sigma(q_1),
\end{equation}
where $\Sigma(0)=0$. This implies that the mass of the field $\varphi$ has to be correctly normalized. Because the VEV of $\normord{\psi^2(x)}$ vanishes, the distribution \eqref{eq:app1_sigma} coincides in some neighborhood of zero with the Fourier transform of $(\Omega|\Adv(\psi^2(x_1);\psi^2(x_2))\Omega)$. Consequently, by the result of Section~\ref{sec:cetral_splitting_massive}, the function $\Sigma:\,\R^4\to\C$ is analytic in some neighborhood of the origin. The function $\Sigma$ is analytic also in the massive spinor and scalar QED. This is not the case in the massless spinor or scalar QED -- for these theories the condition $\Sigma(0)=0$ has to be satisfied in the sense of {\L}ojasiewicz. Note that in the case of the massless $\varphi^3$ theory $\Sigma(q)$ is logarithmically divergent at $q=0$. As a result, it is not possible to define the time-ordered products such that $\Sigma(0)=0$. The reasoning presented below can also be used to show that the weak adiabatic limit does not exist for the second order correction to  $\Gre(\varphi(x_1),\varphi(x_2))$ in the massless $\varphi^3$ theory.

In the EG approach the second order contribution to \eqref{eq:app_green} is defined as the value at $q_1=q_2=0$ in the sense of {\L}ojasiewicz of the Schwartz distributions
\begin{equation}\label{eq:app_adv}
 \int\rd^4 x_1\rd^4 x_2\,f(x_1,x_2)\,(\Omega|\Adv(\F{\mathcal{L}}(q_1),\F{\mathcal{L}}(q_2);\varphi(x_1),\varphi(x_2))\Omega)
\end{equation}
or
\begin{equation}\label{eq:app_ret}
 \int\rd^4 x_1\rd^4 x_2\,f(x_1,x_2)\,(\Omega|\Ret(\F{\mathcal{L}}(q_1),\F{\mathcal{L}}(q_2);\varphi(x_1),\varphi(x_2))\Omega).
\end{equation}
If the weak adiabatic limit exists, then the values at $q_1=q_2=0$ of both of the above distributions exist and coincide. By Theorem \ref{thm:main2} this is certainly the case if the normalization condition \ref{norm:wAL} holds. This condition implies that the self-energy is normalized in such a way that  $\Sigma(0)=0$ and $\partial_\mu \Sigma(0)=0$. Let us redefine the VEV of the time-ordered product \eqref{eq:app1_vev} by adding to it $c\,\delta(x_1-x_2)$ with $c\neq 0$. Note that after this redefinition $\Sigma(0)=c\neq 0$.

The differences between the contributions to \eqref{eq:app_adv} and \eqref{eq:app_ret} before and after the redefinition of \eqref{eq:app1_vev} are given by
\begin{multline}
 d^\pm(q_1,q_2)=\int\mP{k_1}\mP{k_2}\,f(k_1,k_2)
 \\ 
 \left[\frac{\ri}{k_1^2+\ri\zerop}\frac{\ri}{k_2^2+\ri\zerop}
 -(2\pi)\theta(\pm k_1^0)\delta(k_1^2)\,(2\pi)\theta(\pm k_2^0)\delta(k_2^2)\right](2\pi)^4\delta(k_1+k_2+q_1+q_2) ~c,
\end{multline}
where $+$ and $-$ correspond to \eqref{eq:app_adv} and \eqref{eq:app_ret}, respectively. The difference between the contribution to 
\begin{equation}
 \int\rd^4 x_1\rd^4 x_2\,f(x_1,x_2)\,(\Omega|\Dif(\F{\mathcal{L}}(q_1),\F{\mathcal{L}}(q_2);\varphi(x_1),\varphi(x_2))\Omega)
\end{equation}
before and after the redefinition of \eqref{eq:app1_vev} is given by
\begin{multline}
 d(q_1,q_2)
 %=\int \rd^4 x_1\rd^4 x_2\,f(x_1,x_2)\, \Dif(\F{\mathcal{L}}(q_1)\F{\mathcal{L}}(q_2);\varphi(x_1)\varphi(x_2))
 =2\int\mH{0}{k_1}\mH{0}{k_2}\,\F{f}(k_1,k_2)\,(2\pi)^4\delta(k_1+k_2+q_1+q_2) ~c
 \\
 - 2\int\mH{0}{k_1}\mH{0}{k_2}\,\F{f}(k_1,k_2)\,(2\pi)^4\delta(k_1+k_2-q_1-q_2) ~c.
\end{multline}
If $\F{f}(0,0)=0$ then $|d(q_1,q_2)|\leq\const\, |(q_1,q_2)|$. Since $d$ is continuous we have $d(q_1,q_2)=O^\mathrm{dist}(|(q_1,q_2)|)$. Consequently, using Theorem \ref{thm:math_splitting} and the decomposition \eqref{eq:decomposition_weak_massive} one shows that the distributions $d^\pm(q_1,q_2)$ have values at $q_1=q_2=0$ in the sense of {\L}ojasiewicz and these values coincide. As a result, it is enough to consider $f\in\mathcal{S}(\R^8)$ such that $\F{f}=1$ in a neighborhood of $0$. In this case we have $$d(q_1,q_2) =\const\, \sgn(q_1^0+q_2^0) \theta((q_1+q_2)^2)$$ in some neighborhood of $0$. Thus, after the splitting $$d^\pm(q_1,q_2)=\const\, \sgn(q_1^0+q_2^0) \theta((q_1+q_2)^2) \log|q^0_1+q^0_2| + O^\mathrm{dist}(|(q_1,q_2)|^0).$$ It implies that $d^\pm(q_1,q_2)$ do not have values at $q_1=q_2=0$ in the sense of {\L}ojasiewicz. In consequence, the weak adiabatic limit does not exist after the above-mentioned redefinition of the distribution \eqref{eq:app1_vev}.

\printbibliography

\end{document}

\grid